%% file: G195_PDF_arxiv.tex
\documentclass[twocolumn]{aastex631}
\usepackage{subfigure}
\usepackage{amssymb}
\newcommand\CO{$^{12}$CO }
\newcommand\COl{$^{13}$CO }
\newcommand\COll{C$^{18}$O }

\submitjournal{ApJS}

\shorttitle{Gas N-PDFs of molecular clouds in the third quadrant of the Milky Way}
\shortauthors{Ma et al.}

\begin{document}
\defcitealias{Kainulainen2009}{K09}
\title{Gas Column Density Distribution of Molecular Clouds in the Third Quadrant of the Milky Way}

\correspondingauthor{Hongchi Wang}
\email{hcwang@pmo.ac.cn}
\author[0000-0002-8051-5228]{Yuehui Ma}
\affil{Purple Mountain Observatory and Key Laboratory of Radio Astronomy, Chinese Academy of Sciences, 10 Yuanhua Road, Nanjing 210033, China}

\author[0000-0003-0746-7968]{Hongchi Wang}
\affil{Purple Mountain Observatory and Key Laboratory of Radio Astronomy, Chinese Academy of Sciences, 10 Yuanhua Road, Nanjing 210033, China}
\affil{School of Astronomy and Space Science, University of Science and Technology of China, Hefei, Anhui 230026, China}

\author[0000-0002-6388-649X]{Miaomiao Zhang}
\affil{Purple Mountain Observatory and Key Laboratory of Radio Astronomy, Chinese Academy of Sciences, 10 Yuanhua Road, Nanjing 210033, China}

\author[0000-0001-8923-7757]{Chen Wang}
\affil{Purple Mountain Observatory and Key Laboratory of Radio Astronomy, Chinese Academy of Sciences, 10 Yuanhua Road, Nanjing 210033, China}

\author[0000-0003-2549-7247]{Shaobo Zhang}
\affil{Purple Mountain Observatory and Key Laboratory of Radio Astronomy, Chinese Academy of Sciences, 10 Yuanhua Road, Nanjing 210033, China}

\author[0000-0002-7616-666X]{Yao Liu}
\affil{Purple Mountain Observatory and Key Laboratory of Radio Astronomy, Chinese Academy of Sciences, 10 Yuanhua Road, Nanjing 210033, China}

\author[0000-0003-2218-3437]{Chong Li}
\affil{School of Astronomy and Space Science, Nanjing University, 163 Xianlin Avenue, Nanjing 210023, China}
\affil{Key Laboratory of Modern Astronomy and Astrophysics (Nanjing University), Ministry of Education, Nanjing 210023, China}

\author[0000-0002-1022-4249]{Yuqing Zheng}
\affil{Purple Mountain Observatory and Key Laboratory of Radio Astronomy, Chinese Academy of Sciences, 10 Yuanhua Road, Nanjing 210033, China}

\author[0000-0003-0804-9055]{Lixia Yuan}
\affil{Purple Mountain Observatory and Key Laboratory of Radio Astronomy, Chinese Academy of Sciences, 10 Yuanhua Road, Nanjing 210033, China}

\author[0000-0001-7768-7320]{Ji Yang}
\affil{Purple Mountain Observatory and Key Laboratory of Radio Astronomy, Chinese Academy of Sciences, 10 Yuanhua Road, Nanjing 210033, China}

\begin{abstract}
	We have obtained column density maps for an unbiased sample of 120 molecular clouds in the third quadrant of the Milky Way mid-plane (b$\le |5|\arcdeg$) within the galactic longitude range from 195$\arcdeg$ to 225$\arcdeg$, using the high sensitivity \CO and \COl ($J=1-0$) data from the Milky Way Imaging Scroll Painting (MWISP) project. The probability density functions of the molecular hydrogen column density of the clouds, N-PDFs, are fitted with both log-normal (LN) function and log-normal plus power-law (LN+PL) function. The molecular clouds are classified into three categories according to their shapes of N-PDFs, i.e., LN, LN+PL, and UN (unclear), respectively. About 72\% of the molecular clouds fall into the LN category, while 18\% and 10\% into the LN+PL and UN categories, respectively. A power-law scaling relation, $\sigma_s\propto N_{H_2}^{0.44}$, exists between the width of the N-PDF, $\sigma_s$, and the average column density, $N_{H_2}$, of the molecular clouds. However, $\sigma_s$ shows no correlation with the mass of the clouds. A correlation is found between the dispersion of normalized column density, $\sigma_{N/\rm <N>}$, and the sonic Mach number, $\mathcal{M}$, of molecular clouds. Overall, as predicted by numerical simulations, the N-PDFs of the molecular clouds with active star formation activity tend to have N-PDFs with power-law high-density tails.
\end{abstract}


\section{Introduction} \label{sec1}
The hierarchical structure of molecular clouds is dominated by the interplay among turbulence, self-gravity, and magnetic fields \citep{McKee2007}. Turbulence plays a crucial role in the dynamics of molecular clouds. For example, the kinetic energy of turbulence is comparable to gravitational potential energy on large scales, which can support molecular clouds from global collapse \citep{Stahler2004}. The probability distribution function (PDF) of the density of molecular clouds, i.e., the probability of finding the density in the interval of [$\rho,\rho+d\rho$], is a simple but useful tool to study turbulence and the structure of molecular clouds. The shape of the $\rho$-PDF is related to the underlying physical processes that dominate the dynamics of molecular clouds. Numerical simulations suggest that the $\rho$-PDF has the log-normal (LN) form when the molecular cloud is dominated by turbulence, whereas it will develop a power-law (PL) distribution at high-densities when self-gravity becomes dominant \citep{Kritsuk2011, Federrath2013}. The log-normal distribution of $\rho$-PDF can be qualitatively understood as the result of the molecular gas being successively compressed by a series of random and individual shock waves in the turbulent medium \citep{Vazquez-Semadeni1994}. The $\rho$-PDF can not be obtained directly by observations because of the projection effect. Numerical simulations have shown that column density PDFs (N-PDFs) exhibit similar properties as $\rho$-PDF \citep{Federrath2008, Ballesteros-Paredes2011, Ward2014}. For example, \cite{Ballesteros-Paredes2011} suggest that the N-PDFs of molecular clouds evolve from log-normal distributions at early times to log-normal $+$ power-law (LN+PL) distributions at late times.

The observed shapes of N-PDFs rely on accurate measurement of column densities. Three tracers are commonly used for H$_2$ column density, i.e., the dust extinction at near/mid-infrared wavelengths, the dust emission at far-infrared wavelengths, and the molecular line emission at millimeter wavelengths. \cite{Kainulainen2009} (hereafter K09) studied the N-PDFs of 23 nearby (within 200 pc) molecular clouds with dust extinction as the tracer of the column density and found that the N-PDFs evolve from LN to LN+PL when the star-formation activities of the molecular clouds vary from quiescent to active. Other studies using dust emission data have also confirmed this evolutionary trend of N-PDFs \citep{Schneider2013, Schneider2015a, Schneider2015b, Lombardi2015, Alves2017}. \cite{Spilker2021} have made a census of the N-PDFs of 72 molecular clouds within two kpc from the Sun using dust extinction/emission data from the literature, which is by far the most systematic study of N-PDFs of molecular clouds using dust-based tracers, and they found that the N-PDFs of the molecular clouds in their work are not well described by any simple model of LN, PL, or LN+PL.  

Large-scale molecular line surveys conducted toward different regions of the Galactic plane have provided comprehensive knowledge about the distribution and physical properties of the molecular clouds, for example, the CfA-Chile survey that covered the whole Galactic plane \citep{Dame2001, Miville-Deschenes2017}, the FCRAO survey toward the Outer Galaxy in the second quadrant \citep{Heyer2001}, and the Forgotten Quadrant Survey (FQS) toward a part of the third quadrant of the Galactic plane within 220$\arcdeg\le$l$\le$240$\arcdeg$ and $-$2.5$\arcdeg\le$b$\le$0$\arcdeg$, which is partly overlapped with the sky coverage of this work \citep{Benedettini2020}. However, compared with the studies that measure N-PDFs with the dust-based tracers, fewer works investigated the properties of N-PDFs with molecular line emission as the tracer of H$_2$ column density \citep{Goodman2009, Carlhoff2013, Schneider2015b, Schneider2016, Ma2019, Ma2020}. \cite{Goodman2009} suggested that the \COl molecular line emission is under the influence of the critical density for excitation, optical depth, and chemical depletion effect, hence only traces a narrow dynamic range of H$_2$ column density. However, observations from the Milky Way Imaging Scroll Painting (MWISP) project, a high-sensitivity large-scale survey of the \CO, \COl, and \COll $J=1-0$ line emission \citep{Su2019}, reveal that the \COl line is still a good tracer of gas column density on relatively large scales (pc) \citep{Ma2021} since overdense regions with column densities on the order of $\sim10^{23}$ cm$^{-2}$ are generally concentrated in sub-pc scales. \cite{Ma2021} have obtained forty N-PDFs of the molecular clouds in the second quadrant of the Milky Way mid-plane using the \COl $J=1-0$ emission line data from the MWISP project and found that about 78\% of the molecular clouds have log-normal N-PDFs. Moreover, the molecular clouds that have power-law N-PDFs are not necessarily associated with active star-formation, which is different from the results obtained using dust-based tracers. Considering that different tracers of column densities can trace different parts of molecular clouds, and molecular line emission can help to avoid velocity crowding along the line of sight, it is still worthwhile to investigate whether the properties of N-PDFs obtained from the dust-based tracers still exist when using the molecular line tracer.  

The relation between the dispersion of the density PDF, $\sigma_{\ln\rho/\left\langle \rho\right\rangle }$, and the turbulence energy, expressed as the sonic Mach number $\mathcal{M}$, of molecular clouds is also one of the key ingredients in the interpretation of the stellar initial mass function, the core mass function, and the star formation rate in theoretical studies \citep{Krumholz2005, Padoan2011, Federrath2012, Hennebelle2008, Hennebelle2009, Hennebelle2013, Burkhart2018}. For pure hydrodynamical supersonic isothermal turbulent gas, the density dispersion of molecular clouds is linearly scaled with the sonic Mach number, 
\begin{equation}
	\sigma_{\rho/\left\langle \rho\right\rangle } = b \mathcal{M},  
	\label{eq1}
\end{equation}
where $\left\langle \rho\right\rangle $ is the mean of volume density, and b is a constant related to the mixture of the solenoidal (divergence-free) and compressive (curl-free) forcing modes of turbulence \citep{Padoan1997b}. The predicted values of b vary from 1/3 to 1, with 1/3 and 1 corresponding to pure solenoidal and pure compressive forcing mode, respectively. When considering the logarithmic form of the volumn density, the relation given by Eq. \ref{eq1} becomes $\sigma_{\ln \rho/\left\langle \rho\right\rangle}^2 = \ln(1+b^2\mathcal{M}^2)$. However, the $\sigma_{\rho/\left\langle \rho\right\rangle}-\mathcal{M}$ relation and the value of parameter b are hard to be confirmed from observations since the estimation of volume density is tricky. \cite{Kainulainen2017} found a weak correlation between $\sigma_{\ln \rho/\left\langle \rho\right\rangle}$ and $\mathcal{M}$ for fifteen molecular clouds in the solar neighborhood using extinction data. Other efforts seeked clues from the distribution of column densities. For example, \cite{Goodman2009} have found no correlation between $\sigma_s$ and $\mathcal{M}$ in sub-regions in the Perseus GMC, while \cite{Kainulainen2013b} found a linear correlation between $\sigma_{N/{\left\langle N\right\rangle}}$ and $\mathcal{M}$, and got $b = 0.2^{+0.37}_{-0.22}$ for a group of infrared dark clouds (IRDCs). The $b$ value obtained by \cite{Kainulainen2013b} is below the lower limit of the predicted range for non-magnetized isothermal turbulence, which is attributed to the influence of magnetic fields. With the existence of magnetic fields, the correlation between the dispersion of logarithmic density and the Mach number of turbulence is flattened, and the correlation factor b becomes less than 1/3 at $\mathcal{M}$ higher than $\sim7$ \citep{Price2011}. When the density depends strongly on the magnetic field ($B\propto\rho$), the logarithmic density dispersion is nearly independent on $\mathcal{M}$ \citep{Molina2012}. \cite{Brunt2010} developed an analytical method to convert the observed column density dispersions to volume density dispersions, which can be expressed as $\sigma_{N/{\left\langle N\right\rangle}}^2/\sigma^2_{\rho/\left\langle \rho\right\rangle} = R$, where R is determined by the power spectrum of the column density field and lies in the range of [0.03, 0.15].

\cite{Burkhart2012} proposed that the $\mathcal{M}-\sigma_{\ln \rho/\left\langle \rho\right\rangle}$ relation is still applicable for the 2D situation, which is 
\begin{equation}
	\sigma_{s}^2 = A\times \ln(1+b^2\mathcal{M}^2), 
	\label{eq2}
\end{equation}
where $\sigma_s$ is the width of the N-PDF, and the subscript $s$ means $s = \ln\ (N/\left\langle N \right\rangle)$. For a log-normal N-PDF, the above equation becomes
\begin{equation}
\sigma_{N/\left\langle N\right\rangle}^2 = (b^2\mathcal{M}^2+1)^A-1,
\label{eq3}
\end{equation}
using the relation of $\sigma_{N/\left\langle N\right\rangle}^2 = \exp{(\sigma_s^2)}-1$ \citep{Burkhart2012}.
The observed column densities and Mach numbers of the Taurus and IC 5146 clouds are consistent with Eq. \ref{eq2} with b = 0.5, A = 0.16 \citep{Brunt2010A} and b = 0.5, A = 0.12 \citep{Padoan1997}, respectively.

With the extensive sky coverage and high sensitivity, the MWISP survey can provide us with a large sample of molecular clouds \citep{Yuan2021}. In this work, we measure the column densities of the molecular clouds from l = 195$\arcdeg$ to l = 225$\arcdeg$ using the \COl data of the MWISP project and study the shape and statistical properties of the N-PDFs, taking advantage of the fact that molecular clouds in the outer Galaxy suffer less velocity crowding. We also estimate the $\sigma_s-\mathcal{M}$ and $\sigma_{N/\left\langle N\right\rangle}-\mathcal{M}$ relations based on the measured N-PDFs.  

The paper is organized as follows. The observations, data, and the method for the identification of molecular clouds are presented in Section \ref{sec2}. The measurement of the column density and the analysis of the N-PDFs are given in Section \ref{sec3}. We discuss the properties of the observed N-PDFs in Section \ref{sec4} and present our conclusions in Section \ref{sec5}.
 
\section{Data and the Sample of Molecular Clouds} \label{sec2}
\subsection{Observations and Data} \label{sec2.1}

The \CO and \COl data used in this work are part of the MWISP survey. Up to date, Phase I of the MWISP survey has been finished, and the sky area has been covered from $l = 12^{\circ}$ to $l = 230^{\circ}$ within $b=\pm5 \arcdeg$. \cite{Su2019} have given a detailed introduction to the observations and the data reduction processes of the project. We only give a brief description of the data in this work. In the MWISP project, the \CO, \COl, and \COll $J=1-0$ spectra are obtained simultaneously using the nine-beam Superconducting Spectroscopic Array Receiver (SSAR) \citep{Shan2012} equipped on the PMO-13.7m telescope. The half-power beam width (HPBW) of the telescope is around 52$\arcsec$ and 50$\arcsec$ at 110 GHz and 115 GHz, respectively, which correspond to a spatial resolution of $\sim$0.25 pc at the distance of 1 kpc. The survey area is divided into individual cells of 30$\arcmin\times$30$\arcmin$. The final data are regridded into 30$\arcsec\times$30$\arcsec$ pixels in the directions of Galactic longitude and latitude. The spectrometer on the PMO-13.7m telescope has 16,384 velocity channels with a total bandwidth of 1 GHz, which provides a velocity resolution of 0.17 km s$^{-1}$ at 110 GHz. The required sensitivity of the MWISP data is below $\sim$0.5 K per channel at the \CO $J=1-0$ wavelength and below $\sim$0.3 K per channel at the \COl $J=1-0$ and \COll $J=1-0$ wavelength. 

\subsection{The Sample of Molecular Clouds} \label{sec2.2}
\begin{figure*}[!htb]
	\centering
	\includegraphics[trim=2cm 0cm 3cm 0cm, width=\linewidth, clip]{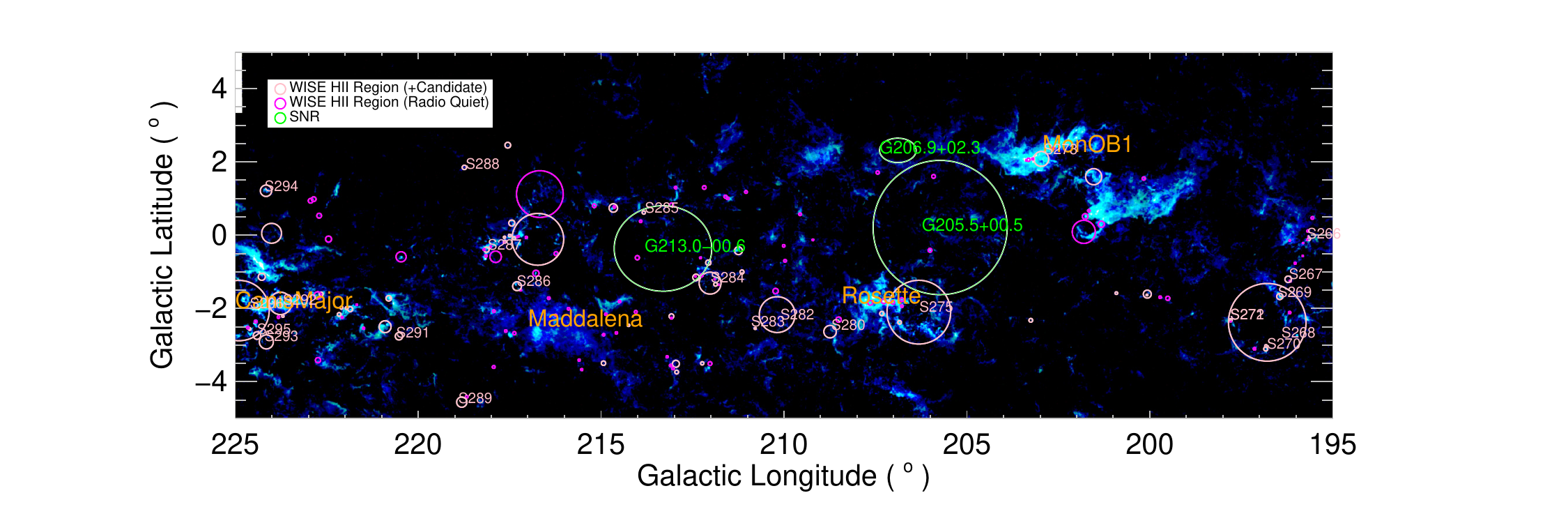}
	\caption{RGB map of the integrated intensity of the \CO, \COl, and \COll $J=1-0$ emission, T$_{mb}$. Red, green, and blue colors correspond to the integrated intensity of the \CO emission in the velocity range from $-$17 to 65 km s$^{-1}$, \COl emission from $-$5 to 55 km s$^{-1}$, and \COll emission from 0 to 33 km s$^{-1}$, respectively. The units of the integrated intensities are K km s$^{-1}$. The \ion{H}{2} regions and \ion{H}{2} region candidates from \cite{Anderson2014} are overlaid as pink circles and magenta circles, respectively, while the supernova remnants from \cite{Green2019} are indicated by the green ellipses. The corresponding names of the \ion{H}{2} regions in the Sharpless' catalog \citep{Sharpless1959} are shown with pink letters, while the names of the supernova remnants are shown in green letters. Some well-known GMCs, like Rosette, Mon OB 1, Canis Major, and Maddalena, are indicated with orange letters.}
	\label{fig1}
\end{figure*}

We use the MWISP data of the outer Galactic plane from l=195${^\circ}$ to l=225${^\circ}$ and within $\pm5^{\circ}$ in this work. Individual cells of the \CO and \COl data are mosaiced into two large datacubes, within which the molecular clouds are identified. Figure \ref{fig1} presents the RGB map of the integrated intensities of the \CO, \COl, and \COll emission. The \ion{H}{2} regions, \ion{H}{2} region candidates, supernova remnants, and the well-known regions, such as the Rosette, Mon OB1, Maddalena, and Canis Major GMCs, are indicated in Figure \ref{fig1}. The figure is similar to that of a forthcoming paper by Wang et al. (2022, in prep) that aims to describe the distribution and basic properties of the three isotopologues of CO molecule in the observed region. Wang et al. (2022, in prep) have built a catalog containing \CO, \COl, and \COll clouds in the third Galactic quadrant which is used in this work. Except for the Maddalena GMC, the other three GMCs are all associated with the \ion{H}{2} regions in the WISE catalog \citep{Anderson2014}. The majority of the \COll emission is detected in the central regions of the four well-known GMCs. Compared with the \CO and \COl emission, the \COll emission is less influenced by optical depth, therefore, being a better tracer than the \COl emission. However, the spatial scales of the regions containing \COll emission are too small to infer the dynamic information of the large-scale turbulence.  

Observationally, molecular clouds are coherent structures composed of contiguous position-position-velocity (PPV) voxels above a given intensity threshold. In the outer Galaxy, the molecular line emission suffers less from velocity crowding, therefore can be easily separated into discrete PPV objects. With the definition in mind, the molecular clouds are identified using a clustering algorithm named the Density-Based Spatial Clustering of Applications with Noise (DBSCAN) \citep{Ester1996}, which is designed to identify clusters of arbitrary shapes. An output cluster, i.e., a molecular cloud, identified by the DBSCAN algorithm, is composed of two kinds of points, ``core'' points and ``border'' points, respectively, and all the points should have brightness above a given threshold. In this work, we adopt a threshold of $T_{mb, sh}=2\sigma_{RMS}$. Voxels with brightness above this threshold are named effective voxels. The types of effective voxels, i.e., ``core'' or ``border'' points, are controlled by two input parameters of the DBSCAN algorithm, i.e., a distance parameter, $\epsilon$, that defines a neighborhood space around an effective voxel, and a number parameter MinPts, that specifies the smallest number of neighborhood effective voxels a core point should have. An effective voxel that has at least MinPts-1 neighborhood effective voxels within a radius of $\epsilon$ is defined as a core point. The border points are those effective voxels that are adjacent to the core points, but do not have enough (MinPts-1) neighborhood voxels. \cite{Yan2020} tested on the effectiveness and robustness of the DBSCAN algorithm against different sets of MinPts$-$1 and $\epsilon$ with the MWISP data and found that the parameter set of  $\epsilon=1$ and $MinPts=4$ is most suitable for the MWISP data. The cloud identification process in this work is the same as in \cite{Yan2020} and is performed separately to the \CO and \COl data. Considering that the emission of \CO is usually more extended than that of \COl in molecular clouds, the \COl emission within one \CO cloud should be dense internal structures of the \CO cloud. Therefore the identified \COl clusters within the boundary of a \CO cloud are considered as one \COl cloud. The clusters should also match the post-selection criterion in Section 2.3 in \cite{Yan2020}, i.e., should contain at least 16 voxels, have peak intensities above 5$\sigma_{RMS}$, have angular sizes larger than 1$\arcmin\times$1$\arcmin$, and occupy at least three velocity channels. Finally, 8946 molecular clouds are identified in the \CO data within the PPV coverage of l = [195$^{\circ}$, 225$^{\circ}$], b = [$-$5$^{\circ}$, 5$^{\circ}$], v = [$-$17, 65] km s$^{-1}$. After mergering the \COl clusters within the boundaries of the \CO clouds, 864 \COl clouds are generated.  
\begin{figure*}[htb!]
	\centering
	\includegraphics[trim=0cm 0cm 0cm 0cm, width= 0.9\linewidth, clip]{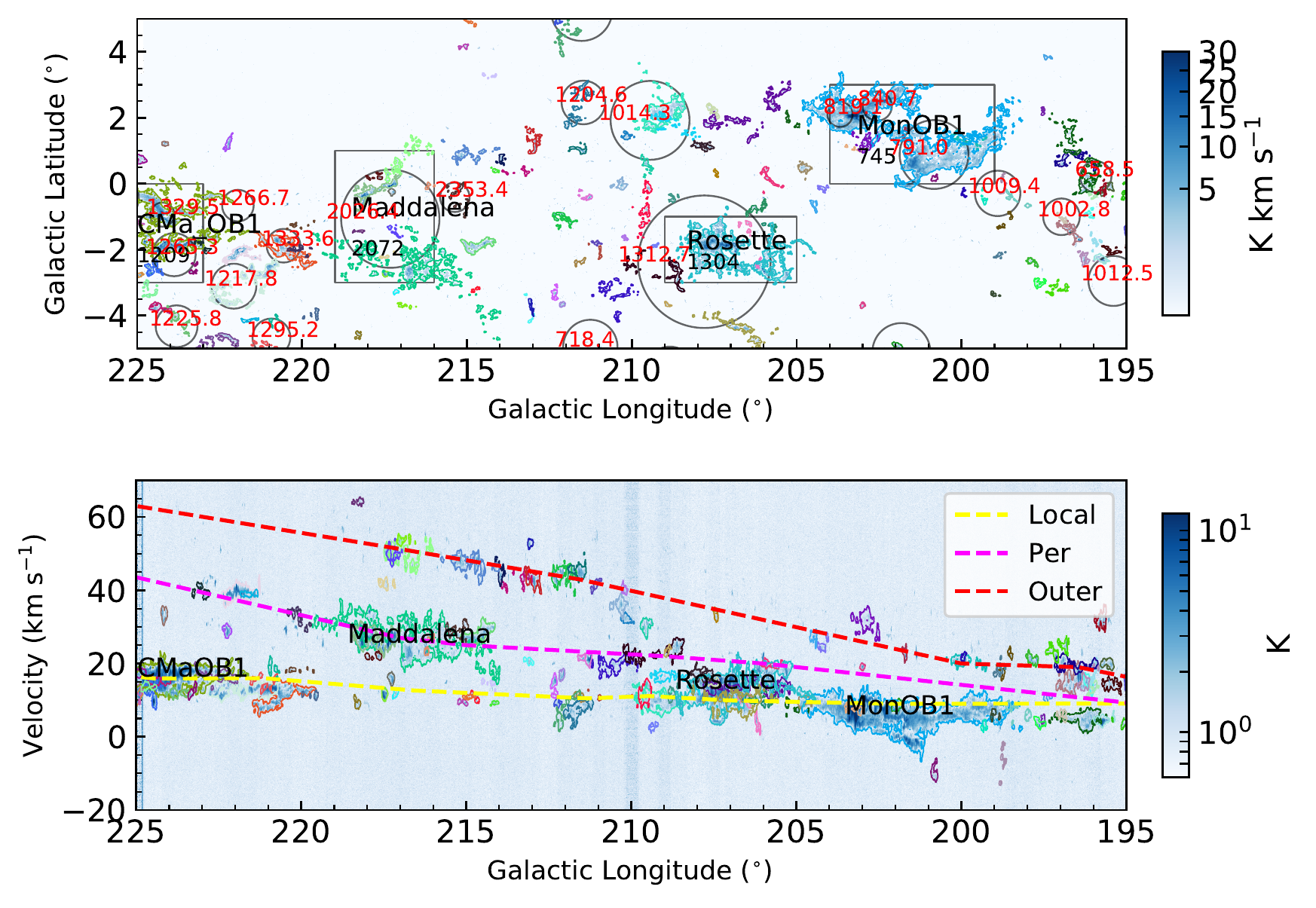}
	\caption{Spatial extents (upper panel) and l-v extents (lower panel) of the molecular clouds for which we have created N-PDFs, with each contour corresponding to an individual \COl cloud. The background image in the upper panel is the integrated intensity map of the \COl emission, T$_{mb}$, in the velocity range from $-$17 to 65 km s$^{-1}$, while in the lower panel we show the peak intensity map extracted along the direction of Galactic latitude. The black numbers and the red numbers in the upper panel are the distances of the molecular clouds measured by \cite{Zucker2019} and \cite{Chen2020}, respectively, while the effective extents of the corresponding molecular clouds are indicated with rectangles and circles. The yellow, magenta, and red dashed lines in the lower panel show the locations of the Local, Perseus, and Outer spiral arms, respectively, which are derived from CO and HI observations \citep{Weaver1970, Cohen1980}.} 
	\label{fig2}
\end{figure*} 
 
For the statistical significance of the N-PDF analysis, we only use a sub-sample of 120 \COl molecular clouds, i.e., the clouds that occupy at least 200 spatial pixels in the Galactic longitude and latitude plane, to build N-PDFs. A complete catalog of all \CO and \COl molecular clouds will be given in a forthcoming paper by Wang et al. (2022, in prep). Figure \ref{fig2} shows the boundaries of the selected sub-sample of the \COl clouds, which are overlaid on the integrated intensity map (upper panel) and the l-v peak intensity map (lower panel) of the \COl emission. 
                                   
\section{Results} \label{sec3}
\subsection{Physical Poperties} \label{sec3.1}
Column densities of the \COl molecular clouds are calculated assuming that the \COl molecules are under the local thermodynamic equilibrium (LTE) condition and that the \CO $J=1-0$ emission is optically thick. Considering the cosmic microwave background radiation (CMB) $T_{\rm bg} = 2.7$ K and the beam filling factor of the \CO emission is unity, the excitation temperature, T$_{\rm ex}$, of the molecular clouds can be derived from the peak intensities of the \CO spectra according to Eq. (1) in \cite{Li2018}. We obtained excitation temperature map for each \COl cloud. The optical depth, $\tau$, of the \COl emission can be derived using Eq. (2) in \cite{Li2018}, providing that the excitation temperature of the \COl emission is the same along the line of sight. Then the \COl column density can be obtained according to the following formula \citep{Li2018}, 

\begin{equation}
\label{f1}
N_{^{13}CO} = 2.42\times10^{14} \frac{1+0.88/{T_{ex}}}{1-e^{-5.29/T_{ex}}}\int \tau (^{13}CO) dv. 
\end{equation}
When the optical depth of the $^{13}$CO emission is small, the integrand $T_{ex}\int\tau(^{13}CO) dv$ on the right side of formula \ref{f1} can be approximated to $\tau_0/(1-e^{-\tau_0})\int T_{\rm mb} dv$ \citep{Pineda2010}, where $\tau_0$ is the peak optical depth of the $^{13}$CO emission. Providing the abundance ratios [$^{12}$C]$/$[$^{13}$C] $=$ 69 \citep{Wilson1999} and [CO]$/$[H] $\sim$ 10$^{-4}$ \citep{Solomon1972, Herbst1973}, the abundance ratio [$H_{2}]/[^{13}$CO] is expected to be $7\times10^{5}$. The column density of molecular hydrogen is converted from the \COl column density with this ratio.   

To get reliable N-PDFs at low column densities, we need to remove the pixels in the column density maps with \COl integrated intensities below three times the noise level of integrated intensity. The noise level is defined as $\sqrt{n}\sigma_{\rm RMS} dv$, where n is the number of velocity channels of a given pixel in a \COl molecular cloud (varying between pixels), $\sigma_{RMS}$ is the RMS noise level of the pixel, and $dv$ is the width of the velocity channel. Although the sub-sample of molecular clouds that we selected for the N-PDF analysis, as described in Section \ref{sec2.2}, have pixel numbers more than 200, the actual pixel numbers used to build N-PDF for some of the clouds are slightly smaller after removing the pixels with relatively low signal-to-noise ratios. Since the molecular clouds are identified above $2\sigma_{RMS}$, and they should occupy at least three velocity channels, we could estimate a reference detection limit of H$_2$ column density for each cloud, which is derived according to Eq. \ref{f1} using the median value of 3$\times$2$\sigma_{RMS} dv$, the median $\rm{T_{ex}}$, and median $\rm {\tau(^{13}CO)}$ of the \COl cloud.  
      
Distances of the clouds are essential for calculating their physical parameters, such as mass and radius. The distances of well-known star-forming regions in this part of the sky area have been measured by \cite{Zucker2019} and \cite{Chen2020} using the method that allocates a molecular cloud a distance corresponding to a ``jump'' in extinction along the line of sight. \cite{Yan2019} also have measured distances of eleven molecular clouds in the third quadrant of the Milky-Way mid-plane using the MWISP data. The sky coverage of their work is 209$.\!\!^{\circ}$75$\leq$l$\leq$219$.\!\!^{\circ}$75, and $\left|\rm{b}\right|\leq5^{\circ}$, which is partly overlapped with the area in this work. The distances of molecular clouds from \cite{Zucker2019} and \cite{Chen2020} are indicated in Figure \ref{fig2}(a) with black numbers and red numbers, respectively, while the spatial extents of the corresponding molecular clouds from these two works are marked with rectangles and circles, respectively. In the sightlines where the molecular clouds from the two works overlap, the measured distances are very close. For example, the distance differences for the CMa OB1, Maddalena, Rosette, and Mon OB1 GMCs between \cite{Zucker2019} and \cite{Chen2020} are all less than 6$\%$. However, the catalogs from \cite{Zucker2019}, \cite{Chen2020}, and \cite{Yan2019} do not provide distances for all the selected molecular clouds in this work. Therefore, we derived the kinematic distances for the clouds in the sub-sample, assuming a flat rotation curve of the Galaxy and adopting the Galactic rotation constants from model A5 of \cite{Reid2014}. The kinematic distances of the CMa OB1, Maddalena, Rosette, and Mon OB1 star-forming regions are 1.25, 2.25, 1.39, and 0.83 kpc, respectively, which are consistent with the distances obtained by \cite{Zucker2019} and \cite{Chen2020} within differences of $\sim$10\%.

For each \COl molecular cloud in the selected sub-sample, maps of the excitation temperature, optical depth, column density, and second-moment are derived. Physical parameters for each clouds, such as position centroid, pixel number, peak excitation temperature, mean H$_2$ column density, radius, mass, median optical depth, virial parameter, and Mach number, are extracted. The intensity-weighted position centroids in the PPV space are calculated according to \cite{Yan2019}. The effective radius of the clouds is calculated according to $R = d\sqrt{\Omega/\pi}$, where $d$ is the distance, and $\Omega$ is the projected angular area. The radius defined this way is possibly a factor of $\sim2$ larger than the definition commonly used in other statistical works of physical properties of molecular clouds, which approximate molecular clouds to elongated gaussian ellipsoids and use the geometric mean of the intensity weighted (or unweighted) FWHMs along two major axes of the clouds as the radii \citep{Heyer2001, Rice2016, Ma2021}. The mass, $M_{LTE}$, of the clouds is derived according to Eq. (5) in \cite{Ma2021}. The surface density is then the ratio of the mass to the area of the clouds. Once the mass, size, and velocity dispersion are derived, the virial parameter can be calculated accrording to $\alpha_{vir} = 5\sigma_v^2R/(GM_{LTE})$, where G is the gravitational constant. The Mach number, $\mathcal{M}$, is defined as the ratio between the non-thermal velocity dispersion and the local sound speed of the cloud. Specifically, we calculated $\mathcal{M}$ with $\mathcal{M} = \sigma_{v_{non}}/c_s$, where $\sigma_{v_{non}}$ is the non-thermal velocity dispersion and $c_s = \sqrt{k_B T_k/(\mu_{H}m_{H})}=0.187\sqrt{T_{kin}/10}$ km s$^{-1}$ is the sound speed \citep{Schneider2013}, where $k_B$ is the Boltzman constant, $\mu_{H} = 2.37$ is the mean molecular weight per free particle \citep{Kauffmann2008}, and $m_{H}$ is the mass of the atomic hydrogen. The kinetic temperature $T_{kin}$ is approximated to the excitation temperature under the LTE condition. The non-thermal velocity dispersion is derived through $\sigma_{v_{non}} = \sqrt{\sigma_{v,^{13}CO}^2-c_{s,^{13}CO}^2}$, where $\sigma_{v,^{13}CO} = \sqrt{\Sigma[T_i(v_i-v_0)^2]/\Sigma T_i}$ is the intensity-weighted veloctiy dispersion, $v_0$ is the intensity-weighted central velocity of the cloud, and $c_{s,^{13}CO}$ is the thermal velocity dispersion of \COl molecule, which is defined as $c_{s,^{13}CO}=\sqrt{k_B T_k/(\mu_{^{13}CO}m_{H})}$ and $\mu_{^{13}CO}=29$ is the relative molecular mass of \COl molecule.

The extracted parameters are listed in Table \ref{tab1}. The molecular clouds are named after their centroid positions and velocities. Specifically, the cloud names start with ``MWISP" and then contain the spatial coordinates of the molecular clouds accurate to three decimal places and the centroid velocities accurate to two decimal places. 

\begin{longrotatetable}
	\begin{deluxetable*}{llrrllrrlrrrrrrrrr}
		\tablewidth{700pt}
		\tabletypesize{\tiny}
		 \tablecaption{Properties of the N-PDFs and Physical Paramteters of the Molecular Clouds\label{tab1}}
		 \tablecolumns{17}
		 \tablehead{
			 \colhead{Name} & \colhead{d} & \colhead{$\mu$} & \colhead{$\sigma_s$} & \colhead{$b_r$} & \colhead{$\alpha$} & \colhead{N$\rm_{pix}$} & \colhead{T$_{\rm{ex, max}}$} & \colhead{$\left\langle N_{H_2}\right\rangle $} & \colhead{R} & \colhead{Mass} & \colhead{Skewness} & \colhead{Kurtosis} & \colhead{$\tau_{13}$} & \colhead{$\mathcal{M}$} &\colhead{N$_{YSO}$} & \colhead{$\alpha_{vir}$} &\colhead{Shape}\\
			& \colhead{(kpc)} &  &  &  &  &  & \colhead{(K)} & \colhead{(cm$^{-2}$)} & \colhead{(pc)} & \colhead{(M$_{\odot}$)} & &  &  & & & & \\
			\colhead{(1)} & \colhead{(2)} & \colhead{(3)} & \colhead{(4)} & \colhead{(5)} & \colhead{(6)} & \colhead{(7)} & \colhead{(8)} & \colhead{(9)} & \colhead{(10)} & \colhead{(11)} & \colhead{(12)} & \colhead{(13)} & \colhead{(14)} & \colhead{(15)} &\colhead{(16)} & \colhead{(17)} &\colhead{(18)}
			}
		 \startdata
		\input{table1.tex}
		\enddata
		 \tablecomments{Column 1: name of the cloud. Column 2: distance of the cloud. Column 3-6: fitted parameters, i.e., $\mu$, $\sigma_s$, $b_r$, $\alpha$, of the N-PDFs in Eqs. \ref{eq5}-\ref{eq6}, where $b_r$ and $\alpha$ are only given for the N-PDFs that have LN+PL shapes. Column 7: pixel number. Column 8: peak excitation temperature. Column 9: average column density. Column 10: radius. Column 11: LTE mass. Column 12: Skewness of the N-PDF. Column 13: Kurtosis of the N-PDF. Column 14: \COl optical depth of the cloud. Column 15: Mach number. Column 16: number of associated YSOs. Column 17: virial parameter. Column 18: category of the N-PDF. The source name is defined under the MWISP standard. According to the MWISP standard for nomenclature, molecular clouds are named after their centroid positions and velocities. Specifically, the names start with ``MWISP" and then contain the spatial coordinates of the molecular clouds accurate to three decimal places and the centroid velocities accurate to two decimal places. The accuracy is set according to the pointing accuracy and the velocity resolution of the PMO-13.7m telescope. The average column density for each cloud is calculated above the reference detection limit of the cloud.}
	\end{deluxetable*}
\end{longrotatetable}
\clearpage

\subsection{Column Density Distribution --- N-PDFs} \label{sec3.2}
\subsubsection{Models and Fitting} \label{sec3.2.1}

An N-PDF is simply a normalized histogram of the distribution of the logarithmic normalized column density, $s = \ln\ (N/\left\langle N\right\rangle)$. The average column density used in the normalization for each N-PDF is calculated using all the effective pixels in the column density map. However, since we fitted each N-PDF above the reference detection limit of $N_{H_2}$ (see details in the next paragraph), the average column densities of the clouds listed in Table \ref{tab1} are calculated above the detection limit of $N_{H_2}$ for both consistency and higher confidence. We build the N-PDFs for the selected molecular clouds with a uniform bin width of $\Delta s = 0.2$, and the count in each bin for an N-PDF is normalized with the product of the total pixel number of the cloud and $\Delta s$. The width of bins is selected in a practical way to have sufficient numbers of bins to describe the shape of the N-PDFs and, at the same time, have a minimum number of bins with zero counts. From the apparent shapes of the N-PDFs and given that some of the molecular clouds in the region are active in star formation, and some are quiescent, we expect to see both turbulence-dominated molecular clouds and self-gravity dominated ones. Hence, we consider the following two models for the fitting of N-PDFs, i.e., a pure log-normal
\begin{equation}
p(s) = Ae^{{-\frac{(s-\mu)^2}{2\sigma_s^2}}}, 
\label{eq5}
\end{equation} 
and a piecewise function of log-normal and power-law 
\begin{eqnarray}
p(s)  & = &
\left\{
\begin{array}{lc}
Ae^{{-\frac{(s-\mu)^2}{2\sigma_s^2}}} & \ \ \ s < b_r \\
Ae^{\alpha s}\times e^{-\alpha b_r}\times e^{-\frac{(br-\mu)^2}{2\sigma_s^2}} & \ \ \ s\ge b_r 
\end{array}     \right.
\label{eq6}
\end{eqnarray}
where $\mu$ and $\sigma_s$ are the mean and dispersion of $s$, $\alpha$ is the power-law index, $b_r$ is the value of $s$ at the break where the N-PDF turns from log-normal to power-law, and A is a constant related to the normalization of the N-PDF. The parameters $\mu$, $\sigma_s$, $\alpha$, and $b_r$ are treated as free in the fitting process.  

\begin{deluxetable*}{cllll}
	\label{tab2}
	\tablecaption{Priors of the parameters used in the fitting processes}
	\tabletypesize{\normalsize}
 	\setlength{\tabcolsep}{10pt}
	\tablehead{\colhead{Parameter} & \colhead{$\mu$} & \colhead{$\sigma_s$} & \colhead{$b_r$} & \colhead{$\alpha$} } 
	\startdata
    Range & $\rm (s_{ref}, s_{max})$ & $(0, \rm 5\times s_{std})$ & $(\mu, \rm s_{max})$ & $(-10, 0)$\\
	\enddata
    \tablecomments{The subscripts ``ref'' represents the value of s corresponding to the reference detection limit of N$_{H_2}$, ``max'' represents the maximum, and ``std'' means the standard deviation.}
\end{deluxetable*}

\begin{figure*}[htb!]
	\centering
	\begin{minipage}[t]{0.8\linewidth}
		\centering
		\includegraphics[width= \linewidth]{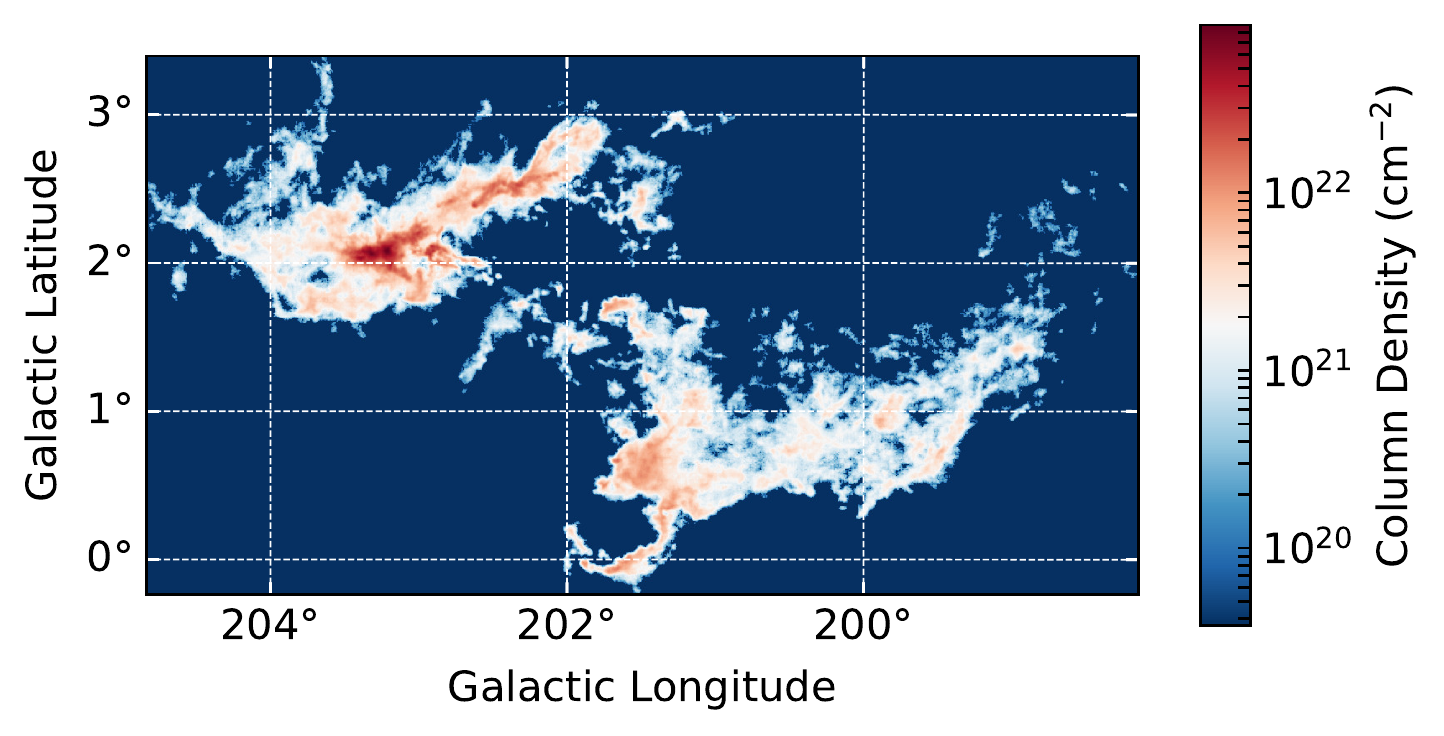}
		\\(a)
	\end{minipage}
	\begin{minipage}[t]{0.47\linewidth}
		\centering
	    \includegraphics[width= \linewidth]{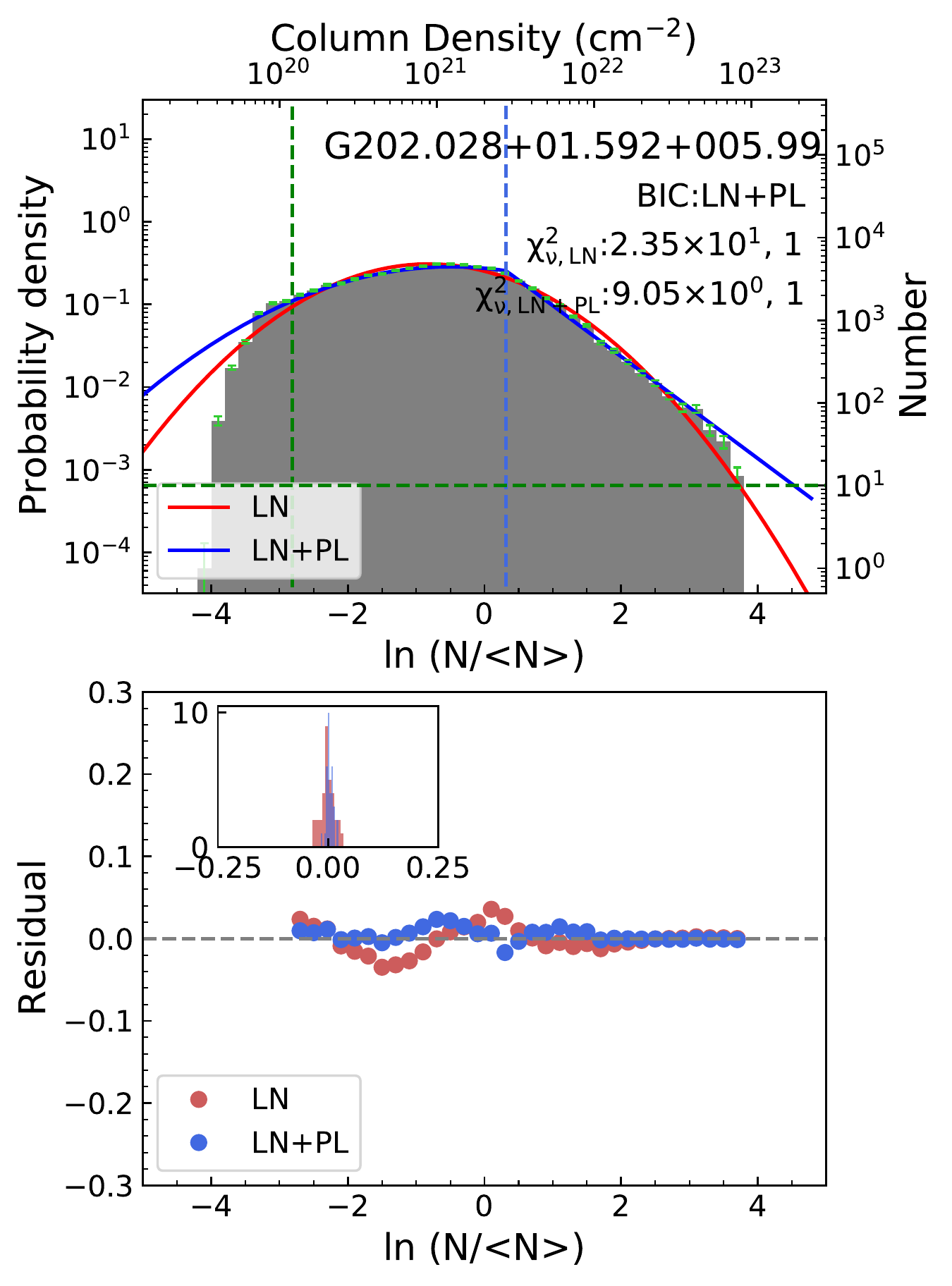}
		\\(b)
	\end{minipage}
	\begin{minipage}[t]{0.51\linewidth}
		\centering 
		\raisebox{0.05\height}{\includegraphics[width= \linewidth]{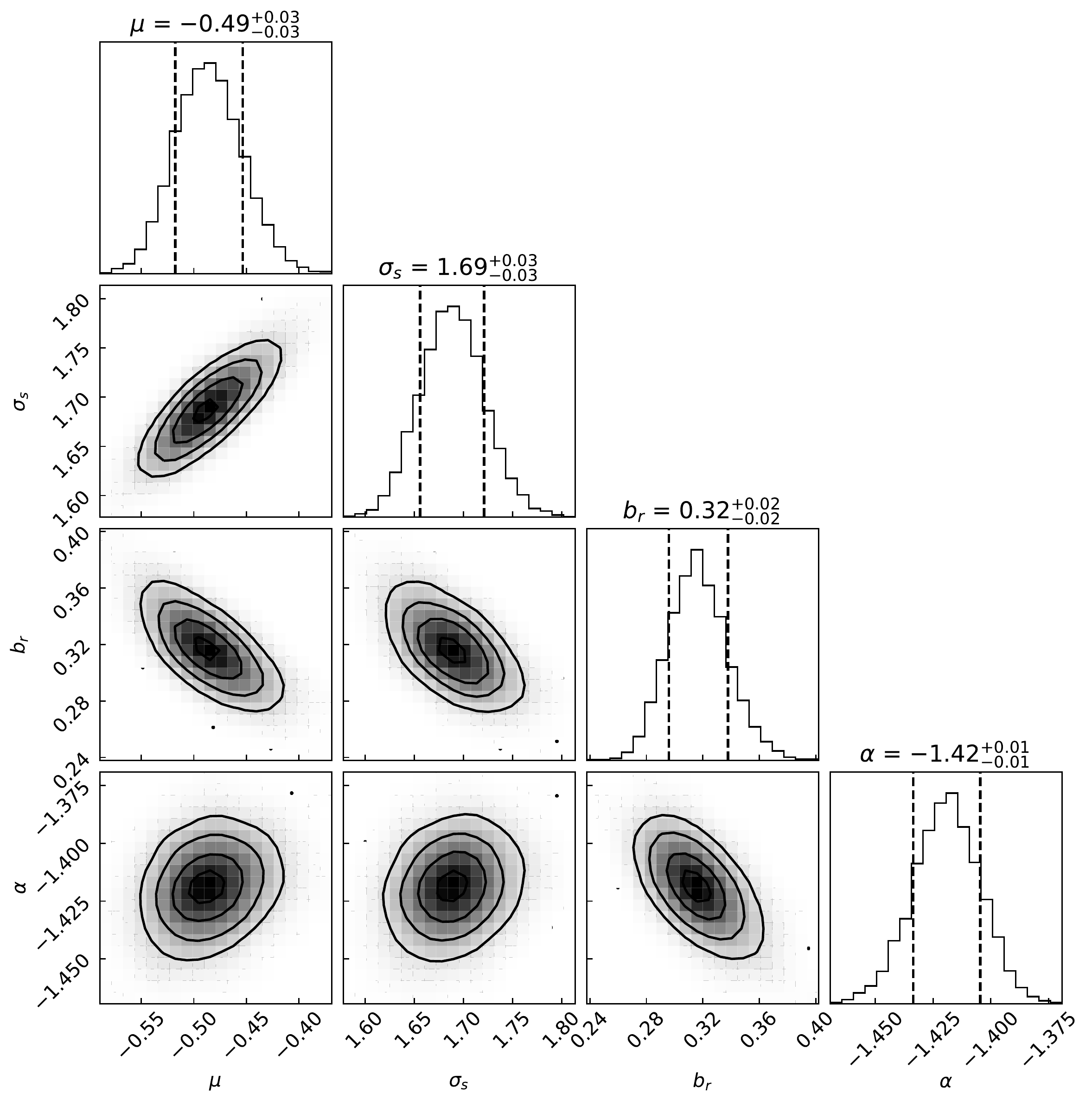}}
		\\(c)
	\end{minipage}
	\caption{An example of the N-PDF fitting process. Panel (a) shows the spatial distribution of the $H_2$ column densities of the Mon OB1 GMC, while panel (b) shows the corresponding N-PDF overlaid with the fitted LN and LN+PL functions. The horizontal green dashed line marks the ten counts level in the statistic bins, while the vertical green dashed line marks the reference detection limit of the H$_2$ column density. The vertical blue dashed line shows the location of $b_r$. The model selected by BIC and the reduced chi-squared, $\chi_{\nu}$, of the LN and LN+PL fittings are shown in the figure. Detailed meaning of number right to $\chi_{\nu}$ is given in the text. The lower subpanel in panel (b) shows the distribution and histograms of the residuals of the two fittings. Panel (c) presents the corner map of the fitted parameters. The two vertical dashed lines, from left to right, in the marginalized distributions of the parameters are the 16$\%$ and 84$\%$ quantiles, respectively. The median values of $\mu$, $\sigma_s$, $b_r$, and $\alpha$ are presented on top of the corresponding histograms. The differences between the two quantiles and the median values are adopted as one-sigma errors of these parameters.} 
	\label{fig3}
\end{figure*}                 

We use the Markov Chain Monte Carlo (MCMC) method for fitting the N-PDFs of the molecular clouds, which can explore the parameter space to find the most probable set of parameters. One advantage of the MCMC fitting is that we do not need to bin the column density data, implying that we may get the N-PDFs for the clouds occupying relatively small projected areas. The python package EMCEE \citep{Foreman-Mackey2013} is used to implement the fitting. Figures \ref{fig3} presents the column density map and N-PDF of the Mon OB1 (G202.028+01.592+005.99) GMC as an example. We fitted each N-PDF of the molecular clouds with both LN and LN+PL models above the reference detection limit of $N_{H_2}$, which is displayed as the green vertical dashed line in Figure \ref{fig3}(b). The MCMC method is based on Bayesian theorem and is an iterating sampling method. In each step of sampling, the EMCEE sampler proposes a model represented by an ensemble of parameters, calculates the posterior probability, accepts or rejects the proposed model given the posterior probability, and moves to the next position in the parameter space according to the stretch move method \citep{Goodman2010,Foreman-Mackey2013}. The posterior probability $p(M|D)$ is calculated through $p(M|D) = p(D|M)p(M)/p(D)$, where $M$ stands for model, $D$ for data, and ``$|$'' for given condition. In our case $p(D)=1$, the prior probability of the model is $p(M)$ = $p(\mu)p(\sigma_s)p(\alpha)p(b_r)$. We adopt uniform priors for the parameters $\mu$, $\sigma_s$, $b_r$, and $\alpha$ within the ranges listed in Table \ref{tab2}. In Table \ref{tab2}, the lower limit of $b_r$ depends on the center, i.e., $\mu$, of the log-normal part of the piecewise function in Eq. \ref{eq6}, which can be seen as a Bayesian hierarchical modeling situation \citep{Sharma2017}. In this situation, $p(b_r) = p(b_r|\mu)p(\mu)$ = 1, when $\mu$ and $b_r$ are in the ranges from Table \ref{tab2}. The n-dimensional (n=4 here) vector that explores the parameter space and extracts samples as above is called a ``walker''. The sampling process is repeated thousands of times for each parameter until the accepted samples cover the delimited parameter space and form a ``Markov Chain''. We used 32 walkers for the fitting in this work. The iterating process is stopped when we get 10,000 accepted samples for each walker. The first 1000 steps of the Markov Chains are discarded for the so-called burn-in period of the sampling process, and then the chains are flattened by extracting samples every 50 steps. The median of the flattened sample of a parameter is considered its best fit, while the 16th and 84th percentiles are considered the lower and upper limit of its one-sigma confidence interval. Figure \ref{fig3}(c) is the corner plot that presents the marginalized distribution of the fitted parameters for the LN+PL model of the Mon OB1 GMC.  

\begin{figure*}[htb!]
	\centering
	\begin{minipage}[t]{0.32\linewidth}
		\centering
		\includegraphics[width= \linewidth]{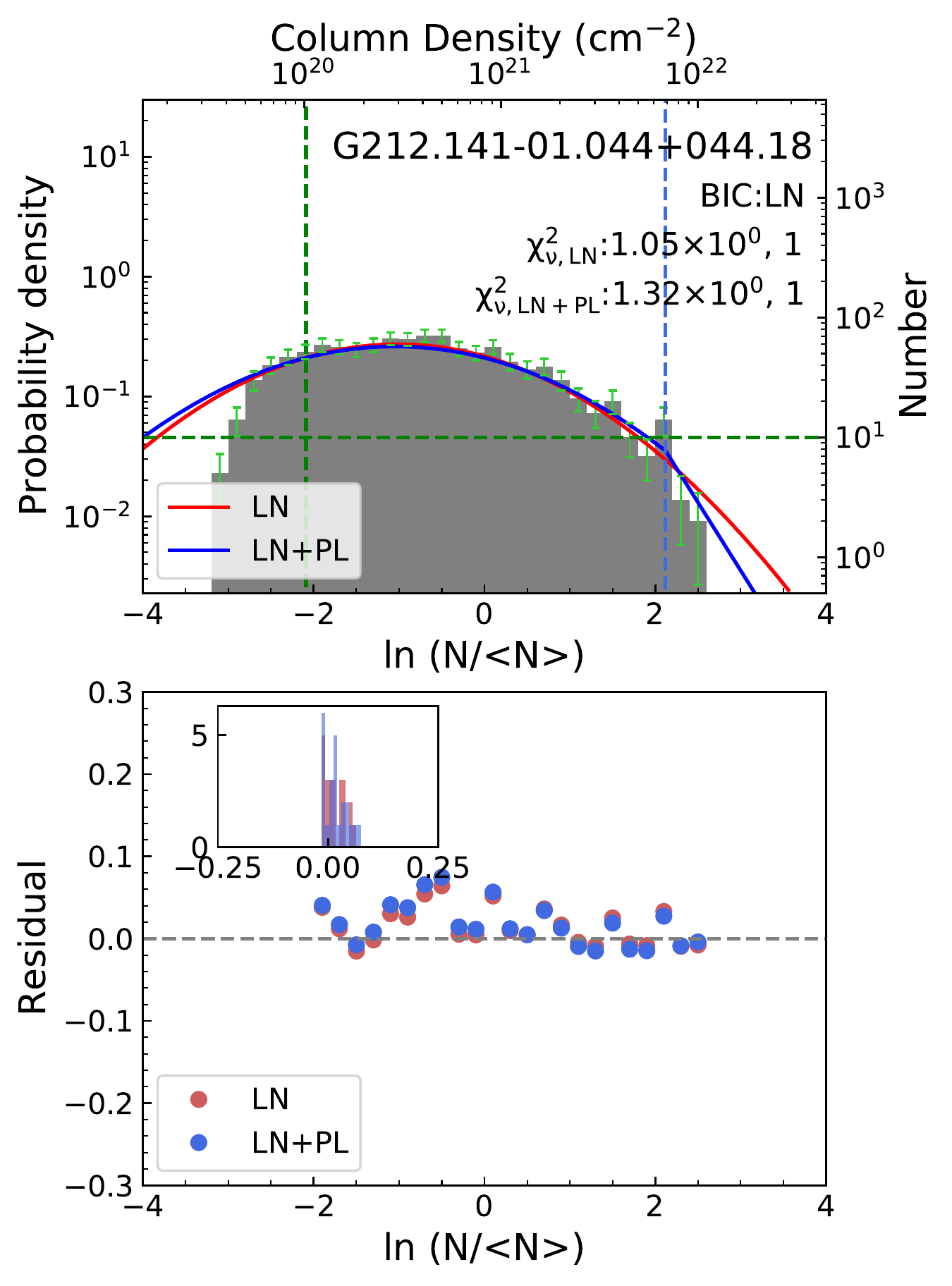}
		\\(a)
	\end{minipage}
	\begin{minipage}[t]{0.32\linewidth}
		\centering
		\includegraphics[width= \linewidth]{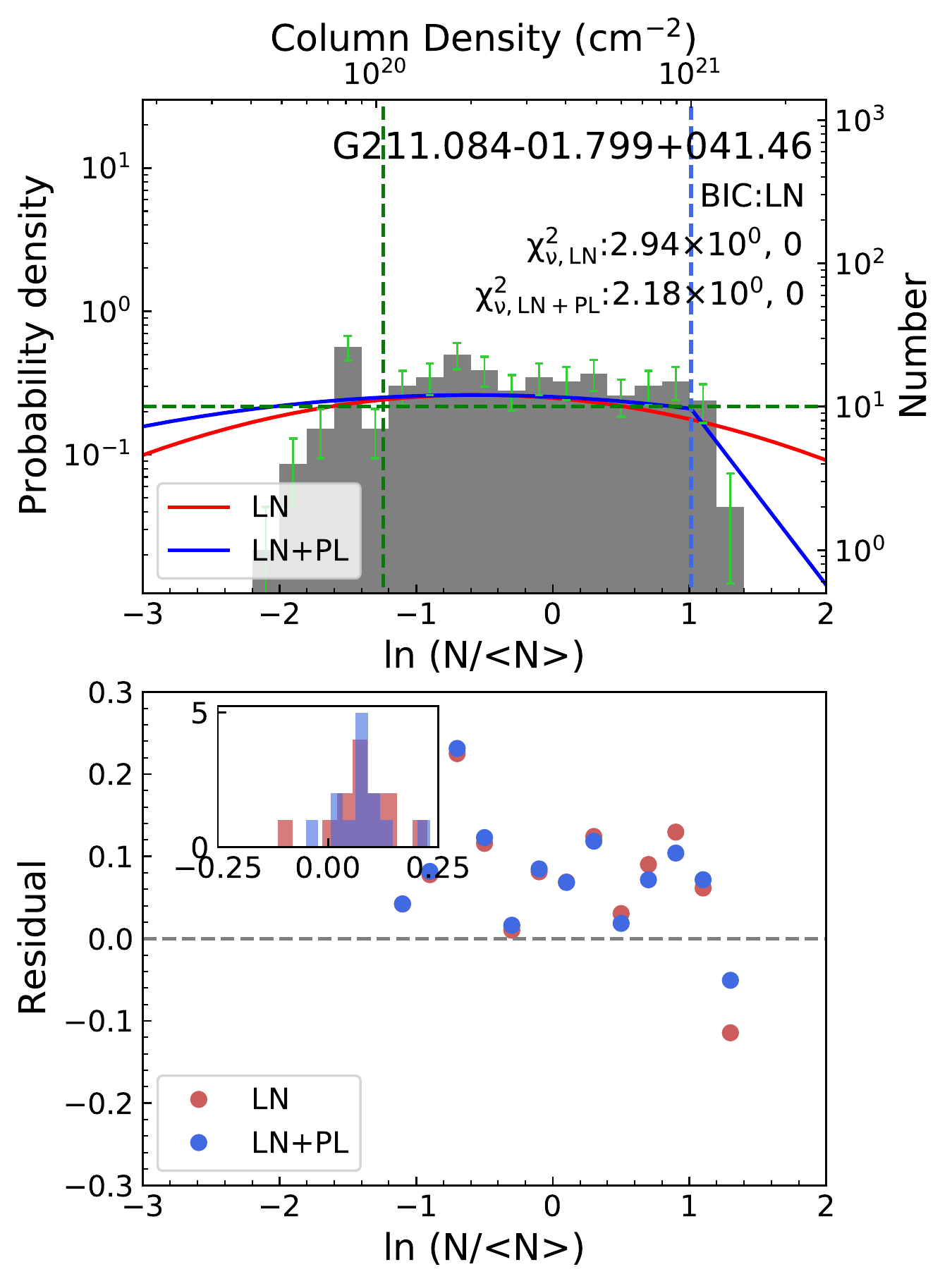}
		\\(b)
	\end{minipage}
	\begin{minipage}[t]{0.32\linewidth}
		\centering 
		\includegraphics[width= \linewidth]{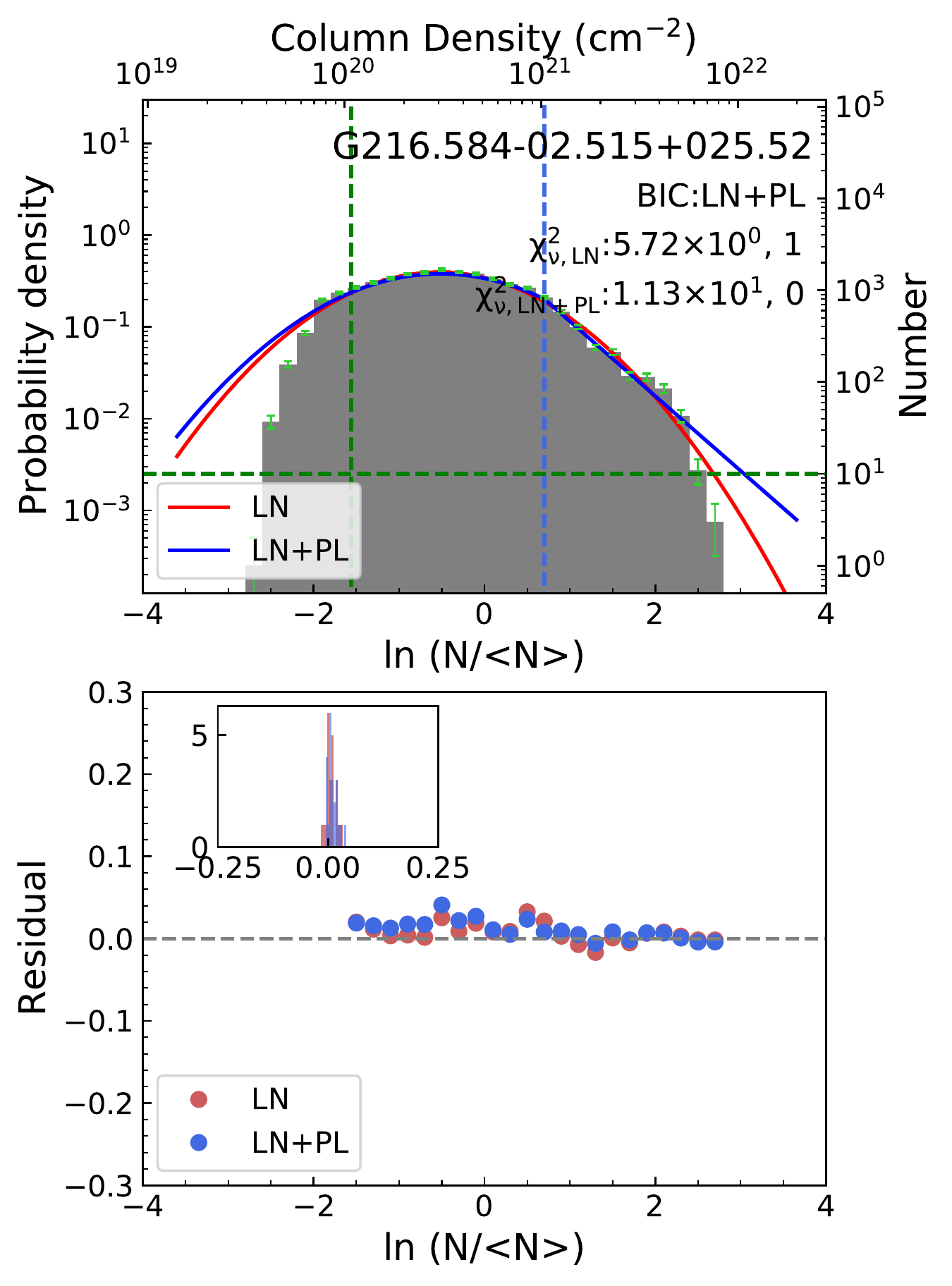}
		\\(c)
	\end{minipage}
	\caption{Examples of N-PDFs of the categories (a) LN, (b) UN (BIC proposing LN), and (c)UN (BIC proposing LN+PL), respectively. The horizontal, vertical green dashed lines and the reduced chi-squared have the same meaning as those in Figure \ref{fig3}. The names of the clouds are given in the upper right corner of panels (a)-(c). The fitted N-PDFs of the LN and the LN+PL models are indicated in red and blue, respectively. The lower panels in subfigures (a)-(c) present the residuals of the fittings for the N-PDFs. The zoom-in histograms in the lower panels show the distributions of the residuals.} 
	\label{fig4}
\end{figure*}     

To classify the N-PDFs into appropriate categories based on their shapes, model selection between the two models of LN and LN+PL is necessary. We choose the model that has a smaller Bayesian Information Criterion (BIC) value \citep{Schwarz1978}. Besides, for a ``good'' fitting, the residuals should be randomly distributed around zero. Therefore, we also checked the residuals and reduced chi-squared, $\chi_{\nu}^2$, of the best fits of the two models for each molecular cloud. The residuals of the fittings are shown below the N-PDFs, as presented in Figure \ref{fig3}(b). If more than 80\% of the residuals of a fitting are above or below zero, or its $\chi_{\nu}^2$ is above 10, we consider the fitting is not good. The model selected using BIC should meet the residual criterion, otherwise we consider the shape of the N-PDF is unclear (UN). Figure \ref{fig3}(b) also shows the values of $\chi_{\nu}^2$ for the best fits of the LN and LN+PL models and shows the model selected using the BIC criterion. The number right to the reduced chi-squared represents whether the fittings meet the 80\% residual criterion, ``1'' for yes and ``0'' for no. The critical value of the ratio 80\% and $\chi_{\nu}^2 = 10$ are selected empirically, as we intend to assign as many N-PDFs with definite shapes as possible. Figure \ref{fig3}(b) is an example of the LN+PL form N-PDFs. Firstly, the BIC criterion prefers the LN+PL model. We can see that the residuals of the LN+PL fitting are randomly distributed around 0 and the reduced chi-squared of the LN+PL fitting is $\sim$9, which is below the critical value 10. Therefore, the cloud G202.028$+$01.592$+$005.99 (Mon OB1) is classified to the LN+PL category. Figure \ref{fig4} shows examples of the LN and the UN categories. In panel (a), the model proposed by the BIC criterion is LN, while the residual distribution and $\chi_{\nu}^2$ all satisfy their criteria. Therefore, the proposed model LN is accepted. In Figures \ref{fig4}(b) and \ref{fig4}(c) (Maddalena), although the BIC criterion selects the LN and LN+PL models, respectively, more than 80\% of the residuals of the LN fitting in Figure \ref{fig4}(b) are greater than 0, and $\chi_{\nu}^2$ of the LN+PL fitting in Figure \ref{fig4}(c) is higher than 10 (11.3). Therefore, the two models in Figures \ref{fig4}(b) and \ref{fig4}(c) are rejected.  

\begin{figure*}[htb!]
	\centering
	\includegraphics[trim=0cm 0cm 0cm 0cm, width= 0.6\linewidth, clip]{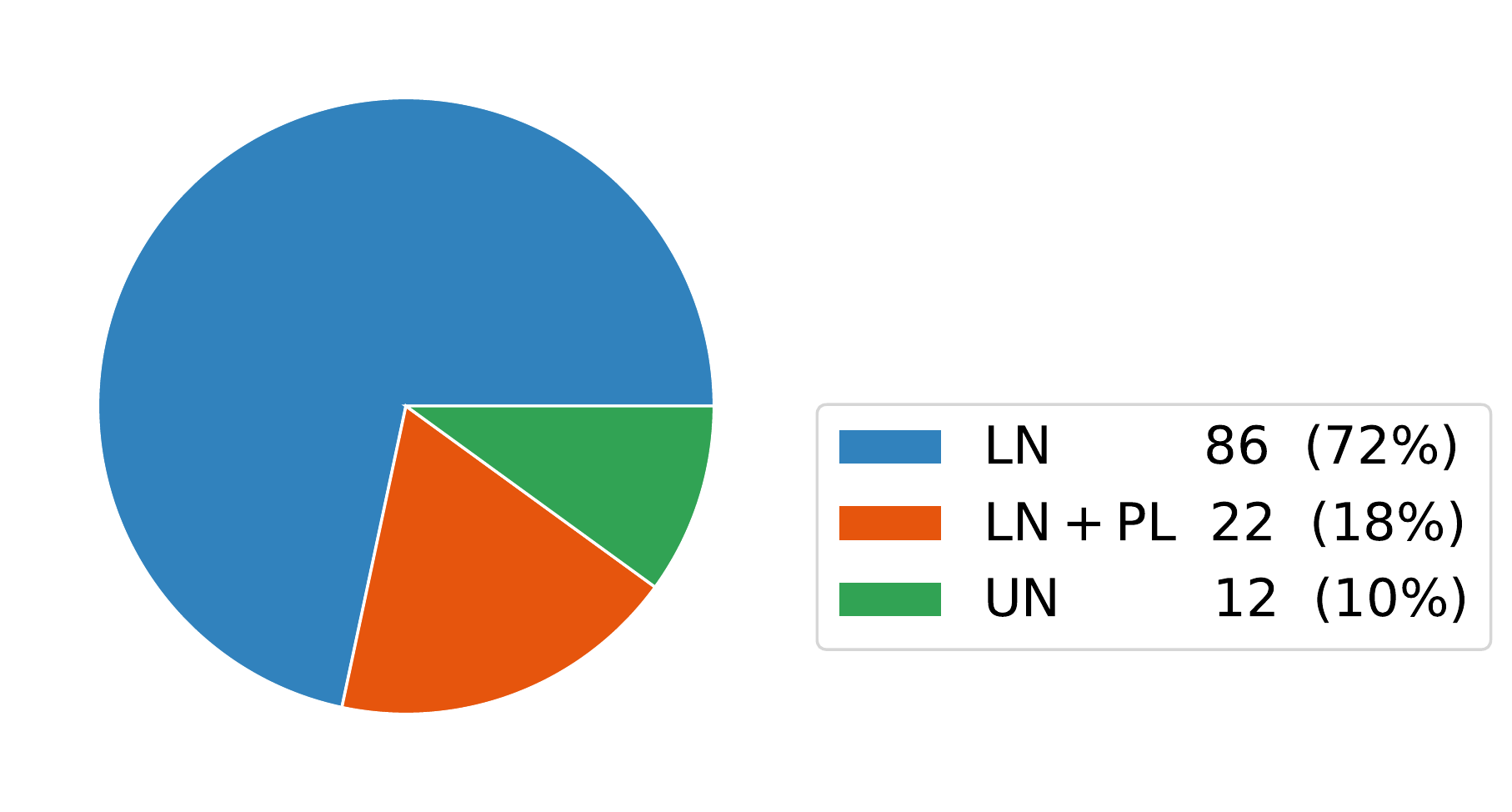}
	\caption{Pie chart of the classification of N-PDFs into LN, LN+PL, and UN.} 
	\label{fig5}
\end{figure*} 
 
In summary, the 120 molecular clouds are classified into three categories, i.e., LN, LN+PL, and UN, respectively. The number and percentage of each category are shown in Figure \ref{fig5}. In Figure \ref{fig5}, 72\% (86) of the N-PDFs have the LN form, 18\% (22) have the LN+PL form, whereas 10\% (12) of the N-PDFs can not be described by any single model. The categories of the N-PDFs and the parameters, $\mu, \sigma_s, b_r, \alpha$, of the adopted fittings for each cloud are listed in Table \ref{tab1}. All the N-PDFs are presented in Figures \ref{fig16}-\ref{fig18} in the Appendix in the order of LN, LN+PL, and UN categories, respectively. The N-PDFs in Figures \ref{fig16}-\ref{fig18} are sorted in ascending order of the spatial pixel numbers of column densities of the clouds. In Figure \ref{fig18}, twelve N-PDFs can not be assigned with definite shapes, among which one is excluded from the fitting because of the high $\chi_{\nu}^2$. Visually, the N-PDFs in the last two panels of Figure \ref{fig18} are well fitted with the LN+PL model. The difference between the statistical results and visual inspection may be caused by the large pixel number and therefore the large dynamical range of the y-axes of the two N-PDFs. The probabilities of the two N-PDFs cover $\sim$3 orders of magnitude. Therefore, in a logarithmic-scale display, even large differences between the data and the model may not be visually evident.   

\subsubsection{Statistics of N-PDF Parameters and Physical Parameters of Molecular Clouds}
 
\begin{figure*}[htb!]
	\centering
	\begin{minipage}[t]{0.38\linewidth}
		\centering
		\includegraphics[width= \linewidth]{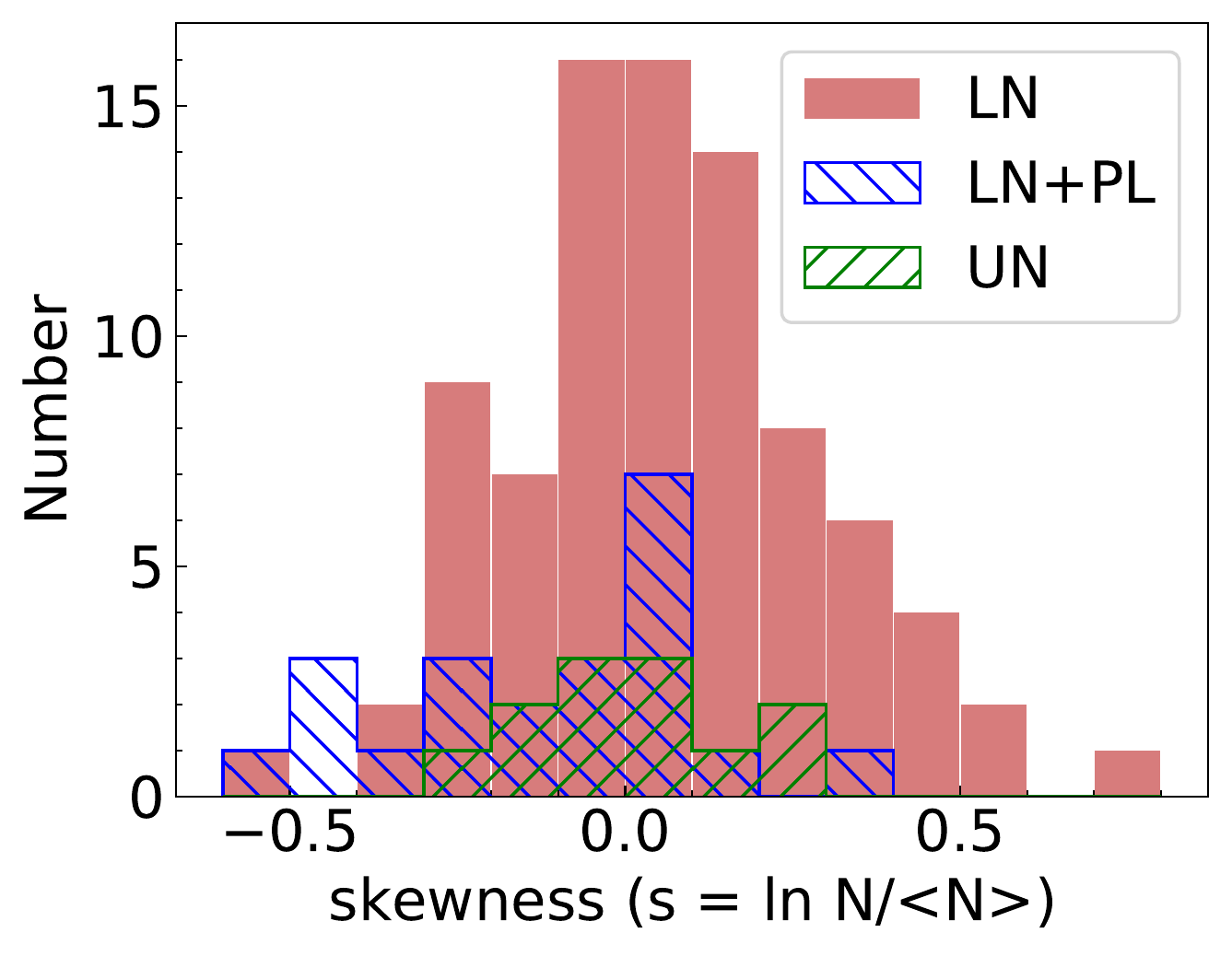}
		\\(a)
	\end{minipage}
	\begin{minipage}[t]{0.38\linewidth}
		\centering
		\includegraphics[width= \linewidth]{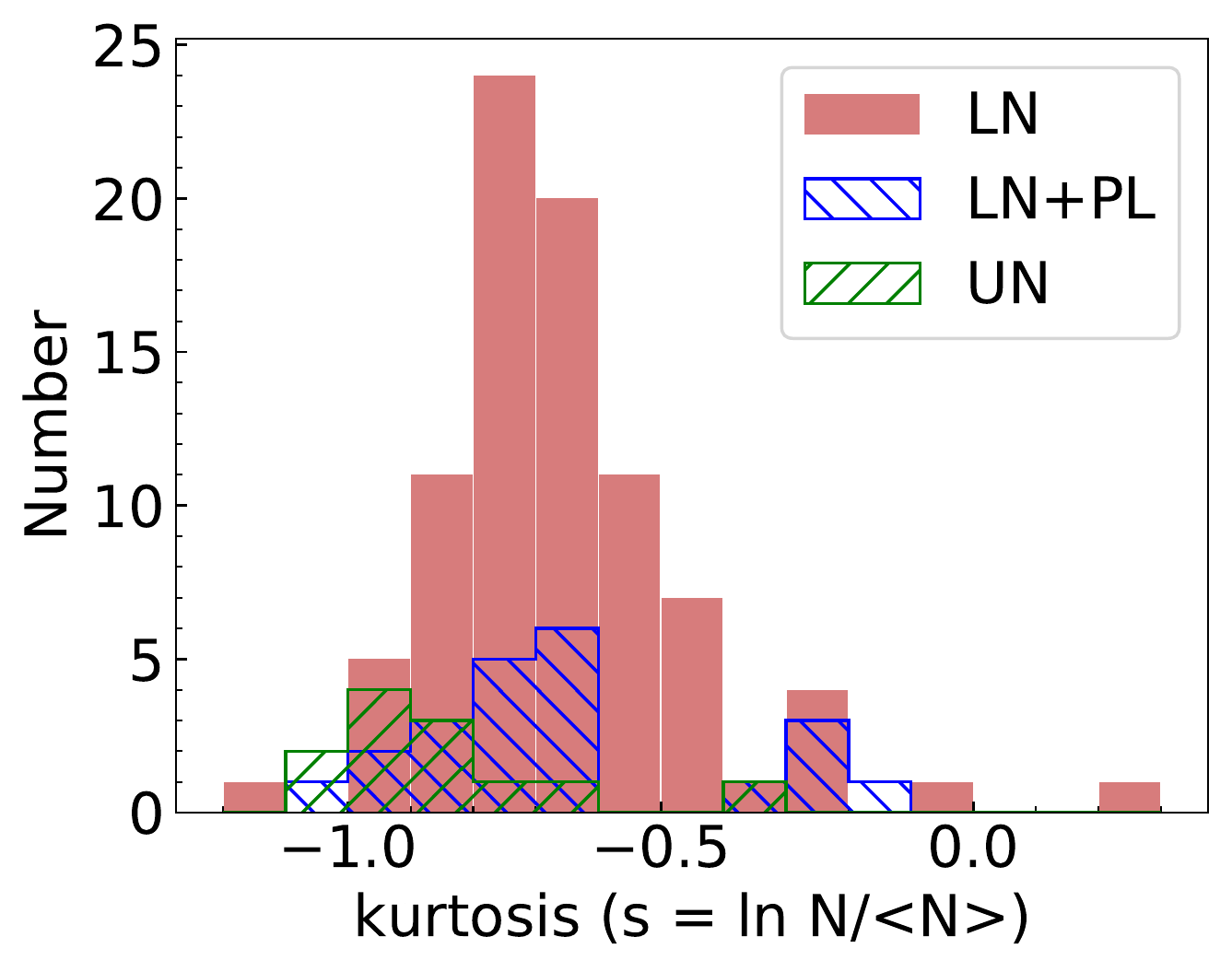}
		\\(b)
	\end{minipage}
	\caption{Distribution of (a) Skewness and (b) Kurtosis of the column densities of the selected clouds.} 
	\label{fig6}
\end{figure*}

In addition to fitting with model distributions, we can also describe the shapes of N-PDFs with the third and fourth moments, i.e., skewness and kurtosis, of $s$. Skewness measures the symmetry property of the distribution of a random variable about its mean, while kurtosis measures the existence of outliers. Figure \ref{fig6} gives the distribution of the skewness and kurtosis of $s$ of the molecular clouds. The excess kurtosis is used, which is defined as the fourth central moment divided by the square of the variance of $s$, and is then subtracted by three \citep{kokoska2000crc}. In this definition, the kurtosis of a normal distribution is zero. The skewness of the molecular clouds in the LN and UN categories are symmetrically distributed around zero. However, the LN+PL fitted molecular clouds tend to have negatively skewed N-PDFs, i.e., right-leaning curves like the N-PDF of cloud G201.963$-$04.904$+$008.54 in Figure \ref{fig17}, which may result from that the PL lines of these N-PDFs are generally below their LN lines. In Figure \ref{fig6}(b), the kurtosises of the LN and LN+PL fitted molecular clouds are concentrated around $-$0.7, while the kurtosises of the molecular clouds in the UN category are relatively more negative. A small kurtosis of $-1$ usually indicates a relatively flat distribution, which is consistent with the situations seen in Figures \ref{fig16}-\ref{fig18}.     

\begin{figure*}[htb!]
	\centering
	\begin{minipage}[t]{0.35\linewidth}
		\centering
		\includegraphics[width= \linewidth]{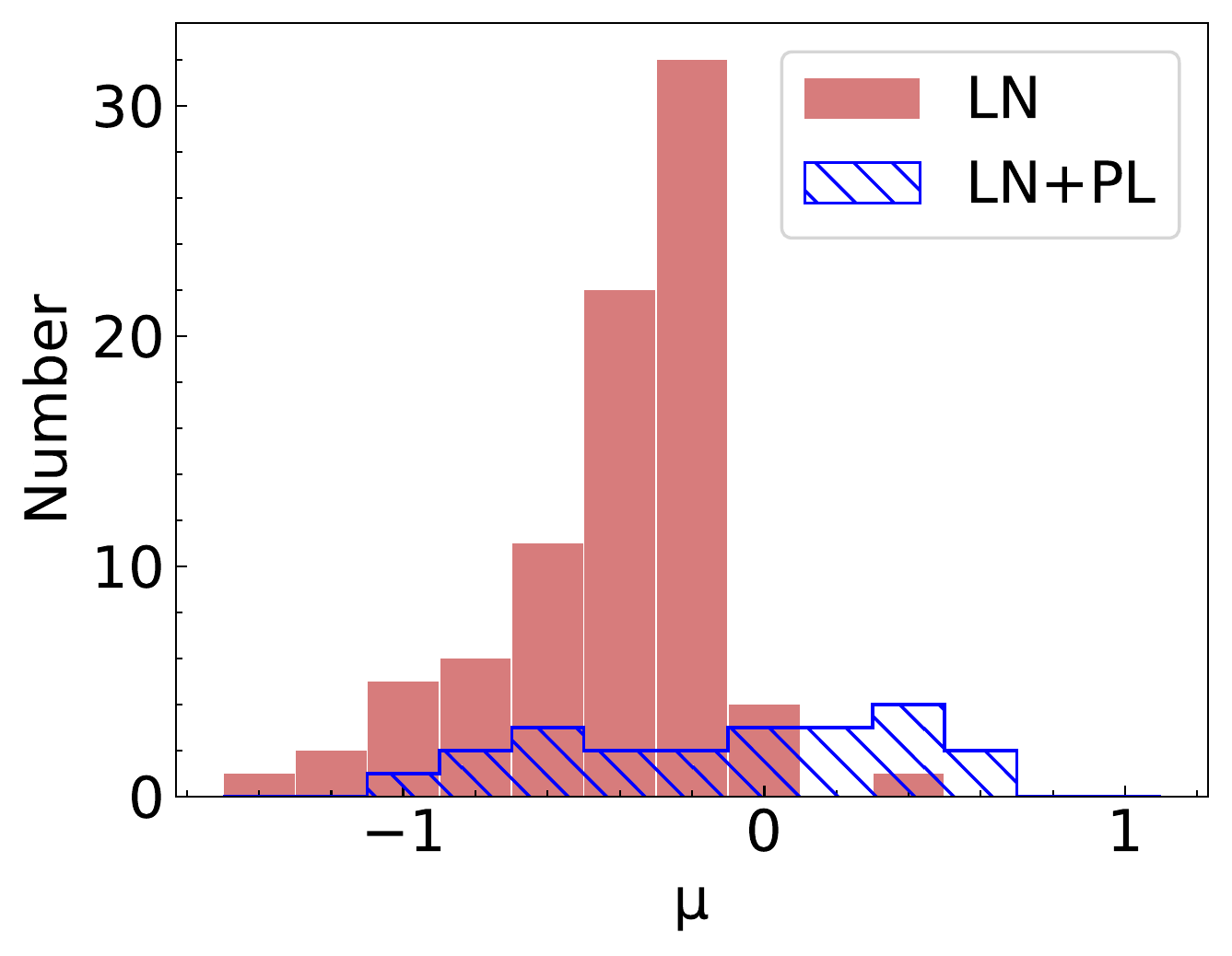}
		\\(a)
	\end{minipage}
	\begin{minipage}[t]{0.35\linewidth}
		\centering
		\includegraphics[width= \linewidth]{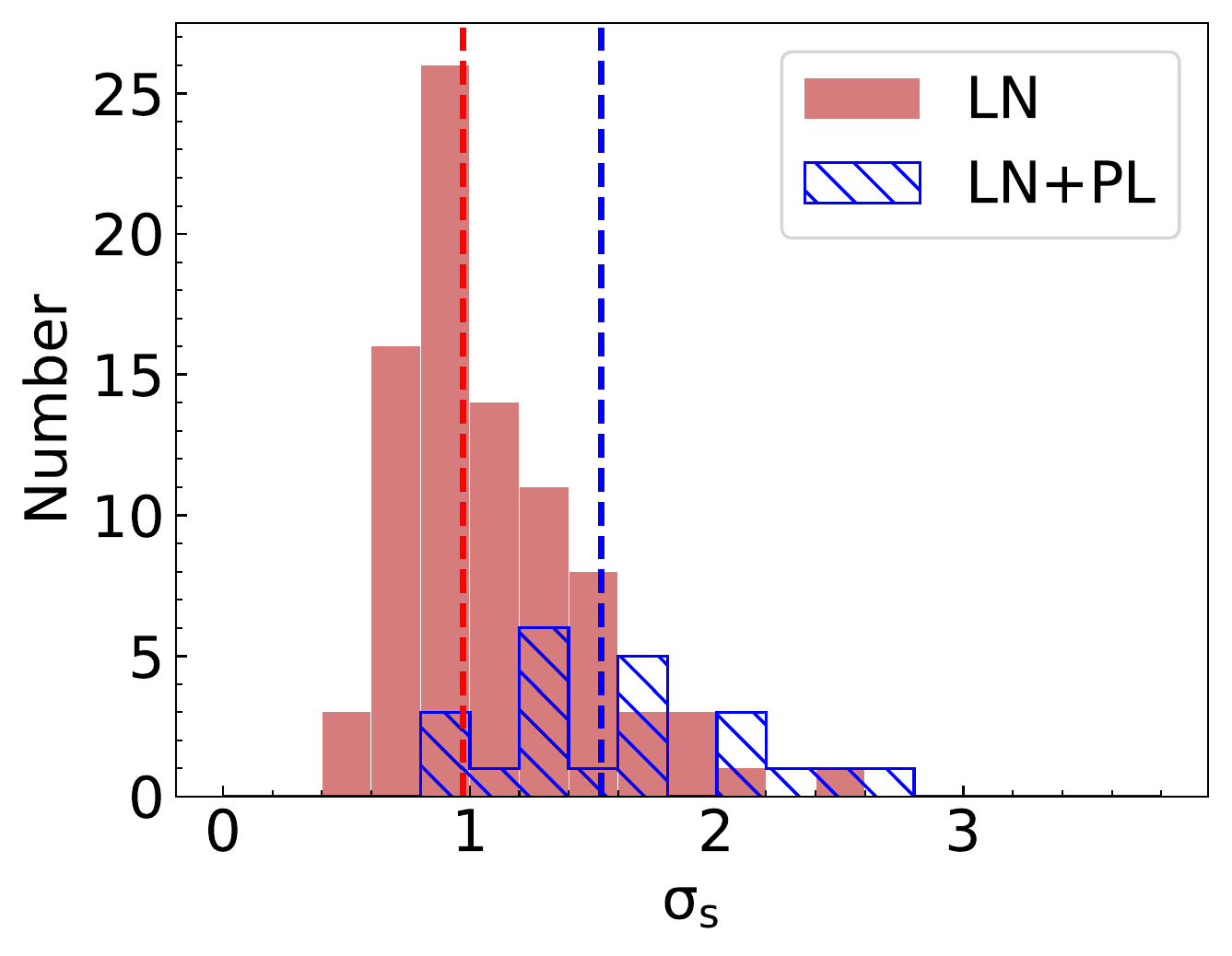}
		\\(b)
	\end{minipage}
	\begin{minipage}[t]{0.37\linewidth}
		\centering
		\includegraphics[width= \linewidth]{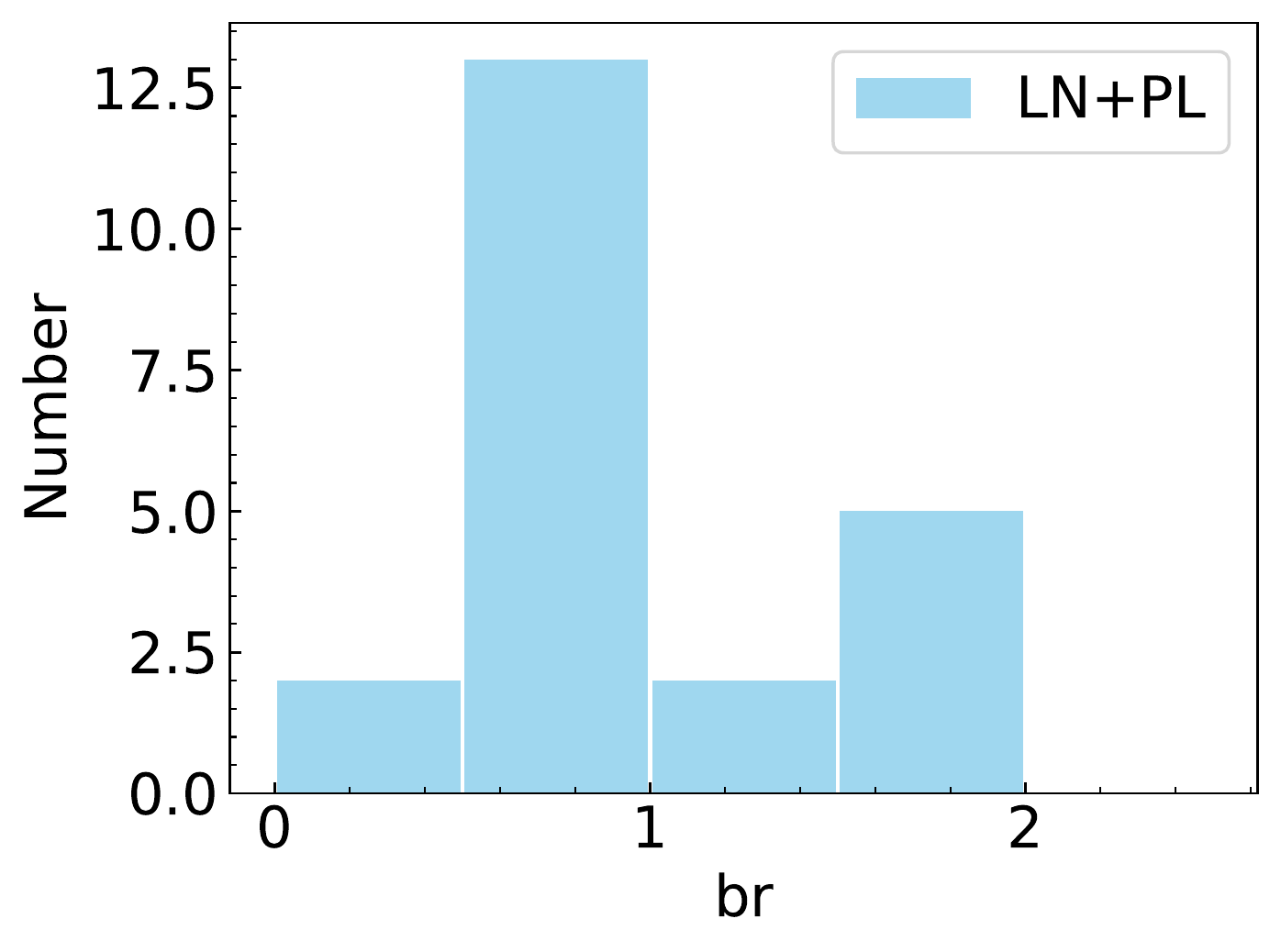}
		\\(c)
	\end{minipage}
	\begin{minipage}[t]{0.345\linewidth}
		\centering
		\includegraphics[width= \linewidth]{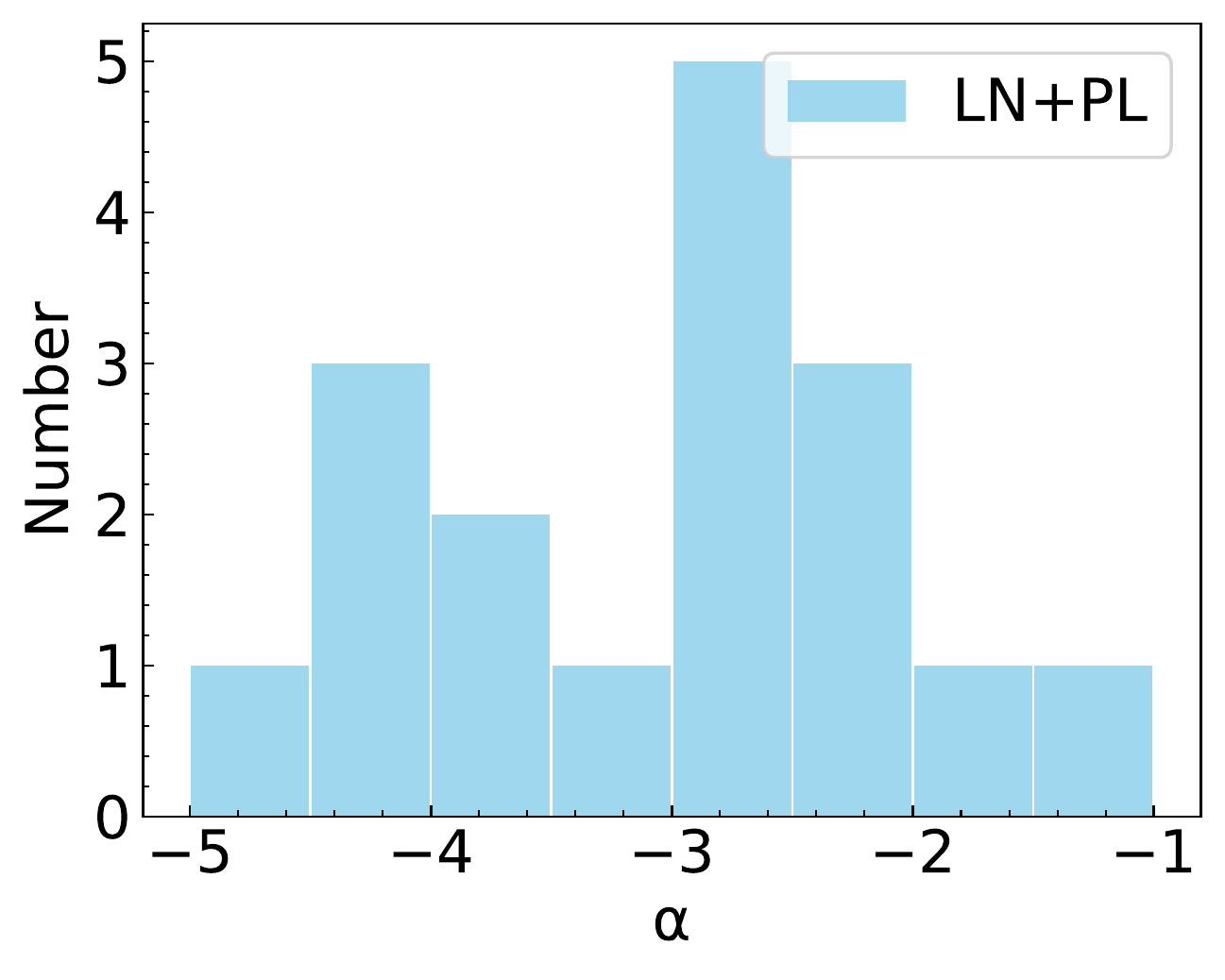}
		\\(d)
	\end{minipage}
	\caption{Distribution of (a) $\mu$, (b) $\sigma_s$, (c) $b_r$, and (d) $\alpha$, respectively, in Eqs. \ref{eq5}-\ref{eq6}. The red dashed line and the blue dashed line in panel (b) represent the median $\sigma_s$ for the LN and LN+PL categories, respectively.} 
	\label{fig7}
\end{figure*}  

Figure \ref{fig7} presents the histograms of the fitted parameters. Parameter $\mu$ of the LN fitted N-PDFs lies within [$-$1.1, 0] with a median value of $\sim-$0.2, while $\sigma_s$ of the LN fitted N-PDFs lies within [0.6, 2.2], with a median value around 1. However, the LN+PL fitted N-PDFs do not show typical values of $\mu$ and $\sigma_s$. Instead, the $\mu$ and $\sigma_s$ of the LN+PL fitted N-PDFs show relatively broad distributions within [$-1$, 0.6] and [0.8, 2.8], respectively. Although having a broad distribution, the median $\sigma_s$, $\sim 1.7$, of N-PDFs in the LN+PL category is greater than that of the LN category. In Figures \ref{fig7}(c) and \ref{fig7}(d), the distribution of $b_r$ is peaked around 0.8, while $\alpha$ is peaked around $-$4.2 and $-$2.8, although the statistics for $b_r$ and $\alpha$ is poor. It is generally thought that molecular gas with column density above $b_r$ is undergoing collapse. The transitional column densities designated by $b_r$ in our sample of clouds lie in the range from $\sim$8$\times10^{20}$ to 3$\times10^{21}$ cm$^{-2}$ without a characteristic value. 

\begin{figure*}[htb!]
	\centering
	\begin{minipage}[t]{0.32\linewidth}
		\centering
		\includegraphics[width= \linewidth]{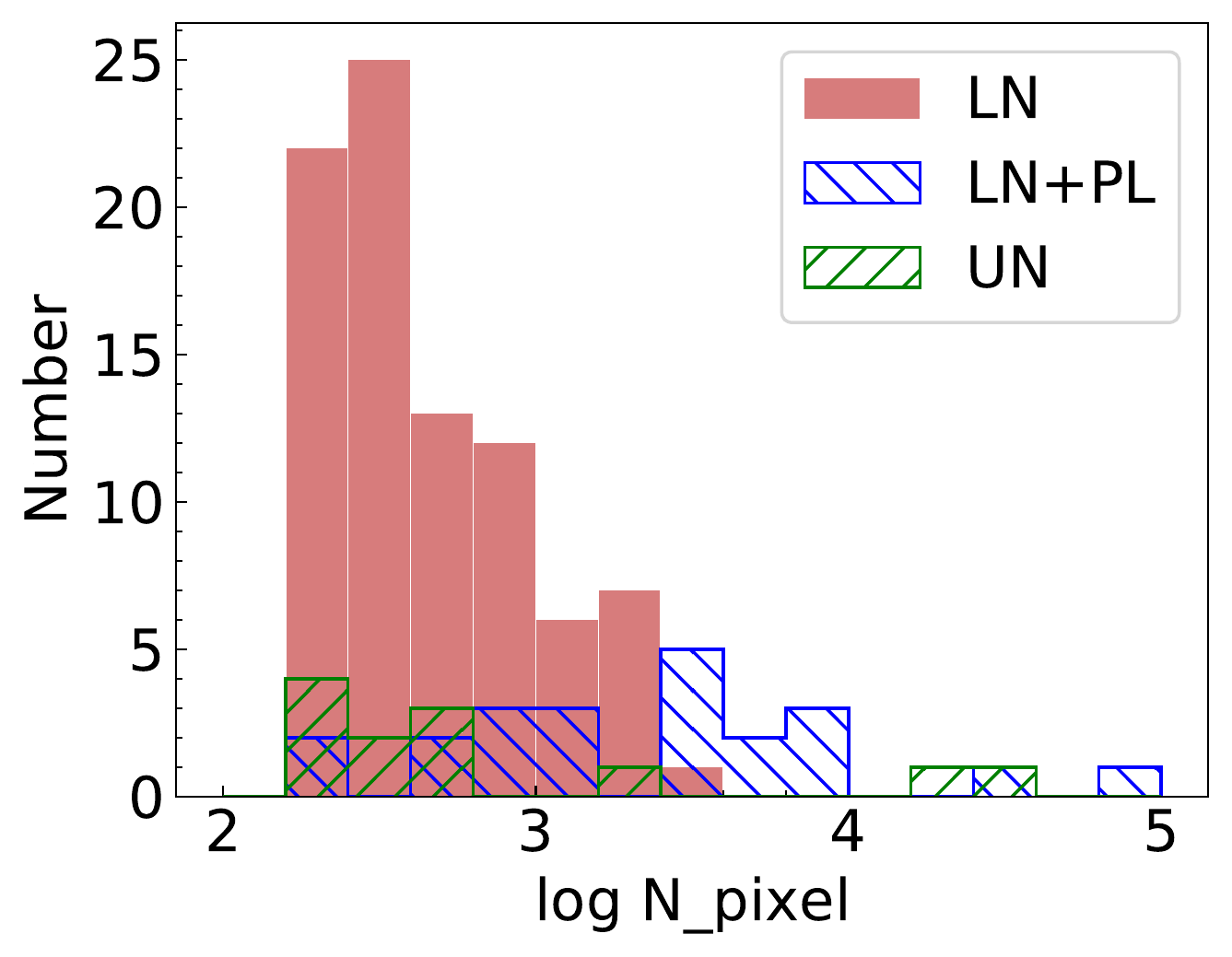}
		\\(a)
	\end{minipage}
	\begin{minipage}[t]{0.32\linewidth}
		\centering
		\includegraphics[width= \linewidth]{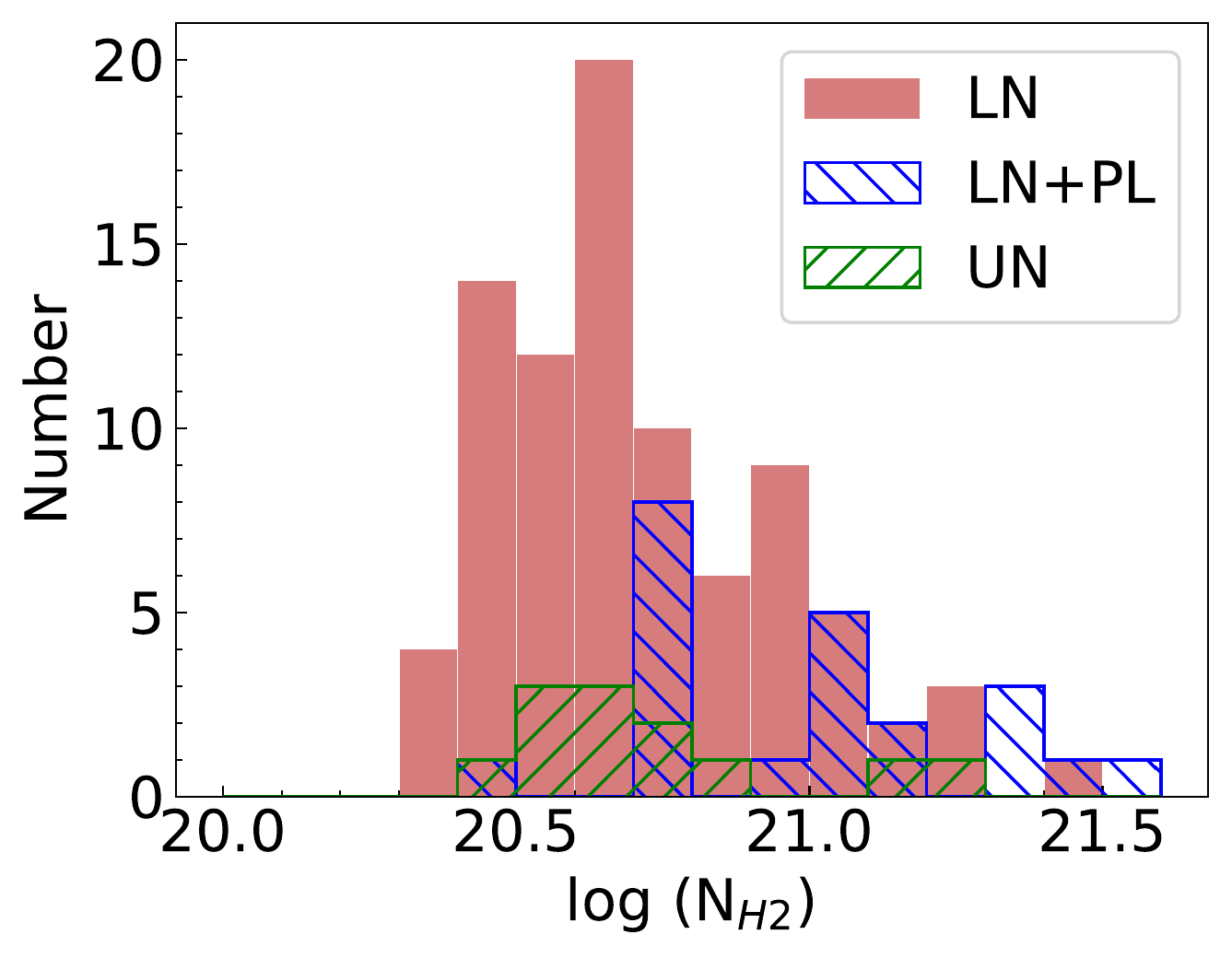}
		\\(b)
	\end{minipage}
	\begin{minipage}[t]{0.32\linewidth}
		\centering
		\includegraphics[width= \linewidth]{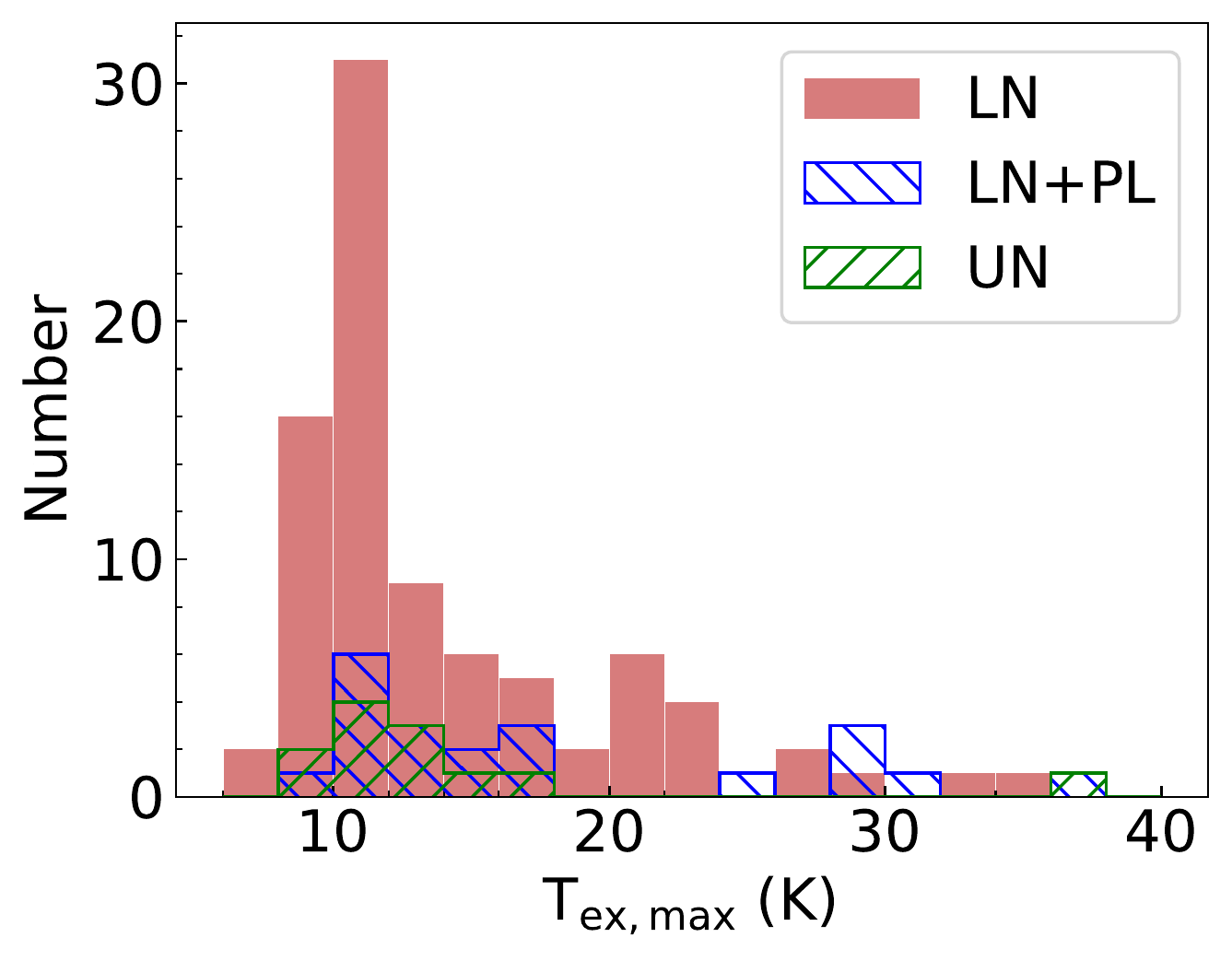}
		\\(c)
	\end{minipage}
	\caption{Histograms of the distance-independent physical parameters, i.e., (a) pixel number, (b) column density, and (c) excitation temperature of the selected clouds, respectively.} 
	\label{fig8}
\end{figure*}  

\begin{figure*}[htb!]
	\centering
	\begin{minipage}[t]{0.32\linewidth}
		\centering
		\includegraphics[width= \linewidth]{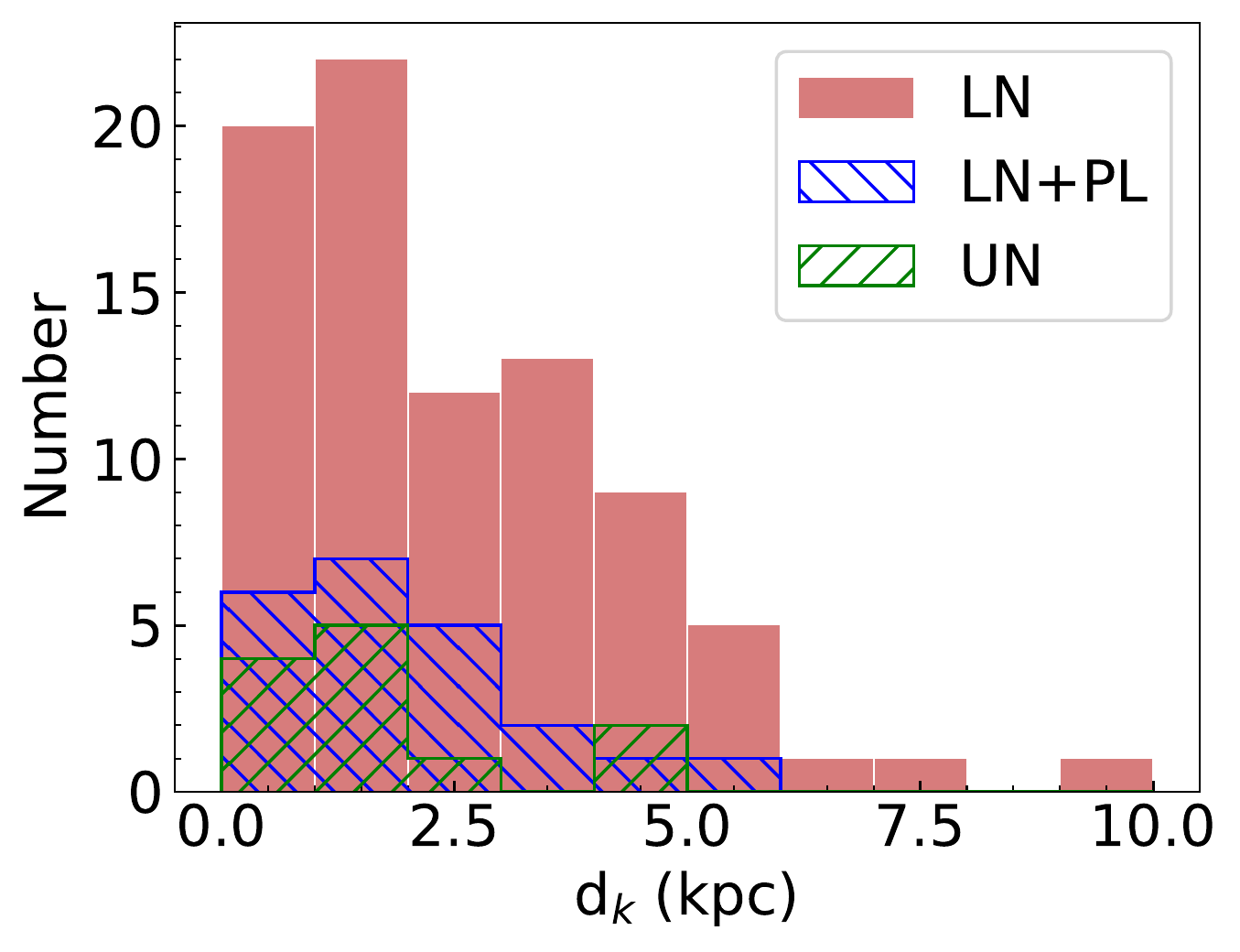}
		\\(a)
	\end{minipage}
	\begin{minipage}[t]{0.32\linewidth}
		\centering
		\includegraphics[width= \linewidth]{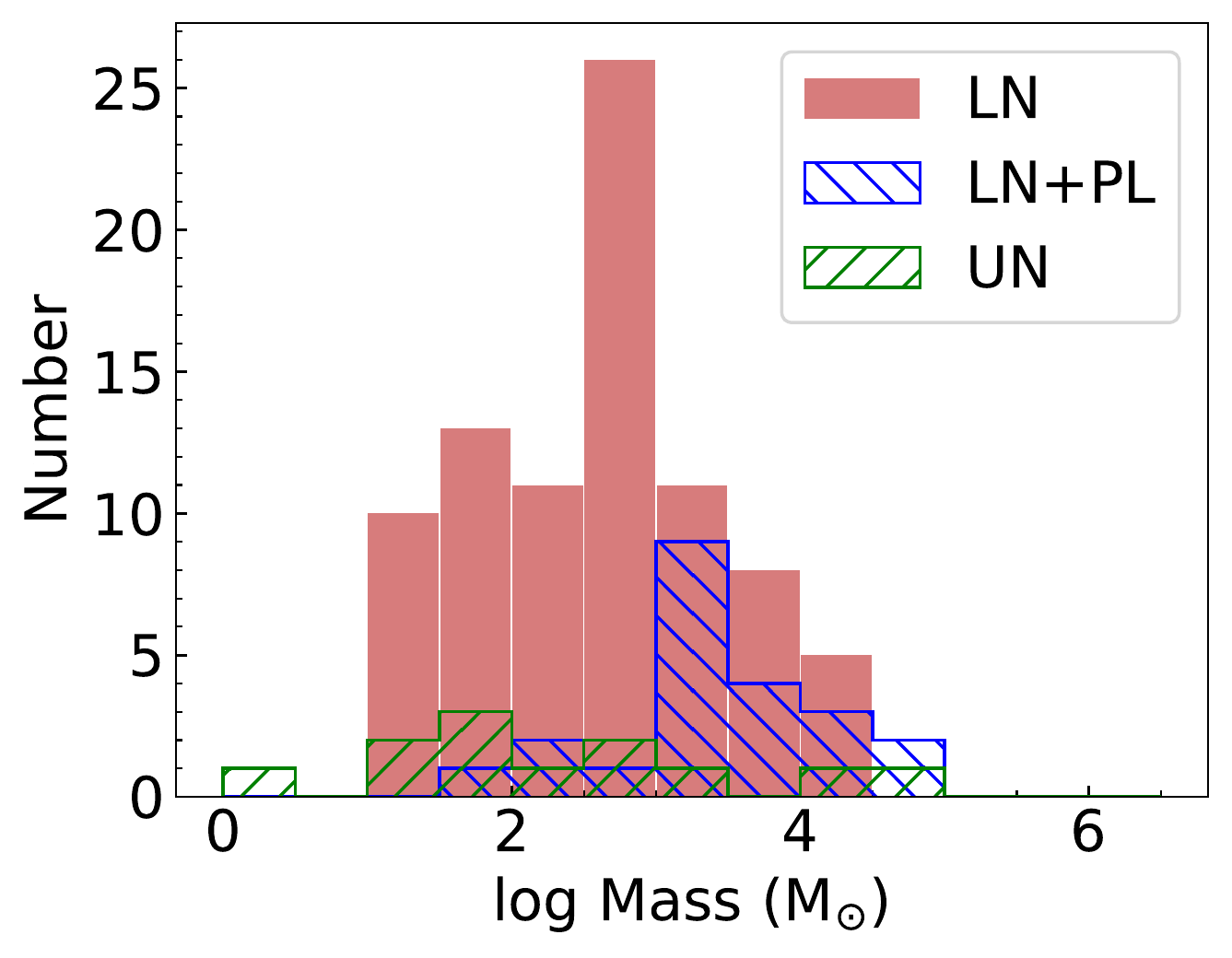}
		\\(b)
	\end{minipage}
	\begin{minipage}[t]{0.32\linewidth}
		\centering
		\includegraphics[width= \linewidth]{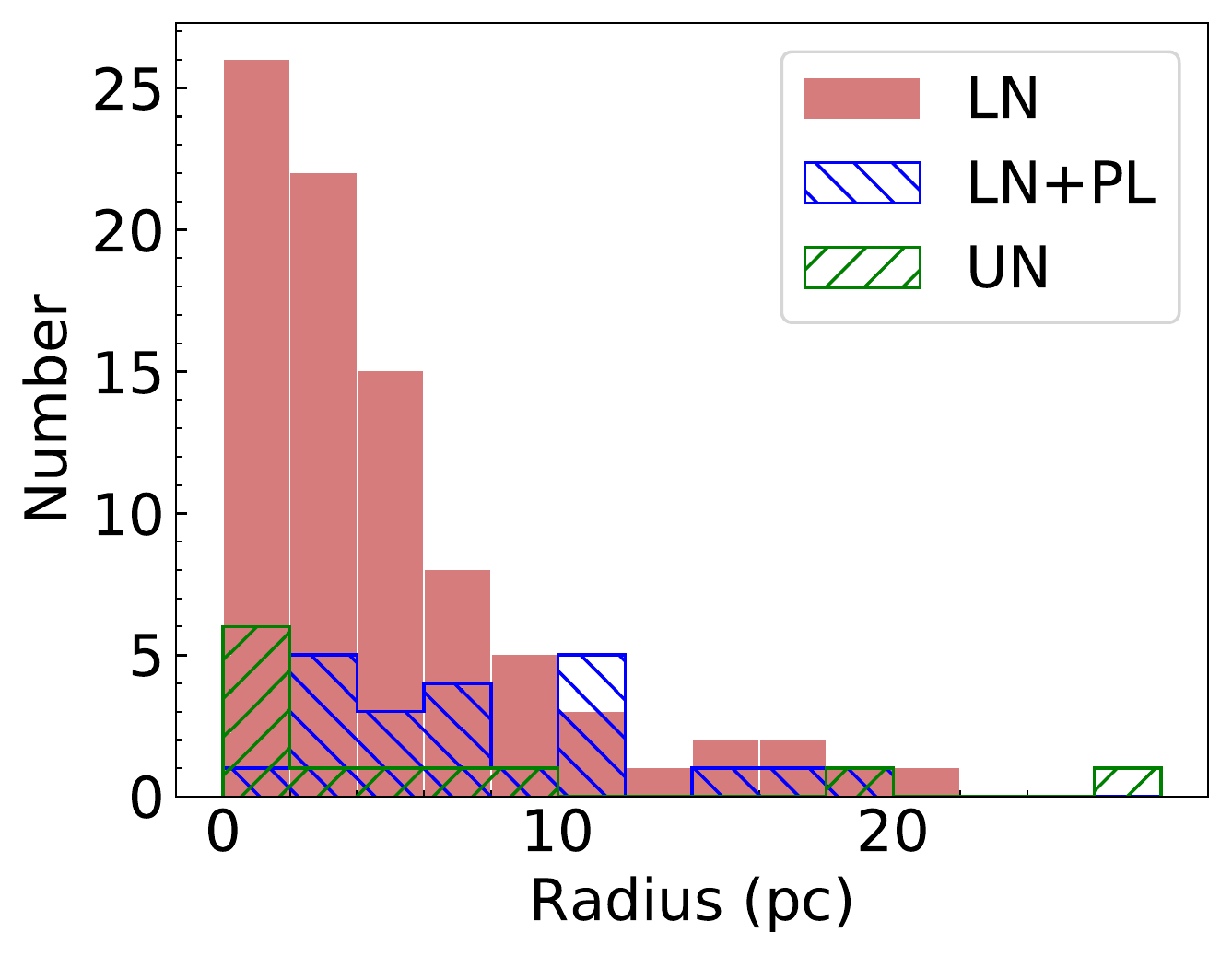}
		\\(c)
	\end{minipage}
	\caption{Histograms of the distance-dependent physical parameters, i.e., (a) kinematic distance, (b) mass, and (c) radius of the selected clouds, respectively.} 
	\label{fig9}
\end{figure*}  

The histograms of the physical parameters independent on distance are shown in Figure \ref{fig8}. In Figure \ref{fig8}, most of the clouds in the UN category have pixel numbers less than $\sim$500. The median pixel numbers of the clouds in the LN catagory is 365. The mean column densities of the molecular clouds in the LN category are concentrated around 5$\times 10^{20}$ cm$^{-2}$, and the mean column densities in the UN category are similar to those in the LN category. Compared to the UN and LN categories, the molecular clouds in the LN+PL category generally have more pixels and higher mean column densities. The clouds in the three categories have similar peak excitation temperatures distributed around $\sim$10 K. Histograms of the kinematic distances and distance-dependent physical parameters of the clouds are shown in Figure \ref{fig9}. The majority of the molecular clouds in the UN category are nearby clouds within 2 kpc, and are small molecular clouds with radius less than $\sim$2 pc. This situation is reasonable because, with a certain sensitivity, we tend to obtain smaller molecular clouds at closer distances. The molecular clouds with LN+PL-shaped N-PDFs are relatively more massive than those in the LN and UN categories, whereas they do not exhibit any preferential spatial scales, i.e., their radii span a wide range from sub-pc to $\sim$20 pc.

\begin{figure*}[htb!]
	\centering
	\begin{minipage}[t]{0.325\linewidth}
		\centering
		\includegraphics[width= \linewidth]{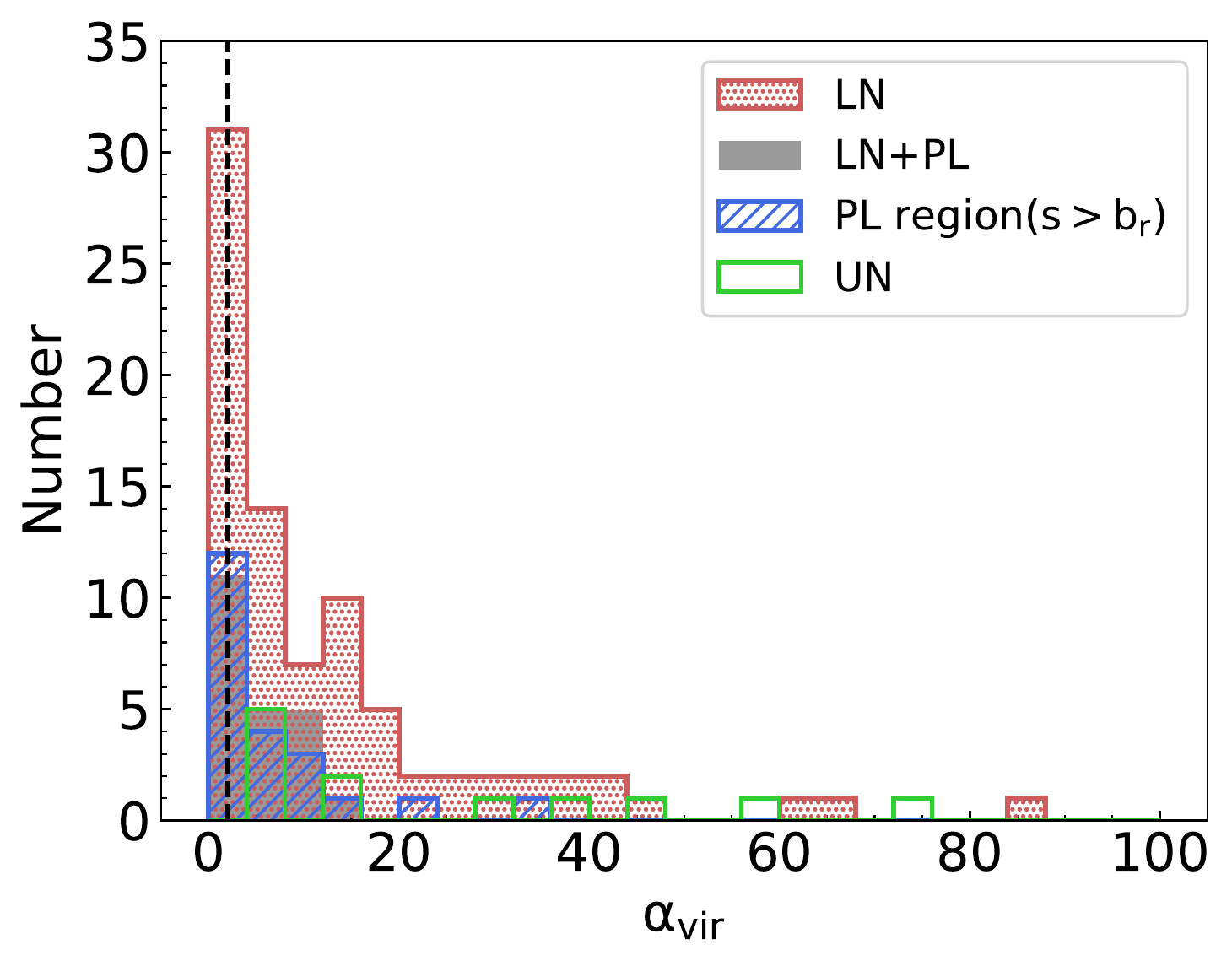}
		\\(a)
	\end{minipage}
	\begin{minipage}[t]{0.33\linewidth}
		\centering
		\includegraphics[width= \linewidth]{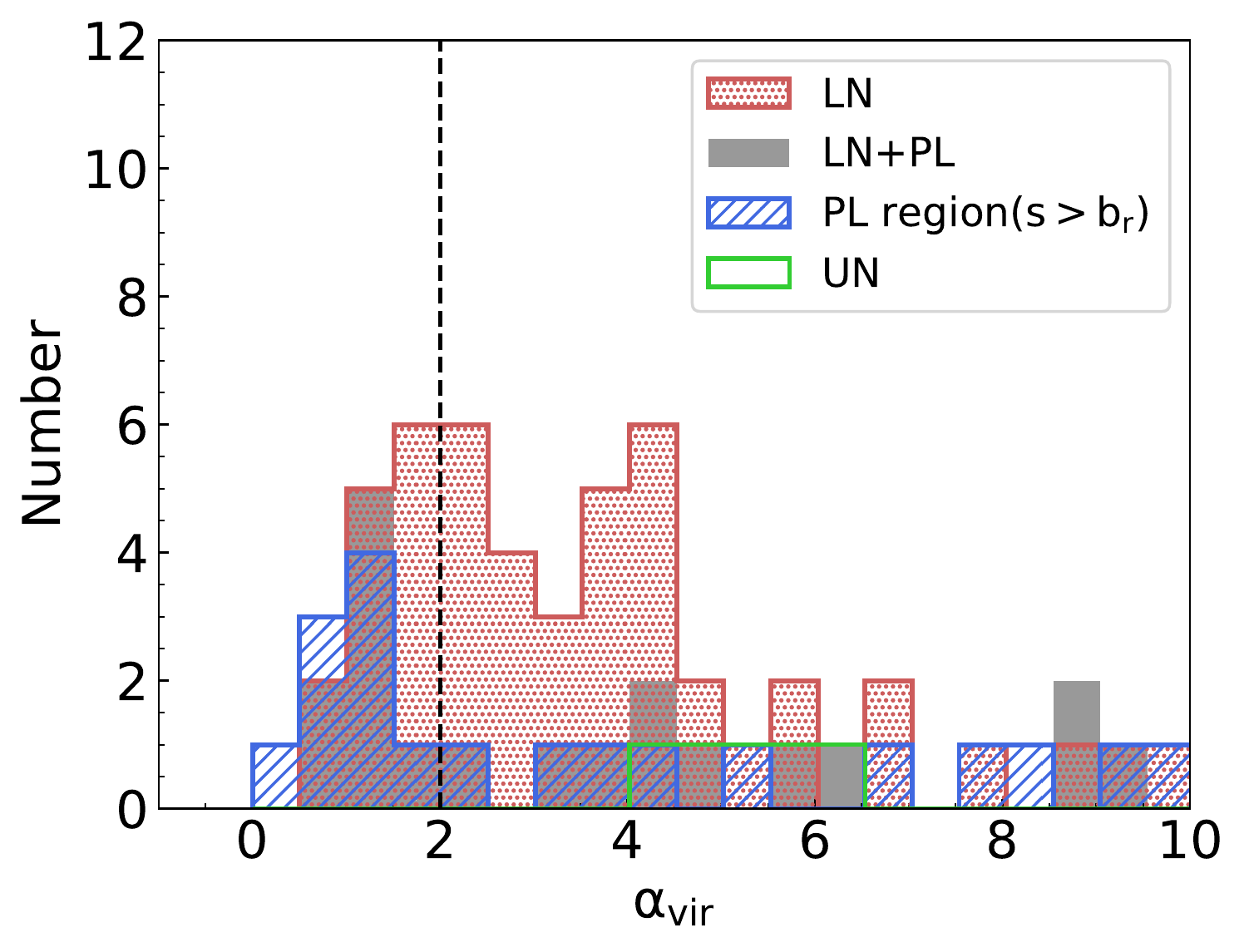}
		\\(b)
	\end{minipage}
	\begin{minipage}[t]{0.315\linewidth}
		\centering
		\includegraphics[width= \linewidth]{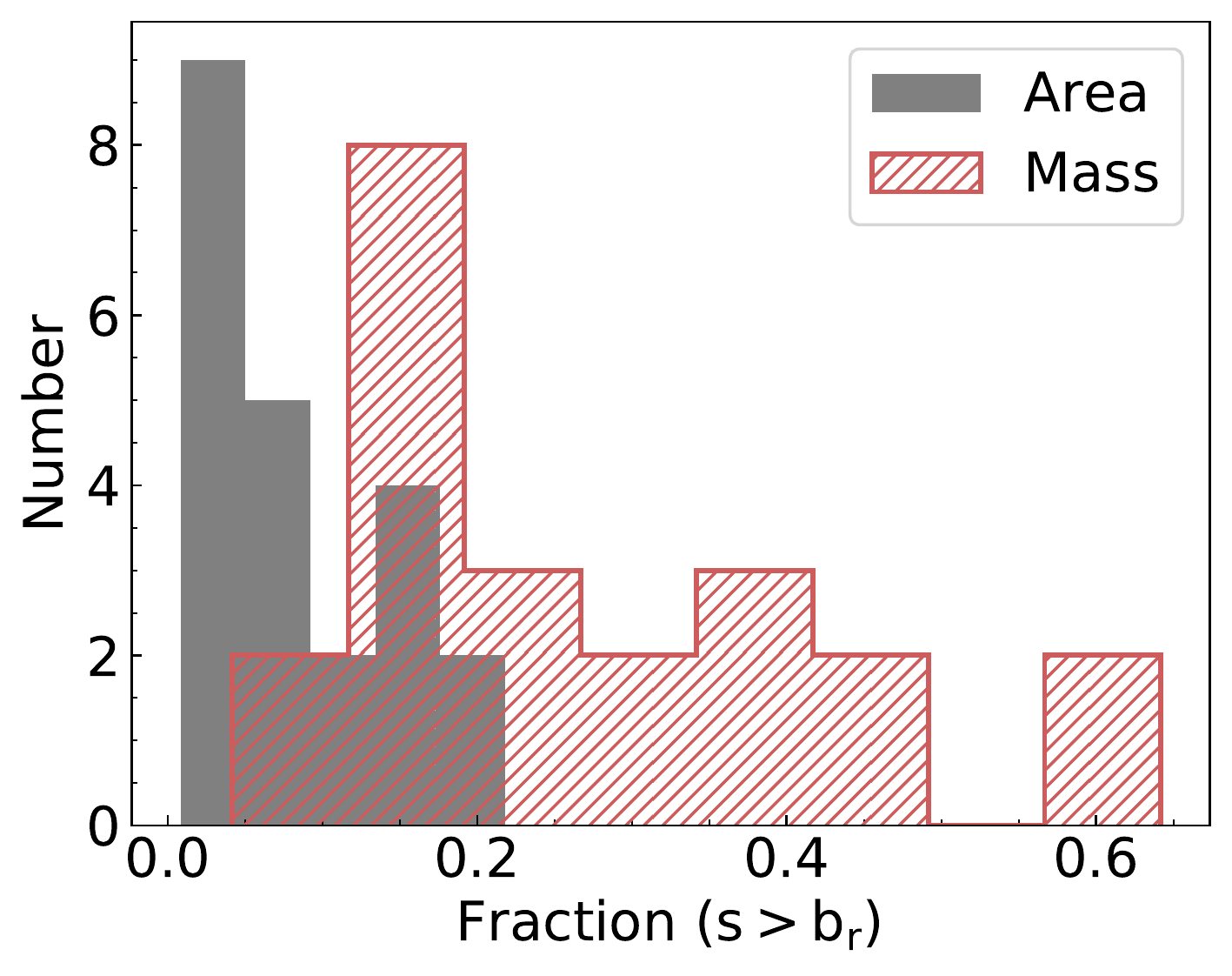}
		\\(c)
	\end{minipage}
	\caption{(a) Histograms of the virial parameters of the clouds and the sub-regions of molecular clouds with column densities in the PL part of the N-PDFs. The vertical dashed line represents $\alpha_{vir}=2$. (b) A zoom-in histogram of the virial parameters within $0<\alpha_{vir}<10$ with bin widths of 0.5. (c) Histograms of the area and mass fraction of the molecular gas of column density in the PL part of the N-PDFs.} 
	\label{fig10}
\end{figure*}  

Figure \ref{fig10}(a) presents the histograms of the virial parameters of the clouds and the regions corresponding to the PL part of the N-PDFs in the LN+PL category. In this figure, the molecular clouds with PL tails are not necessarily gravitationally bound, i.e., with $\alpha_{vir}<2$. The situation is the same when only considering the molecular gas contributing to the PL part of the N-PDFs. The molecular clouds in the LN category generally have larger virial parameters than those in the LN+PL category, and some of them are gravitationally bound. However, when focusing on a narrower range of $\alpha_{vir}$ around $\alpha_{vir} = 2$, for example, $0<\alpha_{vir}<5$, the fractions of self-gravitating molecular clouds in the LN+PL category (57$\%$) and in the PL part sub-regions (69$\%$) are much higher than that of the LN category (33$\%$), as seen in Figure \ref{fig10}(b). All the clouds in the UN category have virial parameters higher than 4, which means they may need external forces to stay in equilibrium. We also calculated the area fraction and the mass fraction of the molecular gas corresponding to the PL tails of the N-PDFs to the whole cloud, as shown in Figure \ref{fig10}(c). Although the molecular gas in the PL part of the N-PDF only accounts for  $\sim 3\%-20\%$ of the area of a molecular cloud, it contains $\sim 10\%-60\%$ of the mass of the molecular cloud.

\subsubsection{Scaling Relations Between Width of N-PDFs and Physical Parameters of the Clouds} \label{sec3.2.3}
As suggested by numerical simulations, the N-PDF is evolved during the formation and evolution of the molecular cloud \citep{Ballesteros-Paredes2011, Federrath2013, Ward2014}. The widths, $\sigma_s$, of the N-PDFs can be considered a measure of the complexity of the internal structure of molecular clouds. Numerical simulations suggest that $\sigma_s$ is related to the sonic Mach number of turbulence. However, the relations between $\sigma_s$ and other physical properties of the molecular clouds are unknown. In this section, we use the observed results to investigate the relationships between $\sigma_s$ and other physical parameters of the molecular clouds.

\begin{figure*}[htb!]
	\centering
	\begin{minipage}[t]{0.45\linewidth}
		\centering
		\includegraphics[width= \linewidth]{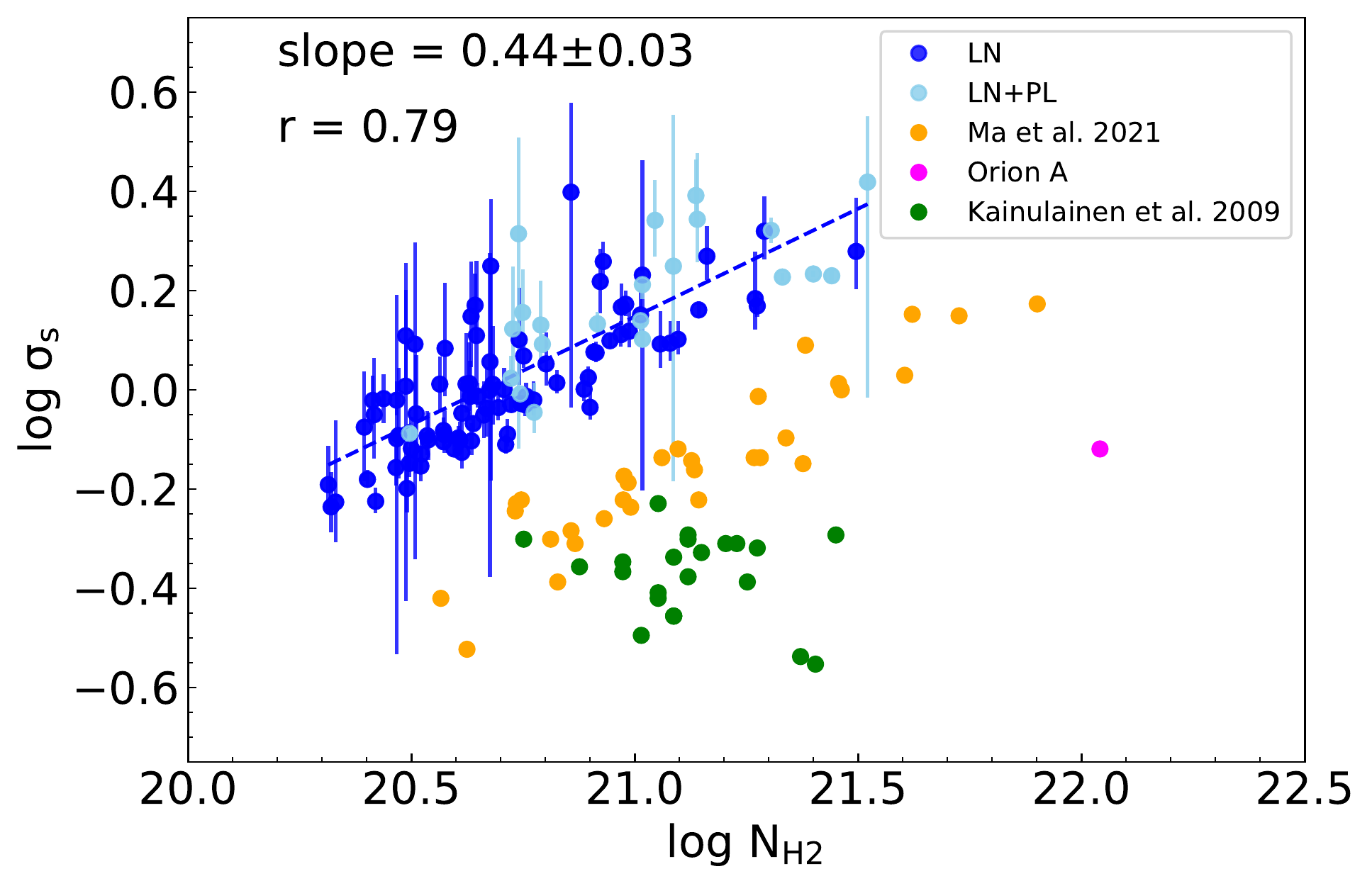}
		\\(a)
	\end{minipage}
	\begin{minipage}[t]{0.44\linewidth}
		\centering
		\includegraphics[width= \linewidth]{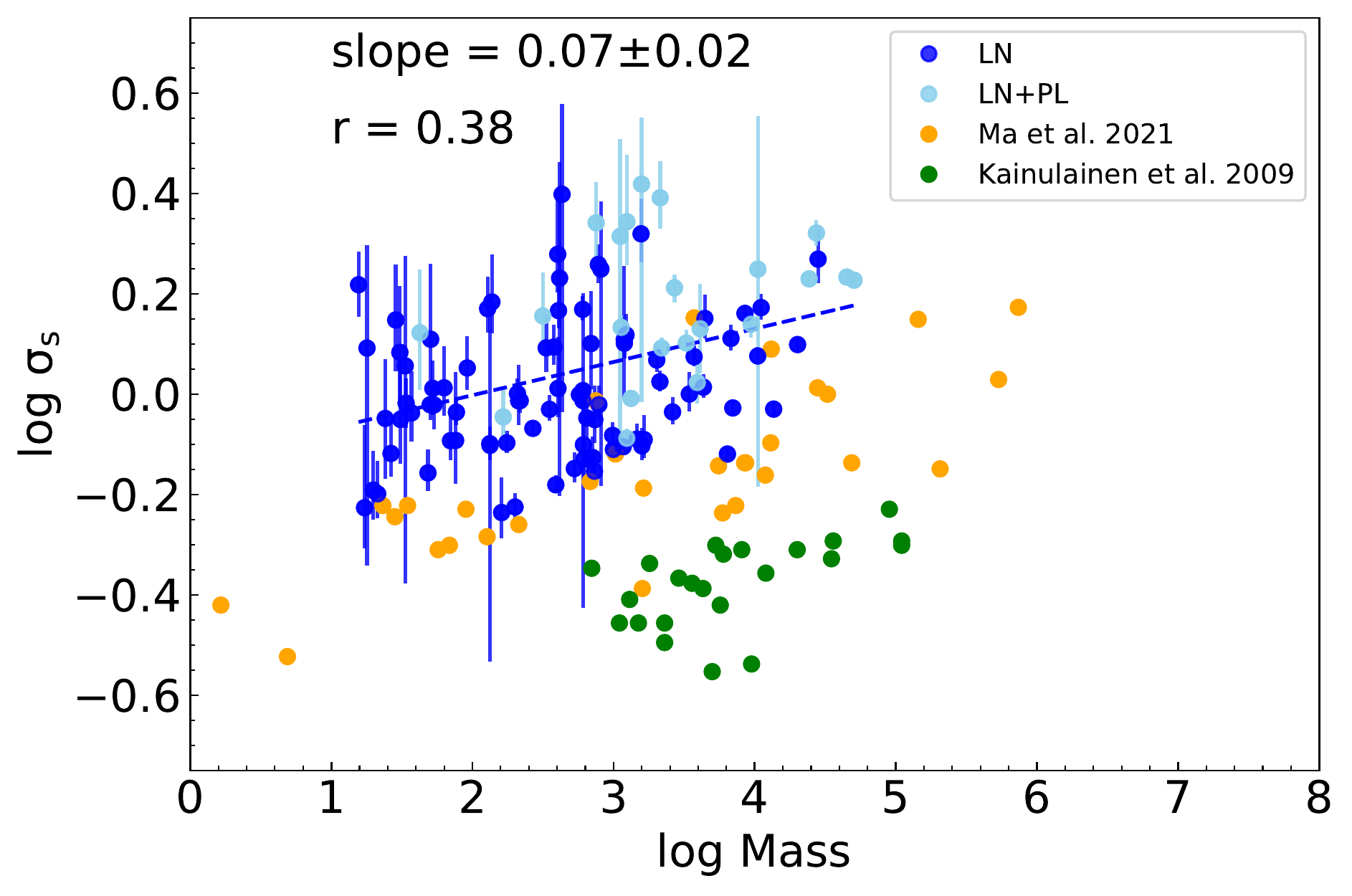}
		\\(b)
	\end{minipage}
	\caption{Variations of the widths of the N-PDFs with (a) the average column densities and (b) mass of the molecular clouds. The dashed lines are the fitted scaling relations between $\log \sigma_s$ and $\log N_{\rm H_2}$ in panel (a), and $\log \sigma_s$ and $\log M$ in panel (b). The blue, light blue, orange, green, and magenta colors correspond to the results obtained in this work, \citetalias{Kainulainen2009}, \cite{Ma2021}, \cite{Ma2020}, respectively. The average column density for each cloud in this work is calculated above the median detection limit of the cloud.} 
	\label{fig11}
\end{figure*}  

\begin{figure*}[htb!]
	\centering
	\begin{minipage}[t]{0.30\linewidth}
		\centering
		\includegraphics[width= \linewidth]{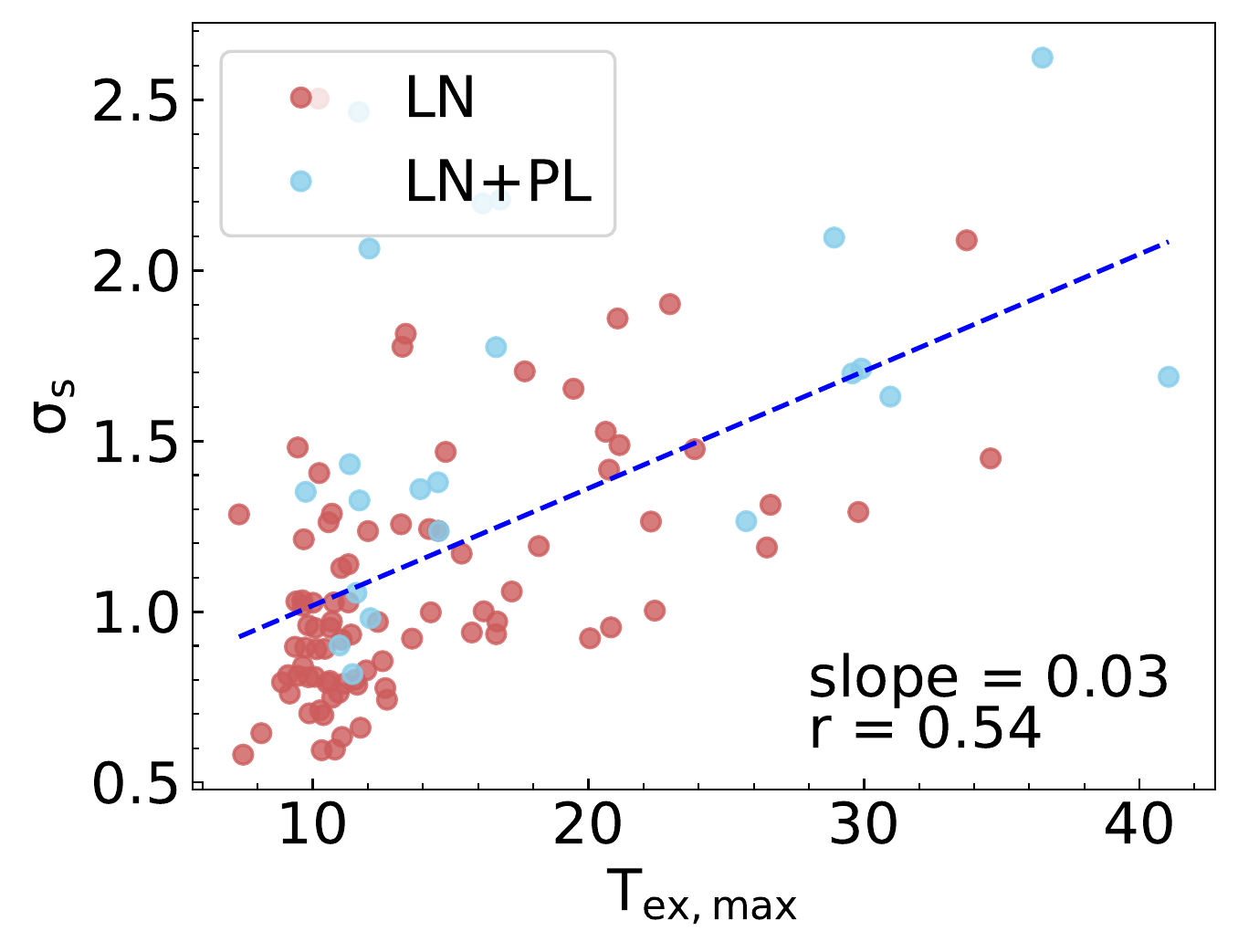}
		\\(a)
	\end{minipage}
	\begin{minipage}[t]{0.31\linewidth}
		\centering
		\includegraphics[width= \linewidth]{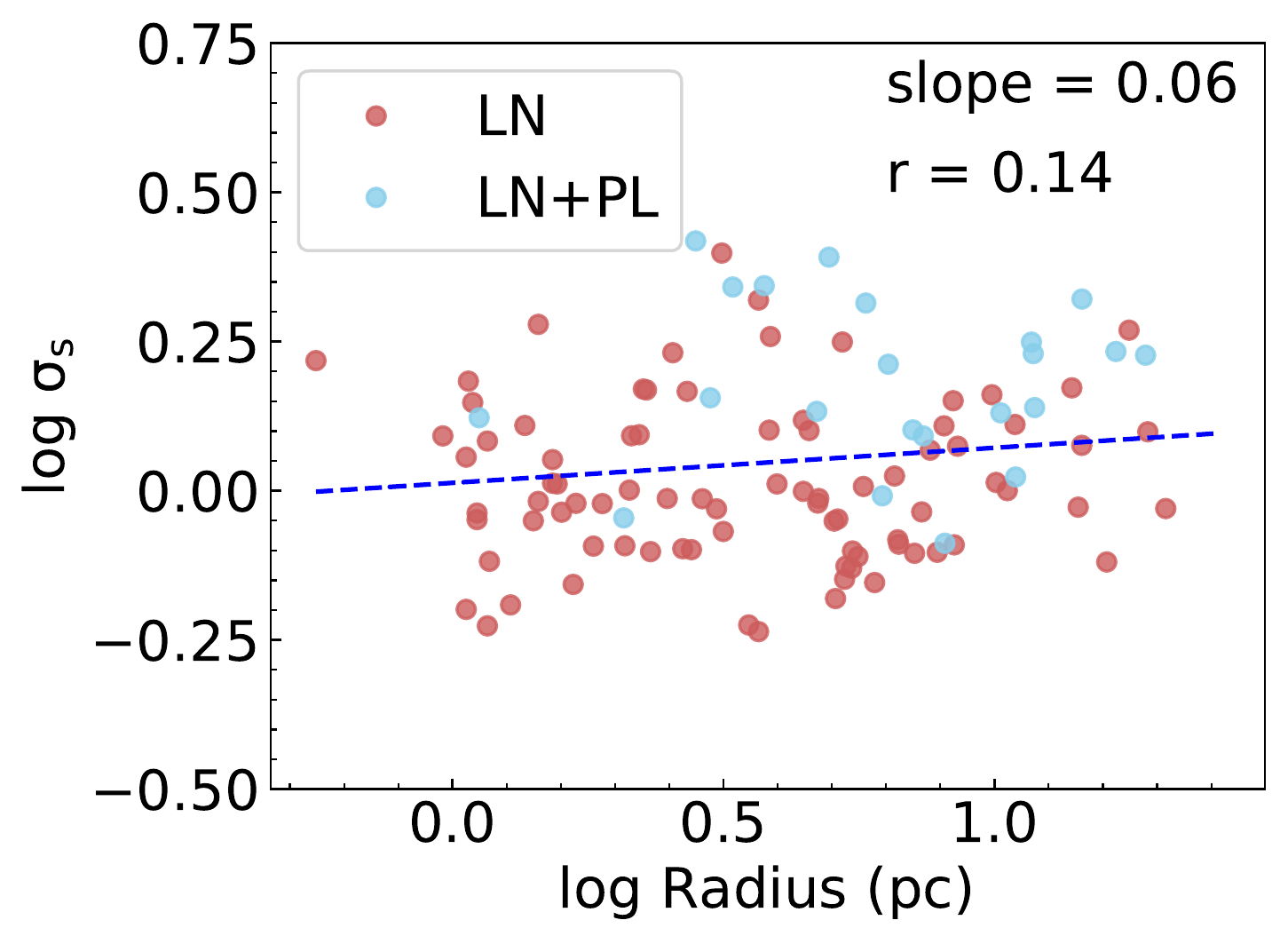}
		\\(b)
	\end{minipage}
	\begin{minipage}[t]{0.33\linewidth}
		\centering
		\includegraphics[width= \linewidth]{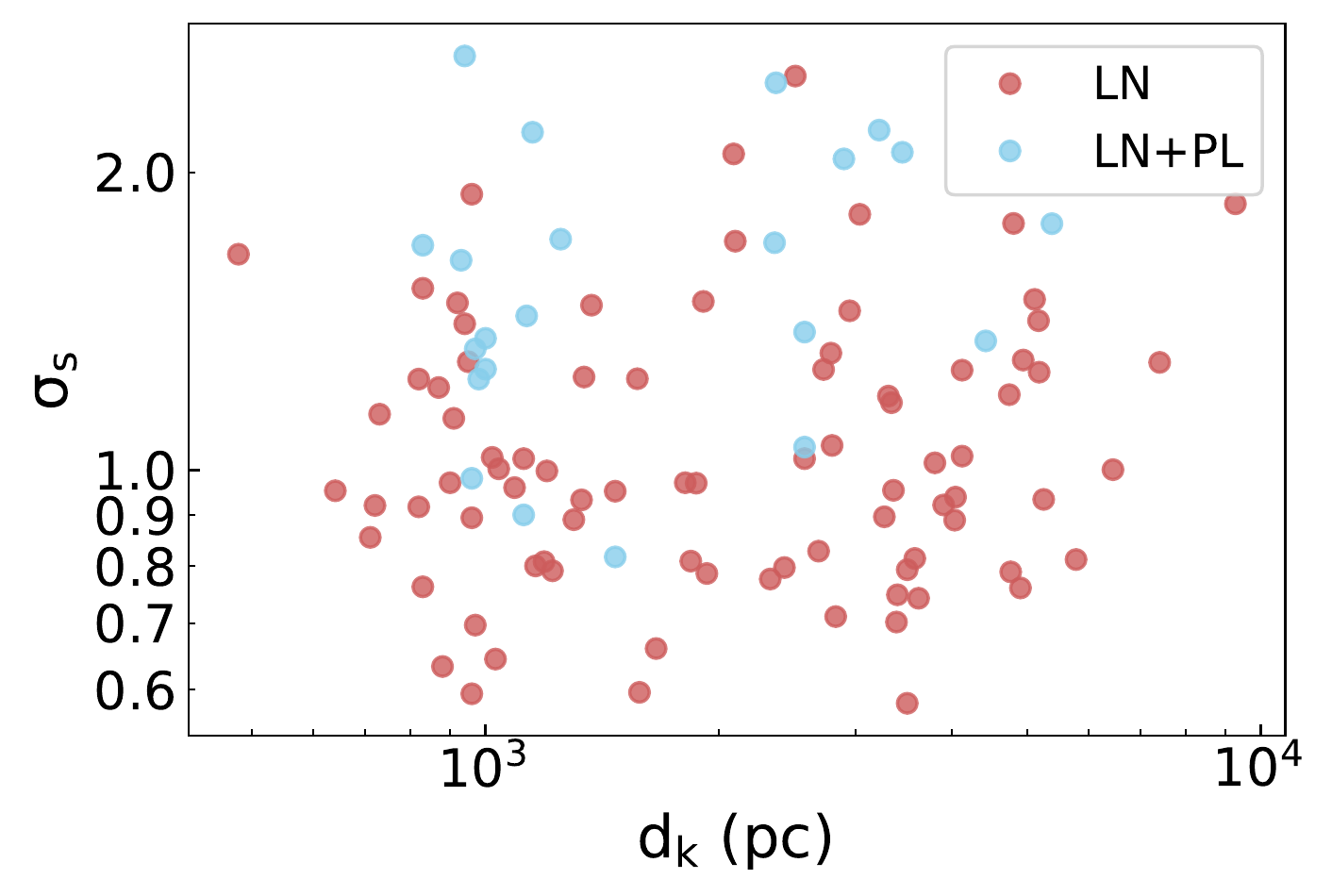}
		\\(c)
	\end{minipage}
	\caption{Variation of the width of the N-PDF with (a) peak excitation temperature, (b) radius, (c) kinematic distance of the clouds. The red color and blue color correspond to the clouds in the LN and LN+PL categories, respectively.} 
	\label{fig12}
\end{figure*}  

Figure \ref{fig11}(a) shows the variation of the widths, $\sigma_s$, of N-PDFs with the average column densities, $\left\langle N_{H_2}\right\rangle $, of the clouds in a log-log space. The $\log\ \sigma_s$ increases linearly with $\log\ \left\langle N_{H_2}\right\rangle $ with a slope of 0.44, indicating a power-law scaling relation $\sigma_s\propto \left\langle N_{\rm H_2}\right\rangle ^{0.44}$. The Pearson correlation coefficient between $\log\ \sigma_s$ and $\log \left\langle N_{\rm H_2}\right\rangle $ is 0.79. For comaprison, we also overlaid the $\log\ \sigma_s-\log \left\langle N_{\rm H_{2}}\right\rangle $ data pairs obtained by \citetalias{Kainulainen2009}, and those obtained by \cite{Ma2020, Ma2021}. The average extinctions of the clouds in \citetalias{Kainulainen2009} are converted to average column densities according to $N_{\rm H_2} = 0.94\times10^{21}Av$ \citep{Bohlin1978}. In Figure \ref{fig11}(a), the results from \cite{Ma2021} are systematically shifted toward smaller $\sigma_s$ and higher $\left\langle N_{\rm H_2}\right\rangle$. However, the scaling relationship between $\sigma_s$ and $\left\langle N_{H_2}\right\rangle$ still exists, and the slope of the scaling resembles what we obtained in this work. The shift of the $\sigma_s$-$\left\langle N_{H_2}\right\rangle$ relation between \citet{Ma2021} and this work may partly result from different criteria of detection threshold of \COl emission used in the two works. The boundaries of the \COl clouds in this work are determined in the PPV space using the DBSCAN algorithm, which means the clouds are defined by 3D contours of $T_{mb}= 2\sigma_{RMS}$. Instead, the boundaries of the \COl clouds in \citep{Ma2021} are determined within the associated \CO clouds using an empirical criterion, which results in a detection threshold of \COl intensity of $\sim$1.5 times higher than the threshold of this work (see Section \ref{sec3.1}). Therefore, the column density maps in this work contain more pixels with low column densities, which lowers the mean column densities and increases the widths of the N-PDFs when compared to \cite{Ma2021}. 

We already noticed in our previous work \citep{Ma2021} that the $\sigma_s$ obtained by \citetalias{Kainulainen2009} are smaller than the values obtained using the \COl data. In Figure \ref{fig11}(a) the $\sigma_s$ values obtained by \citetalias{Kainulainen2009} are distributed within [0.3, 0.7], and are systematically smaller than those of this work. Besides, we do not see any increasing trend of $\sigma_s$ with $\log \left\langle N_{\rm H_2}\right\rangle$. However, \citetalias{Kainulainen2009} fitted their N-PDFs with LN functions in a relatively narrow range of s, [-0.5, 1], around the peaks of the N-PDFs. Power-law-like tails at high densities are common in the N-PDFs of the active star-forming regions in \citetalias{Kainulainen2009}, which means the underlying dispersions of the column densities of the clouds should be larger than the fitted $\sigma_s$ parameters in their work. The magenta point of Orion A GMC is obtained by \cite{Ma2020}, using the \COl data from the pilot survey of the MWISP project. As the nearest high-mass star-forming region, the average column density of the Orion A GMC is much higher than the nearby low-mass star-forming regions in \citetalias{Kainulainen2009} and most of the molecular clouds in the second quadrant in \cite{Ma2021} and in the third quardrant in this work. 

Figure \ref{fig11}(b) shows the relation between $\sigma_s$ and the mass of the clouds. Different from the $\sigma_s-\left\langle N_{\rm H_2}\right\rangle$ relation, the exponent of the fitted $\sigma_s-M_{LTE}$ power-law relation is 0.07, and the correlation coefficient between $\log \sigma_s$ and $\log M_{LTE}$ is small (0.38), indicating $\sigma_s$ is nearly independent of the mass of molecular clouds. The same conclusion also can be drawn from the data pairs from \cite{Ma2021} and \citetalias{Kainulainen2009}. 

Figure \ref{fig12} presents the relation between $\sigma_s$ and (a) the peak excitation temperature, $T_{\rm ex, max}$, (b) radius, and (c) kinematic distance of the molecular clouds. In Figure \ref{fig12}(a), we found a weak correlation between $\sigma_s$ and $\rm T_{ex, max}$ of the clouds. The slope of the linear relation between $\sigma_s$ and $\rm T_{ex, max}$ is 0.03, and the correlation coefficient is 0.54. Generally, when there are no additional heating sources other than the interstellar radiation field, such as feedback from newborn stars, \ion{H}{2} regions, and supernova remnants, the temperature of the molecular cloud is $\sim$10 K. Therefore, the peak excitation temperature of the molecular clouds is partly related to star formation activities. The correlation between $\sigma_s$ and $T_{\rm ex, max}$ is consistent with the fact that high-density gas regions (corresponding to broad N-PDFs) in molecular clouds harbor star-formation. In Figures \ref{fig12}(b) and \ref{fig12}(c), we do not find any correlations between $\sigma_s$ and the radius or kinematic distance of the clouds, which indicates the hierarchy of the internal structures of molecular clouds is not evident at least on the spatial scales from $\sim$1 to $\sim$10 pc, and the complexity of the structure of molecular clouds does not vary with Galactic environments, i.e., Galactocentric distances or locations on/off spiral-arms. 

\section{Discussion} \label{sec4} 
\subsection{Other Relations between the Physical Parameters of the Molecular Clouds}

\begin{figure*}[htb!]
	\centering
	\begin{minipage}[t]{0.33\linewidth}
		\centering
		\includegraphics[width= \linewidth]{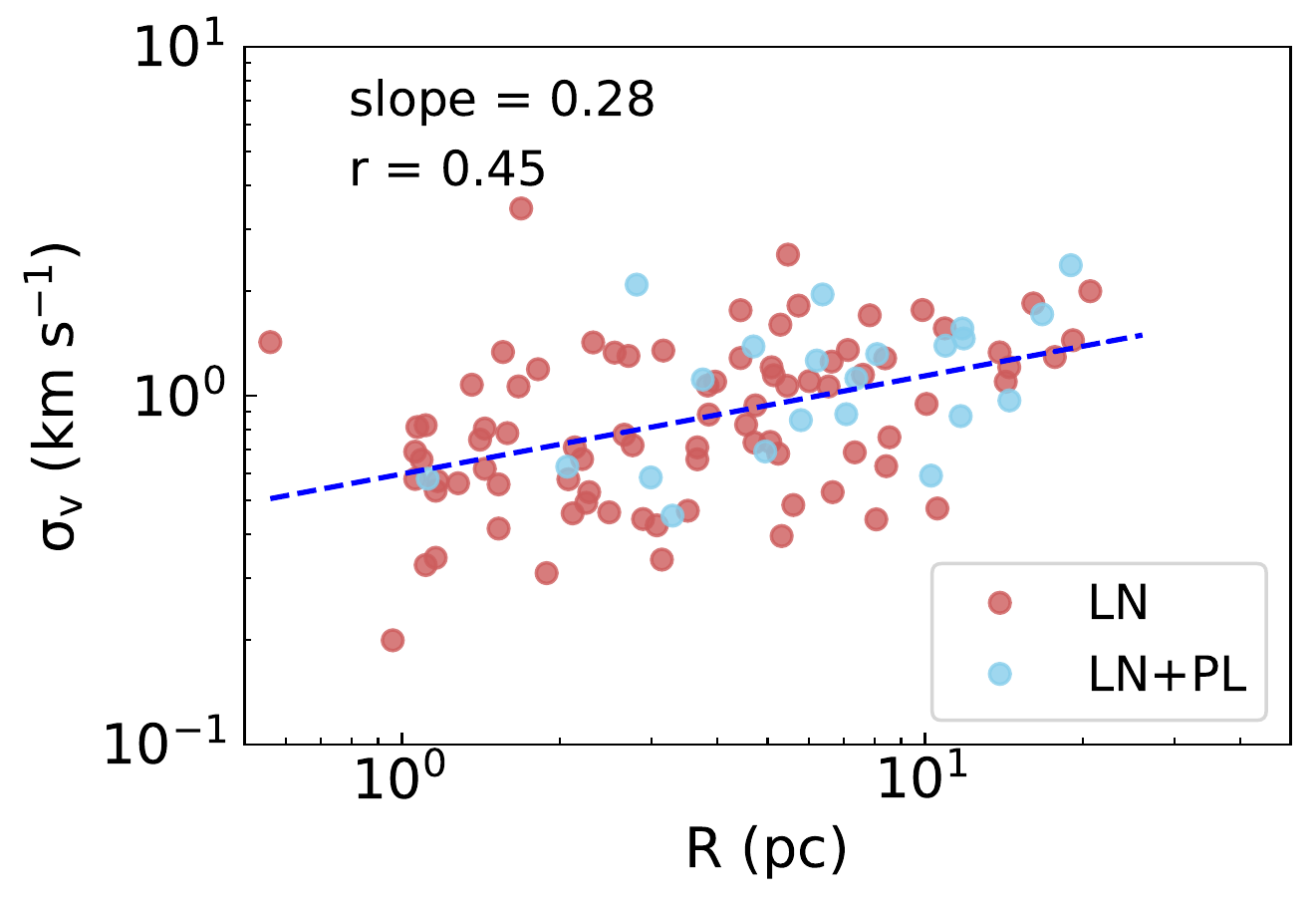}
		\\(a)
	\end{minipage}
	\begin{minipage}[t]{0.32\linewidth}
		\centering
		\includegraphics[width= \linewidth]{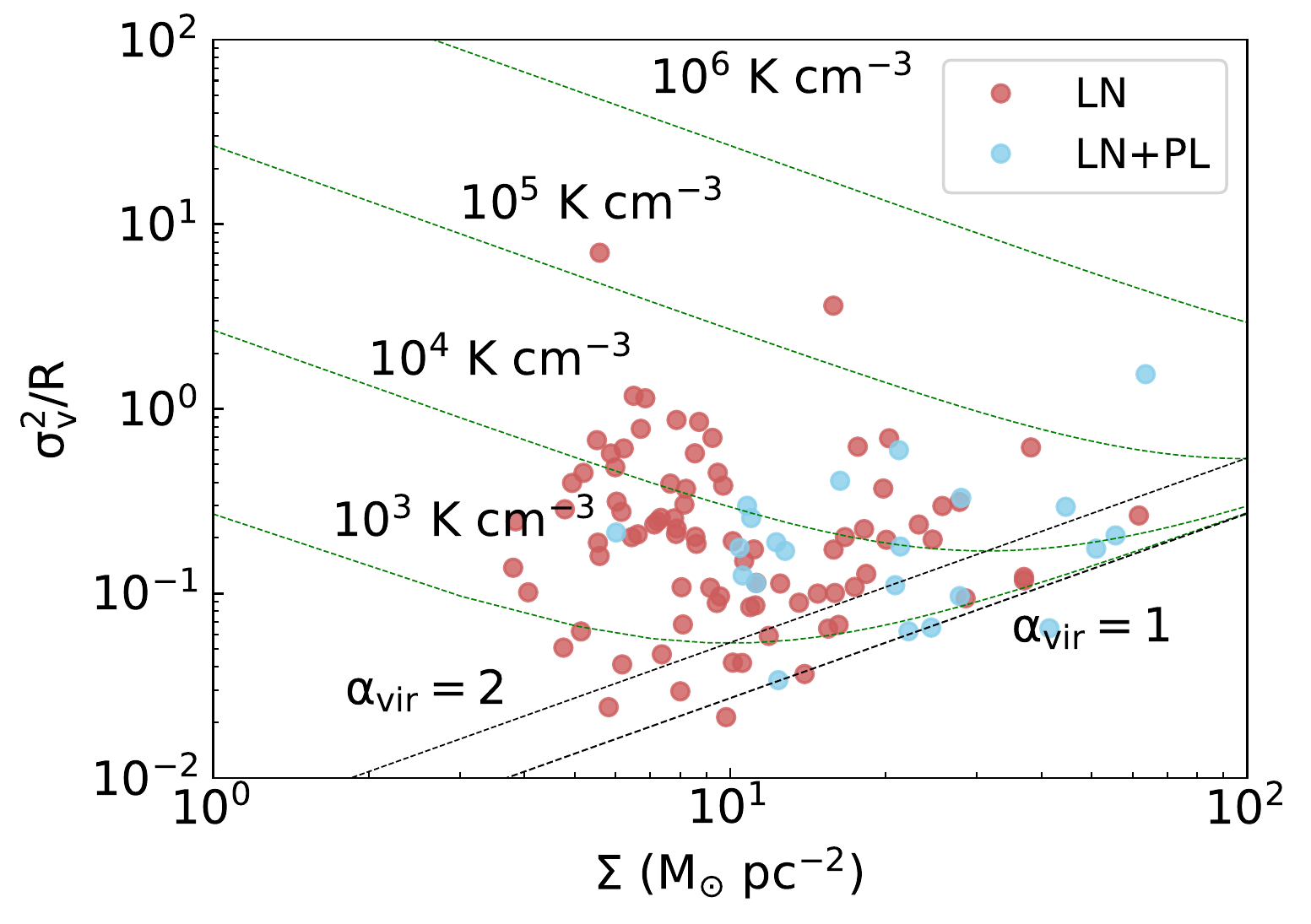}
		\\(b)
	\end{minipage}
	\begin{minipage}[t]{0.32\linewidth}
		\centering
		\includegraphics[width= \linewidth]{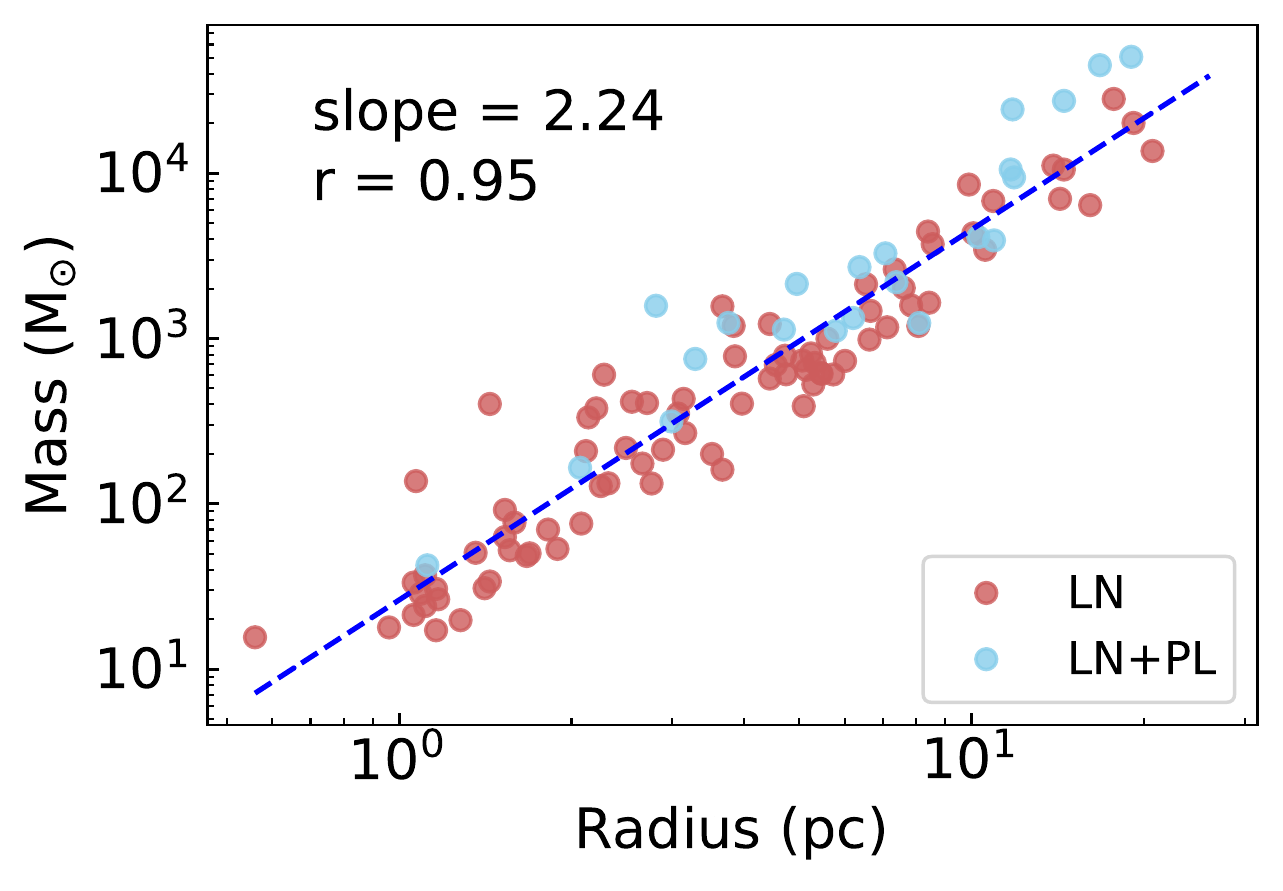}
		\\(c)
	\end{minipage}
	\caption{Relationships between (a) velocity dispersion and radius, (b) the factor $\sigma_v^2/R$ and surface density, and (c) mass and radius of the clouds. The red color and the blue color correspond to the clouds in the LN and LN+PL categories, respectively. The dashed blue lines in panels (a) and (c) are the fitted linear relations for the $\log \sigma_v-\log R$ and $\log M-\log R$ correlations, respectively. } 
	\label{fig13}
\end{figure*}

\cite{Larson1981} found a power-law scaling relation between the velocity dispersion and the size of molecular clouds, $\sigma_v\propto R^{0.38}$, which is the so-called first Larson's relation. Figure \ref{fig13} presents the three widely discussed Larson's relations between the physical parameters of molecular clouds. As presented in Figure \ref{fig13}(a), the scaling exponent of the observed $\sigma_v-R$ relation in this work is only 0.28, and the correlation coefficient between $\ln \sigma_v$ and $\ln R$ is r$=$0.45, which is quite low. No significant difference can be found in the $\sigma_v-R$ relations between the molecular clouds in the LN and LN+PL categories.

\cite{Larson1981} also suggests that molecular clouds tend to be objects under virial equilibrium states, known as the second Larson's relation. Figure \ref{fig13}(b) gives the relationship between the factor $\sigma_v^2/R$ and the surface density $\Sigma$ of the clouds. The reference relations for the molecular clouds that are under the virial equilibrium ($\alpha_{vir}=$ 1) state or are marginally gravitationally bound (with a virial parameter of $\alpha_{vir}=$ 2) are indicated in the figure. If considering external pressure for the dynamics of the clouds in addition to gravity and internal pressure, equilibrium states can be expressed with the green dotted lines in Figure \ref{fig13}(b) \citep{Field2011}. Molecular clouds with the same virial parameter will distribute along lines parallel to the dashed lines of $\alpha_{vir}=1, 2$. We can see from Figure \ref{fig13}(b) that the shapes of the N-PDFs are not related to the gravitational states of the molecular clouds. Most of the clouds in the LN and LN+PL categories are distributed above the marginally bound line, indicating they may need external pressure to stay at equilibrium states. However, the molecular clouds in the LN category do not exhibit typical virial parameters, i.e., distributed along the directions parallel to the $\alpha_{vir} = 1, 2$ lines. In contrast, the molecular clouds in the LN+PL category tend to have higher surface densities and to be distributed along the directions parallel to the $\alpha_{vir} = 1, 2$ lines, which may indicate different importance of self-gravity between the two categories.  

The third Larson's relation suggests that molecular clouds have nearly constant surface density, which can be expressed as $Mass\propto R^2$. Figure \ref{fig13}(c) shows the $Mass-R$ relation of the clouds, which is well fitted by a power-law function with an exponent of 2.24. An exponent in the $Mass-R$ scaling relation larger than two indicates larger molecular clouds have higher column densities. From Figure \ref{fig13}(c), we can see that the clouds both in the LN and the LN+PL categories show the same $Mass-R$ relation. According to the above discussion, molecular clouds, whether having PL tails, exhibit the same behavior regarding the first and third Larson's relations while they are slightly different in gravitational states.

In Section \ref{sec3.2.3}, we obtain a power-law scaling relation between $\sigma_s$ and $\left\langle N_{\rm H_2}\right\rangle$ of the molecular clouds, which has not been reported so far. The relation suggests that the complexity of the internal structure of a molecular cloud is correlated with its mean column density. The form of the scaling relation $\sigma_s\propto \left\langle N_{\rm H_2}\right\rangle^{0.44}$ is reminiscent of the first Larson's relation, i.e., $\sigma_v\propto R^{0.38}$ (or $\sigma_v\propto R^{0.5}$) \citep{Larson1981, Solomon1987}. In rough deduction, the parameters $\sigma_s$ and $N_{H_2}$ can be related if the following relations, $\sigma_s\propto \mathcal{M}^{\beta_1}$, $\sigma_v\propto R^{\beta_2}$, and $N_{H_2}\propto R^{\beta_3}$ can be established, specifically, $\sigma_s\propto\mathcal{M}^{\beta_1}\propto \sigma_v^{\beta_1}\propto{(R^{\beta_2})^{\beta_1}\propto[(N_{H_2}^{1/{\beta_3}})^{\beta_2}]^{\beta1}} = N_{H_2}^{\beta_1\beta_2/\beta_3}$. The results in this work imply $\beta_1\beta_2/\beta_3 = 0.44$, i.e., $\beta_1 = 0.44\beta_3/\beta_2$. From previous discussions in this section, we have got $\sigma_v\propto R^{0.28}$ and $Mass\propto R^{2.2}$, which indicate $\beta_2 = 0.28$ and $\beta_3 = 0.2$, respectively. Therefore, the relation $\sigma_s\propto \left\langle N_{\rm H_2}\right\rangle^{0.44}$ implies an observed scaling relation between the width of the N-PDF, $\sigma_s$, and the Mach number, $\mathcal{M}$, of the molecular clouds, which is $\sigma_s\propto\mathcal{M}^{0.31}$. If we further consider the more common Larson's relation $\sigma_v\propto R^{0.5}$, the observed $\sigma_s-\mathcal{M}$ relation becomes $\sigma_s\propto\mathcal{M}^{0.18}$. Although the analytic form of the observed $\sigma_s-\mathcal{M}$ relation is different from that predicted by numerical simulations (Eq.\ref{eq2}), their behaviors are similar, as shown in Figure \ref{fig14}. In the lower panel of Figure \ref{fig14}, the grey dashed line and grey solid line represent the observed $\sigma_s-\mathcal{M}$ relation with exponents of 0.3 and 0.18, respectively, and other lines represent the relations in the form of Eq. \ref{eq2} with different values of A and b. The grey solid line lies very close to the best fit of our data of Eq. \ref{eq2}. Therefore, the observed $\sigma_s-\left\langle N_{\rm H_2}\right\rangle$ relation may have the same physical origin as the predicted $\sigma_s-\mathcal{M}$ relation, i.e., successive compression on the density and velocity structure of the interstellar medium by shock waves \citep{Molina2012}. We can conduct a similar deduction of the relation $\sigma_s\propto Mass^{\beta_4}$, adopting the scaling relations of $\sigma_s\propto \mathcal{M}^{0.31}$, $\sigma_v\propto R^{0.28}$, and $Mass\propto R^{2.2}$, and found $\beta_4\sim 0.04$, which is consistent with the observed results in Figure \ref{fig10} within errors. Therefore, the width of the N-PDFs, $\sigma_s$, has very weak, if any, dependency on the mass of the molecular clouds.   

\subsection{Relation between turbulence energy and the width of N-PDF} \label{sec4.2}

\begin{figure*}[htb!]
	\centering
	\includegraphics[width= 0.5\linewidth]{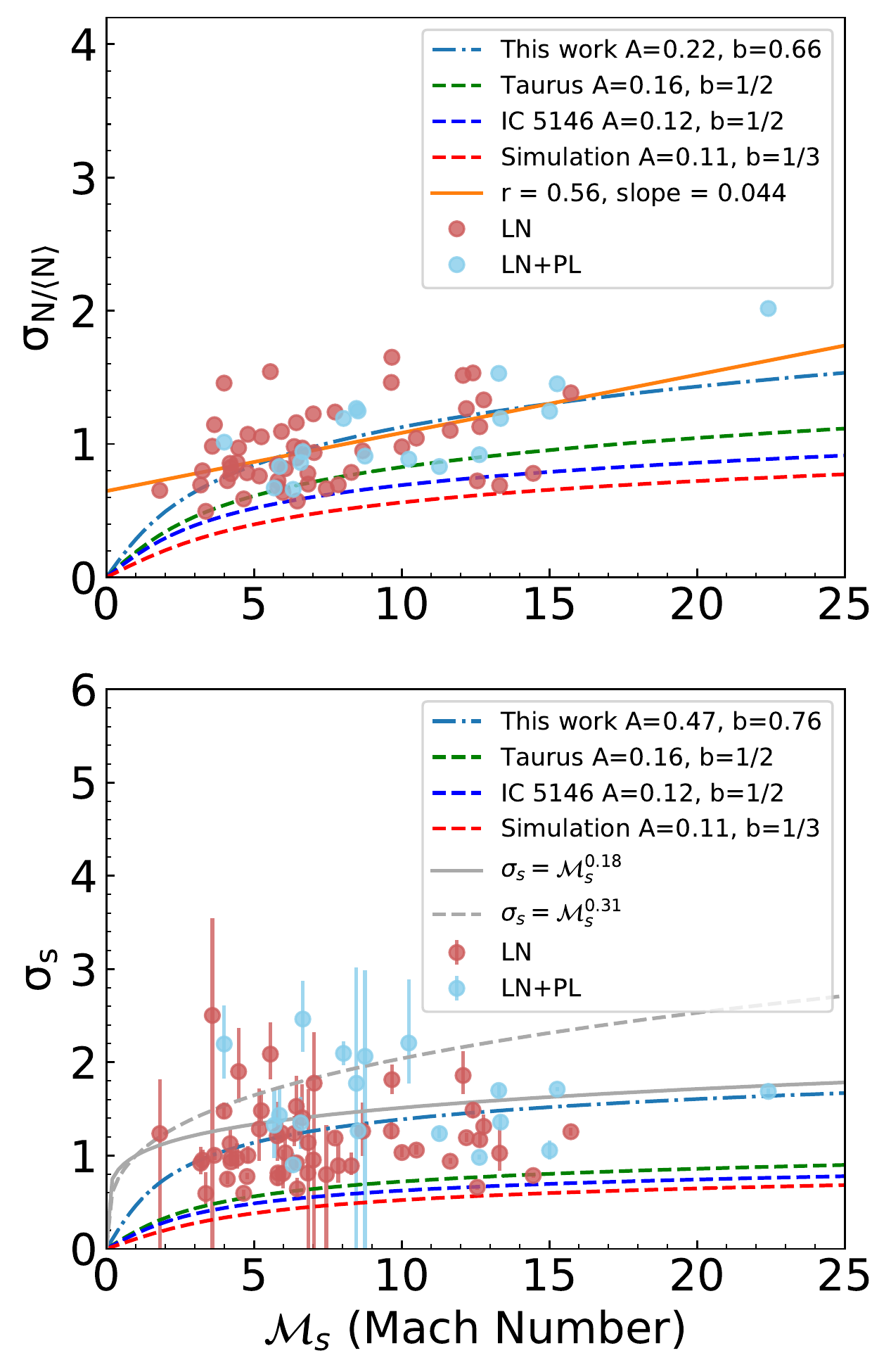}
	\caption{Relationship between the dispersion of the normalized column density, $\sigma_{N/\left\langle N\right\rangle}$, and sonic Mach number, $\rm{\mathcal{M}}$, of the clouds (upper panel). The lower panel presents the relationship between $\sigma_s$ and $\mathcal{M}$. The green, blue, and red dashed curves show the case [A = 0.16, b = 1/2], [A = 0.12, b = 1/2], and [A = 0.11, b = 1/3], corresponding to the Taurus, IC 5146, and simulated molecular clouds, respectively \citep{Burkhart2018}. The dark blue dot-dashed lines in the upper and lower panels are the best fit of the relation $\sigma_{N/\left\langle N\right\rangle } = \sqrt{(b^2\mathcal{M}^2+1)^A-1}$ and $\sigma_s = \sqrt{A\ln (1+b^2\mathcal{M}^2)}$, respectively, while the orange straight line is the best fit of the linear relation of $\sigma_{N/\left\langle N\right\rangle}- \mathcal{M}$. The grey solid and dashed curves represent the relations $\sigma_s= \mathcal{M}^{0.18}$ and $\sigma_s= \mathcal{M}^{0.31}$, respectively.} 
	\label{fig14}
\end{figure*}

The $\sigma_{N/\left\langle N\right\rangle}-\mathcal{M}$ and the $\sigma_s-\mathcal{M}$ relations derived in this work are presented in Figure \ref{fig14}. Molecular clouds in this work are defined as contiguous PPV structures. Therefore, large-scale velocity gradients between substructures within a GMC may cause multi-peaks in the \COl average spectrum of the GMC, which may result in an overestimation of the velocity dispersion, and then $\mathcal{M}$. We checked the average spectra of the \CO and \COl emission of the clouds, and the clouds with multi-peaks in the \COl spectra are excluded in the investigation of the $\sigma_{N/\left\langle N\right\rangle}-\mathcal{M}$ and $\sigma_s-\mathcal{M}$ relations. Fifty three LN molecular clouds and eighteen LN+PL molecular clouds are preserved in Figure \ref{fig14}.  

\begin{figure*}[htb!]
	\centering
	\begin{minipage}[t]{0.3\linewidth}
		\centering
		\includegraphics[width= \linewidth]{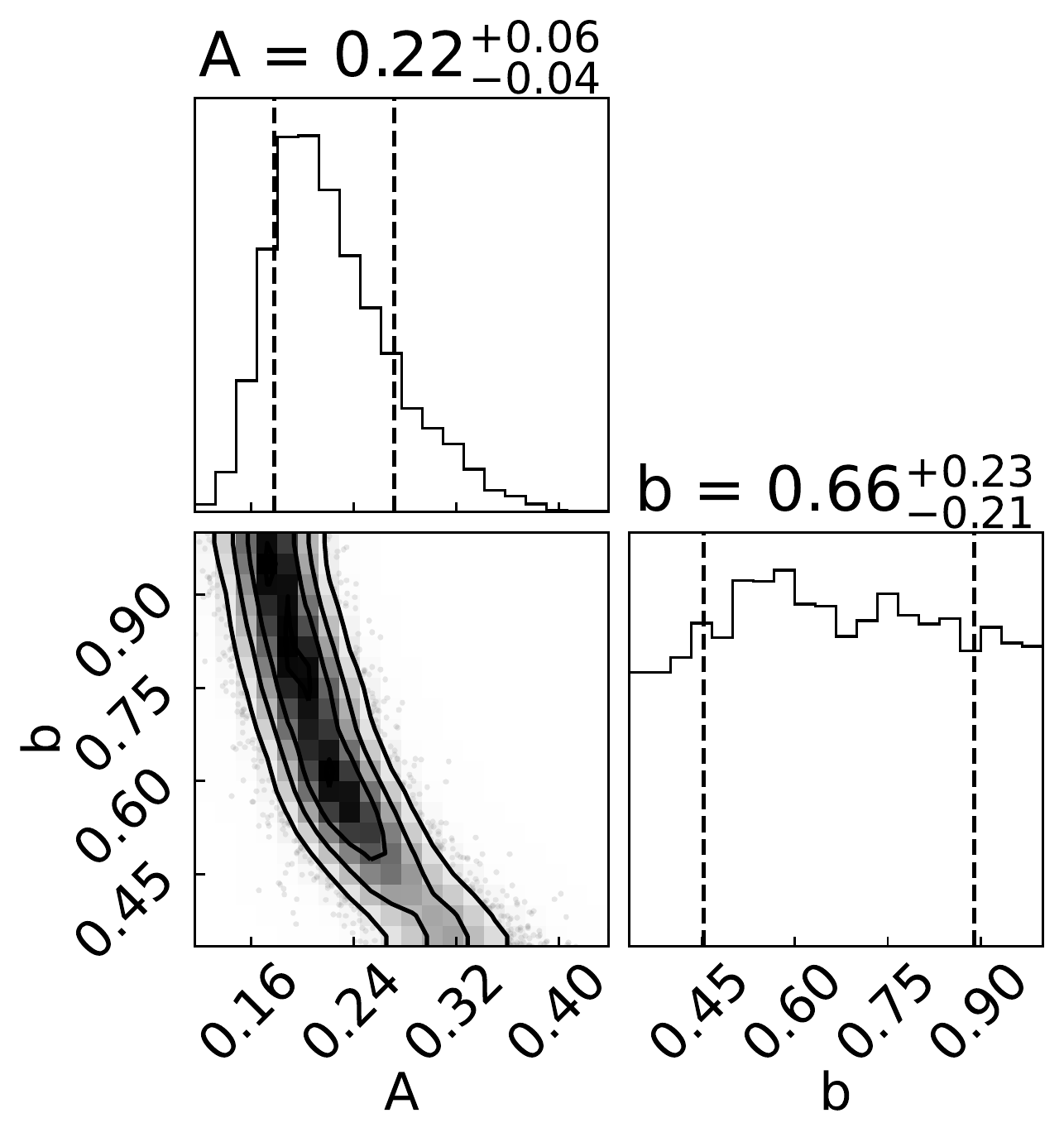}
		\\(a)
	\end{minipage}
	\begin{minipage}[t]{0.3\linewidth}
		\centering
		\includegraphics[width= \linewidth]{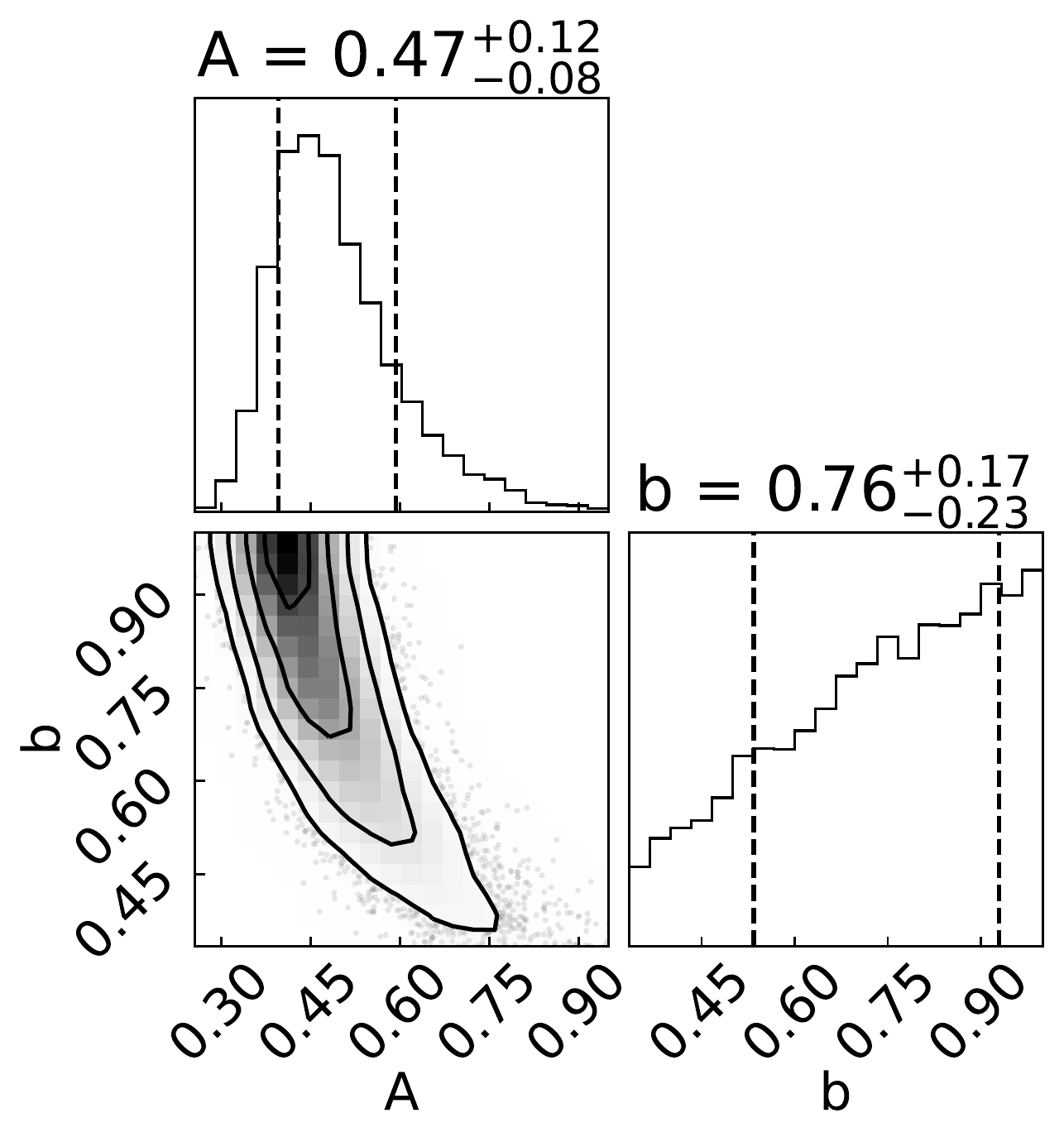}
		\\(b)
	\end{minipage}
	\caption{Corner maps of the fitted parameters in the relationships of (a) $\sigma_{N/\left\langle N\right\rangle } = \sqrt{(b^2\mathcal{M}^2+1)^A-1}$ and (b) $\sigma_s = \sqrt{A\ln (1+b^2\mathcal{M}^2)}$, respectively. The numbers in the marginalized distributions present the median value and the 16th and 84th percentiles of the parameters A and b. The vertical lines represent the 16th and 84th percentiles.} 
	\label{fig15}
\end{figure*}  

In Figure \ref{fig14}, although the values of $\sigma_{N/\left\langle N\right\rangle}$ are comparable to those of the IRDCs and the nearby star-forming regions listed in table 3 in \cite{Kainulainen2013b}, the sonic Mach numbers of the molecular clouds in this work are smaller by a factor around two. The molecular clouds in the LN and LN+PL categories do not show significant differences in the $\sigma_{N/\left\langle N\right\rangle}-\mathcal{M}$ relation. The orange solid-line is the best fit for the linear relation between $\sigma_{N/\left\langle N\right\rangle }$ and $\mathcal{M}$ for all the selected clouds. The fitted slope is 0.044$\pm 0.01$, which is similar to the slope obtained by \cite{Kainulainen2013b}, and the Pearson correlation coefficient is $0.56$. The p-value of the correlation coefficient is nearly zero under the null hypothesis that the slope of the $\sigma_{N/\left\langle N\right\rangle}-\mathcal{M}$ relation is zero, which indicates that we can not rule out a correlation between the two quantities. Our results extended the  $\sigma_{N/\left\langle N\right\rangle }-\mathcal{M}$ relation for IRDCs obtained by \cite{Kainulainen2013b}, to molecular clouds with relatively low column densities, on the order of $\sim10^{20}-10^{21}$ cm$^{-2}$, and smaller Mach numbers.

We also fitted the $\sigma_{s}-\mathcal{M}$ and $\sigma_{N/\left\langle N\right\rangle}-\mathcal{M}$ relations in the forms of Eqs. \ref{eq2} and \ref{eq3}, using the MCMC approach with a uniform prior limitation of $0<A<1$ and $1/3\le b \le 1$. The dark blue dot-dashed lines in Figure \ref{fig14} represent the best fittings of our data. The dashed lines with green, blue, and red colors represent the relations for the Taurus, IC 5146 GMCs, and the simulated clouds in \cite{Burkhart2012}, respectively. The majority of the molecular clouds in the upper panel in Figure \ref{fig14} lie above these three known relations. The best fits of the two forms of the $\sigma_{N/\left\langle N\right\rangle}-\mathcal{M}$ relation, i.e., the form of Eq. \ref{eq3} (dark blue dot-dashed line) and the linear form (orange solid line), are very consistent in the range of $\mathcal{M}=[3, 18]$, and they describe the majority of the clouds. The clouds in this work also lie above the known $\sigma_{s}-\mathcal{M}$ relations in the lower panel of Figure \ref{fig14}. Nevertheless, here the correlation between $\mathcal{M}$ and $\sigma_s$ is relatively weak and $\sigma_s$ has large scatter. Figure \ref{fig15} shows the corner maps of the fitted A and b parameters for the $\sigma_{N/\left\langle N\right\rangle} -\mathcal{M}$ (panel a) and $\sigma_{s}-\mathcal{M}$ (panel b) relations. For the first relation, we obtain $A = 0.22^{+0.06}_{-0.04}$ and $b = 0.66^{+0.23}_{-0.21}$, while for the second relation, we obtain $A = 0.47^{+0.12}_{-0.08}$ and $b = 0.76^{+0.17}_{-0.23}$. The values of parameters $A$ and $b$ obtained from the two forms are consistent within errors and are greater than those in the Taurus and IC 5146 molecular clouds. The parameter $A$ in the two forms has a relatively concentrated distribution near its expected value, whereas the parameter $b$ shows broad distributions without any preferred values. As suggested by \cite{Burkhart2012}, like the forcing parameter $b$, the coefficient $A$ may also depend on the driving of the turbulence. For the column density dispersion of pure solenoidal and pure compressive turbulence, the $A$ parameter in numerical simulations is 0.14 and 0.6, respectively \citep{Federrath2010}. The values of $A$ in Figure \ref{fig15} lie between 0.14 and 0.6. The values of $b$ obtained from our data are consistent with the more compressive turbulent in numerical simulations \citep{Menon2020}. However, the relatively weak correlation between $\mathcal{M}$ and $\sigma_s$ and broad distribution of $b$ may indicate various mixtures of solenoidal and compressive forcing modes in the observed molecular clouds. The properties of the molecular clouds in the outer Galaxy that we investigate in this work may be different from the nearby molecular clouds like Taurus and IC 5146 in terms of turbulent forcing. 

\subsection{Ongoing Star Formation and the Shapes of N-PDFs}\label{sec4.3}
Both numerical simulations and Herschel observations have found that the shapes of the N-PDFs are related to star formation activities of molecular clouds (\citetalias{Kainulainen2009}; \citealt{Kainulainen2013a, Federrath2013}), and the high-density power-law portion of N-PDFs is considered to correspond to the collapsing molecular gas which is dominated by gravity \citep{Girichidis2014, Jaupart2020}. In this section, we discuss the relation between the shapes of N-PDFs measured by \COl line emission and the ongoing star formation activities of the clouds. The true physical environments of molecular clouds are usually complicated. We counted the Class I, Flat spectrum, and Class II young stellar objects (YSOs) within the boundary of each molecular cloud and simply assume that these YSOs are associated with the clouds. The YSOs are identified from the AllWISE data release using the photometric classification scheme of \cite{Koenig2014}. The obtained YSO catalog is further supplemented by the catalogs from \cite{Marton2016} and \cite{Marton2019}. The catalog from \cite{Marton2016} is obtained using the AllWISE data release together with the support vector machine method, while the catalog from \cite{Marton2019} is generated using machine learning techniques from the GAIA DR2 and AllWISE database and combined with Planck measurements. The total number of associated YSOs for each cloud is listed in Column 17 in Table \ref{tab1}. Molecular clouds containing more than ten YSOs are considered to harbor ongoing star formation. 

\begin{deluxetable*}{crrr}
	\label{tab3}
	\tablecaption{Fraction of molecular clouds with ongoing star formation} 
	\tabletypesize{\normalsize}
 	\setlength{\tabcolsep}{20pt}
	\tablehead{
	\colhead{Shape of N-PDF} & \colhead{LN} & \colhead{LN+PL} & \colhead{UN}
	}
	\startdata
	Fraction & 27.9\% & 59.1\% & 16.7\%
	\enddata
\end{deluxetable*}

Table \ref{tab3} gives the percentage of the molecular clouds with ongoing star formation in each category. About 60\% of the molecular clouds in the LN+PL category hosts ongoing star formation, while this percentage is only 28\% and 17\% in the LN and UN categories, respectively. The fractions are statistically consistent with the prediction of numerical simulations and the results from dust-based observations, i.e., molecular clouds with active star formation activities tend to have power-law tails in their N-PDFs at the high column density end.     

\section{Conclusion} \label{sec5}
We have studied the properties of N-PDFs with an unbiased sample of 120 molecular clouds in the third quadrant of the Galactic mid-plane. The molecular clouds are identified from the \COl data from the MWISP survey using the DBSCAN clustering algorithm. The N-PDFs are fitted with LN and LN+PL functions and classified into three categories, i.e., LN, LN+PL, and UN, according to the BIC model selection criterion and the distribution of residuals. The statistics of the fitted model parameters of the N-PDFs and the physical properties of the selected molecular clouds like peak excitation temperature, optical depth, kinematic distance, mass, and radius are presented. Relations between N-PDF parameters and the physical parameters of the molecular clouds are discussed. We also investigated the relationship between the shapes of N-PDFs and the star formation activities of the molecular clouds. The main results are summarized as follows. 

1. About 72\%, 18\%, and 10\%, of the 120 selected molecular clouds in the third quadrant of the Milky Way within l = [195$\arcdeg$, 225$\arcdeg$], and b = [$-$5$\arcdeg$, 5$\arcdeg$], are classified into the LN, LN+PL, and UN categories, respectively, indicating that most of the molecular clouds are dominated by turbulence. The results are significantly different from those obtained using the dust-based tracers of H$_2$ column density, for which power-law tails in N-PDFs are common.

2. The widths of the N-PDFs, $\sigma_s$, lying in the range from $\sim$0.6 to 2, are independent of the sizes and distances of molecular clouds, indicating that the hierarchy of the structures of molecular clouds is not evident on scales of $\sim$1-10 pc, and the complexity of molecular clouds is irrelevant to the Galactic environment.

3. The widths of the N-PDFs, $\sigma_s$, are scaled with the mean column densities of the molecular clouds, whereas independent of the masses of the molecular clouds.

4. A correlation is found between the column density fluctuation, $\sigma_{N/\left\langle N\right\rangle}$, and the Mach number, $\mathcal{M}$, of the molecular clouds, which extends the $\sigma_{N/\left\langle N\right\rangle }-\mathcal{M}$ relation found in IRDCs to molecular clouds with relatively lower column densities and smaller Mach numbers. The width, $\sigma_s$, of N-PDFs and $\mathcal{M}$ of the molecular clouds shows relatively weak correlation based on the \COl line emission data in this work. The distributions of $A$ and $b$ parameters in the $\sigma_{N/\left\langle N\right\rangle }-\mathcal{M}$ and $\sigma_s-\mathcal{M}$ relations suggest complex mixture of turbulent forcing modes in the molecular clouds in this work.

5. Consistent with the predictions of numerical simulations and the observations using dust-based tracers for column density, 60\% of the molecular clouds with LN+PL N-PDFs contain ongoing star formation, while this percentage is only 28\% and 17\% for the molecular clouds in the LN and UN categories, respectively. 

\begin{acknowledgments}
This research made use of the data from the Milky Way Imaging Scroll Painting (MWISP) project, which is a multi-line survey in \CO/\COl/\COll along the northern galactic plane with PMO-13.7m telescope. We are grateful to all the members of the MWISP working group, particularly the staff members at PMO-13.7m telescope, for their long-term support. MWISP was sponsored by National Key R$\&$D Program of China with grant 2017YFA0402701 and CAS Key Research Program of Frontier Sciences with grant QYZDJ-SSW-SLH047. J. Y. is supported by National Natural Science Foundation of China through grant 12041305. H. W. and Y. M. acknowledge the support by NSFC grant 11973091. M. Zhang is supported by the National Natural Science Foundation of China (grants No. 12073079). Y. L. and Y. M. acknowledge financial supports by NSFC grant 11973090 and the Natural Science Foundation of Jiangsu Province of China (Grant No. BK20181513). We thank the staffs of the MWISP scientific group for their valuable suggestions. Y. M. thanks Yang Su and Fujun Du for their helpful discussions. We thank the anonymous referee for his/her constructive suggestions that help to improve this manuscript. This work makes use of the SIMBAD database, operated at CDS, Strasbourg, France. 
\end{acknowledgments}

\appendix
Figures \ref{fig16} to \ref{fig18} present the N-PDFs of the molecular clouds in the LN, LN+PL, and UN categories, respectively. 
\input{pdf_residual_LN.tex}

\clearpage
\input{pdf_residual_LN+PL.tex}

\clearpage
\input{pdf_residual_NC.tex}

\clearpage
% \clearpage
% \input{figset16.tex}
% \input{figset17.tex}
% \input{figset18.tex}
% \clearpage
\bibliography{G195_PDF_arxiv.bbl}
\bibliographystyle{aasjournal}

\end{document}

%% file: table1.tex
    MWISP G195.019$-$01.386$+$015.12&   3.22&  0.566&  2.207&  0.993& $-$6.047&     202&   16.80&   1.38$\times10^{21}$&    3.76&    1242& $-$0.430& $-$0.790&   0.320&  10.03&       0&    4.34&   LN$+$PL\\
    MWISP G195.097$-$00.279$+$003.58&   0.83& $-$0.220&  0.762&...&...&     297&   10.94&   3.16$\times10^{20}$&    1.17&      26& $-$0.050& $-$0.730&   0.230&   5.51&       0&   16.63&      LN\\
    MWISP G195.413$-$02.030$+$014.24&   2.90&  0.119&  2.064&  0.847& $-$3.610&     592&   12.05&   5.49$\times10^{20}$&    5.79&    1113& $-$0.090& $-$0.900&   0.270&   8.53&       3&    4.37&   LN$+$PL\\
    MWISP G195.552$-$00.136$+$020.02&   4.42&  0.149&  1.351&  0.546& $-$2.714&     801&    9.74&   6.16$\times10^{20}$&   10.27&    4106& $-$0.120& $-$0.630&   0.470&   6.31&      35&    1.01&   LN$+$PL\\
    MWISP G195.729$-$00.158$+$032.03&   9.28& $-$1.426&  1.859&...&...&     541&   21.06&   1.45$\times10^{21}$&   17.71&   28114&  0.380& $-$0.520&   0.260&  11.87&      30&    1.22&      LN\\
    MWISP G195.753$-$02.284$+$004.42&   0.92& $-$0.796&  1.476&...&...&     914&   23.86&   1.88$\times10^{21}$&    2.28&     604&  0.040& $-$0.690&   0.200&   3.59&      35&    1.22&      LN\\
    MWISP G195.940$-$01.742$+$018.39&   3.80& $-$0.163&  1.017&...&...&     338&    9.64&   3.07$\times10^{20}$&    5.73&     606& $-$0.210& $-$0.950&   0.250&  19.16&       0&   35.96&      LN\\
    MWISP G196.009$-$00.292$+$013.37&   2.58& $-$0.321&  1.027&...&...&     351&   10.76&   4.19$\times10^{20}$&    3.97&     403& $-$0.120& $-$0.740&   0.320&  11.07&       2&   13.70&      LN\\
    MWISP G196.132$-$00.901$+$014.49&   2.79& $-$0.777&  1.313&...&...&     376&   26.61&   9.72$\times10^{20}$&    4.44&    1224&  0.140& $-$0.660&   0.260&  12.57&      12&    6.91&      LN\\
    MWISP G196.275$-$02.285$+$014.69&   2.80& $-$0.397&  1.059&...&...&     811&   17.22&   7.87$\times10^{20}$&    6.54&    2129& $-$0.090& $-$0.710&   0.270&  10.29&      20&    4.02&      LN\\
    MWISP G196.292$+$00.336$+$004.84&   0.96& $-$0.384&  0.981&  1.508& $-$7.967&    6213&   12.09&   5.54$\times10^{20}$&    6.21&    1330&  0.090& $-$0.620&   0.270&  12.43&       3&    8.60&   LN$+$PL\\
    MWISP G196.590$-$01.527$+$015.67&   2.95& $-$1.098&  1.449&...&...&    1668&   34.60&   1.39$\times10^{21}$&    9.89&    8533&  0.400& $-$0.220&   0.220&  16.43&      45&    4.16&      LN\\
    MWISP G196.669$+$00.683$+$020.14&   4.04& $-$0.415&  0.939&...&...&    1854&   15.77&   5.56$\times10^{20}$&   14.27&    6996&  0.200& $-$0.360&   0.330&  11.44&      16&    2.84&      LN\\
    MWISP G196.935$-$01.070$+$005.97&   1.09& $-$0.586&  0.960&...&...&     261&    9.83&   2.74$\times10^{20}$&    1.44&      33&  0.390& $-$0.490&   0.240&   8.39&       0&   32.04&      LN\\
    MWISP G197.015$-$03.153$+$023.88&   4.94& $-$0.819&  1.292&...&...&     725&   29.80&   9.31$\times10^{20}$&   10.91&    6788&  0.340& $-$0.410&   0.270&  15.29&      31&    4.54&      LN\\
    MWISP G197.412$+$03.829$+$010.68&   1.84& $-$0.218&  0.809&...&...&     189&   10.07&   2.96$\times10^{20}$&    2.08&      76&  0.010& $-$0.780&   0.250&   5.70&       0&   10.52&      LN\\
    MWISP G197.528$+$02.181$+$004.77&   0.88& $-$0.101&  0.633&...&...&     215&   11.06&   3.09$\times10^{20}$&    1.06&      21& $-$0.230& $-$0.640&   0.230&   5.32&       0&   19.23&      LN\\
    MWISP G197.653$-$03.026$+$003.82&   0.72& $-$0.366&  0.921&...&...&     725&   13.60&   4.95$\times10^{20}$&    1.59&      77&  0.170& $-$0.500&   0.200&   6.87&       1&   14.62&      LN\\
    MWISP G197.851$-$02.375$+$021.12&   3.90& $-$0.213&  0.922&...&...&     526&   20.06&   7.94$\times10^{20}$&    7.34&    2617& $-$0.240& $-$0.770&   0.290&   6.17&      14&    1.54&      LN\\
    MWISP G198.392$-$00.430$+$007.48&   1.22& $-$0.228&  0.791&...&...&     535&   10.52&   4.06$\times10^{20}$&    2.32&     133& $-$0.030& $-$0.640&   0.270&  14.59&       0&   40.69&      LN\\
    MWISP G198.650$-$01.128$+$020.30&   3.50& $-$0.225&  0.793&...&...&     363&    8.88&   3.45$\times10^{20}$&    5.47&     611&  0.100& $-$0.700&   0.350&  27.80&       0&   66.74&      LN\\
    MWISP G198.753$+$00.018$-$006.28&   ...& $-$0.258&  0.841&...&...&     191&    9.64&   2.48$\times10^{20}$&    ...&       ...& $-$0.010& $-$0.780&   0.290&  35.15&       0&    ...&      LN\\
    MWISP G198.855$-$02.159$+$018.35&   3.04& $-$1.735&  1.814&...&...&     239&   13.37&   8.50$\times10^{20}$&    3.86&     780&  0.740& $-$0.080&   0.390&   9.45&       4&    4.46&      LN\\
    MWISP G199.025$-$03.341$+$006.06&   0.96& $-$0.309&  0.895&...&...&     199&    9.72&   3.24$\times10^{20}$&    1.11&      24&  0.140& $-$0.660&   0.230&   8.34&       0&   36.20&      LN\\
    MWISP G199.307$-$00.478$+$003.23&   0.59&...&...&...&...&     281&   11.60&   4.05$\times10^{20}$&    0.81&      16& $-$0.160& $-$0.990&   0.230&   6.96&       1&   31.68&       UN\\
    MWISP G199.942$-$00.275$+$021.69&   3.50& $-$0.044&  0.581&...&...&     163&    7.47&   2.09$\times10^{20}$&    3.67&     161& $-$0.030& $-$0.800&   0.320&   8.20&       0&   13.36&      LN\\
    MWISP G199.973$+$01.326$+$020.78&   3.31& $-$0.641&  1.188&...&...&     992&   26.48&   8.19$\times10^{20}$&    8.55&    3721&  0.160& $-$0.650&   0.310&   7.50&      26&    1.54&      LN\\
    MWISP G200.820$+$00.158$-$010.00&   ...& $-$0.221&  1.026&...&...&     335&   10.00&   4.81$\times10^{20}$&    ...&       ...& $-$0.260& $-$0.790&   0.280&  13.11&       0&    ...&      LN\\
    MWISP G201.707$+$01.444$+$022.86&   3.36& $-$0.351&  0.954&...&...&     293&   20.82&   5.94$\times10^{20}$&    4.72&     786&  0.080& $-$0.500&   0.210&   6.75&      12&    3.75&      LN\\
    MWISP G201.963$-$04.904$+$008.54&   1.12&  0.089&  0.901&  0.817& $-$7.109&     507&   10.97&   5.95$\times10^{20}$&    2.07&     165& $-$0.580& $-$0.280&   0.280&   6.05&       1&    5.69&   LN$+$PL\\
    MWISP G202.028$+$01.592$+$005.99&   0.83& $-$0.486&  1.688&  0.316& $-$1.419&   77828&   41.06&   2.14$\times10^{21}$&   19.00&   50552&  0.060& $-$0.280&   0.300&  22.18&    1031&    2.44&   LN$+$PL\\
    MWISP G202.854$+$01.263$+$031.72&   4.90& $-$0.225&  0.760&...&...&    1605&    9.16&   3.94$\times10^{20}$&   16.11&    6407& $-$0.070& $-$0.600&   0.330&  20.68&      18&    9.85&      LN\\
    MWISP G203.004$-$03.703$+$019.15&   2.51& $-$0.961&  2.503&...&...&     233&   10.21&   7.20$\times10^{20}$&    3.14&     431&  0.080& $-$1.170&   0.520&   3.15&       1&    0.97&      LN\\
    MWISP G203.261$+$01.059$+$007.41&   0.94& $-$0.665&  1.406&...&...&     199&   10.23&   4.30$\times10^{20}$&    1.09&      28&  0.400& $-$0.810&   0.250&   6.37&       0&   18.91&      LN\\
    MWISP G204.189$-$00.143$+$024.79&   3.27& $-$0.260&  0.897&...&...&     367&    9.35&   4.10$\times10^{20}$&    5.14&     646&  0.050& $-$0.880&   0.370&  12.31&       8&   12.07&      LN\\
    MWISP G204.470$-$00.652$+$027.17&   3.62& $-$0.185&  0.742&...&...&     337&   12.69&   3.33$\times10^{20}$&    5.45&     618& $-$0.110& $-$0.710&   0.280&  11.13&       2&   11.64&      LN\\
    MWISP G204.790$+$00.474$+$009.58&   1.13&  0.184&  1.432&  0.582& $-$2.555&    1040&   11.34&   5.61$\times10^{20}$&    2.99&     315& $-$0.180& $-$0.710&   0.280&   5.57&       0&    3.73&   LN$+$PL\\
    MWISP G205.111$+$01.868$+$008.31&   0.97& $-$0.143&  0.697&...&...&     439&   10.38&   2.92$\times10^{20}$&    1.67&      48&  0.050& $-$0.750&   0.290&  11.35&       2&   45.04&      LN\\
    MWISP G205.839$-$01.276$+$009.98&   1.12& $-$0.607&  1.027&...&...&     289&   11.29&   3.66$\times10^{20}$&    1.56&      52&  0.380& $-$0.250&   0.270&  13.39&       2&   61.47&      LN\\
    MWISP G205.866$+$00.227$+$019.93&   2.33& $-$0.191&  0.776&...&...&     858&   12.63&   5.14$\times10^{20}$&    5.60&     997& $-$0.120& $-$0.710&   0.310&   4.43&       2&    1.53&      LN\\
    MWISP G206.185$-$02.423$+$003.91&   0.48& $-$1.245&  1.653&...&...&     202&   19.46&   8.37$\times10^{20}$&    0.56&      15&  0.430& $-$0.460&   0.290&  14.51&      18&   84.05&      LN\\
    MWISP G206.244$-$00.741$+$010.46&   1.16& $-$0.304&  0.800&...&...&     782&   11.53&   4.03$\times10^{20}$&    2.66&     175&  0.160& $-$0.280&   0.270&   7.52&      14&   10.53&      LN\\
    MWISP G206.477$+$02.119$+$013.33&   1.47& $-$0.252&  0.817&  1.243& $-$5.737&    4509&   11.44&   3.13$\times10^{20}$&    8.10&    1240&  0.080& $-$0.670&   0.230&  13.36&       0&   13.11&   LN$+$PL\\
    MWISP G206.511$-$01.267$+$010.06&   1.10&...&...&...&...&     471&   10.68&   4.15$\times10^{20}$&    1.96&      96& $-$0.160& $-$0.950&   0.290&   7.23&       3&   12.18&       UN\\
    MWISP G206.698$-$04.407$+$009.26&   1.00& $-$0.637&  1.359&  1.500& $-$4.053&    3277&   13.90&   8.25$\times10^{20}$&    4.70&    1132&  0.170& $-$0.670&   0.220&  13.14&       4&    9.20&   LN$+$PL\\
    MWISP G207.156$-$01.984$+$012.95&   1.39&...&...&...&...&   28623&   37.35&   1.91$\times10^{21}$&   19.30&   44954&  0.220& $-$0.630&   0.280&  27.10&    1097&    4.16&       UN\\
    MWISP G207.436$+$00.327$+$033.13&   4.12& $-$0.663&  1.262&...&...&     181&   10.57&   5.51$\times10^{20}$&    4.55&     692&  0.030& $-$0.730&   0.310&   8.45&       5&    5.20&      LN\\
    MWISP G207.526$+$02.216$+$014.76&   1.58& $-$0.061&  0.596&...&...&     739&   10.80&   2.63$\times10^{20}$&    3.52&     200& $-$0.210& $-$0.760&   0.200&   4.30&       0&    4.46&      LN\\
    MWISP G207.823$+$01.124$+$016.92&   1.81& $-$0.359&  0.971&...&...&    1017&   10.69&   4.45$\times10^{20}$&    4.74&     607&  0.040& $-$0.770&   0.290&   9.64&       2&    7.95&      LN\\
    MWISP G208.761$-$00.442$+$011.70&   1.19& $-$0.245&  0.808&...&...&     349&    9.84&   3.43$\times10^{20}$&    1.82&      69& $-$0.010& $-$0.780&   0.240&  12.23&       0&   42.88&      LN\\
    MWISP G208.863$-$02.722$+$024.02&   2.58& $-$0.513&  1.055&  0.838& $-$2.181&    2672&   11.58&   5.28$\times10^{20}$&   10.94&    3921&  0.330& $-$0.130&   0.330&  14.80&      38&    6.26&   LN$+$PL\\
    MWISP G208.887$+$01.874$+$016.12&   1.66& $-$0.177&  0.660&...&...&    1394&   11.73&   2.52$\times10^{20}$&    5.09&     389&  0.110& $-$0.410&   0.220&  12.35&       3&   21.97&      LN\\
    MWISP G208.927$-$03.661$+$022.94&   2.43& $-$0.014&  0.797&...&...&     192&   10.62&   2.93$\times10^{20}$&    2.76&     132& $-$0.150& $-$0.900&   0.230&   7.20&       0&   12.54&      LN\\
    MWISP G209.003$+$02.193$+$009.67&   0.98& $-$0.078&  1.236&  0.926& $-$4.547&    8434&   14.57&   6.21$\times10^{20}$&    7.39&    2191& $-$0.270& $-$0.710&   0.250&  11.07&      16&    4.91&   LN$+$PL\\
    MWISP G209.338$-$01.896$+$002.74&   0.33&...&...&...&...&     171&    9.30&   2.65$\times10^{20}$&    0.35&       2&  0.110& $-$0.820&   0.240&   6.24&       1&   75.93&       UN\\
    MWISP G209.565$+$00.605$+$030.11&   3.34& $-$0.602&  1.170&...&...&     770&   15.40&   5.64$\times10^{20}$&    7.61&    2022&  0.150& $-$0.770&   0.320&  12.44&      13&    5.74&      LN\\
    MWISP G209.566$-$00.565$+$012.12&   1.20& $-$0.444&  0.998&...&...&    2033&   14.28&   4.73$\times10^{20}$&    4.44&     572&  0.200& $-$0.590&   0.310&  19.09&       6&   27.79&      LN\\
    MWISP G210.159$+$01.433$+$032.33&   3.58& $-$0.188&  0.814&...&...&     514&    9.10&   5.19$\times10^{20}$&    6.66&    1469& $-$0.280& $-$0.720&   0.400&   5.52&       3&    1.47&      LN\\
    MWISP G210.221$-$01.632$+$040.24&   4.80& $-$0.905&  1.776&...&...&     177&   13.25&   4.76$\times10^{20}$&    5.24&     812&  0.110& $-$0.950&   0.240&   6.78&       5&    3.47&      LN\\
    MWISP G210.327$-$00.047$+$036.13&   4.12& $-$0.414&  1.033&...&...&     888&    9.62&   6.69$\times10^{20}$&   10.07&    4329& $-$0.010& $-$0.670&   0.310&   9.79&       6&    2.42&      LN\\
    MWISP G210.544$-$03.342$+$019.58&   1.93& $-$0.268&  0.786&...&...&    2024&   11.61&   3.73$\times10^{20}$&    7.12&    1169&  0.070& $-$0.680&   0.280&  14.25&       2&   12.91&      LN\\
    MWISP G210.981$-$02.064$+$010.80&   1.03& $-$0.214&  0.644&...&...&     228&    8.13&   2.06$\times10^{20}$&    1.28&      19&  0.090& $-$0.540&   0.380&   6.20&       0&   23.53&      LN\\
    MWISP G211.084$-$01.799$+$041.46&   4.85&...&...&...&...&     231&   12.41&   4.29$\times10^{20}$&    6.05&     930&  0.040& $-$1.070&   0.180&   8.52&       2&    6.33&       UN\\
    MWISP G211.309$-$00.403$+$021.33&   2.09& $-$1.503&  2.088&...&...&     458&   33.73&   1.95$\times10^{21}$&    3.67&    1565&  0.270& $-$0.800&   0.230&   5.25&      24&    1.17&      LN\\
    MWISP G211.558$-$02.368$+$010.94&   1.02& $-$0.278&  1.030&...&...&     335&    9.40&   4.26$\times10^{20}$&    1.53&      63& $-$0.030& $-$0.660&   0.360&   5.78&       2&    8.74&      LN\\
    MWISP G211.568$+$01.022$+$045.05&   5.38& $-$0.291&  1.775&  0.872& $-$2.368&     702&   16.65&   1.22$\times10^{21}$&   11.70&   10528& $-$0.040& $-$0.730&   0.300&   8.24&      10&    0.98&   LN$+$PL\\
    MWISP G211.671$+$02.298$+$007.35&   0.71& $-$0.335&  0.855&...&...&    2938&   12.53&   4.35$\times10^{20}$&    3.16&     268&  0.160& $-$0.550&   0.260&  13.53&       0&   24.76&      LN\\
    MWISP G211.994$+$04.394$+$006.70&   0.64& $-$0.497&  0.953&...&...&    1037&   10.63&   2.93$\times10^{20}$&    1.69&      50&  0.260& $-$0.600&   0.240&  36.12&       0&  462.19&      LN\\
    MWISP G212.098$-$03.656$+$016.83&   1.57& $-$0.346&  1.237&...&...&     277&   14.54&   1.14$\times10^{21}$&    2.14&     333& $-$0.260& $-$0.760&   0.270&   6.09&       0&    3.77&      LN\\
    MWISP G212.141$-$01.044$+$044.18&   5.11& $-$1.016&  1.488&...&...&    1097&   21.13&   9.54$\times10^{20}$&   13.89&   11111&  0.300& $-$0.560&   0.210&  12.21&      61&    2.57&      LN\\
    MWISP G212.241$+$04.926$+$009.29&   0.87& $-$0.445&  1.212&...&...&     265&    9.67&   3.76$\times10^{20}$&    1.16&      30&  0.020& $-$0.990&   0.290&   5.50&       0&   12.59&      LN\\
    MWISP G212.331$-$03.313$+$014.88&   1.37& $-$0.830&  1.468&...&...&     580&   14.82&   9.33$\times10^{20}$&    2.71&     407&  0.240& $-$0.790&   0.250&  11.72&       3&   13.03&      LN\\
    MWISP G212.966$+$01.223$+$042.73&   4.74& $-$0.660&  1.192&...&...&    1388&   18.20&   8.10$\times10^{20}$&   14.49&   10515&  0.200& $-$0.640&   0.300&  11.98&      23&    2.33&      LN\\
    MWISP G213.069$-$03.666$+$052.82&   6.45& $-$0.475&  1.001&...&...&     398&   16.20&   5.10$\times10^{20}$&   10.56&    3441&  0.240& $-$0.450&   0.260&   4.46&       4&    0.80&      LN\\
    MWISP G213.093$-$04.027$+$020.63&   1.91& $-$0.985&  1.481&...&...&     207&    9.45&   4.39$\times10^{20}$&    2.25&     128&  0.520& $-$0.560&   0.330&   4.94&       1&    4.96&      LN\\
    MWISP G213.358$+$00.364$+$043.28&   4.76& $-$0.195&  0.789&...&...&     403&   11.13&   4.31$\times10^{20}$&    7.84&    1586&  0.000& $-$0.740&   0.270&  17.57&       3&   16.55&      LN\\
    MWISP G213.966$+$00.763$+$044.10&   4.80&...&...&...&...&     431&   12.27&   7.54$\times10^{20}$&    8.18&    3071& $-$0.040& $-$1.010&   0.340&  12.63&       9&    4.53&       UN\\
    MWISP G214.135$+$00.385$+$009.23&   0.82& $-$0.384&  1.236&...&...&     202&   12.00&   3.22$\times10^{20}$&    0.96&      17& $-$0.010& $-$0.980&   0.240&   0.68&       0&    2.45&      LN\\
    MWISP G214.300$+$03.277$+$014.69&   1.30& $-$0.325&  0.891&...&...&     175&   10.43&   2.61$\times10^{20}$&    1.41&      30&  0.070& $-$0.670&   0.220&   7.61&       0&   29.57&      LN\\
    MWISP G214.689$-$01.865$+$027.63&   2.58&  0.353&  1.379&  0.516& $-$2.656&    3138&   14.54&   1.03$\times10^{21}$&   11.86&    9430& $-$0.410& $-$0.270&   0.340&  14.11&      40&    3.10&   LN$+$PL\\
    MWISP G214.739$-$03.208$+$010.74&   0.93&...&...&...&...&     182&   11.01&   3.92$\times10^{20}$&    1.03&      24& $-$0.080& $-$0.800&   0.270&   9.41&       0&   45.54&       UN\\
    MWISP G214.903$+$04.893$+$011.00&   0.96& $-$0.024&  0.594&...&...&     218&   10.32&   2.14$\times10^{20}$&    1.16&      17& $-$0.180& $-$0.860&   0.190&   2.90&       0&    9.20&      LN\\
    MWISP G214.979$+$00.850$+$047.55&   5.18& $-$0.720&  1.256&...&...&    2037&   13.20&   8.80$\times10^{20}$&   19.18&   20125&  0.220& $-$0.610&   0.350&  15.52&      55&    2.30&      LN\\
    MWISP G215.081$+$04.151$+$009.43&   0.82& $-$0.114&  0.918&...&...&     273&   11.05&   4.66$\times10^{20}$&    1.11&      37& $-$0.310& $-$0.600&   0.320&   2.70&       0&    3.71&      LN\\
    MWISP G215.242$-$00.376$+$028.93&   2.69& $-$0.273&  0.828&...&...&     902&   11.94&   3.73$\times10^{20}$&    6.63&     985&  0.080& $-$0.730&   0.260&  12.83&       3&   12.22&      LN\\
    MWISP G216.184$-$00.070$+$023.77&   2.10& $-$0.716&  1.704&...&...&     219&   17.69&   1.04$\times10^{21}$&    2.55&     414&  0.000& $-$0.830&   0.240&  11.27&       2&   12.59&      LN\\
    MWISP G216.584$-$02.515$+$025.52&   2.25&...&...&...&...&   19899&   16.32&   5.76$\times10^{20}$&   26.04&   24393&  0.280& $-$0.310&   0.310&  37.70&      71&   15.06&       UN\\
    MWISP G216.941$-$03.700$+$053.55&   5.78& $-$0.205&  0.812&...&...&     316&    9.46&   3.75$\times10^{20}$&    8.43&    1647& $-$0.080& $-$0.600&   0.300&   6.56&       1&    2.34&      LN\\
    MWISP G217.055$+$00.622$+$050.10&   5.25& $-$0.435&  0.934&...&...&    2309&   16.65&   5.29$\times10^{20}$&   20.70&   13620&  0.300& $-$0.290&   0.300&  20.75&      44&    7.00&      LN\\
    MWISP G217.154$-$03.279$+$011.67&   0.97&  0.380&  1.326&  0.743& $-$5.699&     198&   11.69&   5.33$\times10^{20}$&    1.12&      42& $-$0.440& $-$0.840&   0.270&   5.39&       0&   10.20&   LN$+$PL\\
    MWISP G217.258$-$00.253$+$015.94&   1.34& $-$0.473&  1.242&...&...&     405&   14.22&   1.20$\times10^{21}$&    2.21&     378& $-$0.020& $-$0.590&   0.270&   5.65&      14&    2.94&      LN\\
    MWISP G217.271$-$01.410$+$049.84&   5.17& $-$1.169&  1.416&...&...&     391&   20.75&   1.03$\times10^{21}$&    8.39&    4435&  0.590&  0.250&   0.250&  12.06&      13&    3.59&      LN\\
    MWISP G217.434$-$02.202$+$041.72&   4.03& $-$0.286&  0.890&...&...&     234&   10.13&   4.59$\times10^{20}$&    5.06&     736& $-$0.040& $-$0.720&   0.390&   8.06&       2&    4.33&      LN\\
    MWISP G217.485$-$02.736$+$011.62&   0.96&  0.356&  1.901&...&...&     336&   22.96&   3.13$\times10^{21}$&    1.44&     402& $-$0.520& $-$0.780&   0.290&   4.12&       6&    1.59&      LN\\
    MWISP G217.742$+$00.196$+$022.01&   1.87& $-$0.396&  0.970&...&...&     354&   12.36&   4.28$\times10^{20}$&    2.89&     212&  0.150& $-$0.610&   0.280&   4.06&       0&    3.08&      LN\\
    MWISP G217.803$-$00.246$+$027.03&   2.36& $-$0.946&  1.698&  1.882& $-$4.058&    3709&   29.58&   2.76$\times10^{21}$&   11.79&   24300&  0.070& $-$0.740&   0.280&  13.08&     211&    1.37&   LN$+$PL\\
    MWISP G218.250$-$01.440$+$064.17&   7.41& $-$0.501&  1.285&...&...&     176&    7.32&   3.07$\times10^{20}$&    8.07&    1191&  0.190& $-$0.840&   0.400&   4.87&       0&    1.53&      LN\\
    MWISP G218.296$-$04.581$+$009.03&   0.73& $-$0.145&  1.139&...&...&     312&   11.29&   4.74$\times10^{20}$&    1.06&      33& $-$0.310& $-$0.890&   0.250&   6.58&       0&   17.52&      LN\\
    MWISP G219.597$-$03.936$+$011.39&   0.91& $-$0.271&  1.128&...&...&     421&   11.03&   6.33$\times10^{20}$&    1.53&      91& $-$0.170& $-$0.860&   0.400&   3.81&       1&    3.34&      LN\\
    MWISP G220.155$-$01.985$+$032.80&   2.83& $-$0.209&  0.711&...&...&     519&   10.26&   3.13$\times10^{20}$&    5.29&     527&  0.160& $-$0.580&   0.290&  16.99&       4&   29.69&      LN\\
    MWISP G220.200$-$01.255$+$018.24&   1.47& $-$0.568&  0.952&...&...&     245&   10.09&   2.59$\times10^{20}$&    1.89&      53&  0.410& $-$0.510&   0.260&   2.77&       0&    3.94&      LN\\
    MWISP G220.607$-$01.924$+$011.77&   0.93&  0.075&  1.630&  0.265& $-$1.653&    6971&   30.96&   1.04$\times10^{21}$&    6.37&    2703&  0.020& $-$0.310&   0.300&  18.27&      97&   10.39&   LN$+$PL\\
    MWISP G220.878$-$04.172$+$018.83&   1.50&...&...&...&...&     280&    9.95&   3.85$\times10^{20}$&    2.06&      99&  0.010& $-$0.980&   0.280&  13.45&       0&   38.99&       UN\\
    MWISP G221.180$-$04.865$+$014.69&   1.15&...&...&...&...&     439&   10.56&   3.44$\times10^{20}$&    1.98&      82& $-$0.240& $-$0.780&   0.310&  15.28&       0&   57.51&       UN\\
    MWISP G221.570$-$02.857$+$012.81&   1.00& $-$0.561&  1.265&  1.464& $-$2.968&    7426&   25.73&   1.04$\times10^{21}$&    7.07&    3275&  0.060& $-$0.670&   0.330&   8.29&     132&    1.96&   LN$+$PL\\
    MWISP G221.578$-$04.142$+$039.01&   3.40& $-$0.145&  0.748&...&...&     363&   10.70&   4.10$\times10^{20}$&    5.32&     712& $-$0.210& $-$0.670&   0.320&   3.71&       2&    1.35&      LN\\
    MWISP G221.933$-$02.107$+$039.51&   3.45&  0.478&  2.096&  0.599& $-$2.099&    2624&   28.92&   2.02$\times10^{21}$&   14.50&   27370& $-$0.270& $-$0.640&   0.280&   7.79&      74&    0.58&   LN$+$PL\\
    MWISP G222.233$+$01.220$+$028.83&   2.37&  0.641&  2.464&  0.922& $-$3.113&     647&   11.67&   1.37$\times10^{21}$&    4.95&    2139& $-$0.370& $-$0.910&   0.410&   6.39&       7&    1.28&   LN$+$PL\\
    MWISP G222.290$-$04.709$+$016.40&   1.27&...&...&...&...&    1672&   12.05&   5.68$\times10^{20}$&    4.26&     629&  0.080& $-$0.970&   0.330&   8.35&       3&    5.30&       UN\\
    MWISP G222.991$-$01.921$+$039.47&   3.39& $-$0.159&  0.702&...&...&     466&    9.87&   3.32$\times10^{20}$&    6.01&     732& $-$0.050& $-$0.680&   0.280&  11.47&       6&   11.53&      LN\\
    MWISP G223.299$-$02.601$+$012.39&   0.95& $-$0.433&  1.287&...&...&     306&   10.69&   4.42$\times10^{20}$&    1.36&      50& $-$0.030& $-$0.890&   0.320&  10.96&       0&   36.13&      LN\\
    MWISP G223.750$-$04.108$+$011.87&   0.90& $-$0.278&  0.971&...&...&    1139&   16.69&   5.71$\times10^{20}$&    2.49&     217& $-$0.070& $-$0.870&   0.220&   3.84&       5&    2.84&      LN\\
    MWISP G224.065$-$01.303$+$016.32&   1.25& $-$0.854&  1.712&  1.727& $-$4.000&   26660&   29.91&   2.51$\times10^{21}$&   16.75&   44993&  0.020& $-$0.800&   0.300&  15.06&     818&    1.26&   LN$+$PL\\
    MWISP G224.166$+$01.218$+$033.30&   2.73& $-$0.683&  1.264&...&...&     294&   22.27&   1.25$\times10^{21}$&    3.84&    1192&  0.030& $-$0.470&   0.200&   9.43&      12&    4.27&      LN\\
    MWISP G224.370$-$03.770$+$015.18&   1.15&  0.358&  2.196&  0.841& $-$2.927&    1216&   16.16&   1.11$\times10^{21}$&    3.29&     752& $-$0.230& $-$0.810&   0.300&   3.60&      26&    1.04&   LN$+$PL\\
    MWISP G224.530$-$02.548$+$012.35&   0.94& $-$0.775&  2.623&  1.591& $-$4.319&    1325&   36.48&   3.32$\times10^{21}$&    2.81&    1578& $-$0.080& $-$1.030&   0.200&  14.77&     109&    8.93&   LN$+$PL\\
    MWISP G224.653$-$03.173$+$013.74&   1.04& $-$0.420&  1.003&...&...&     619&   22.41&   7.70$\times10^{20}$&    2.12&     208&  0.100& $-$0.420&   0.170&   3.23&      15&    2.50&      LN\\
    MWISP G224.890$+$00.737$+$017.38&   1.33& $-$0.227&  0.933&...&...&     791&   11.39&   5.68$\times10^{20}$&    3.07&     351& $-$0.240& $-$0.610&   0.290&   3.84&       0&    1.83&      LN\\
    MWISP G224.918$-$01.217$+$012.58&   0.95&...&...&...&...&     228&   14.42&   1.52$\times10^{21}$&    1.18&     135& $-$0.050& $-$0.880&   0.420&   7.58&       0&    6.02&       UN\\
    MWISP G224.918$-$02.765$+$010.99&   0.83& $-$0.561&  1.527&...&...&     249&   20.63&   1.86$\times10^{21}$&    1.07&     137& $-$0.130& $-$0.740&   0.230&   6.16&      15&    6.00&      LN\\

%% file: pdf_residual_LN.tex
\begin{figure}
\subfigure{\includegraphics[trim=0cm 0cm 0cm 0cm, width= 0.23\linewidth, clip]{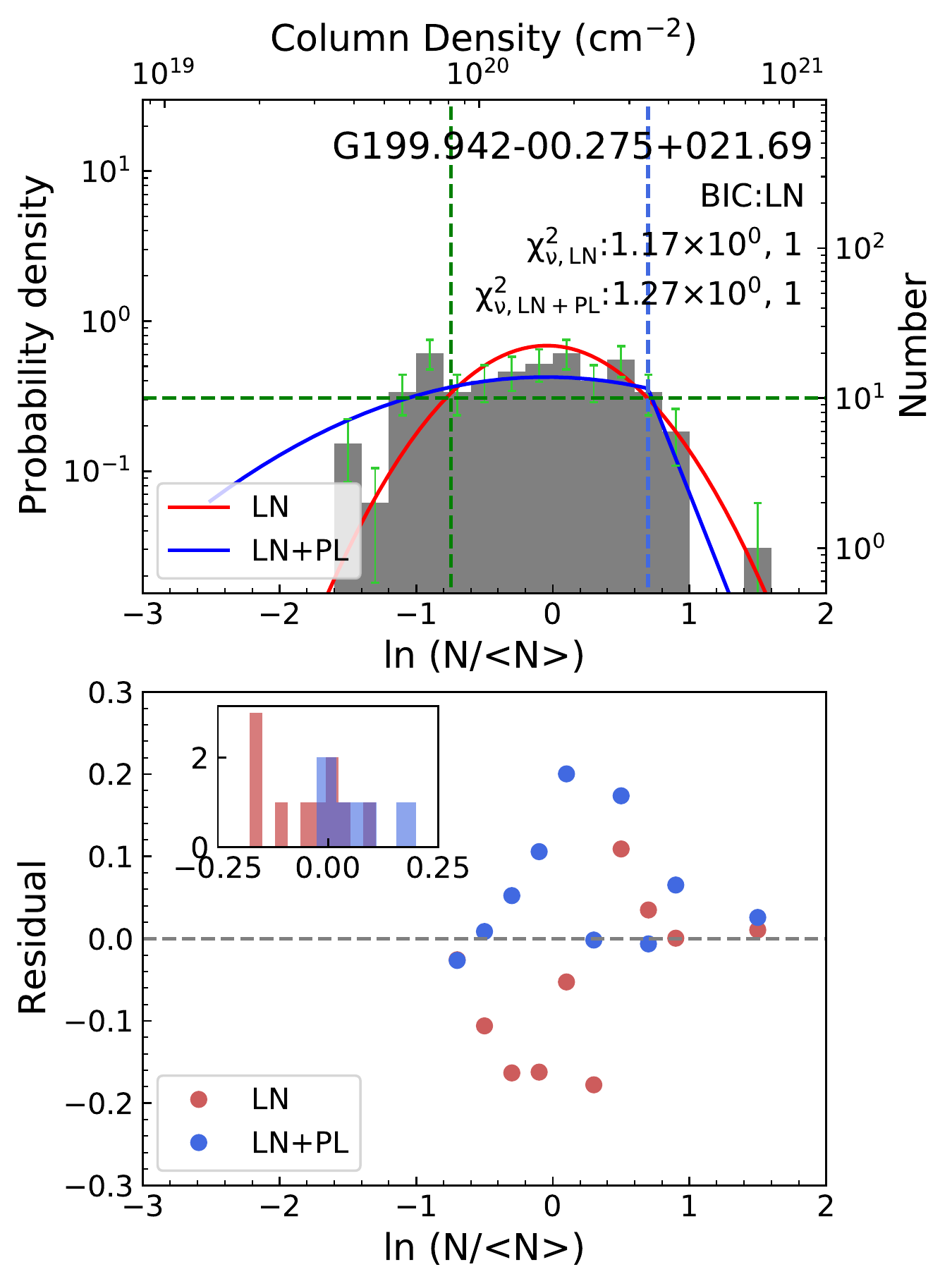}}
\subfigure{\includegraphics[trim=0cm 0cm 0cm 0cm, width= 0.23\linewidth, clip]{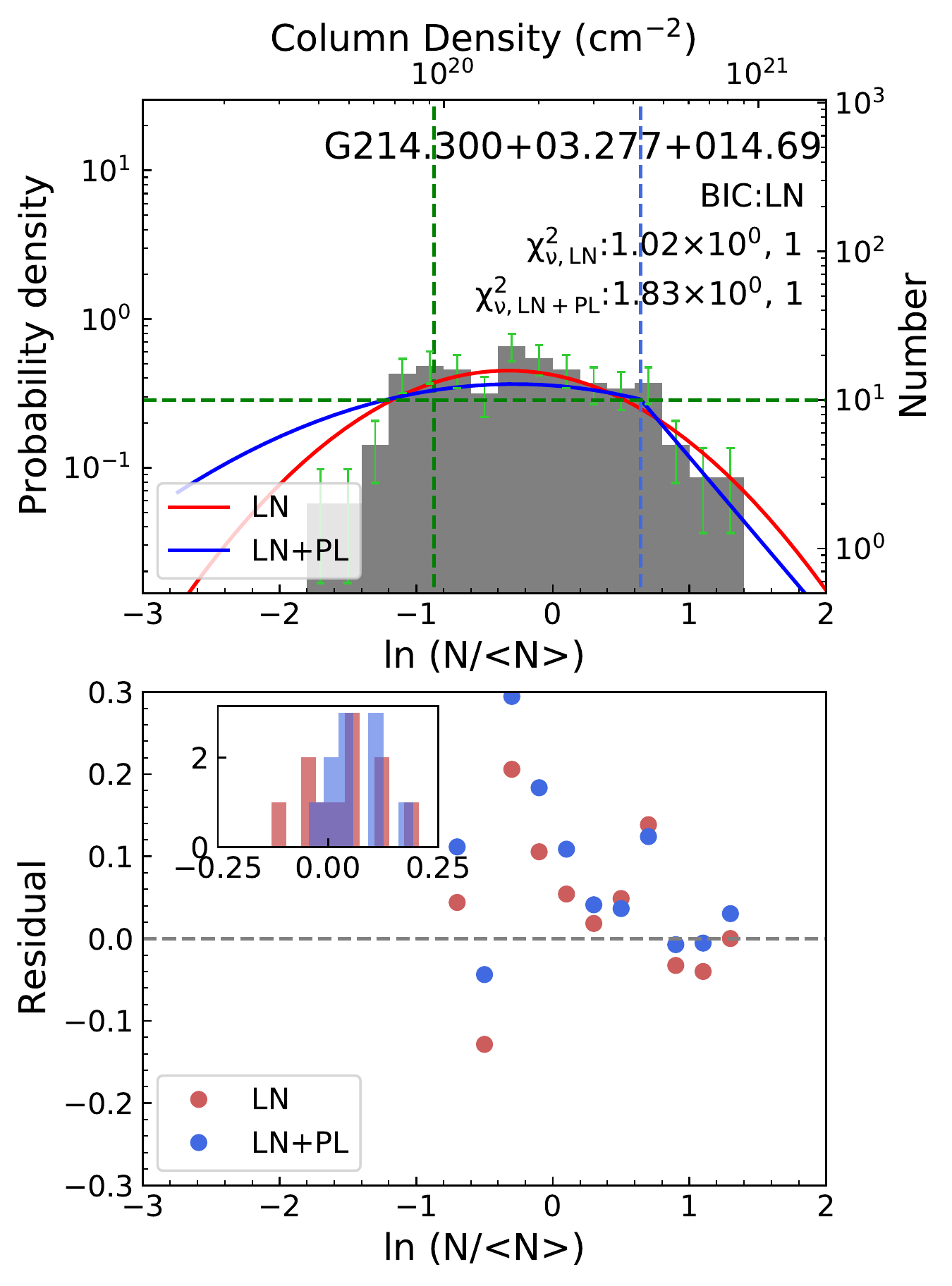}}
\subfigure{\includegraphics[trim=0cm 0cm 0cm 0cm, width= 0.23\linewidth, clip]{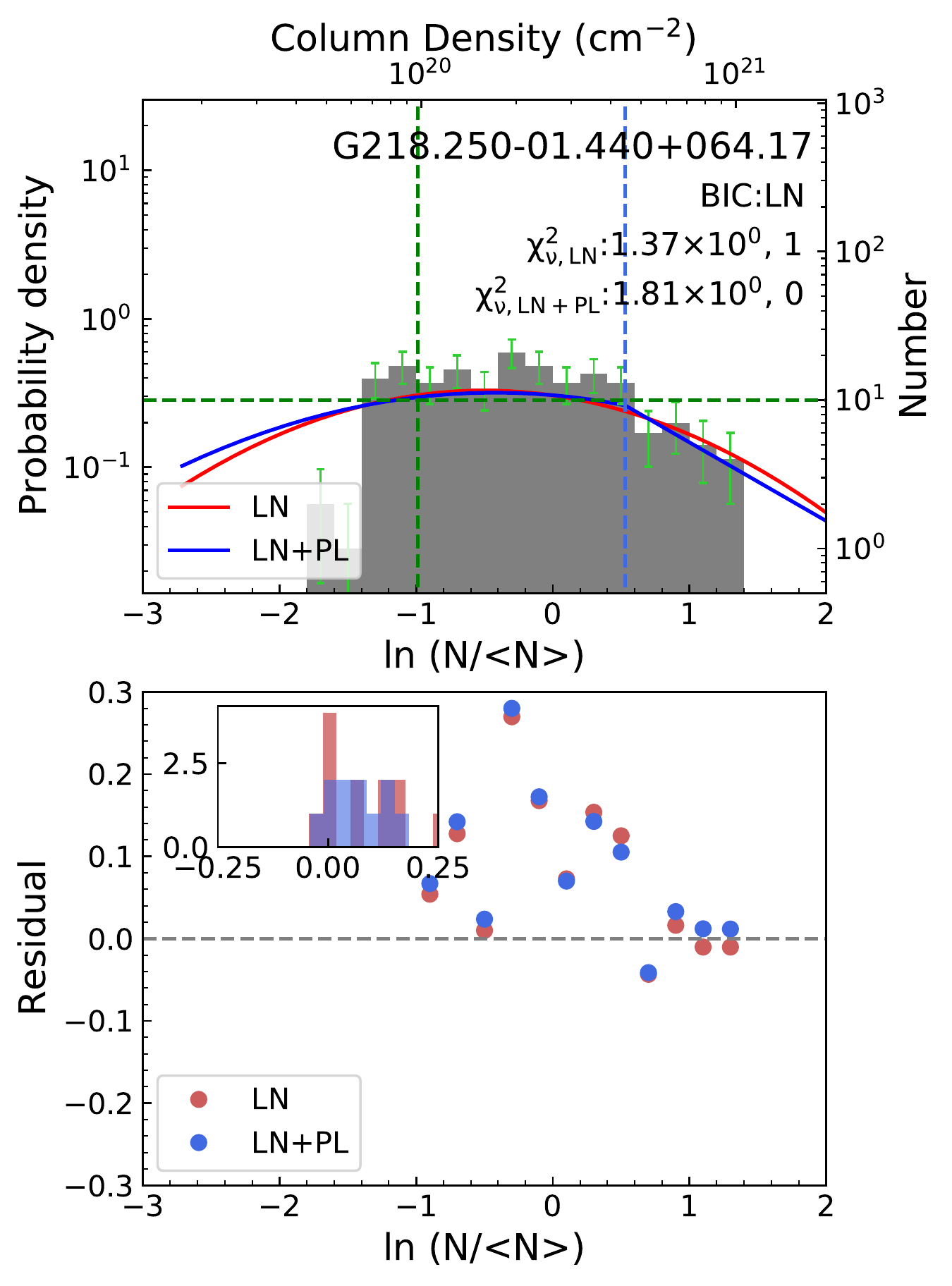}}
\subfigure{\includegraphics[trim=0cm 0cm 0cm 0cm, width= 0.23\linewidth, clip]{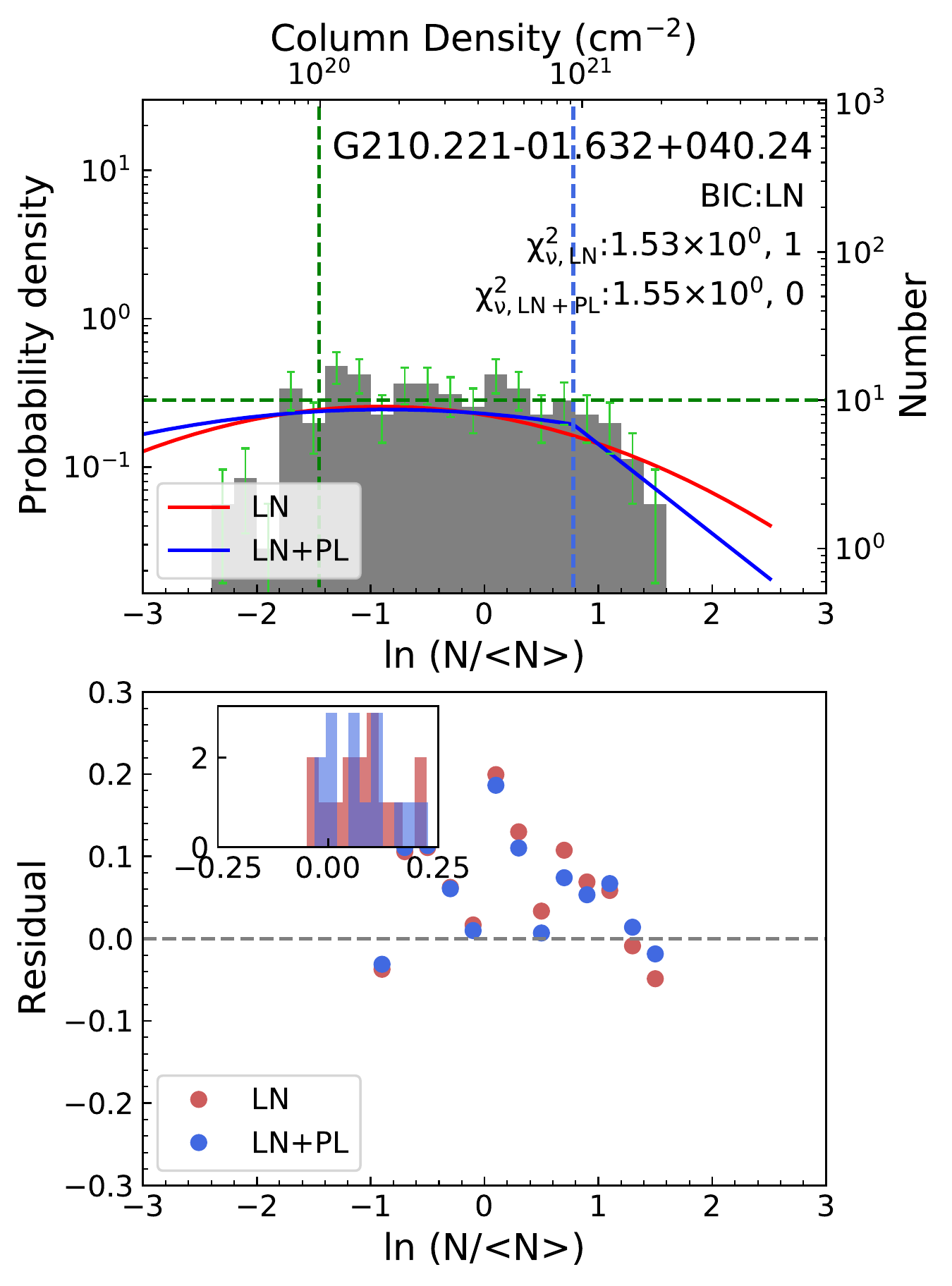}}

\subfigure{\includegraphics[trim=0cm 0cm 0cm 0cm, width= 0.23\linewidth, clip]{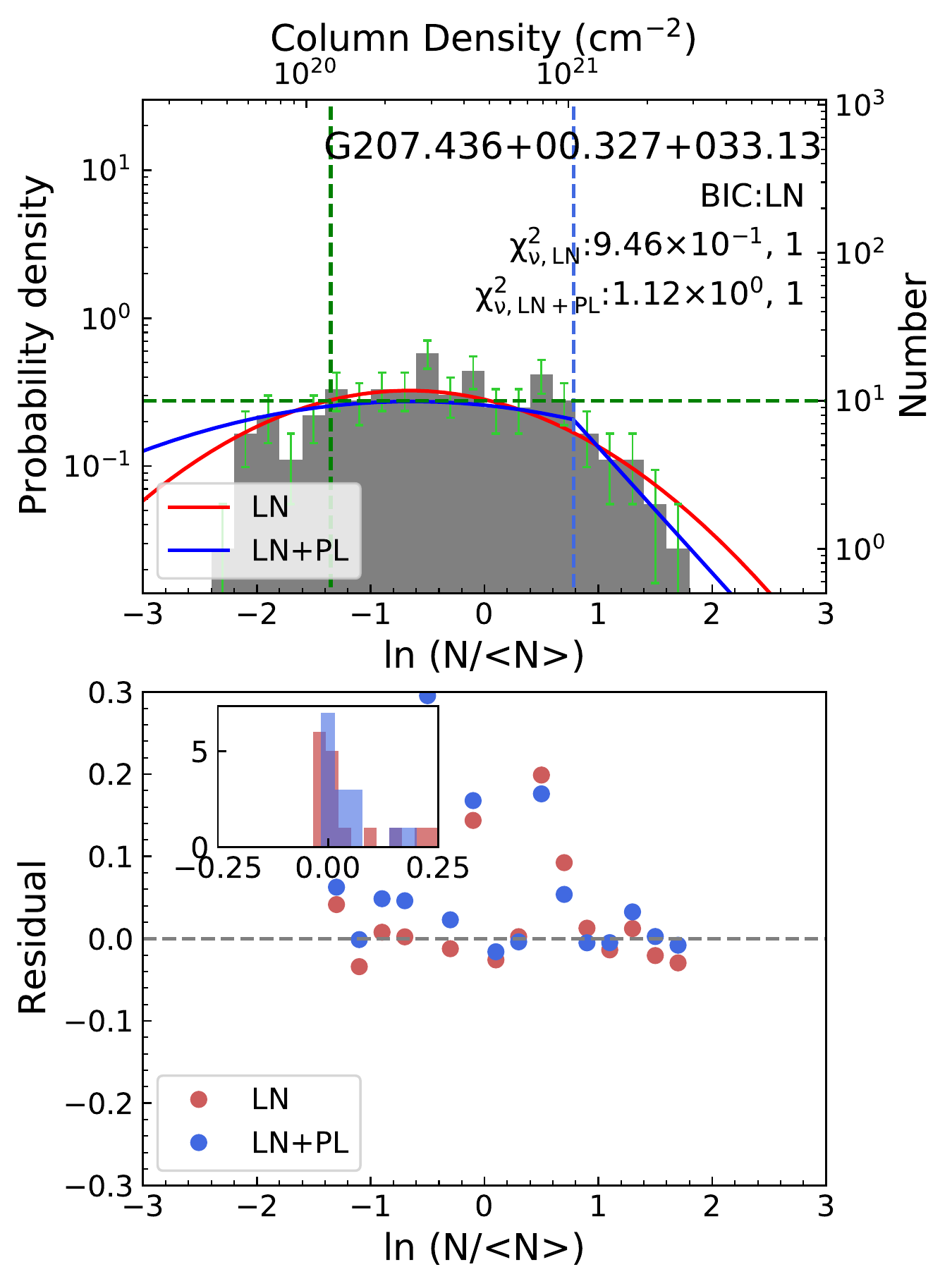}}
\subfigure{\includegraphics[trim=0cm 0cm 0cm 0cm, width= 0.23\linewidth, clip]{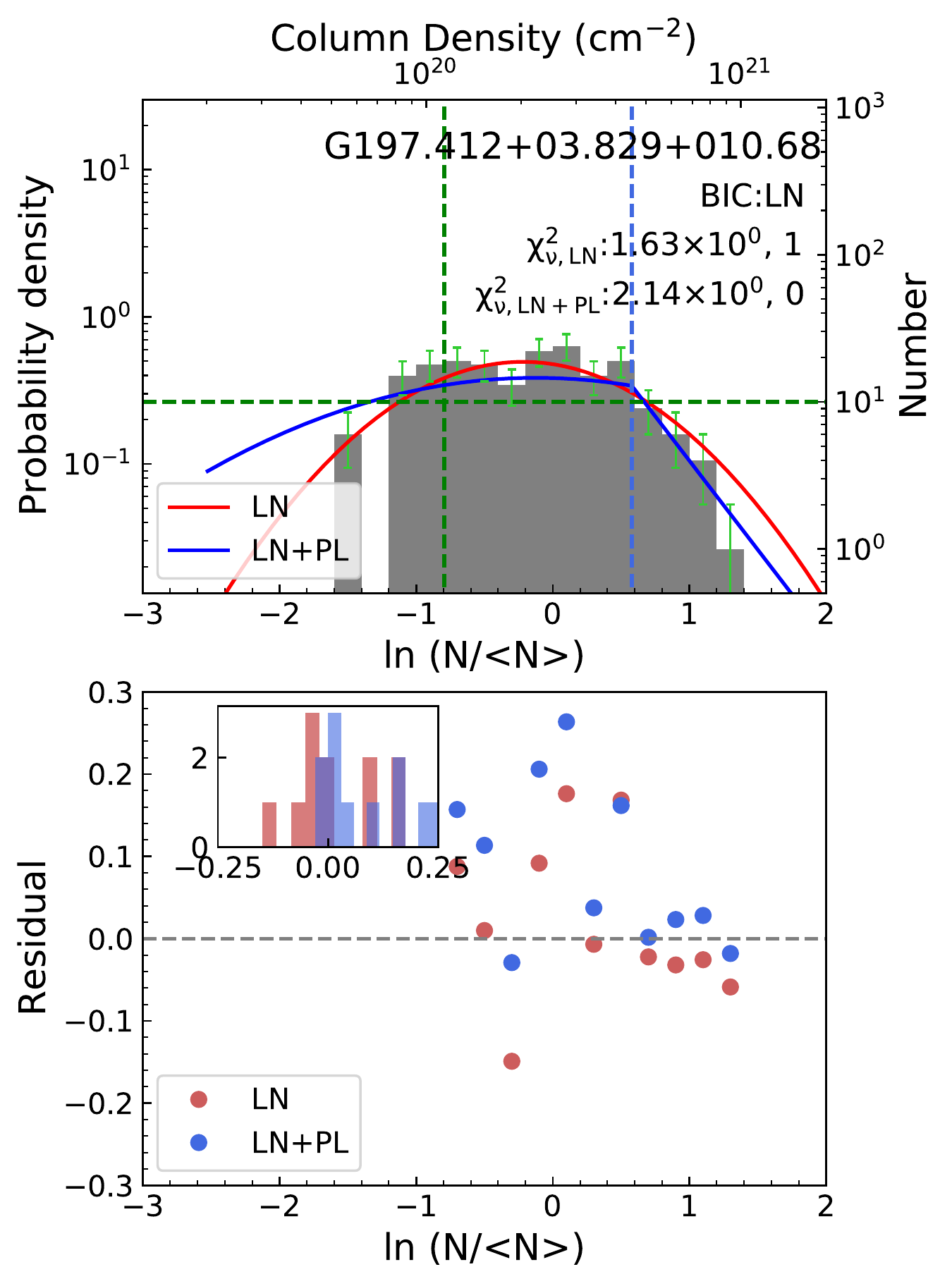}}
\subfigure{\includegraphics[trim=0cm 0cm 0cm 0cm, width= 0.23\linewidth, clip]{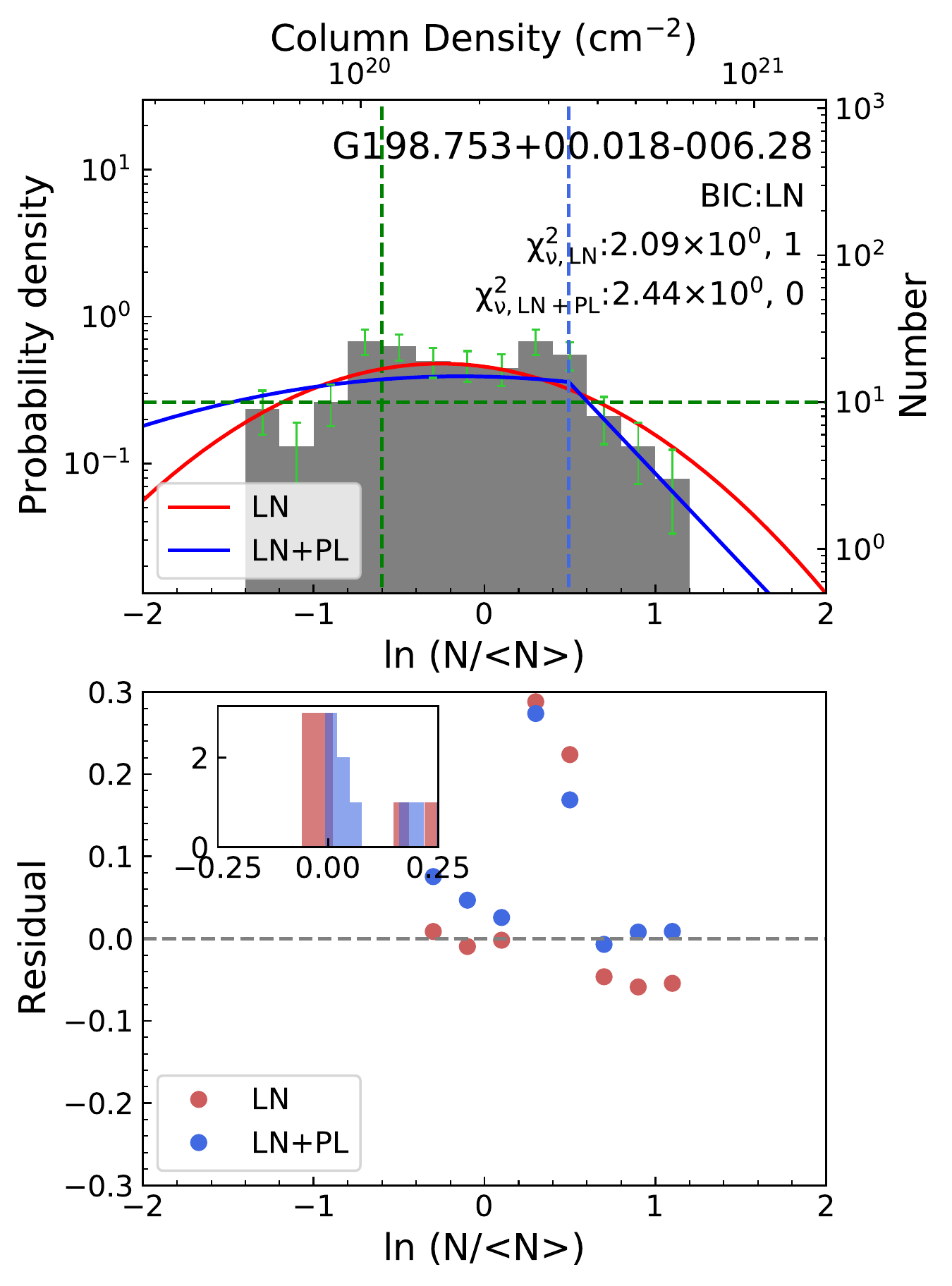}}
\subfigure{\includegraphics[trim=0cm 0cm 0cm 0cm, width= 0.23\linewidth, clip]{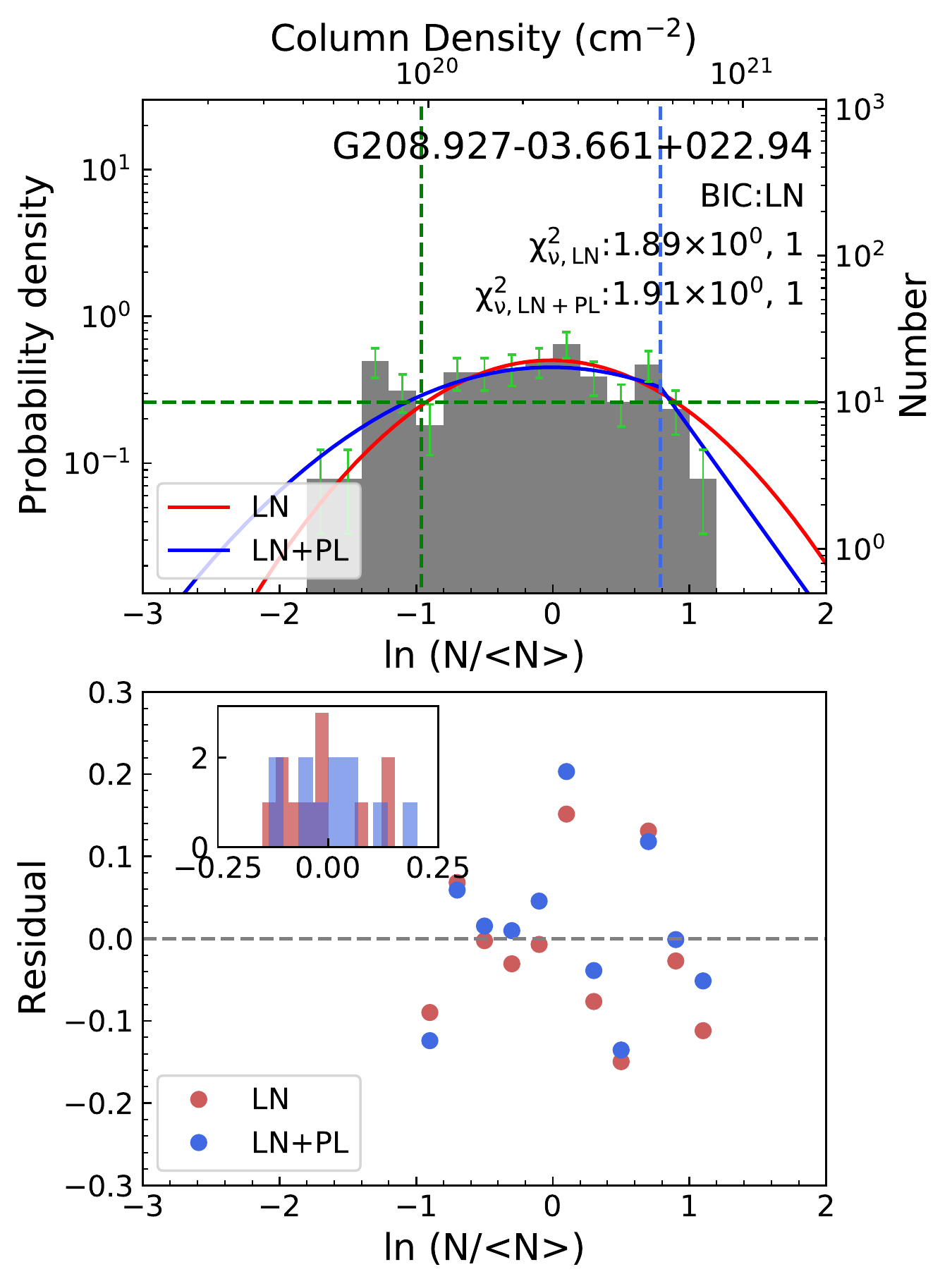}}

\subfigure{\includegraphics[trim=0cm 0cm 0cm 0cm, width= 0.23\linewidth, clip]{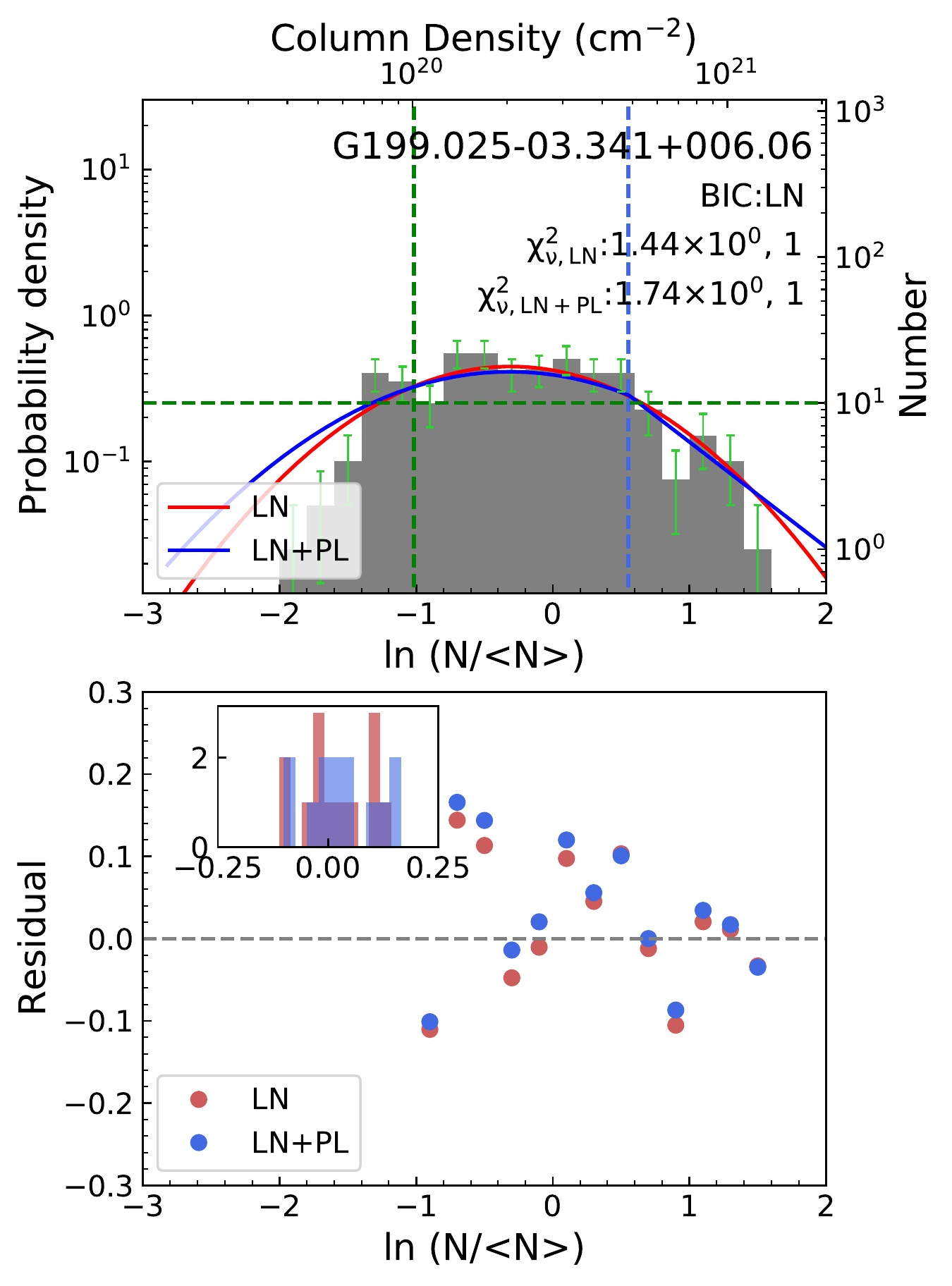}}
\subfigure{\includegraphics[trim=0cm 0cm 0cm 0cm, width= 0.23\linewidth, clip]{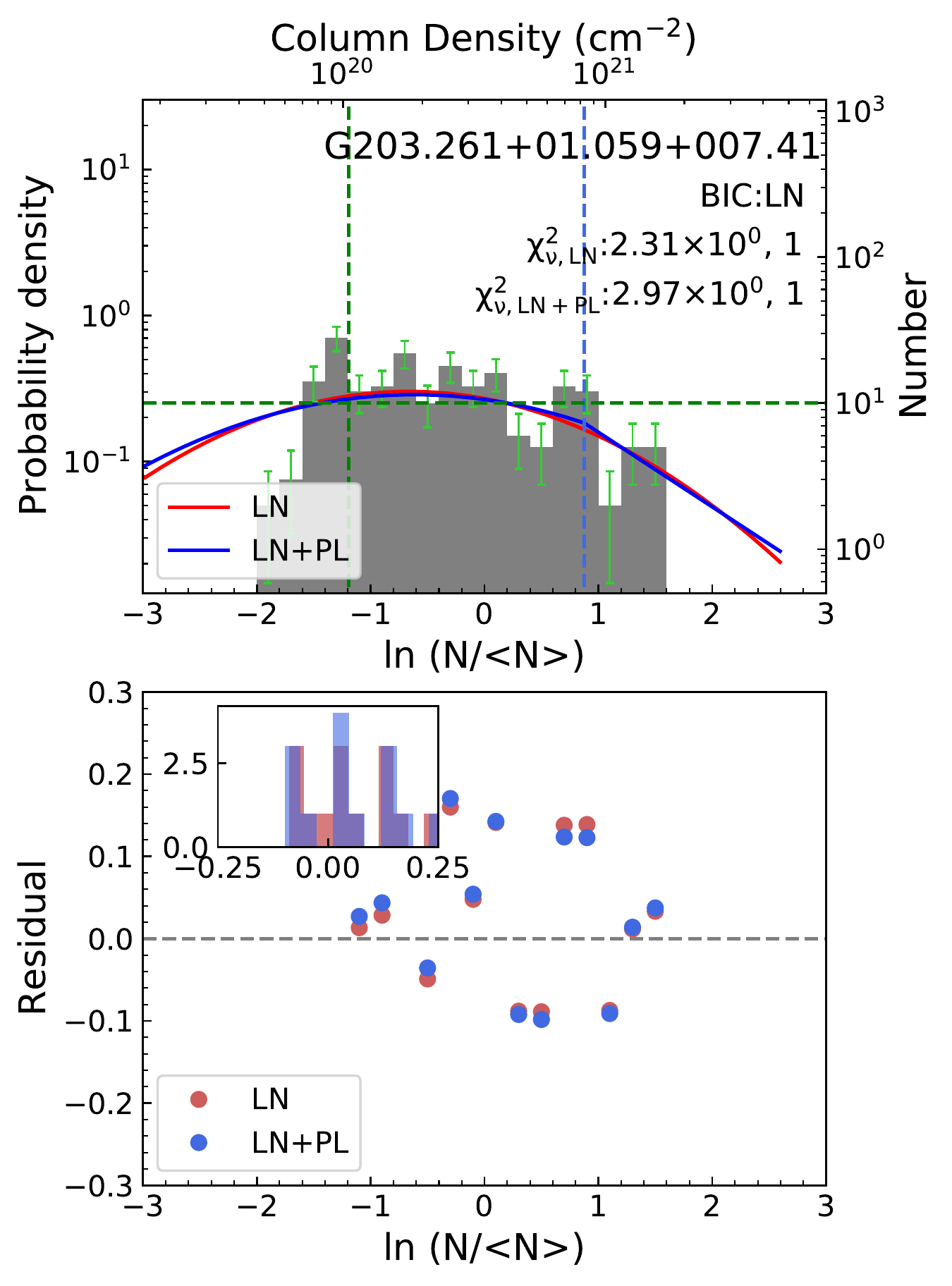}}
\subfigure{\includegraphics[trim=0cm 0cm 0cm 0cm, width= 0.23\linewidth, clip]{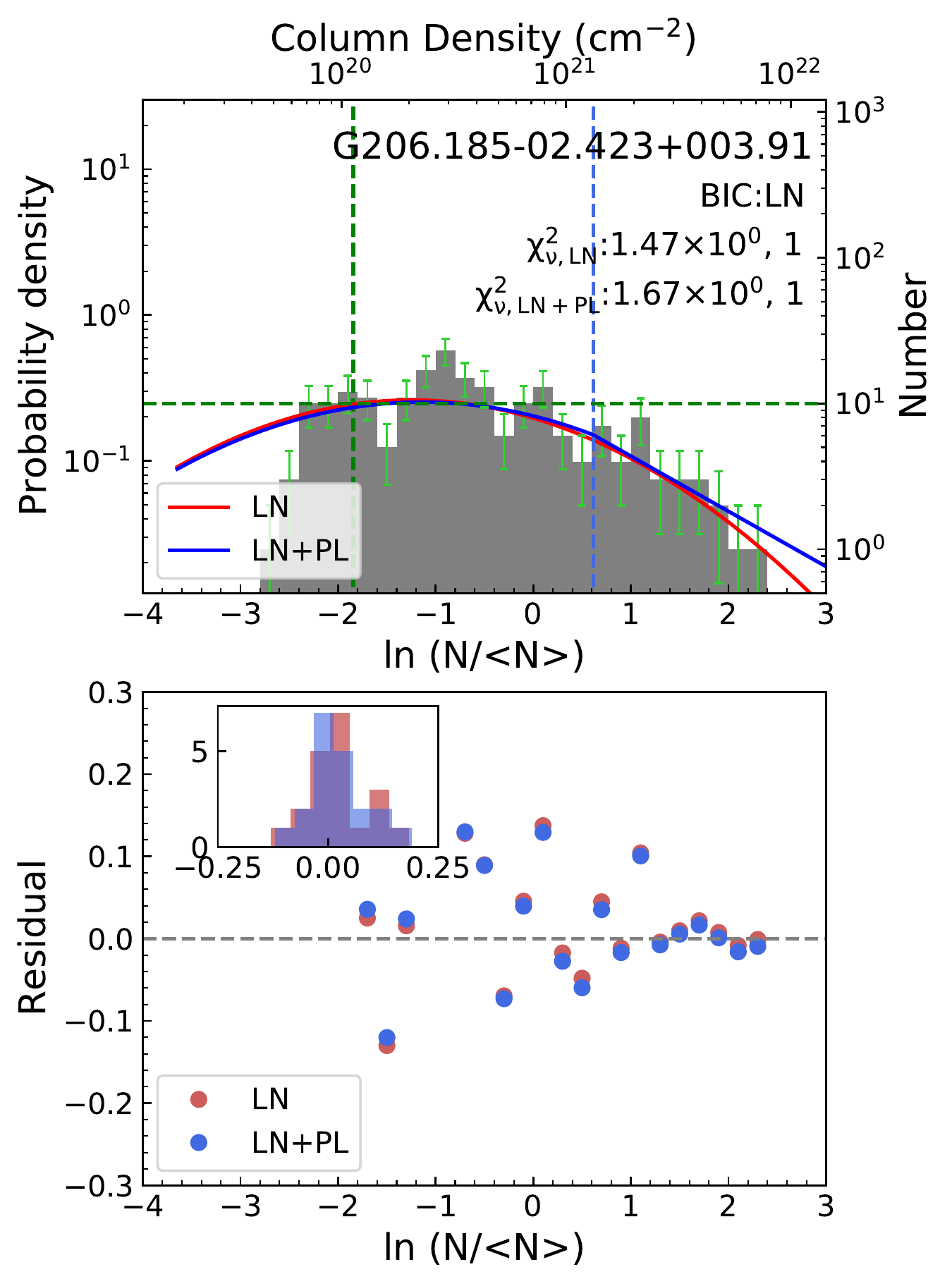}}
\subfigure{\includegraphics[trim=0cm 0cm 0cm 0cm, width= 0.23\linewidth, clip]{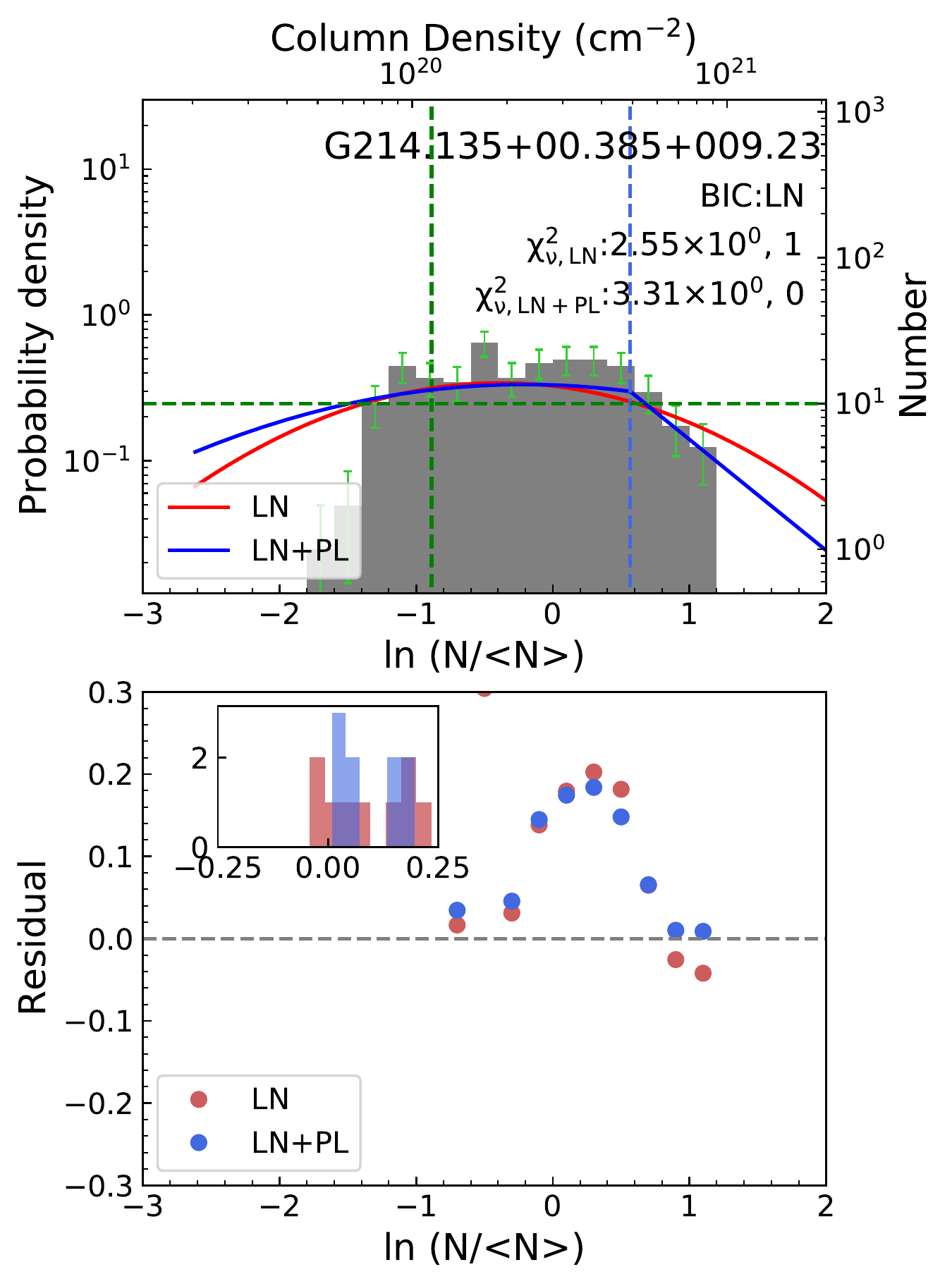}}

\subfigure{\includegraphics[trim=0cm 0cm 0cm 0cm, width= 0.23\linewidth, clip]{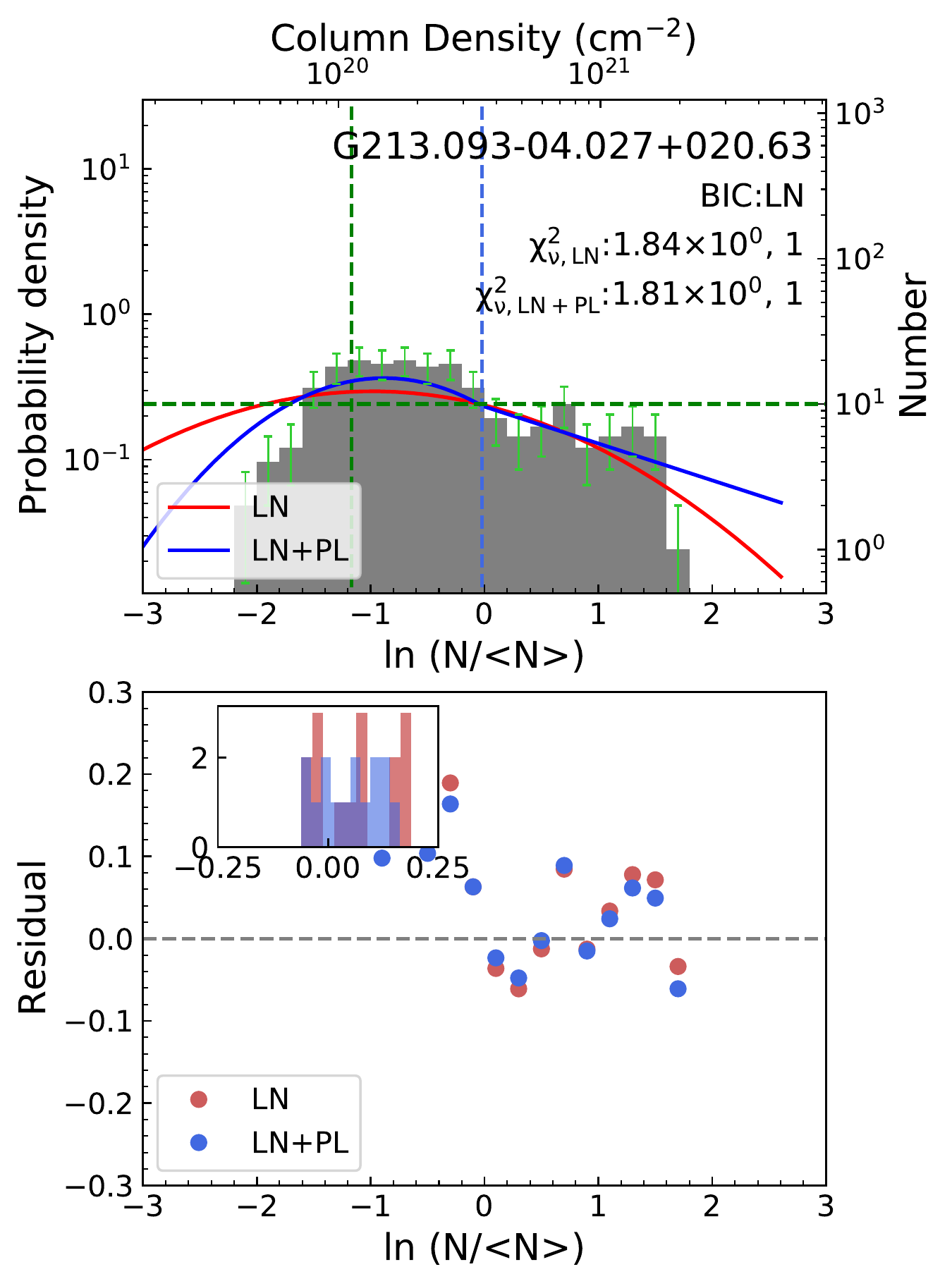}}
\subfigure{\includegraphics[trim=0cm 0cm 0cm 0cm, width= 0.23\linewidth, clip]{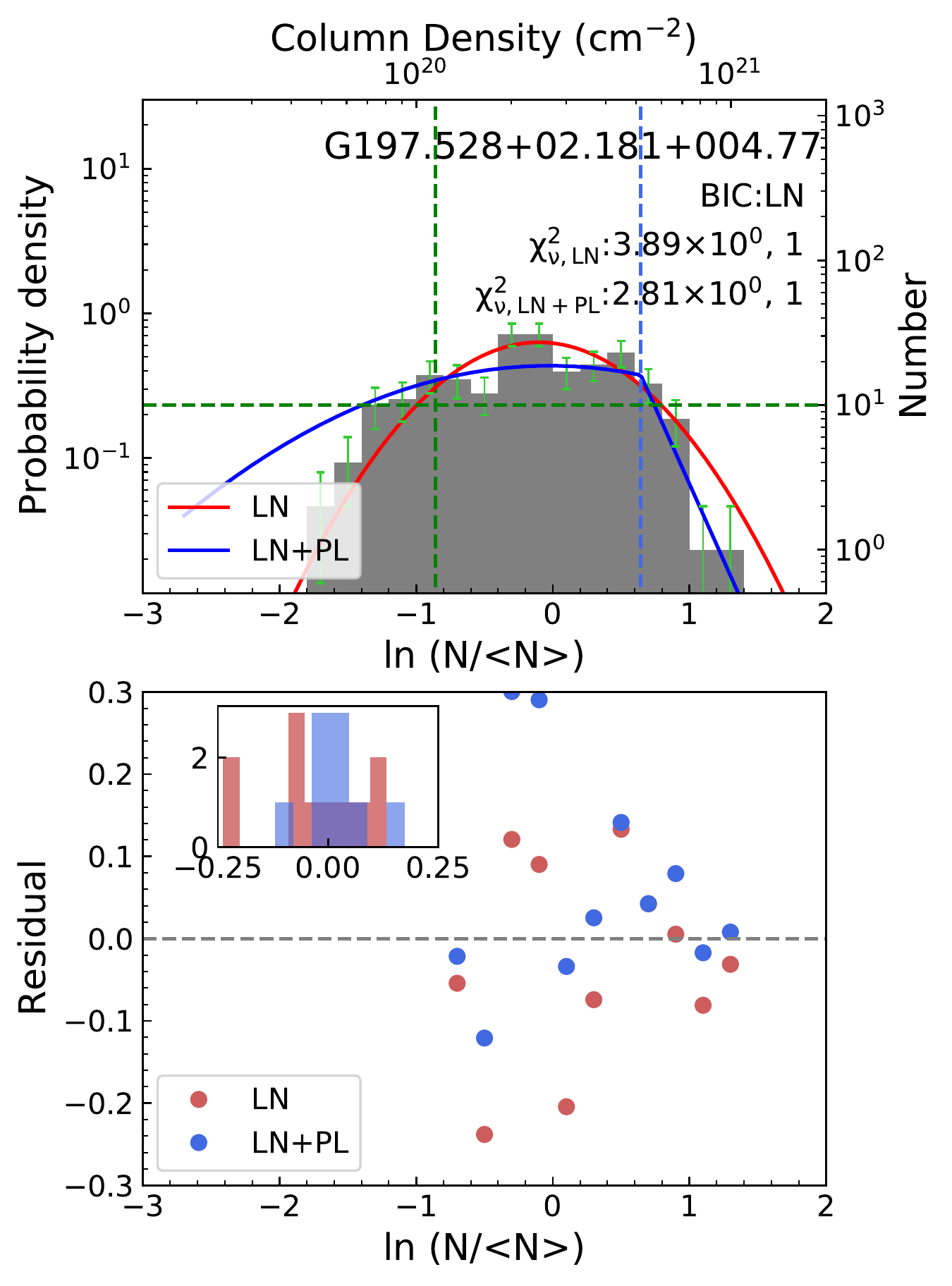}}
\subfigure{\includegraphics[trim=0cm 0cm 0cm 0cm, width= 0.23\linewidth, clip]{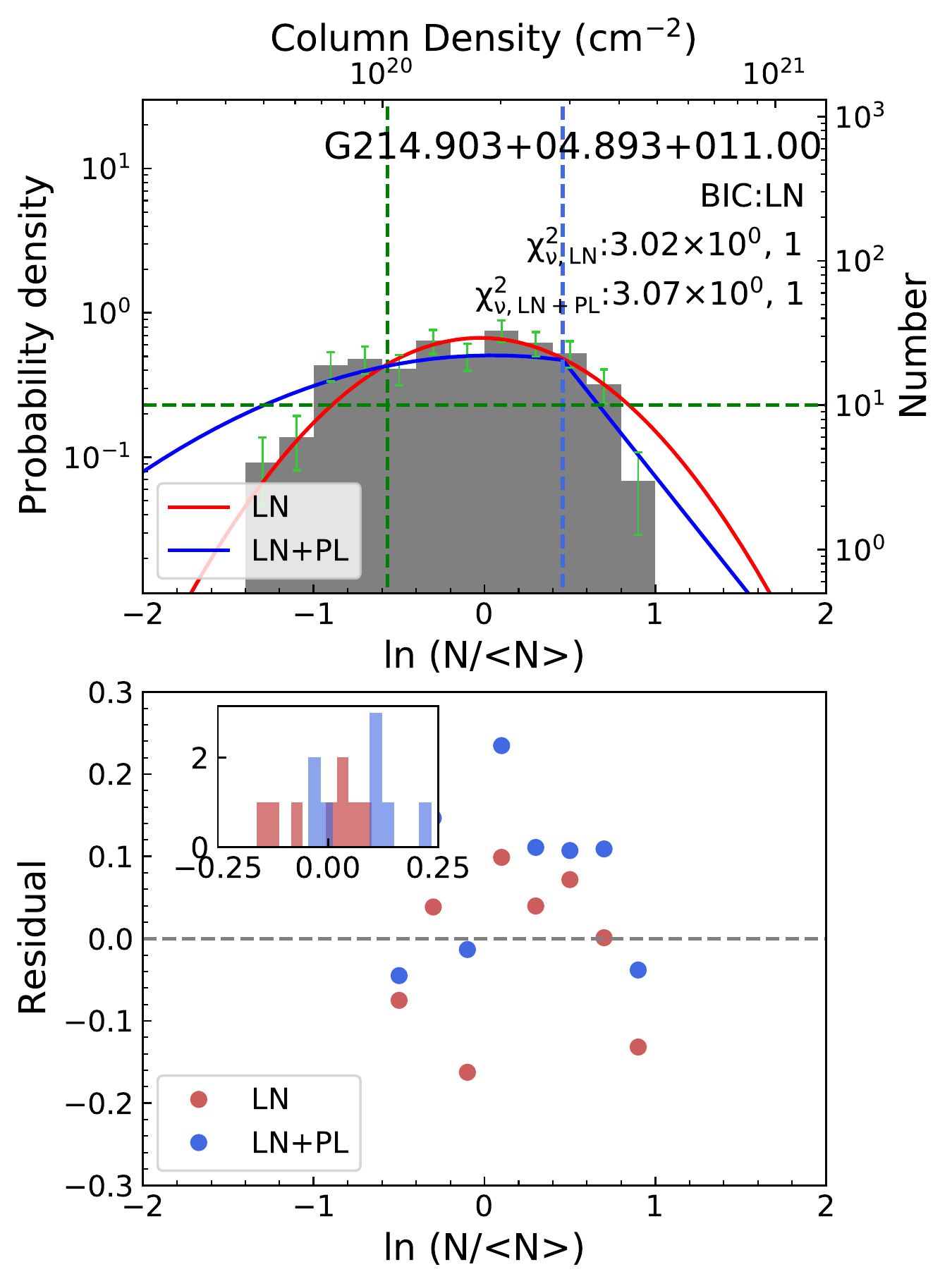}}
\subfigure{\includegraphics[trim=0cm 0cm 0cm 0cm, width= 0.23\linewidth, clip]{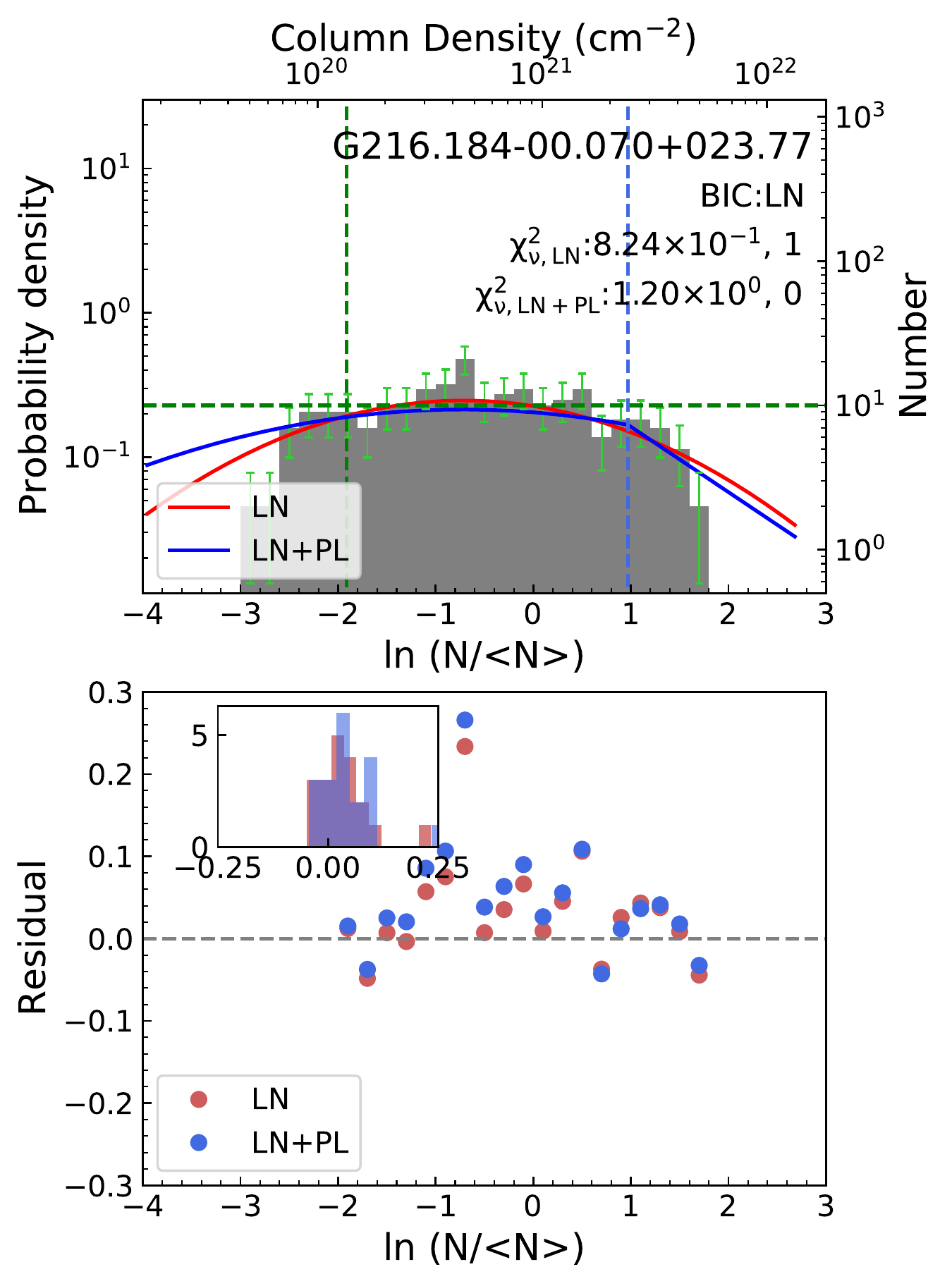}}

\end{figure}
\begin{figure}
\subfigure{\includegraphics[trim=0cm 0cm 0cm 0cm, width= 0.23\linewidth, clip]{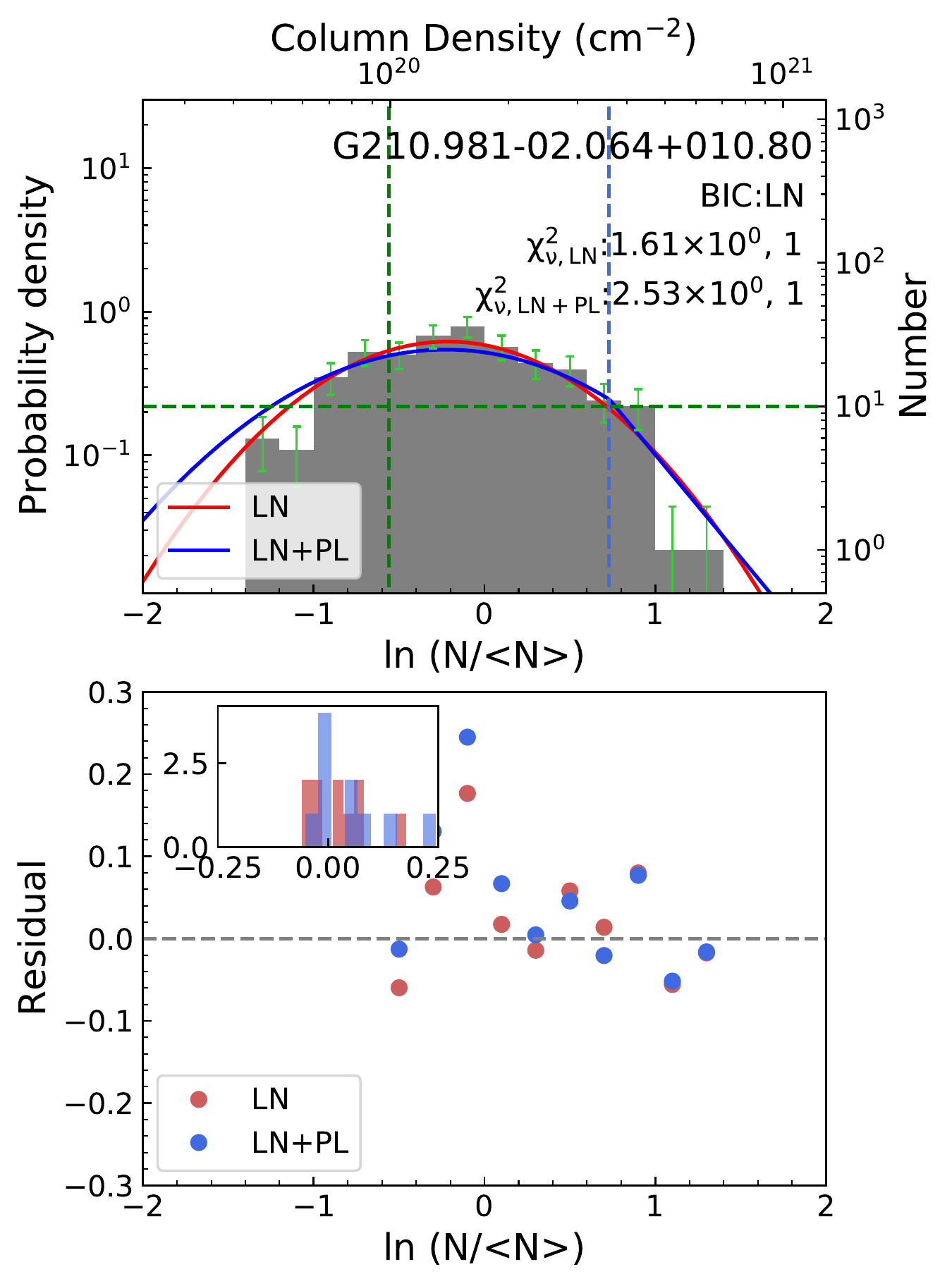}}
\subfigure{\includegraphics[trim=0cm 0cm 0cm 0cm, width= 0.23\linewidth, clip]{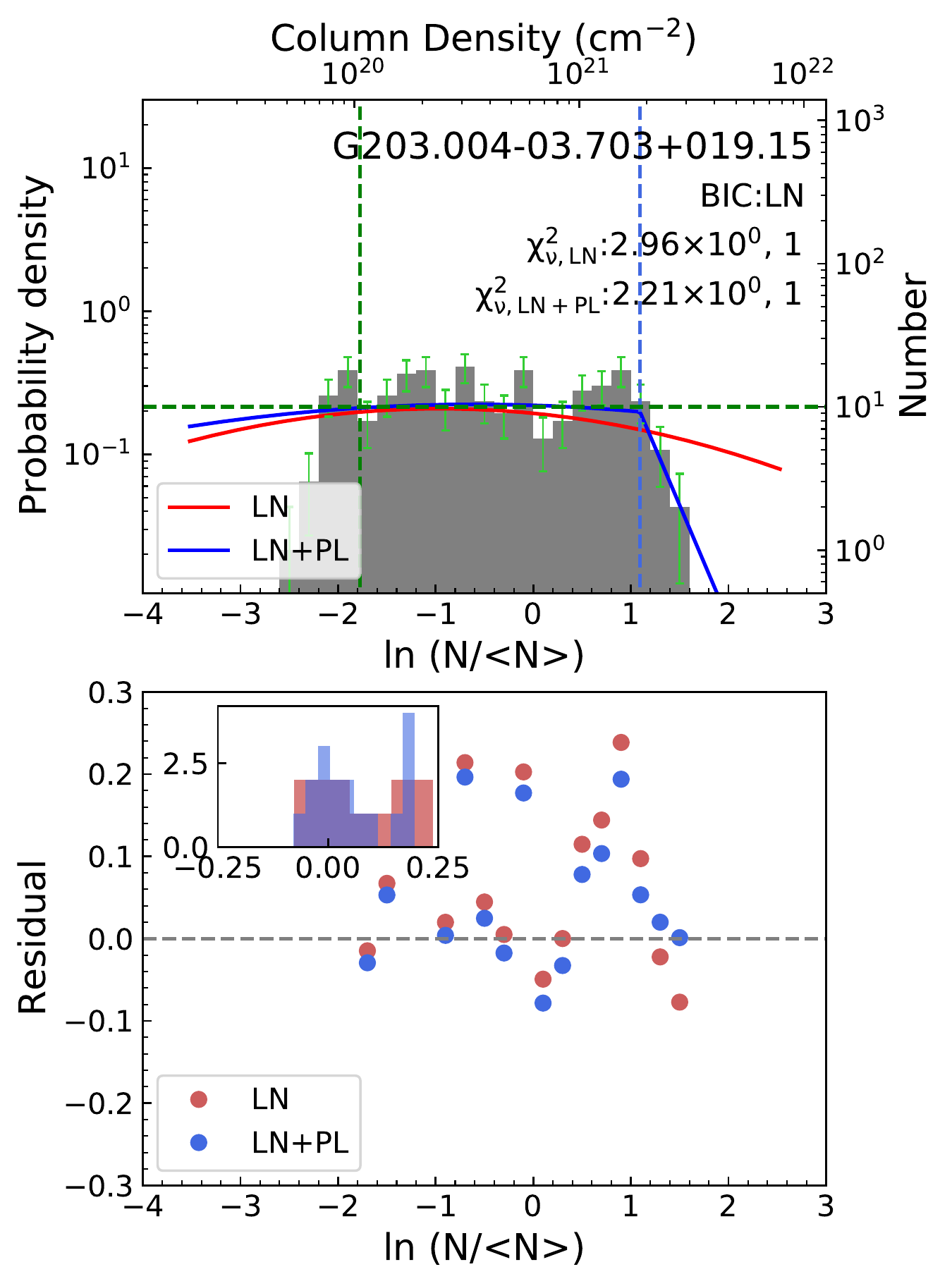}}
\subfigure{\includegraphics[trim=0cm 0cm 0cm 0cm, width= 0.23\linewidth, clip]{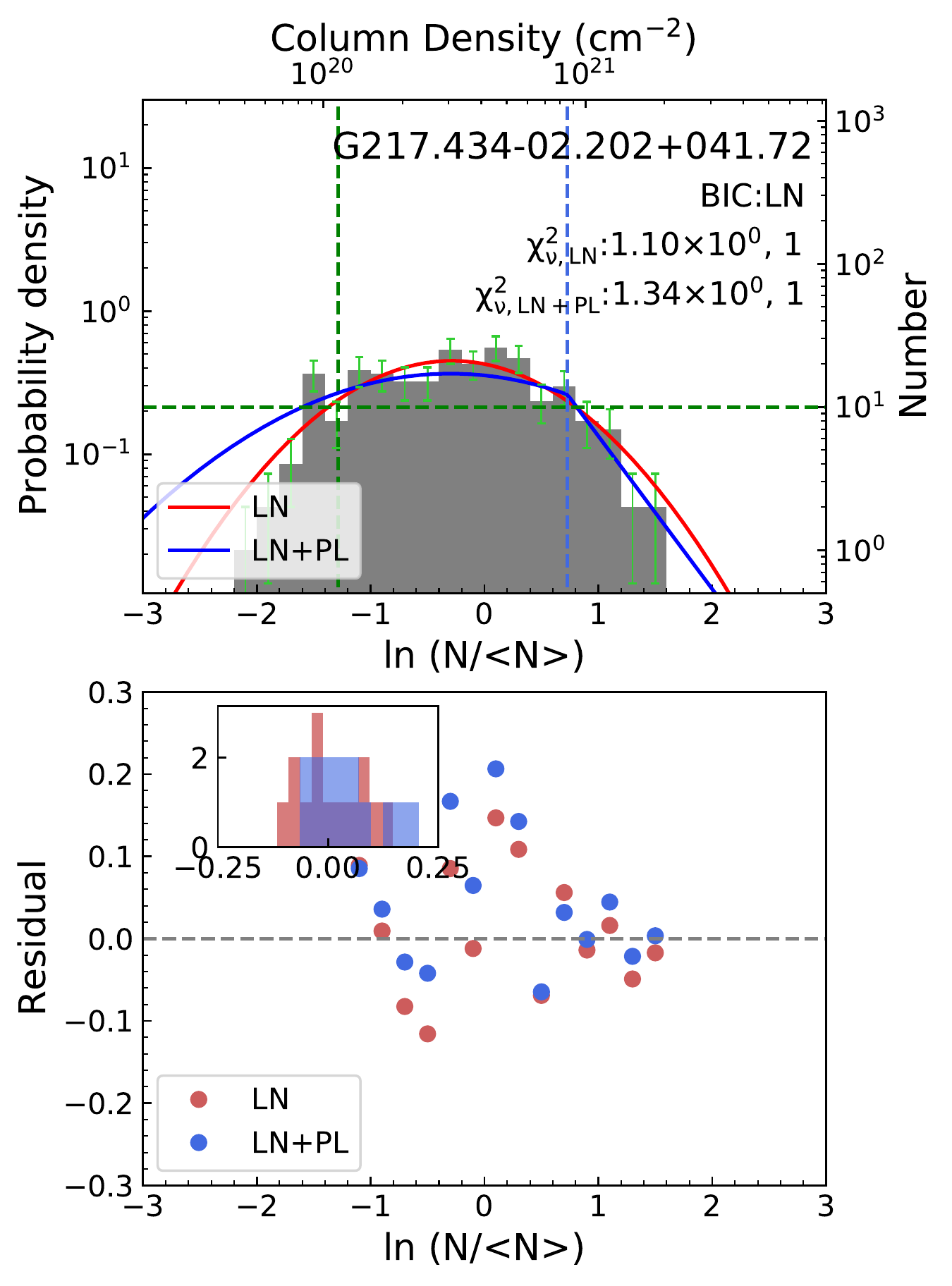}}
\subfigure{\includegraphics[trim=0cm 0cm 0cm 0cm, width= 0.23\linewidth, clip]{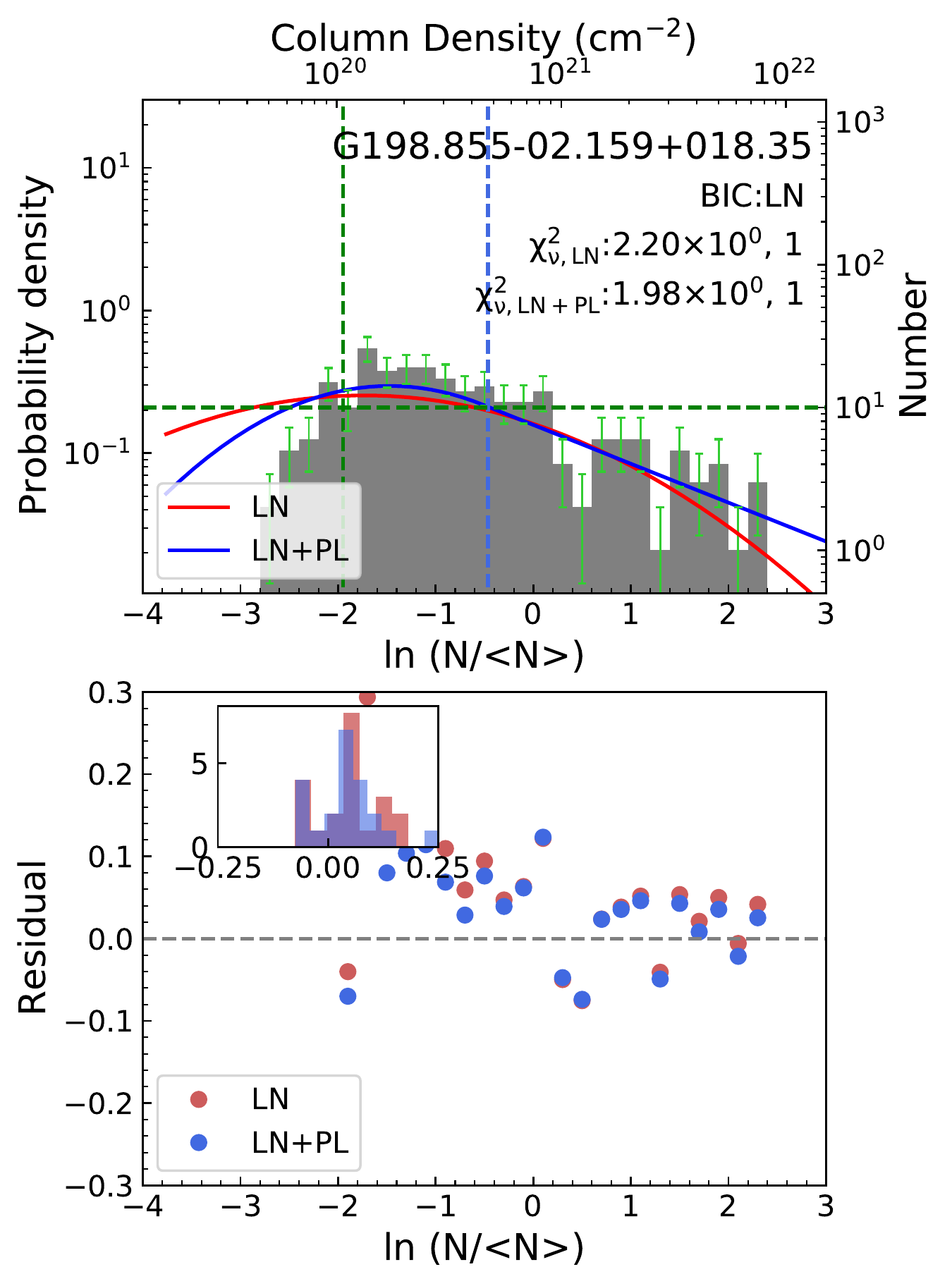}}

\subfigure{\includegraphics[trim=0cm 0cm 0cm 0cm, width= 0.23\linewidth, clip]{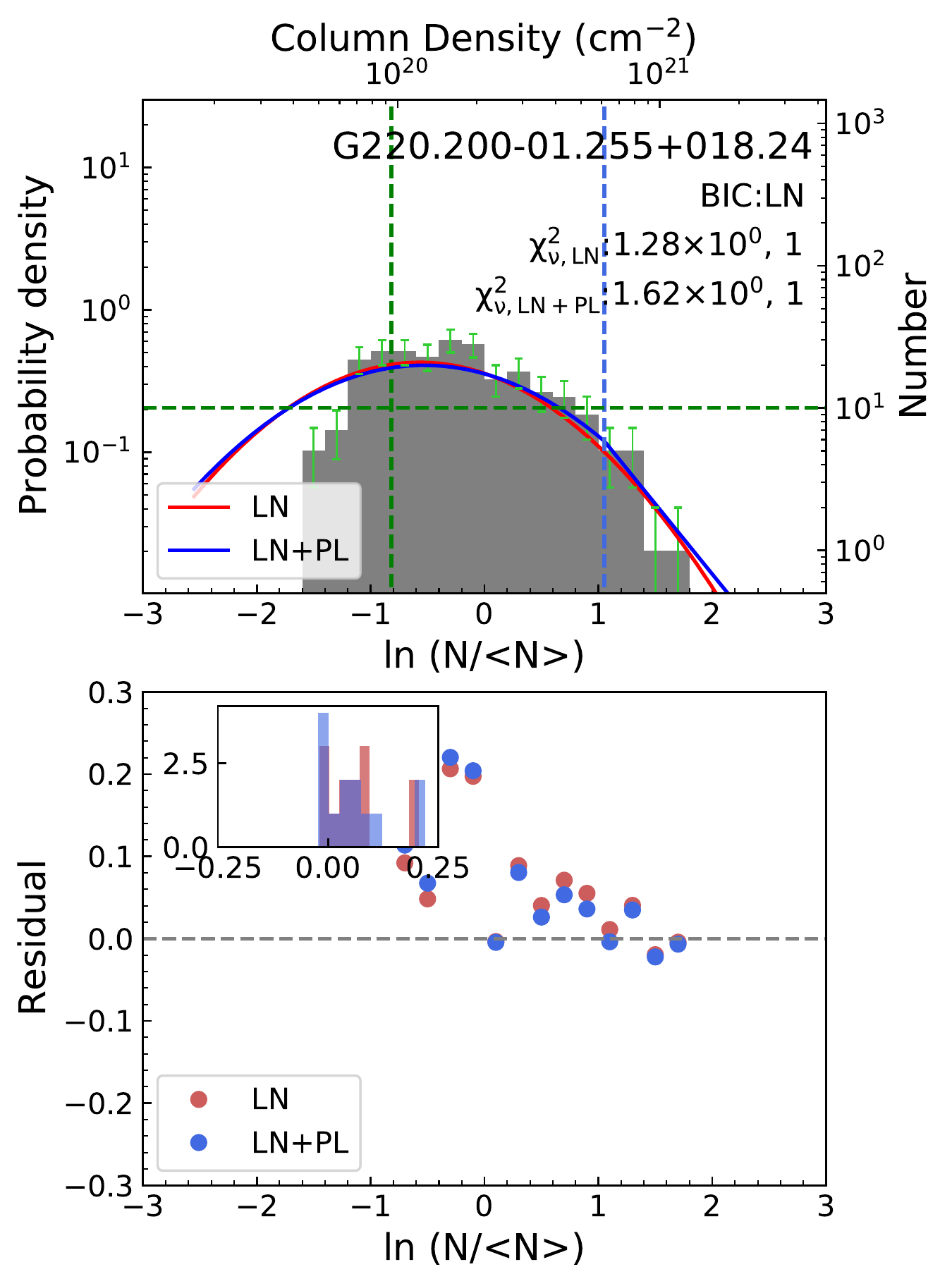}}
\subfigure{\includegraphics[trim=0cm 0cm 0cm 0cm, width= 0.23\linewidth, clip]{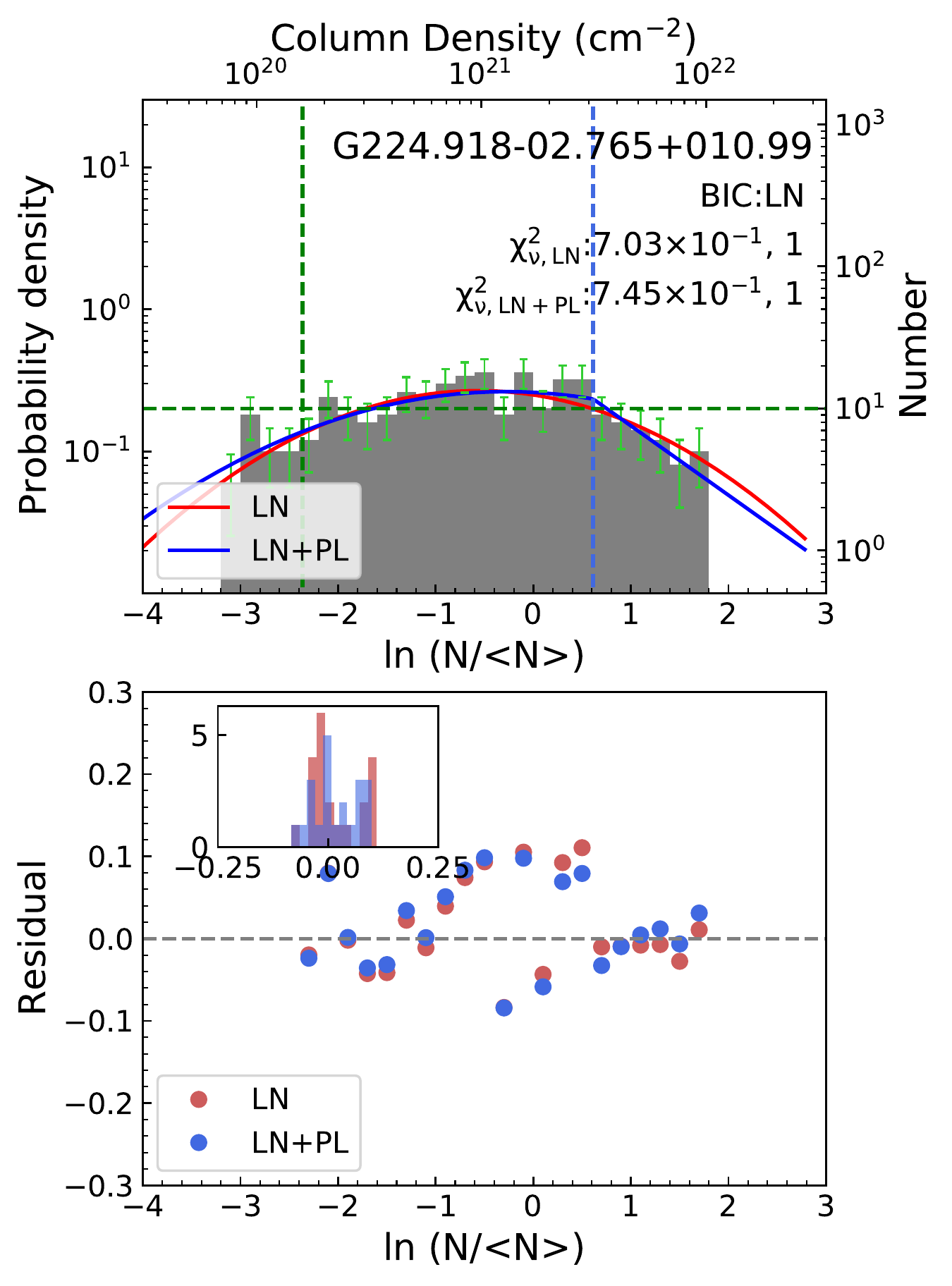}}
\subfigure{\includegraphics[trim=0cm 0cm 0cm 0cm, width= 0.23\linewidth, clip]{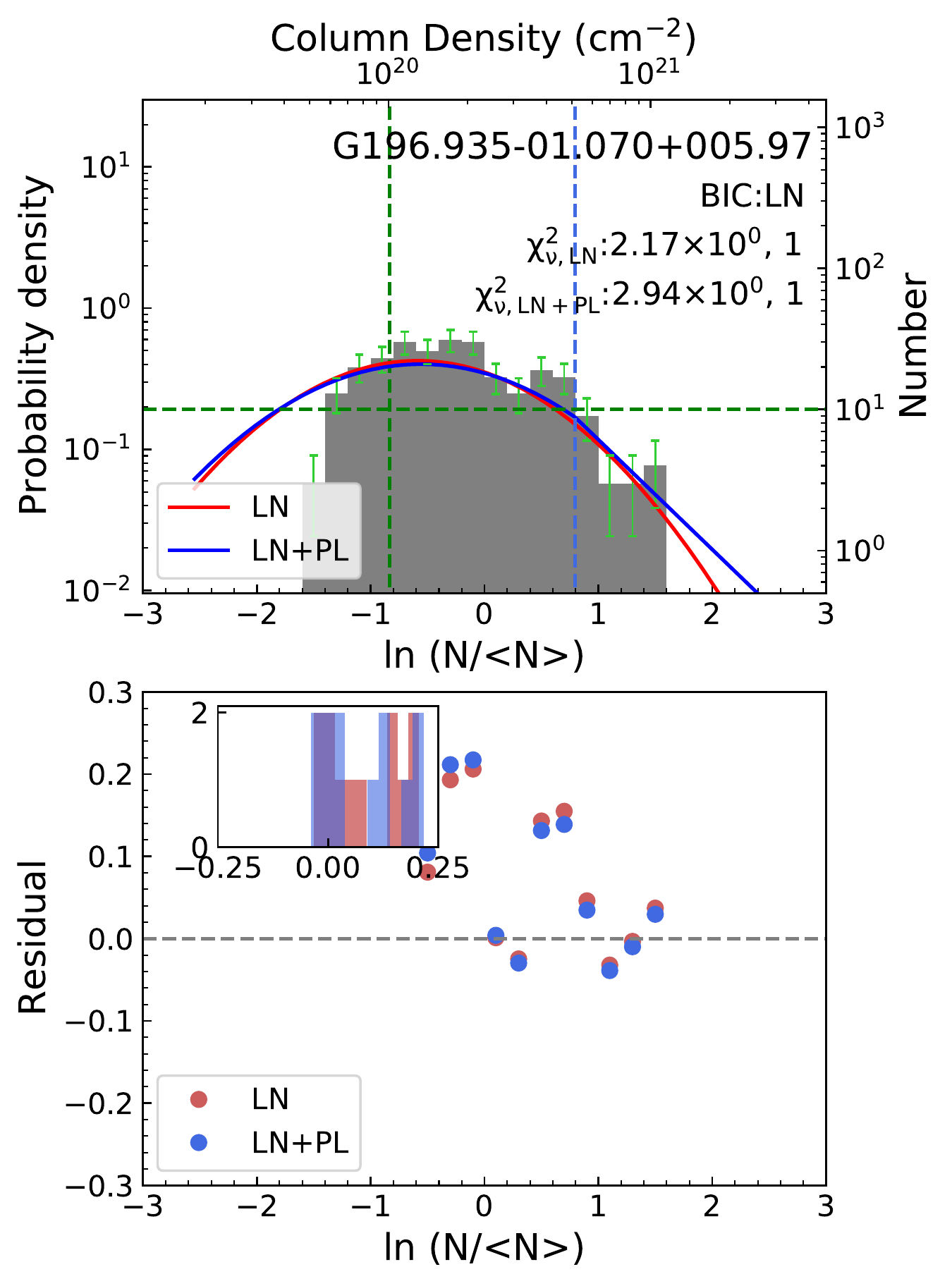}}
\subfigure{\includegraphics[trim=0cm 0cm 0cm 0cm, width= 0.23\linewidth, clip]{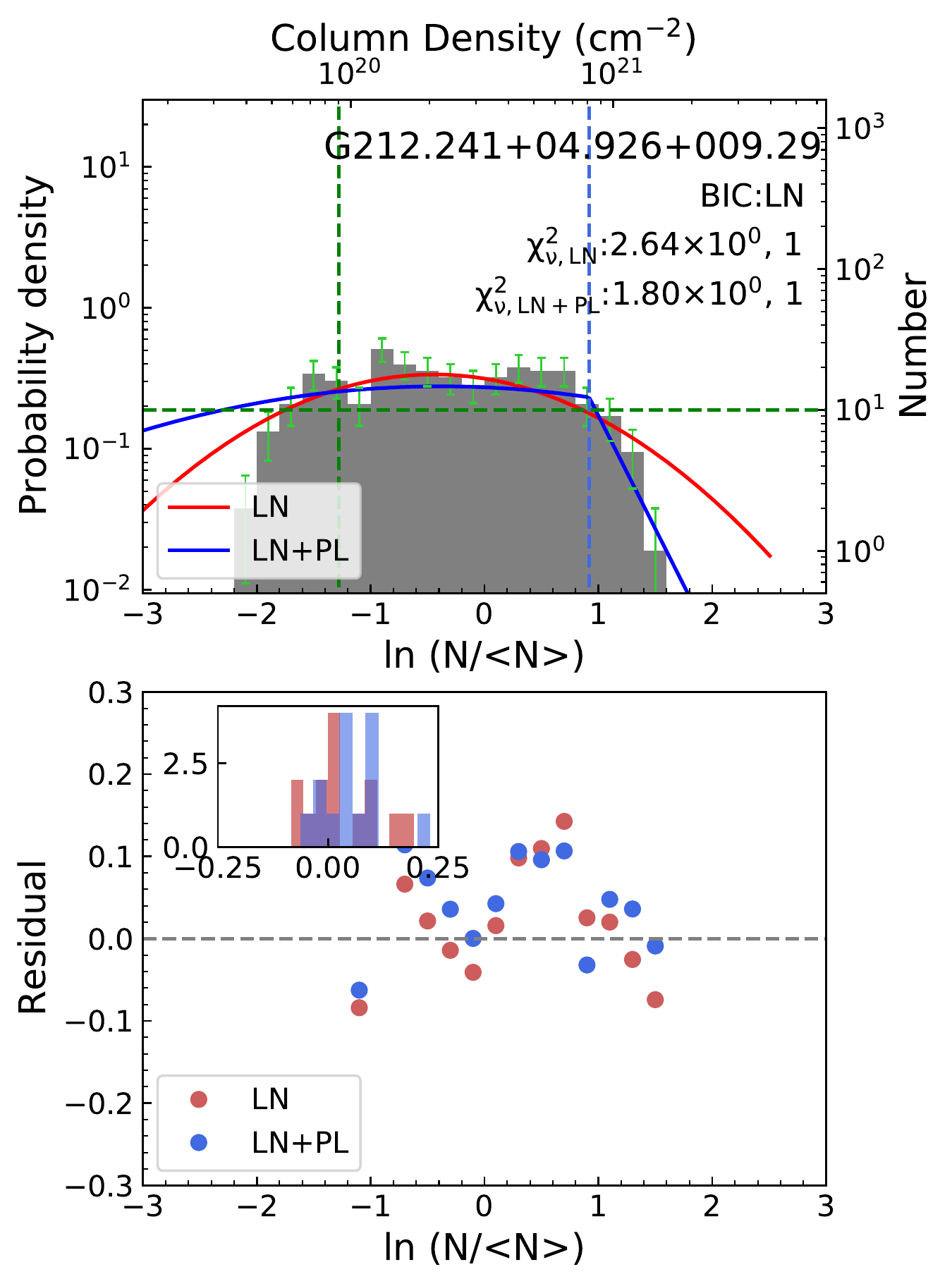}}

\subfigure{\includegraphics[trim=0cm 0cm 0cm 0cm, width= 0.23\linewidth, clip]{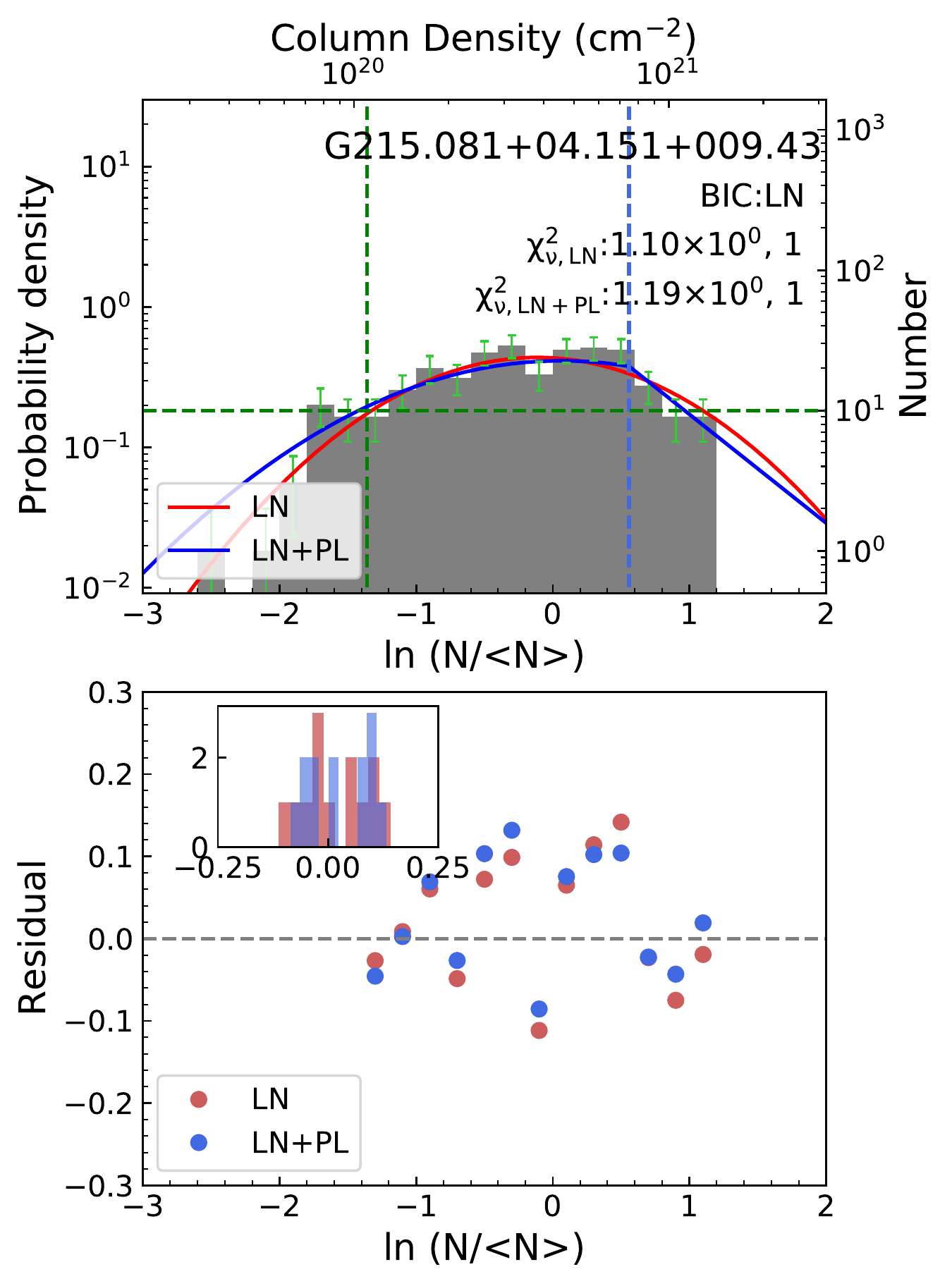}}
\subfigure{\includegraphics[trim=0cm 0cm 0cm 0cm, width= 0.23\linewidth, clip]{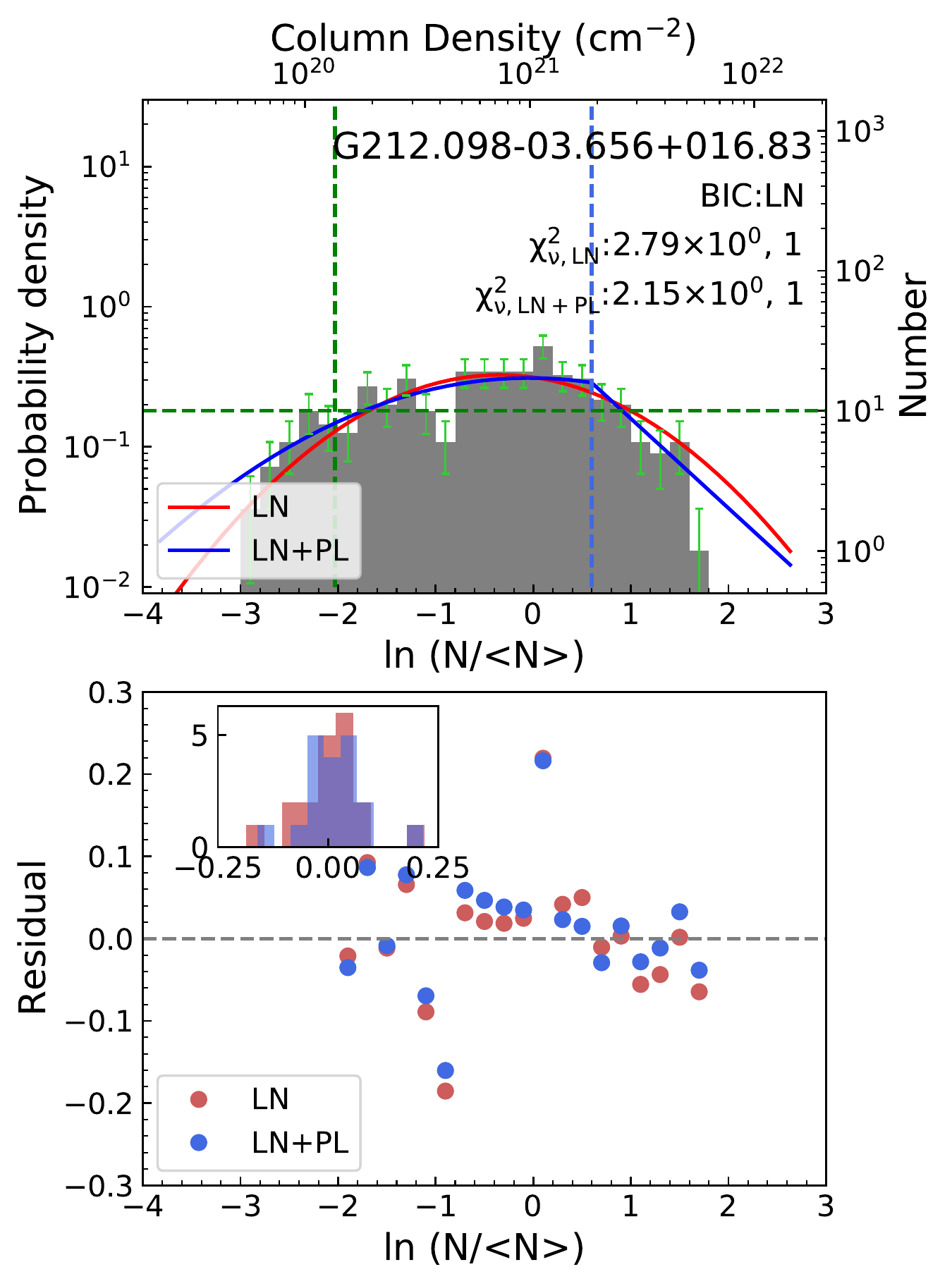}}
\subfigure{\includegraphics[trim=0cm 0cm 0cm 0cm, width= 0.23\linewidth, clip]{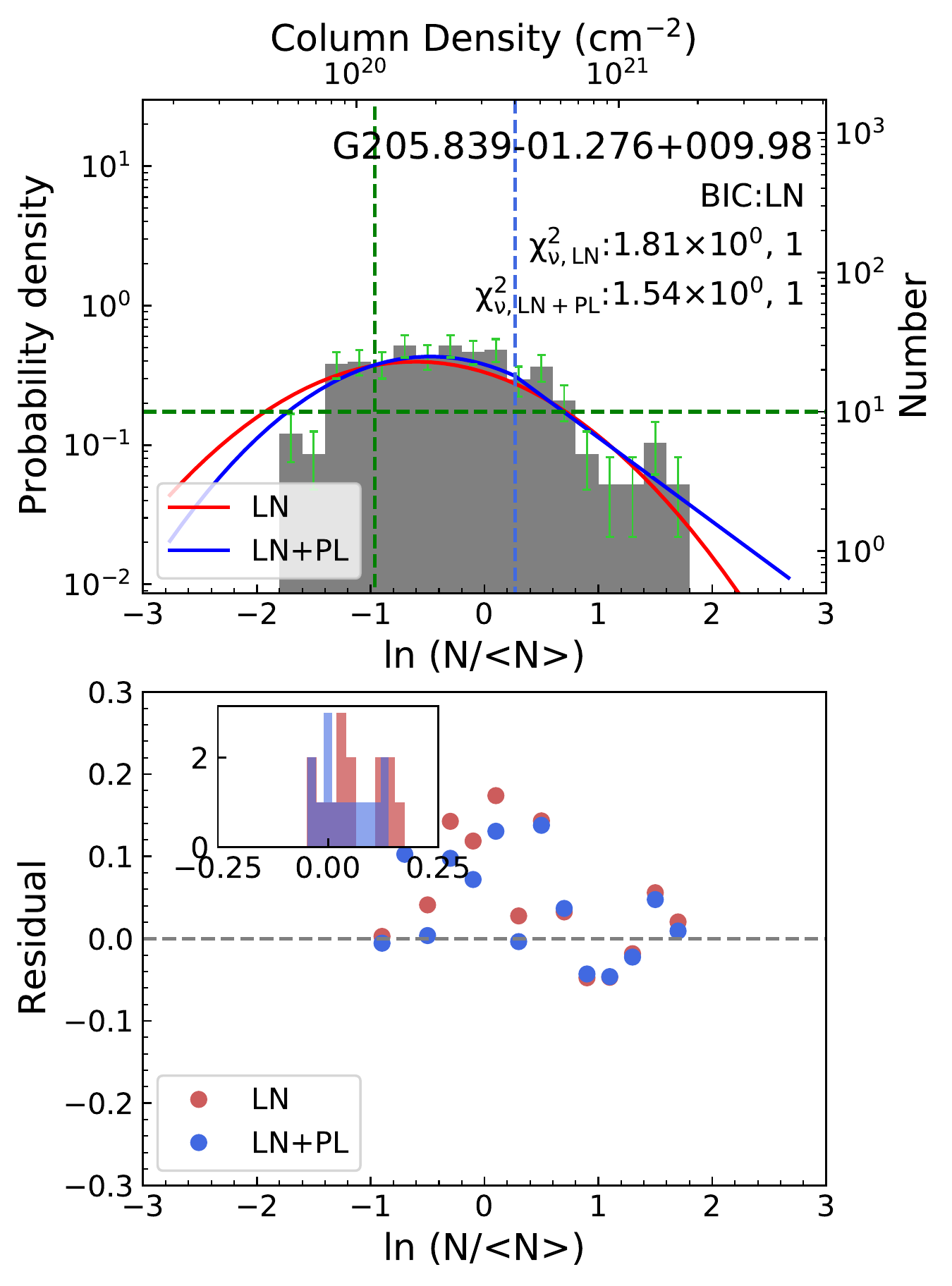}}
\subfigure{\includegraphics[trim=0cm 0cm 0cm 0cm, width= 0.23\linewidth, clip]{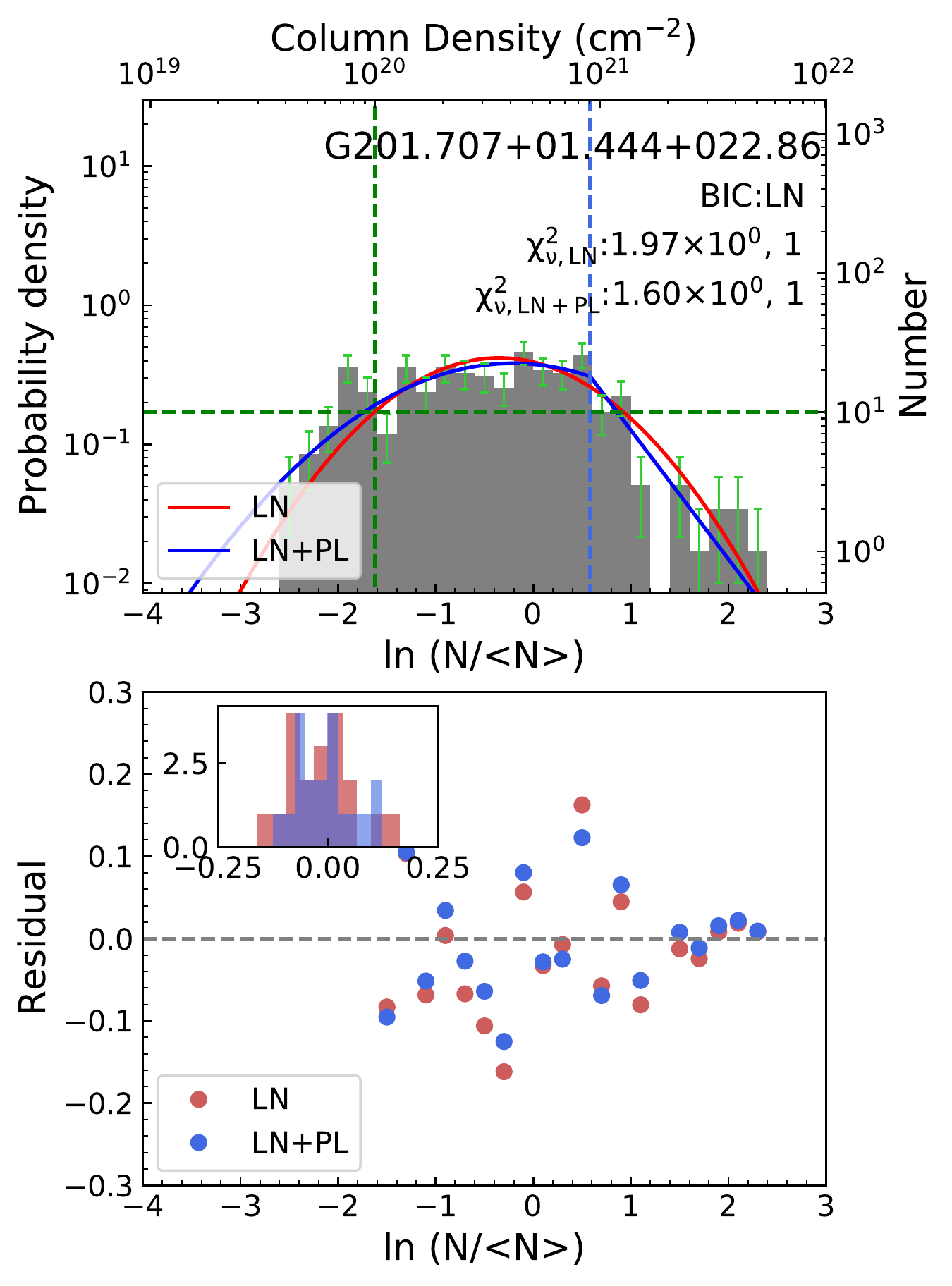}}

\subfigure{\includegraphics[trim=0cm 0cm 0cm 0cm, width= 0.23\linewidth, clip]{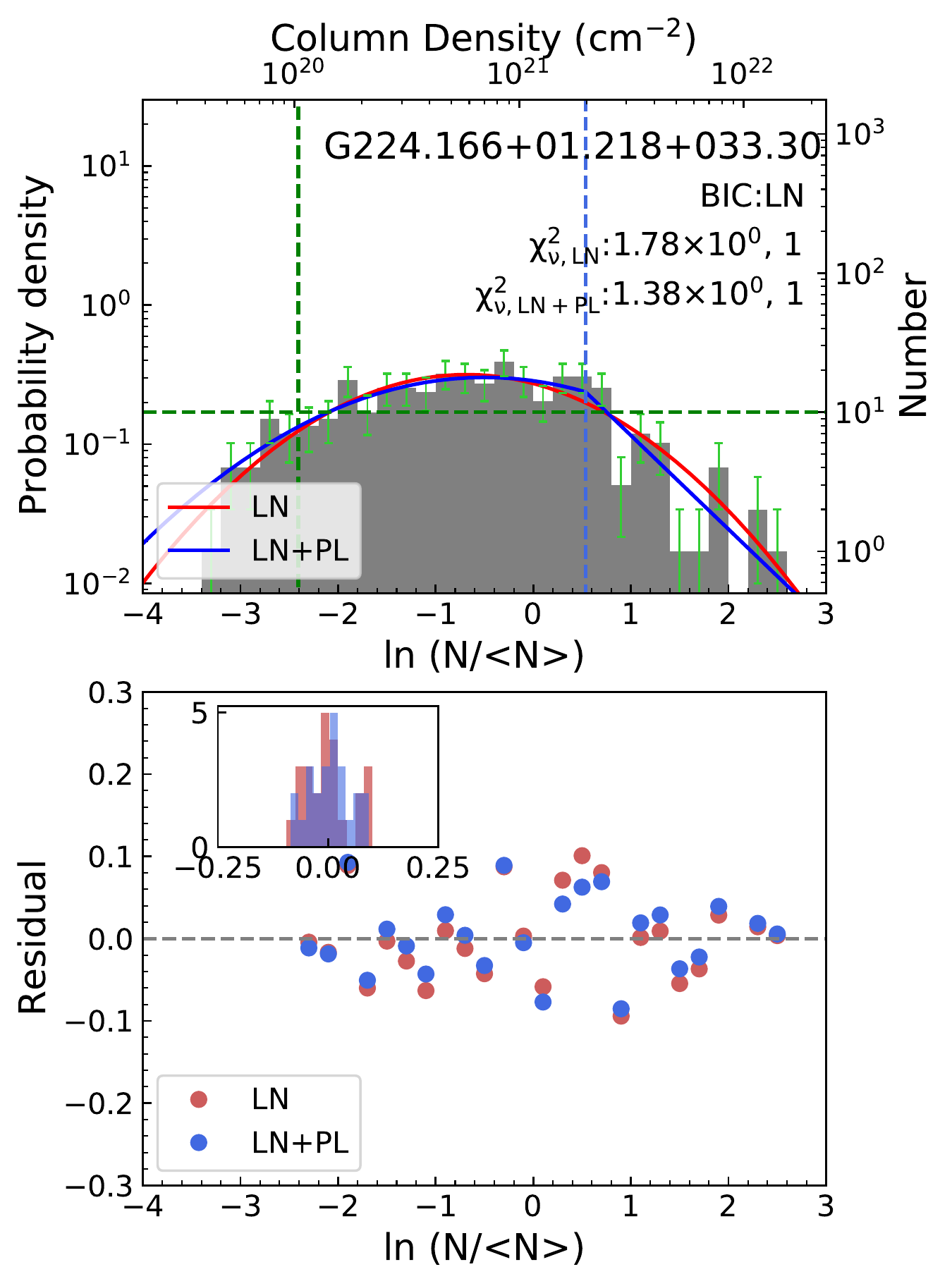}}
\subfigure{\includegraphics[trim=0cm 0cm 0cm 0cm, width= 0.23\linewidth, clip]{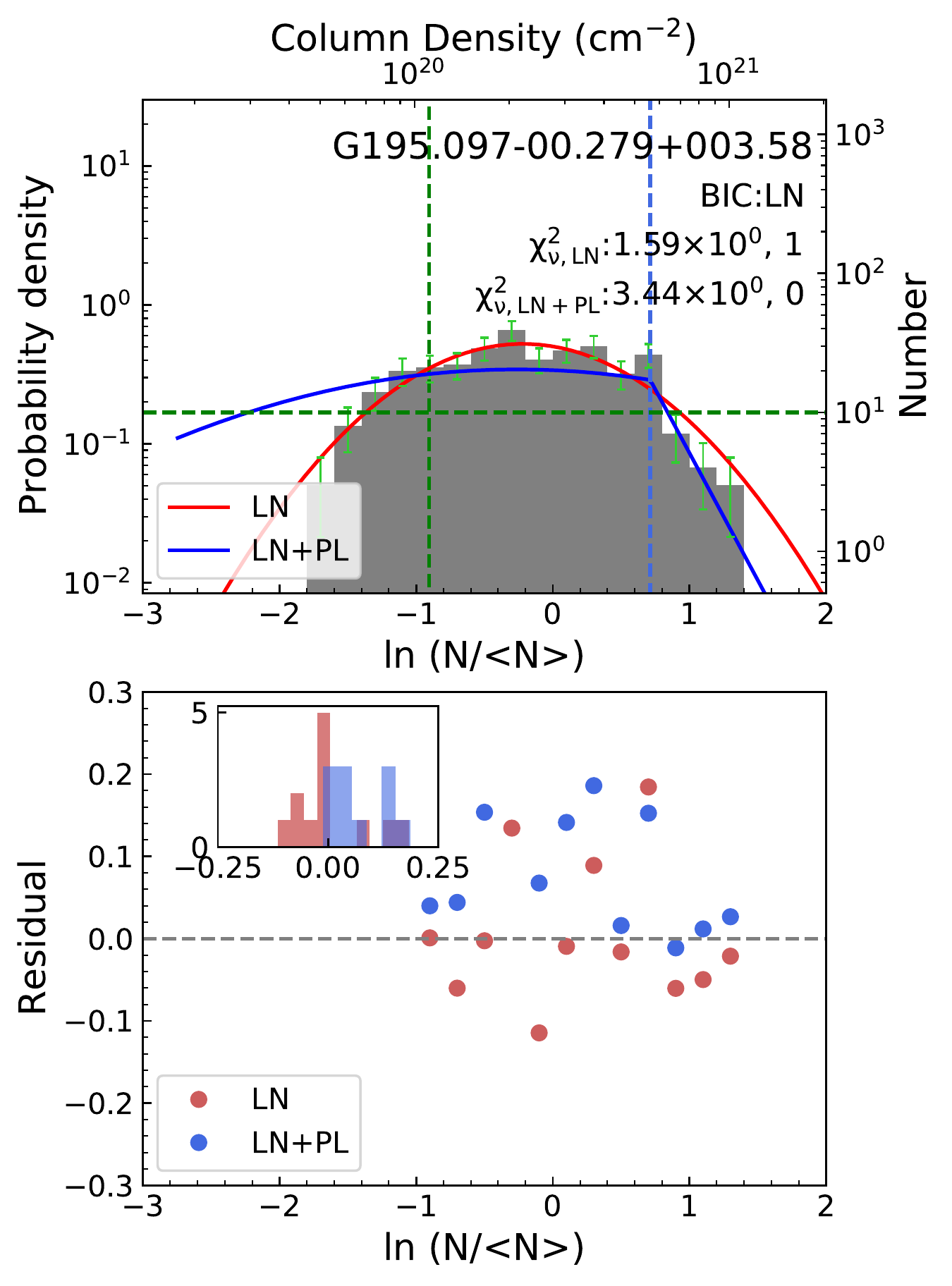}}
\subfigure{\includegraphics[trim=0cm 0cm 0cm 0cm, width= 0.23\linewidth, clip]{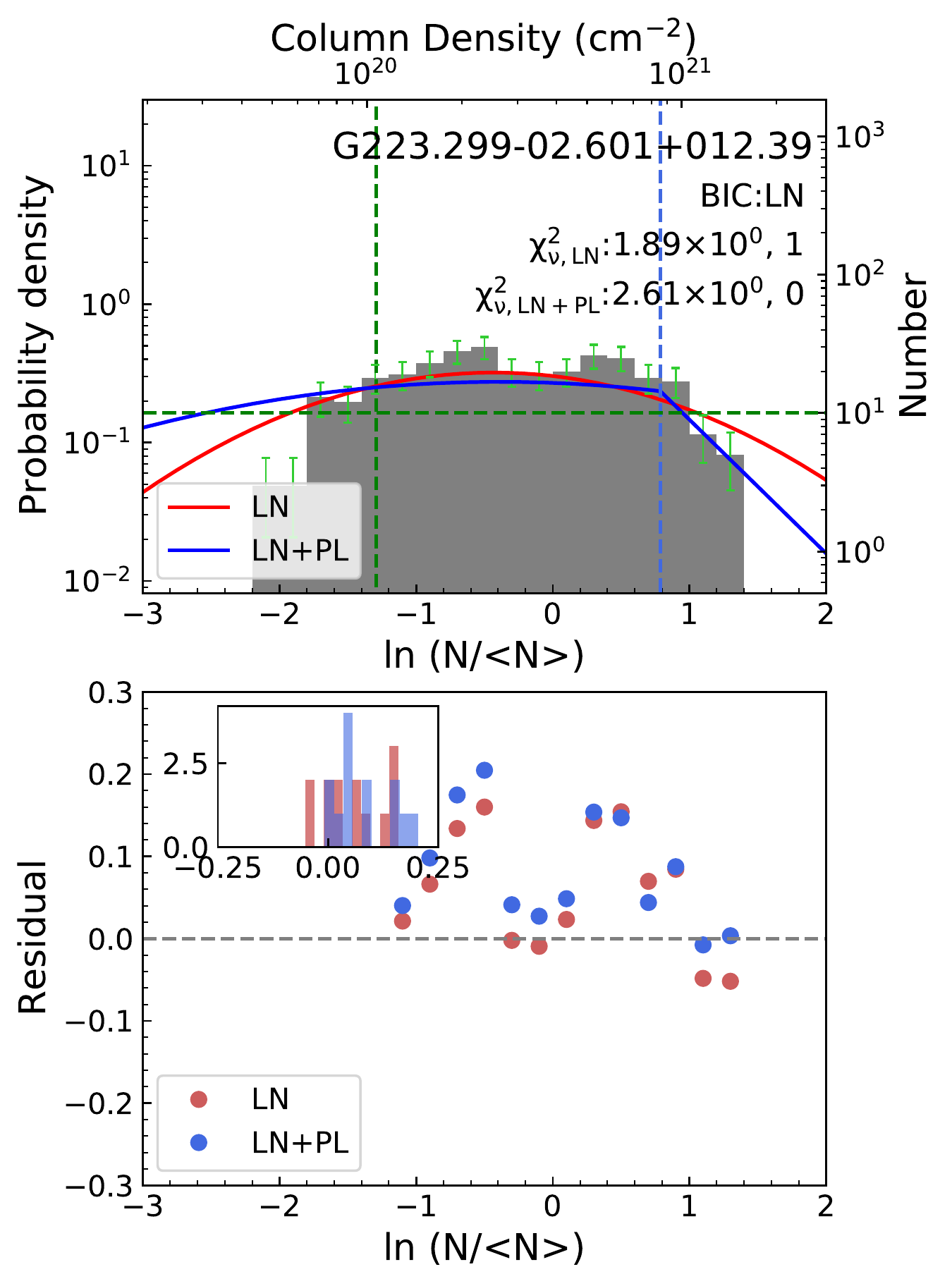}}
\subfigure{\includegraphics[trim=0cm 0cm 0cm 0cm, width= 0.23\linewidth, clip]{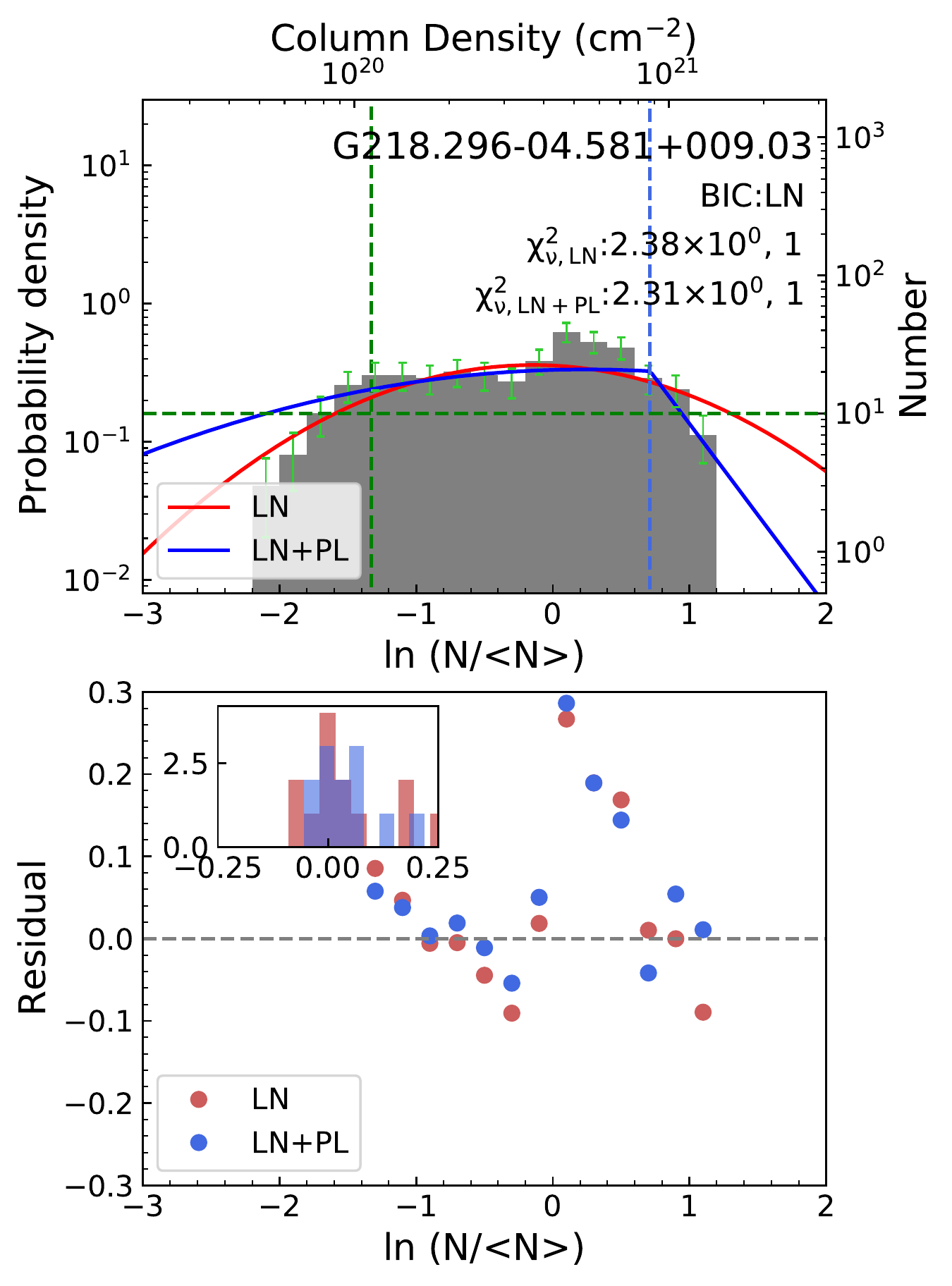}}

\end{figure}
\begin{figure}
\subfigure{\includegraphics[trim=0cm 0cm 0cm 0cm, width= 0.23\linewidth, clip]{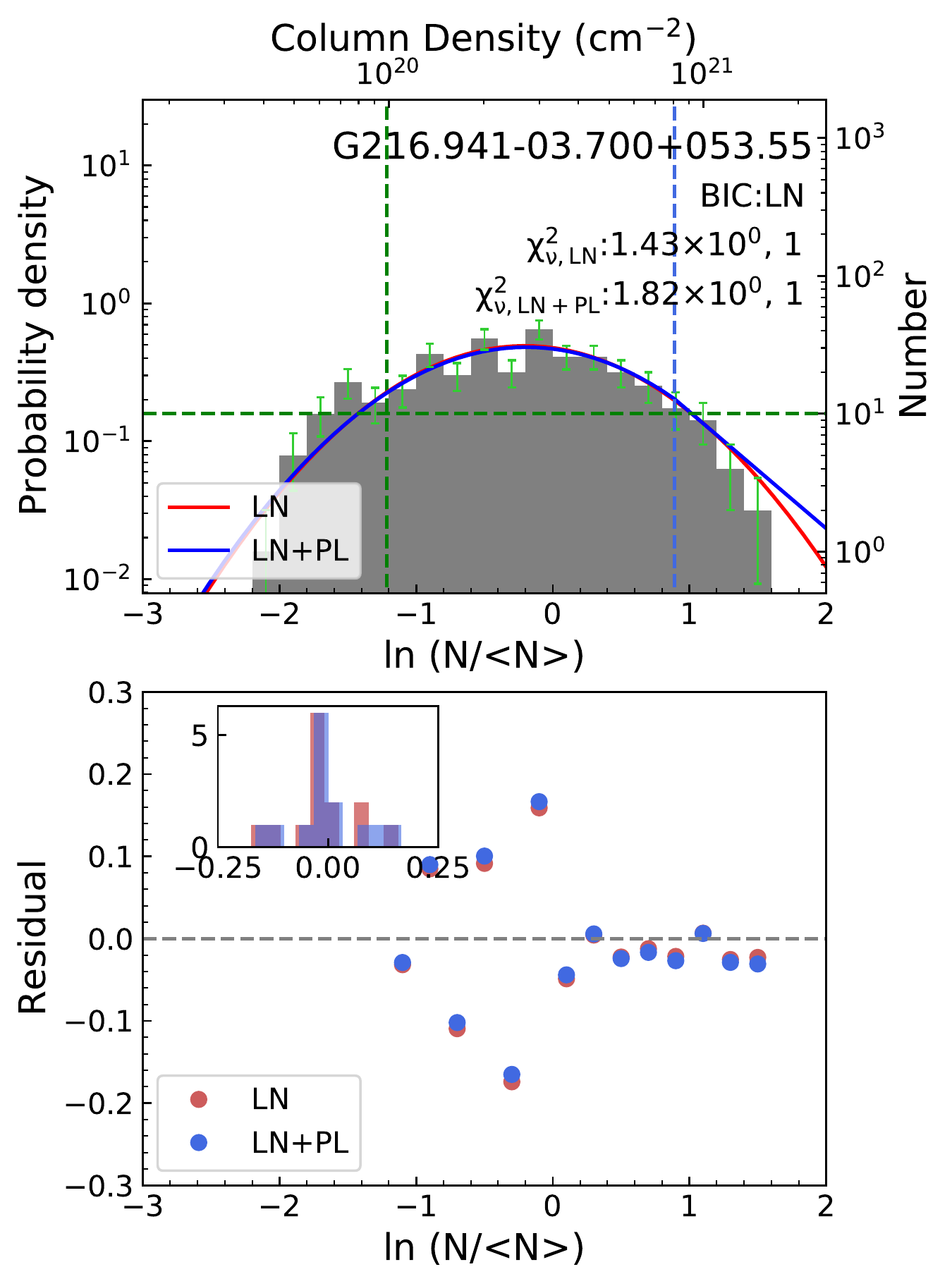}}
\subfigure{\includegraphics[trim=0cm 0cm 0cm 0cm, width= 0.23\linewidth, clip]{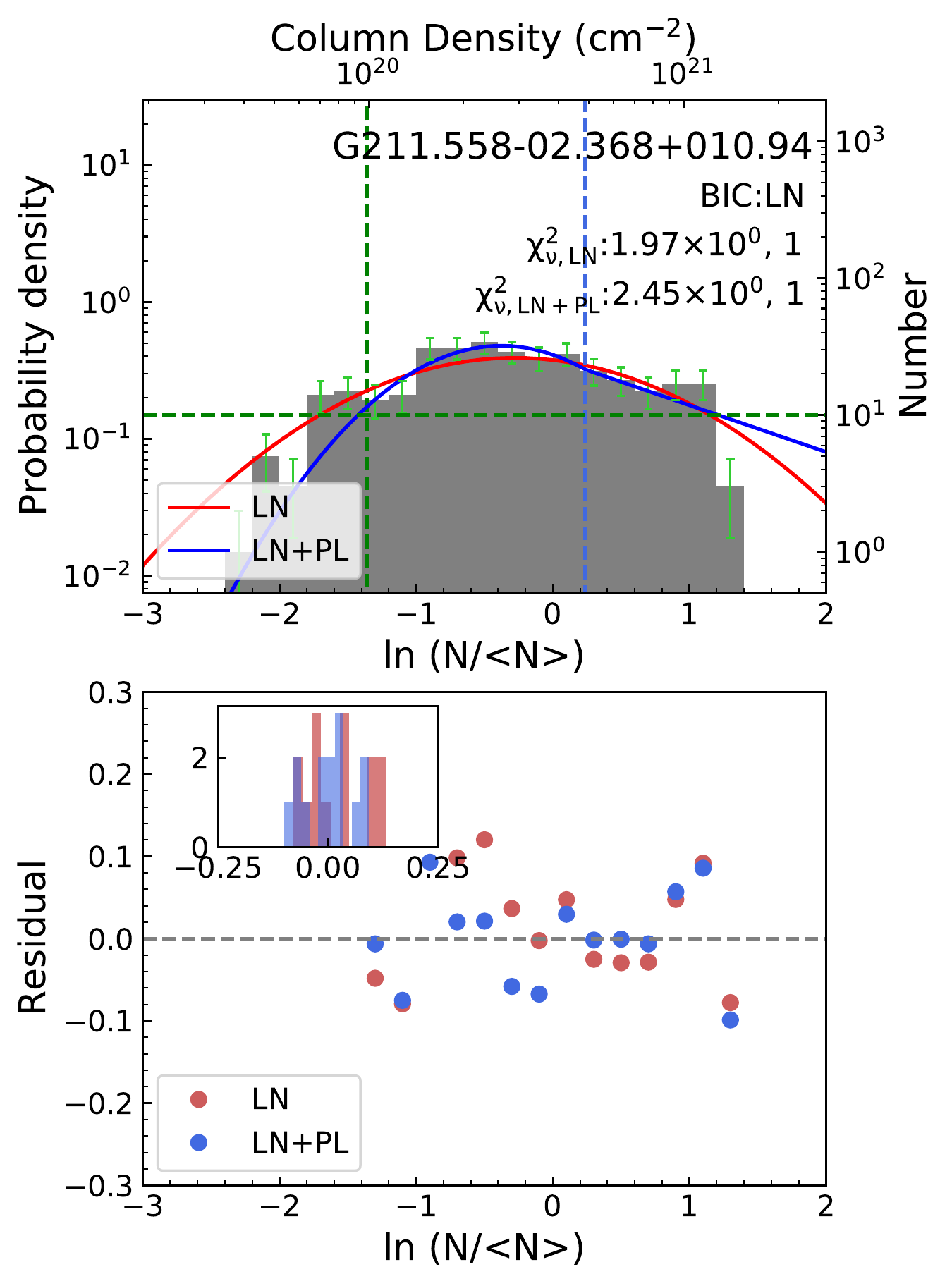}}
\subfigure{\includegraphics[trim=0cm 0cm 0cm 0cm, width= 0.23\linewidth, clip]{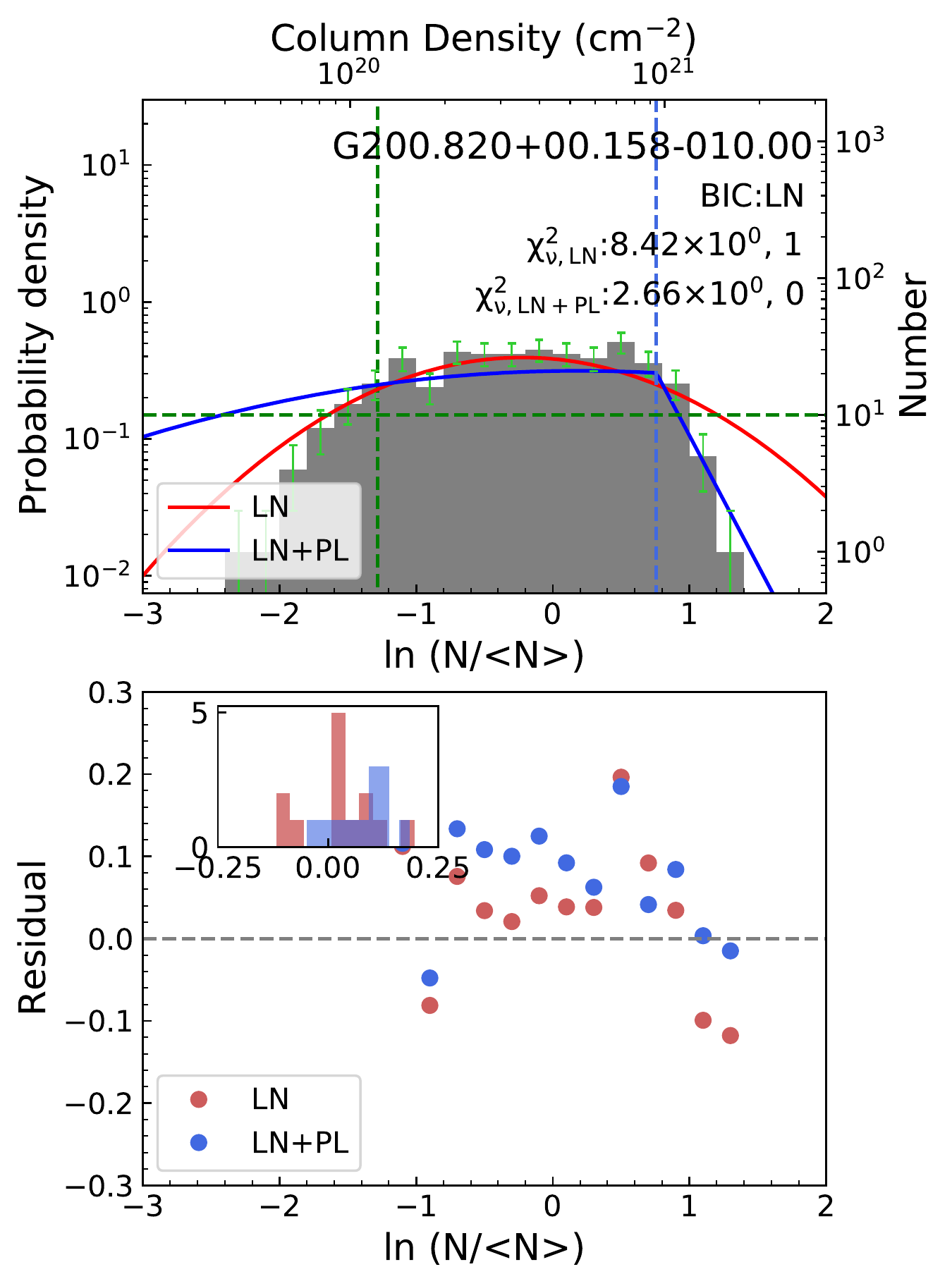}}
\subfigure{\includegraphics[trim=0cm 0cm 0cm 0cm, width= 0.23\linewidth, clip]{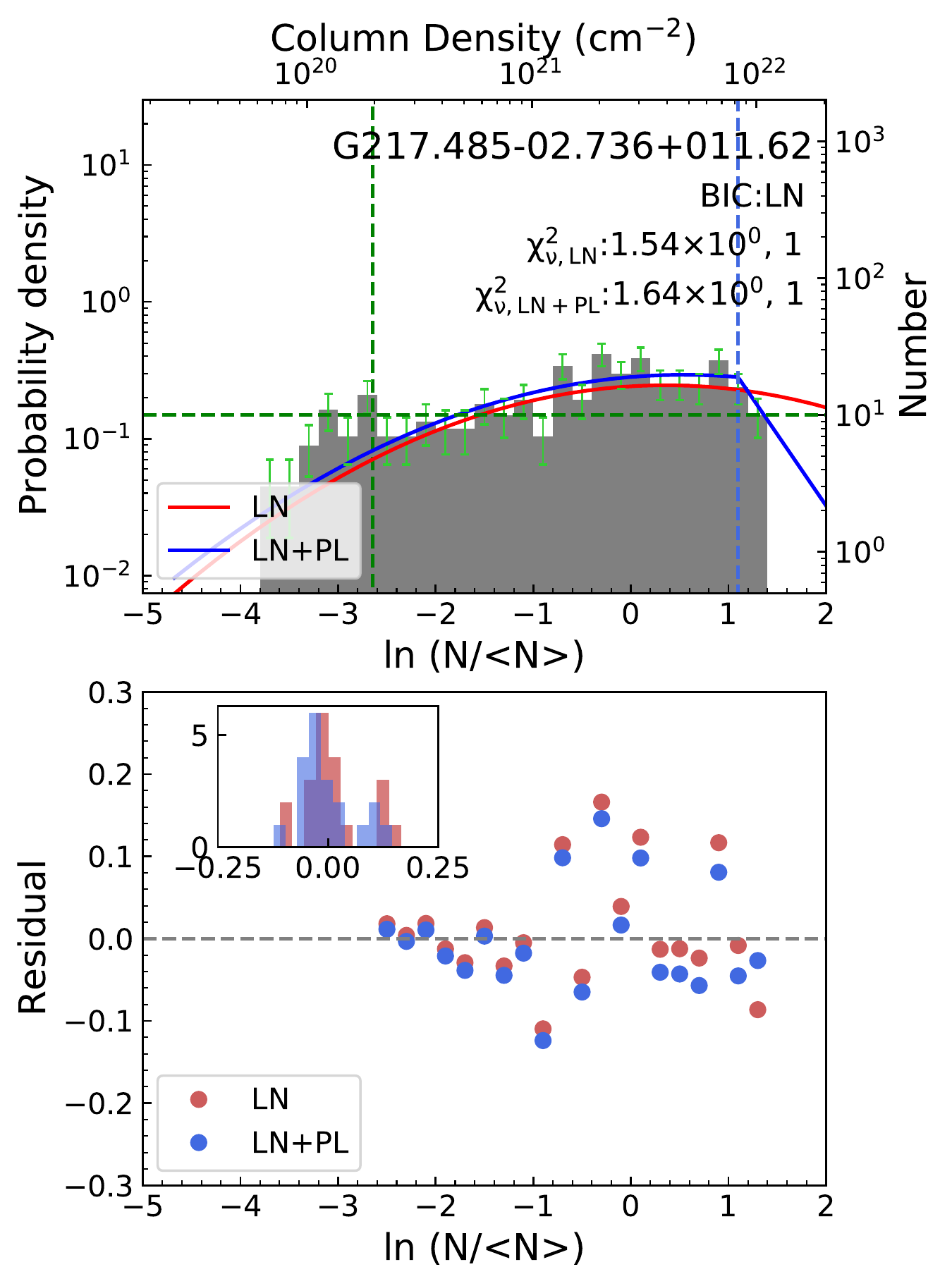}}

\subfigure{\includegraphics[trim=0cm 0cm 0cm 0cm, width= 0.23\linewidth, clip]{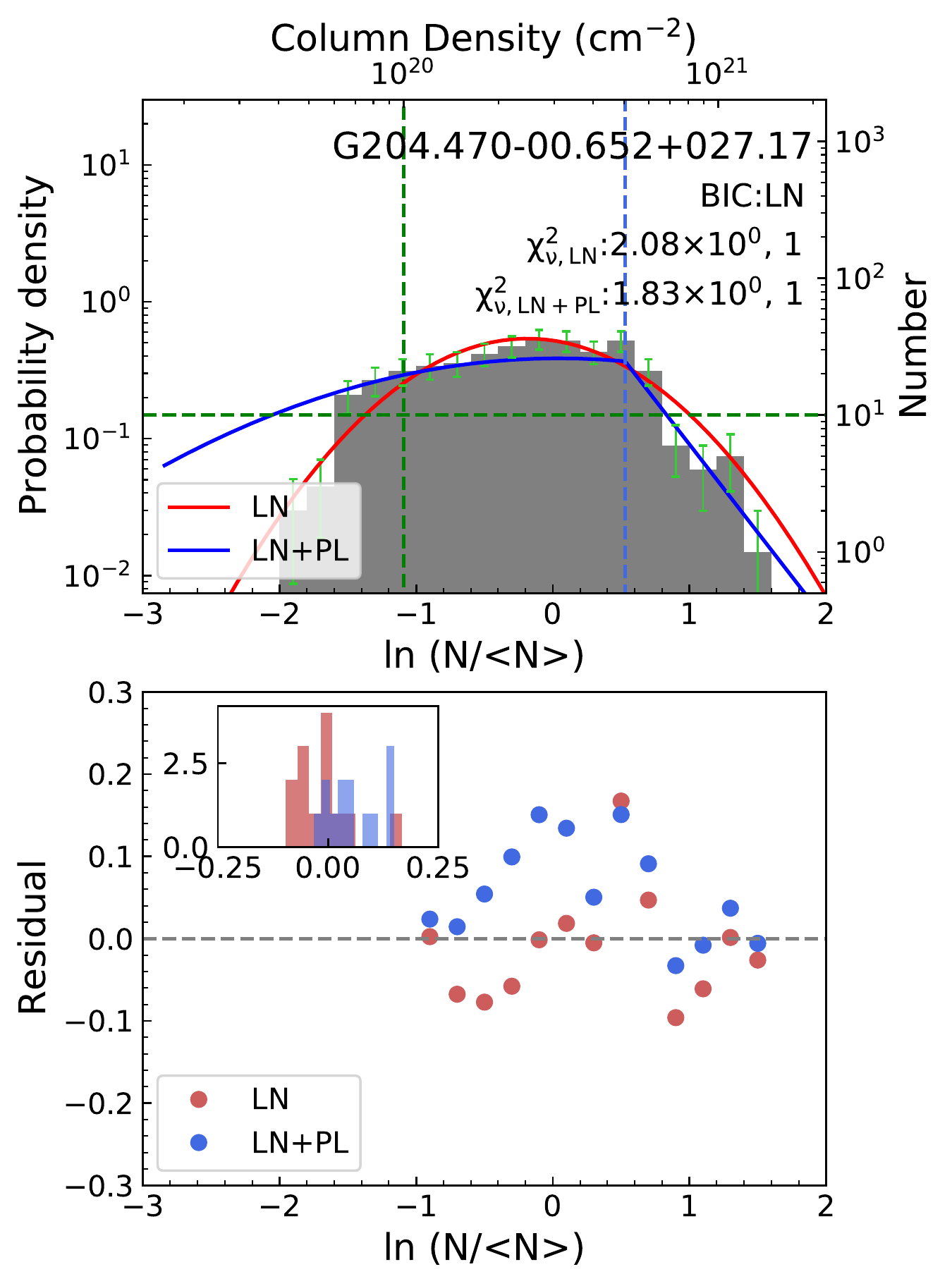}}
\subfigure{\includegraphics[trim=0cm 0cm 0cm 0cm, width= 0.23\linewidth, clip]{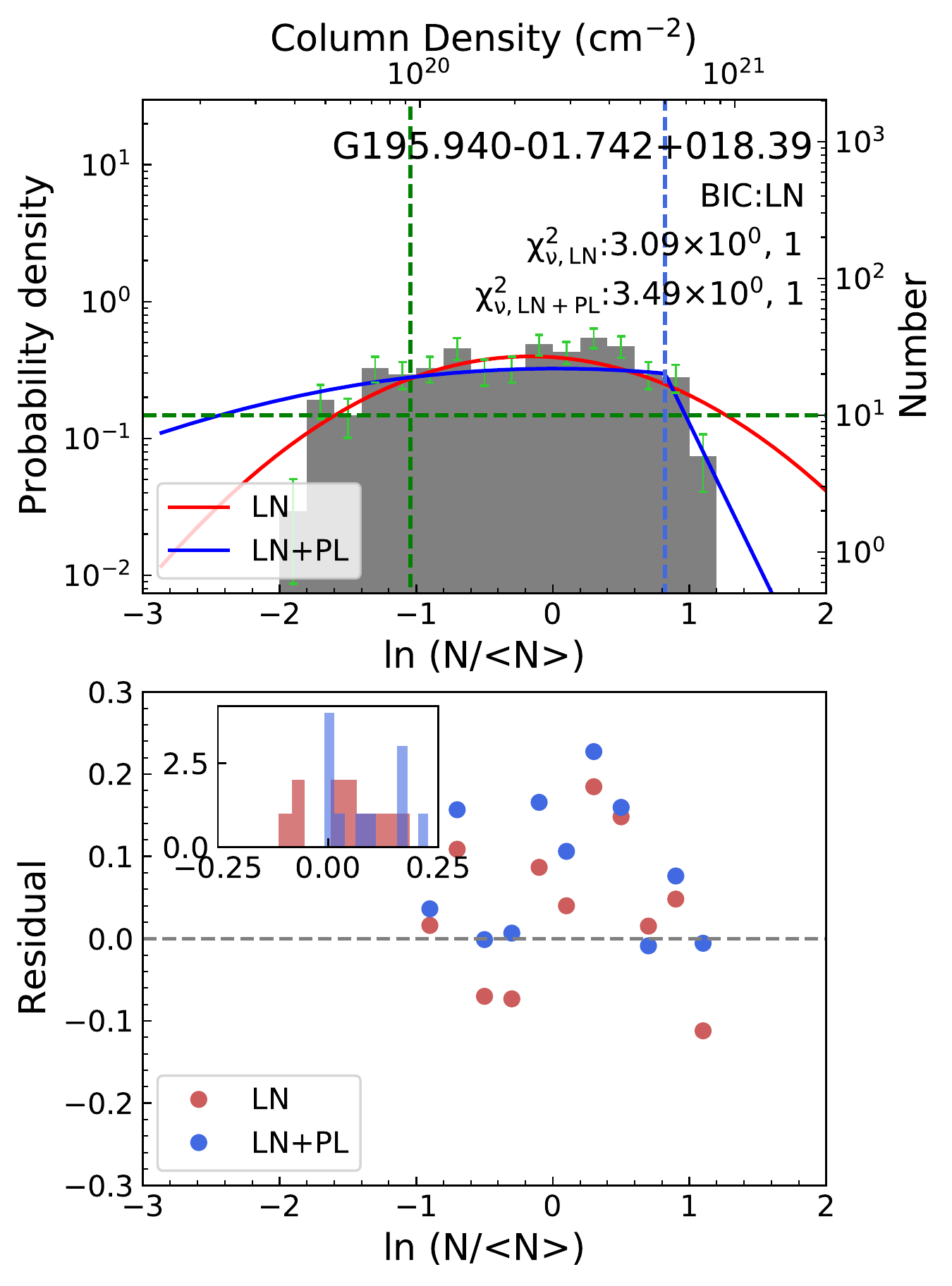}}
\subfigure{\includegraphics[trim=0cm 0cm 0cm 0cm, width= 0.23\linewidth, clip]{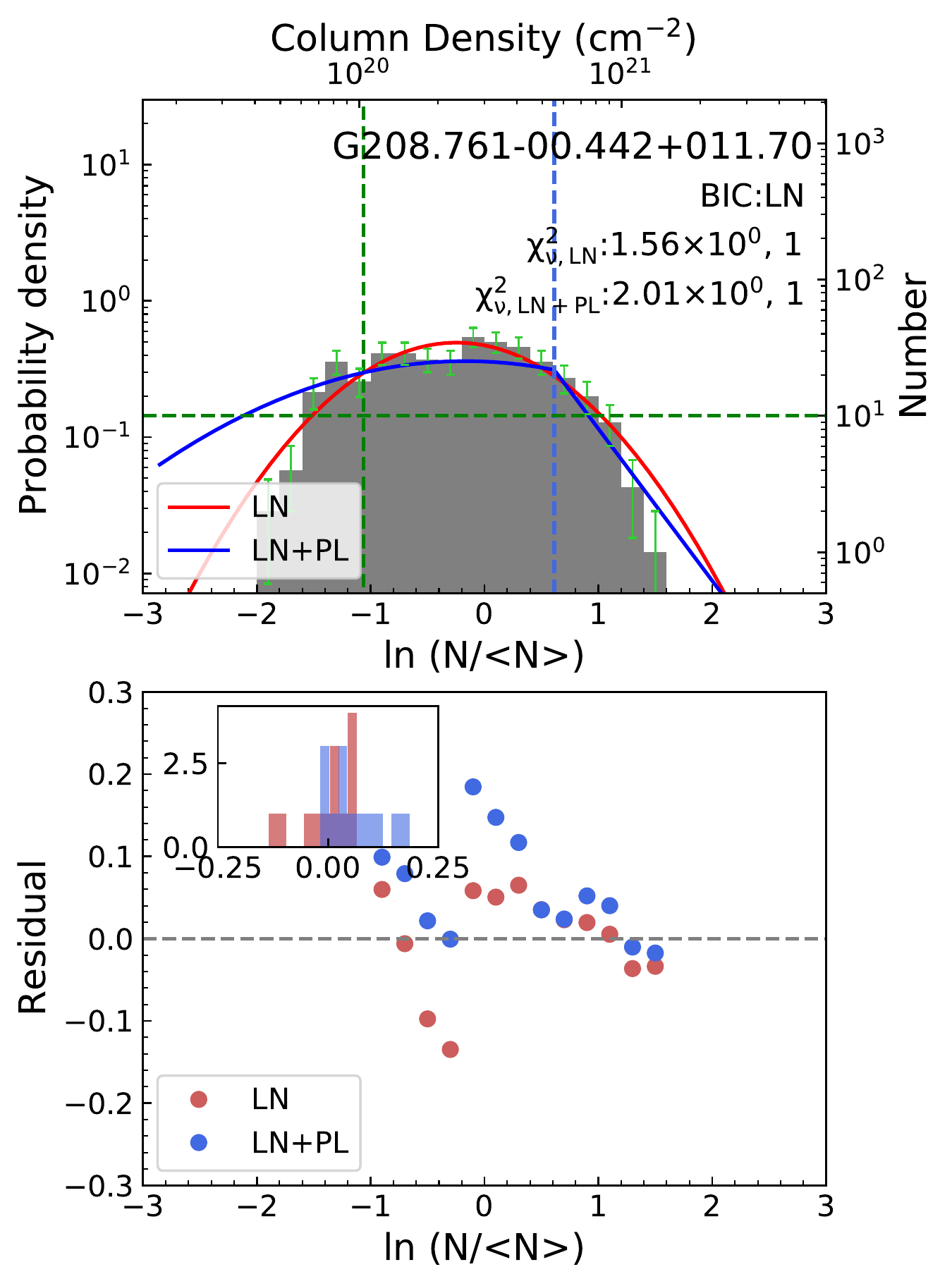}}
\subfigure{\includegraphics[trim=0cm 0cm 0cm 0cm, width= 0.23\linewidth, clip]{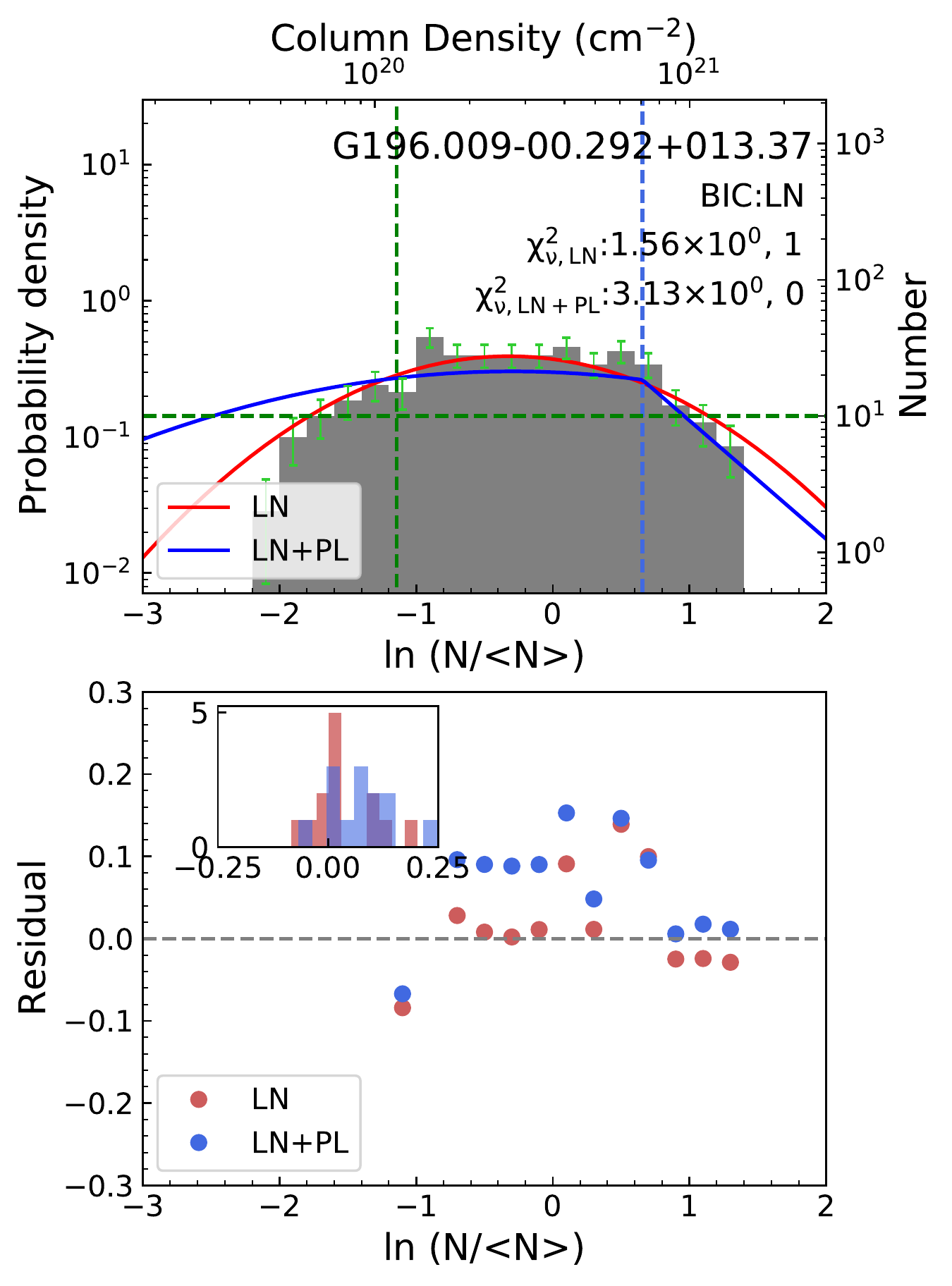}}

\subfigure{\includegraphics[trim=0cm 0cm 0cm 0cm, width= 0.23\linewidth, clip]{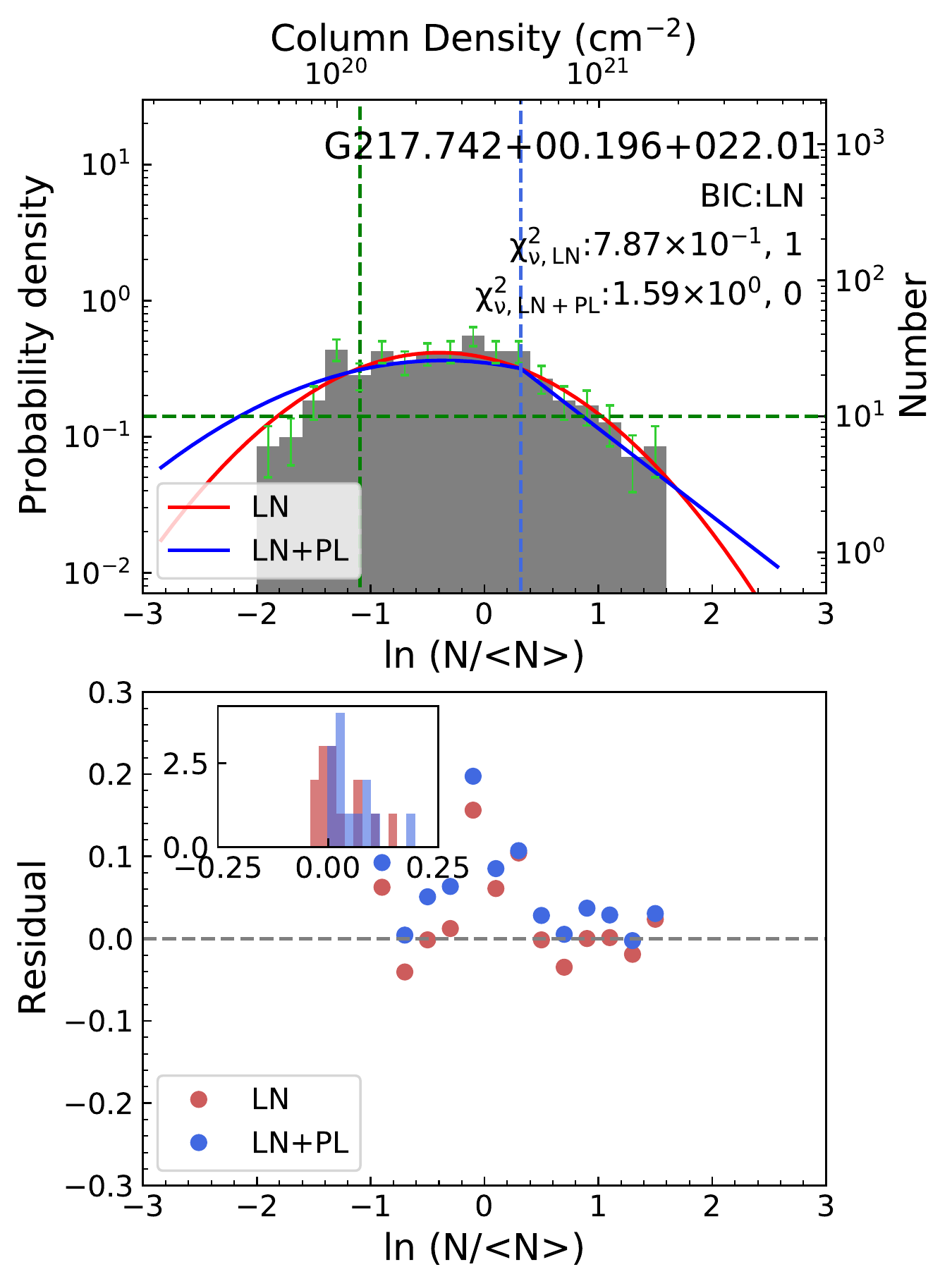}}
\subfigure{\includegraphics[trim=0cm 0cm 0cm 0cm, width= 0.23\linewidth, clip]{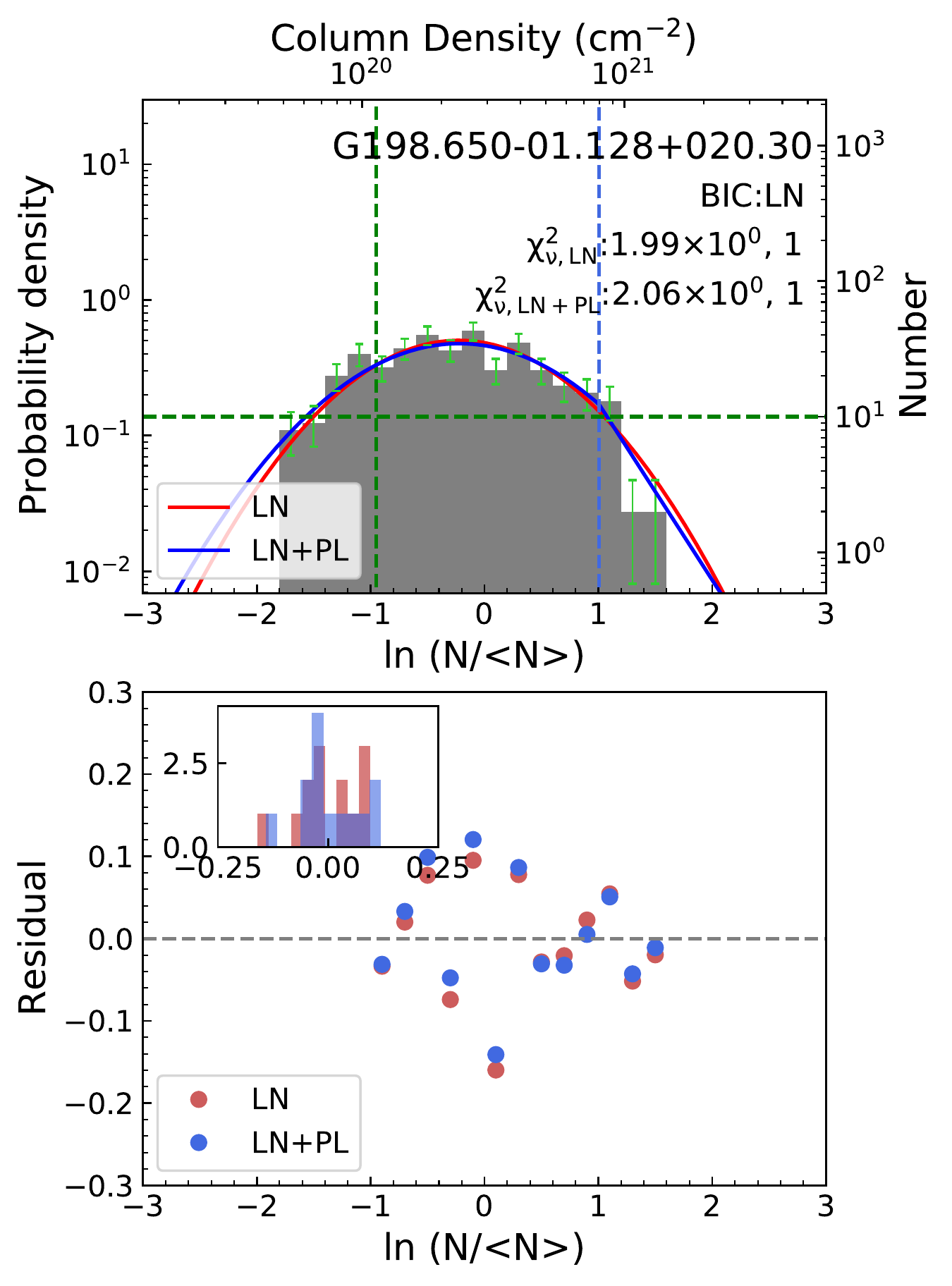}}
\subfigure{\includegraphics[trim=0cm 0cm 0cm 0cm, width= 0.23\linewidth, clip]{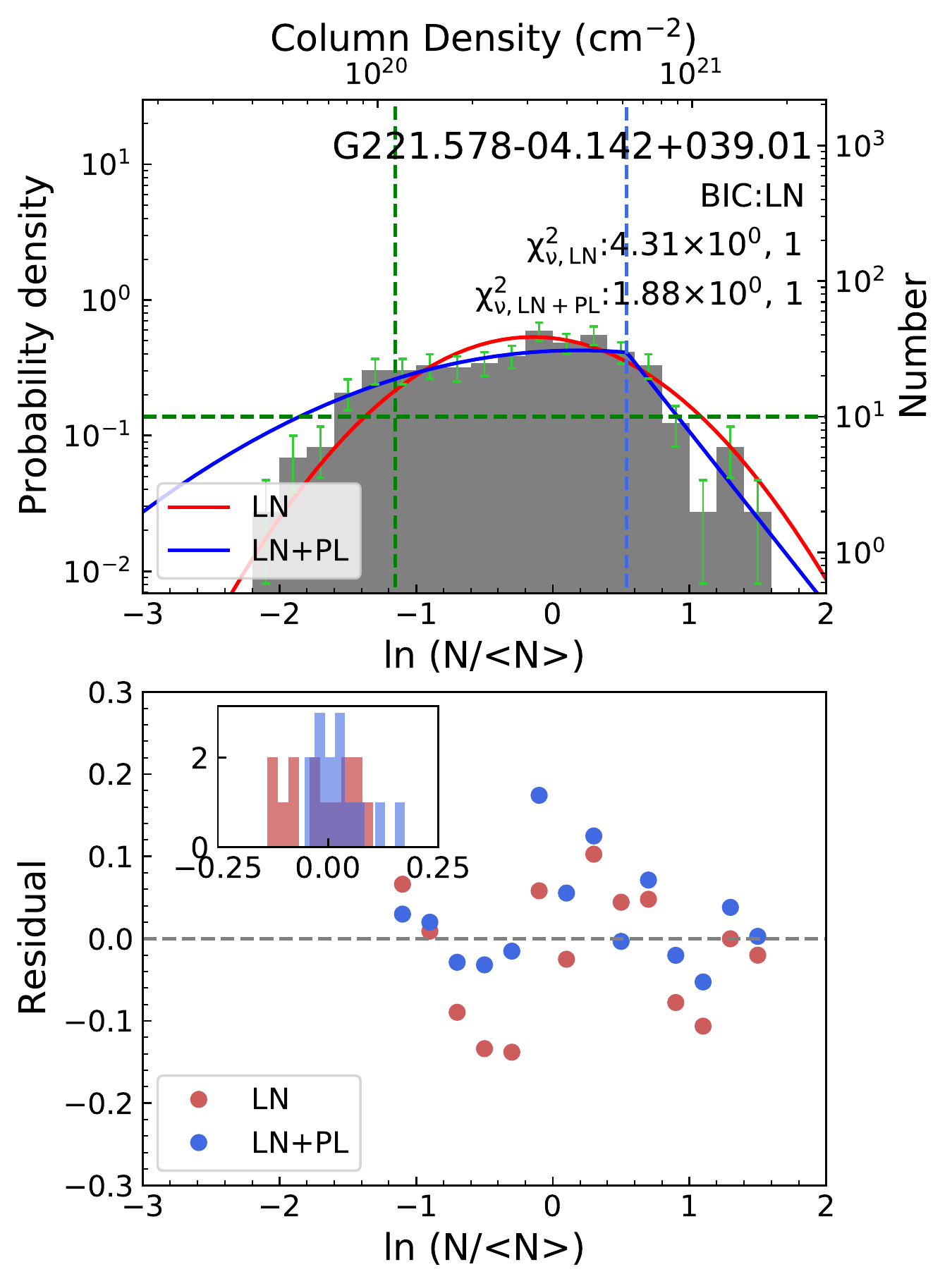}}
\subfigure{\includegraphics[trim=0cm 0cm 0cm 0cm, width= 0.23\linewidth, clip]{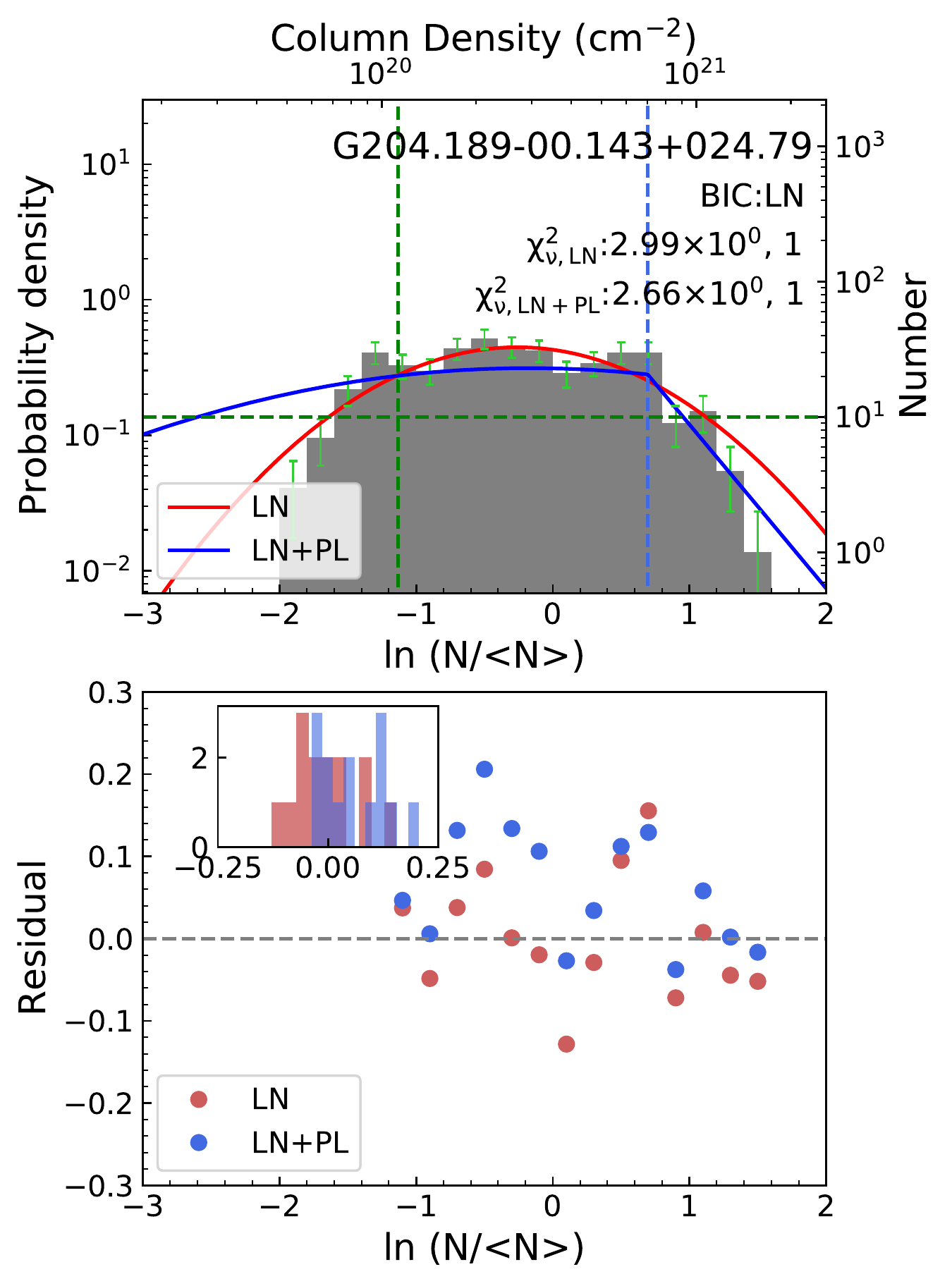}}

\subfigure{\includegraphics[trim=0cm 0cm 0cm 0cm, width= 0.23\linewidth, clip]{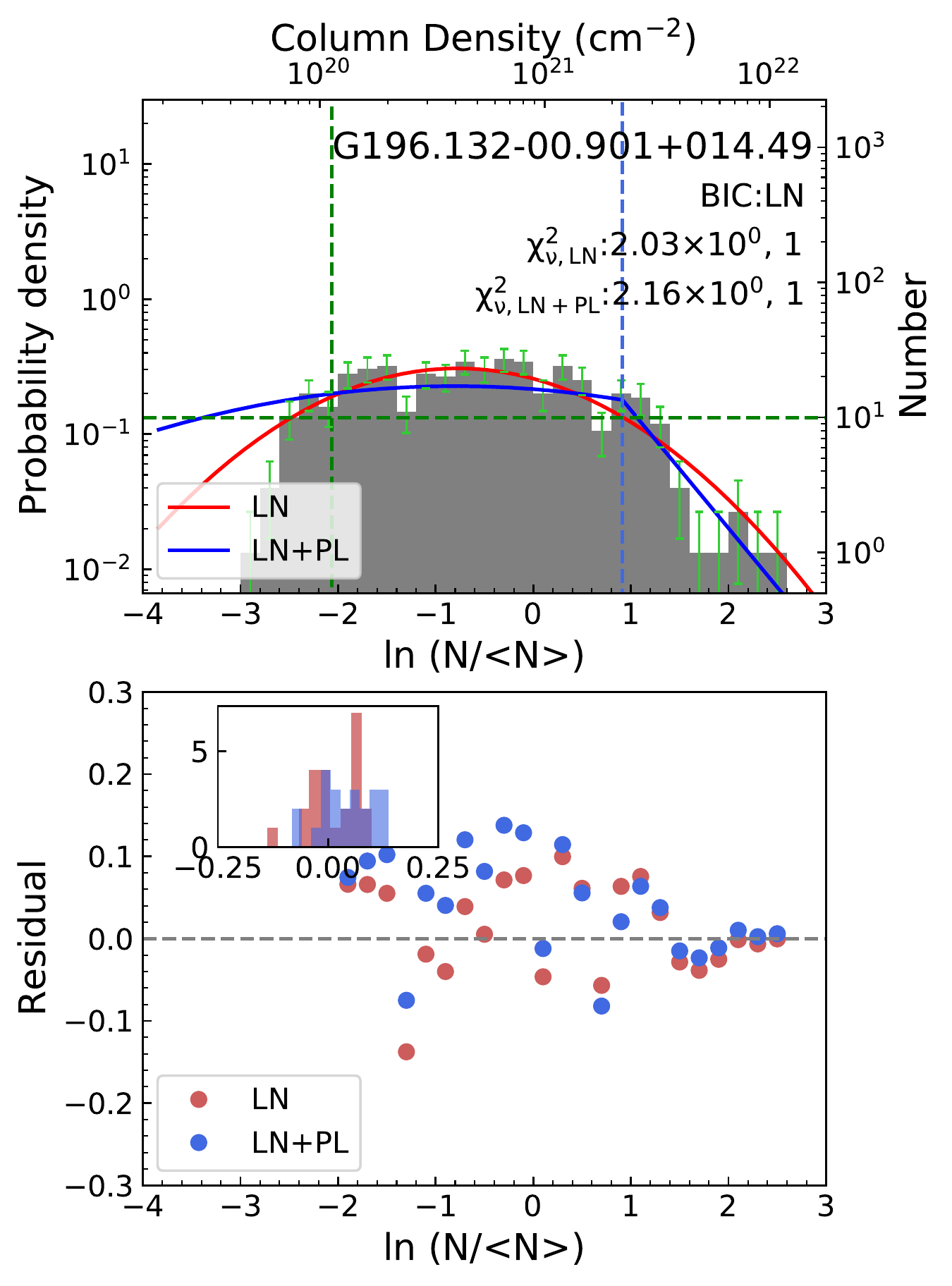}}
\subfigure{\includegraphics[trim=0cm 0cm 0cm 0cm, width= 0.23\linewidth, clip]{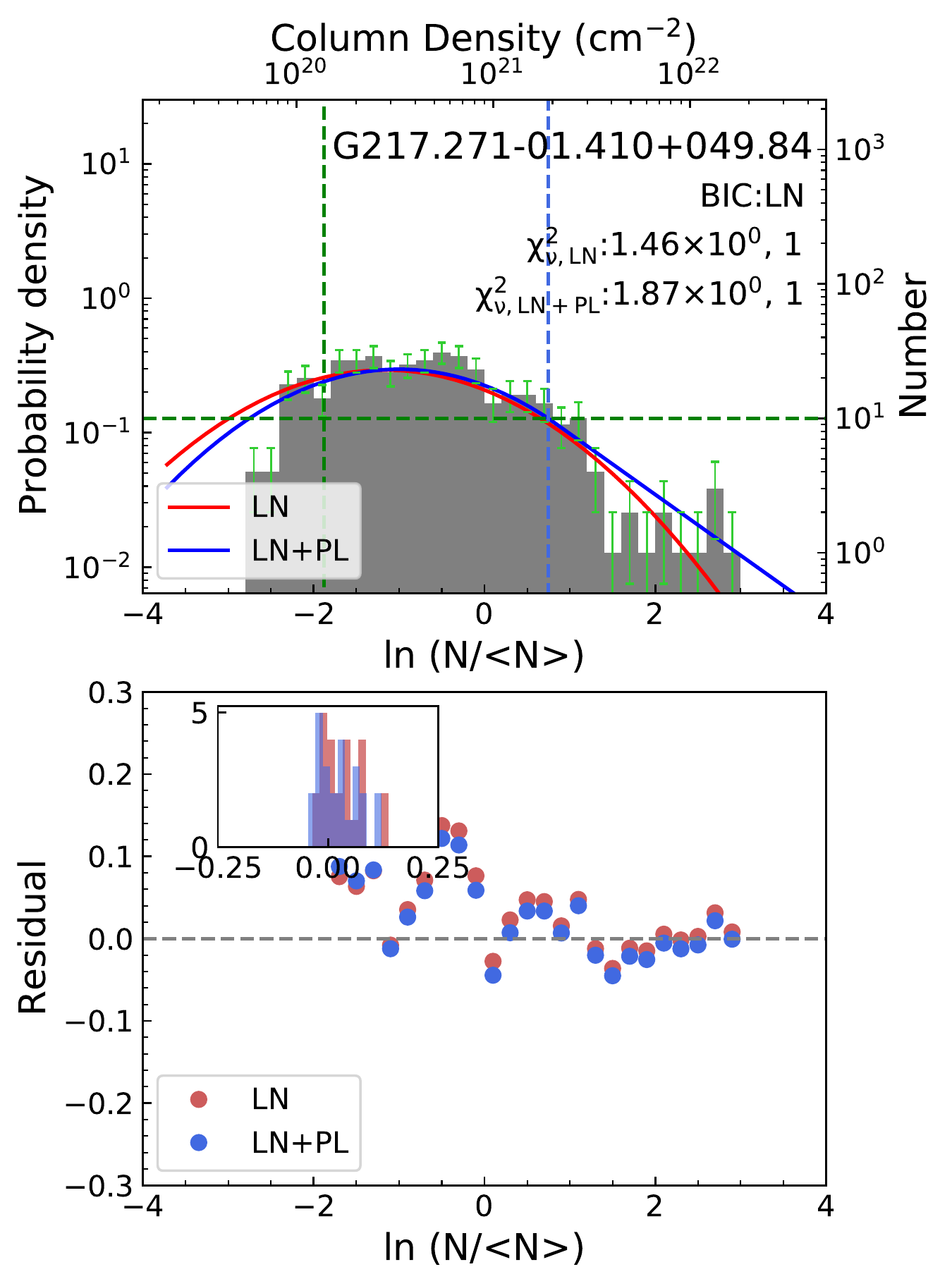}}
\subfigure{\includegraphics[trim=0cm 0cm 0cm 0cm, width= 0.23\linewidth, clip]{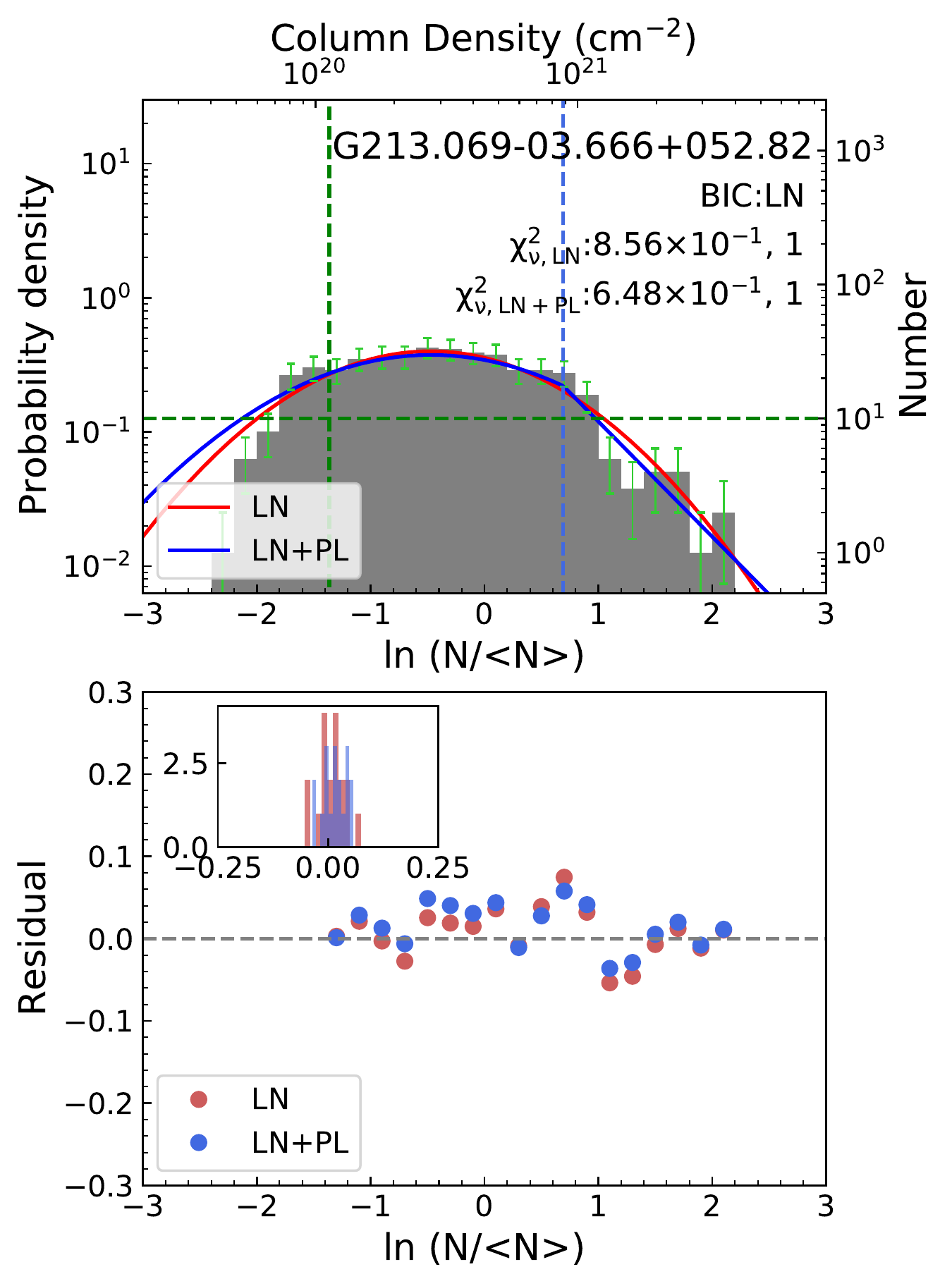}}
\subfigure{\includegraphics[trim=0cm 0cm 0cm 0cm, width= 0.23\linewidth, clip]{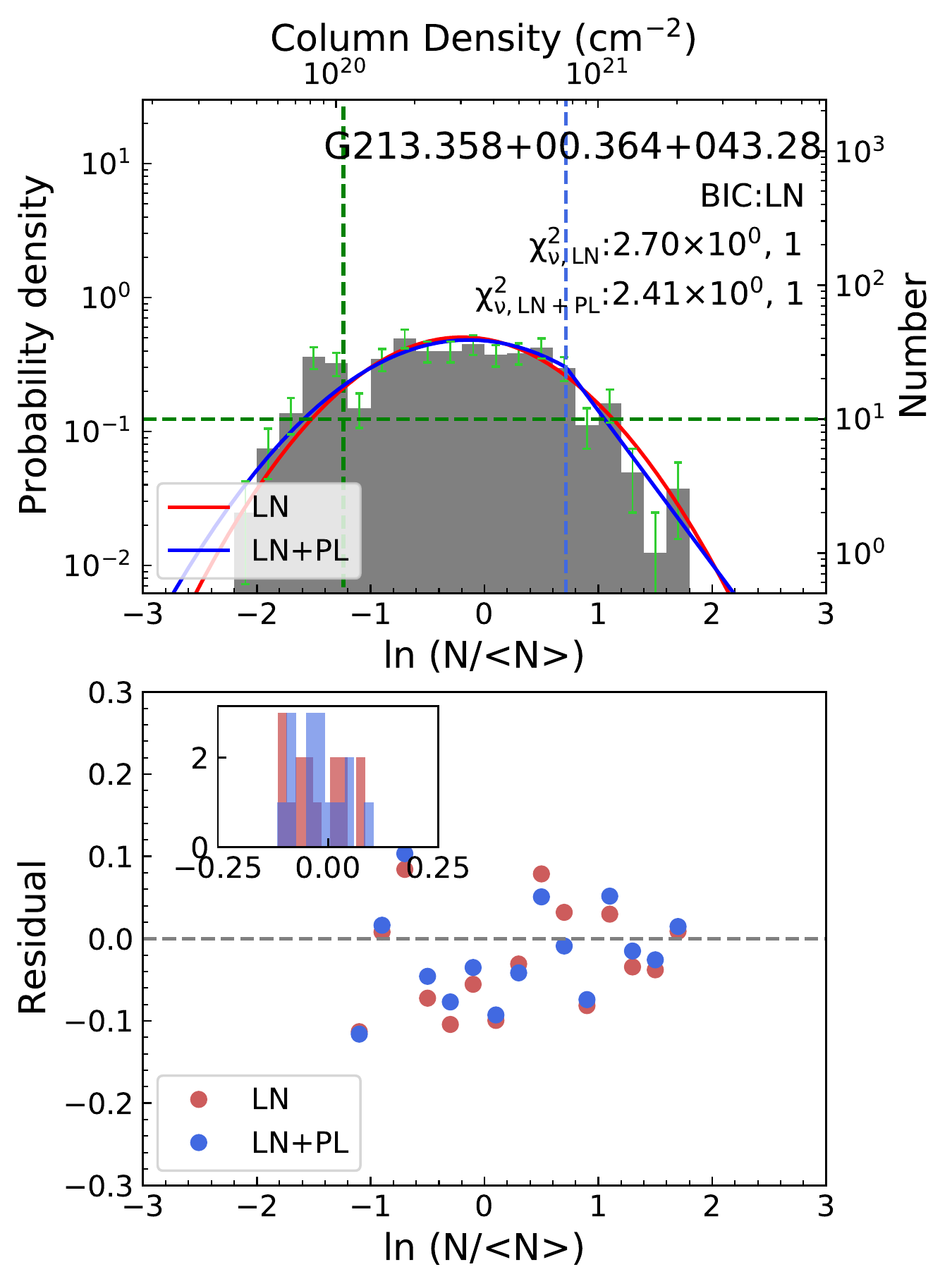}}

\end{figure}
\begin{figure}
\subfigure{\includegraphics[trim=0cm 0cm 0cm 0cm, width= 0.23\linewidth, clip]{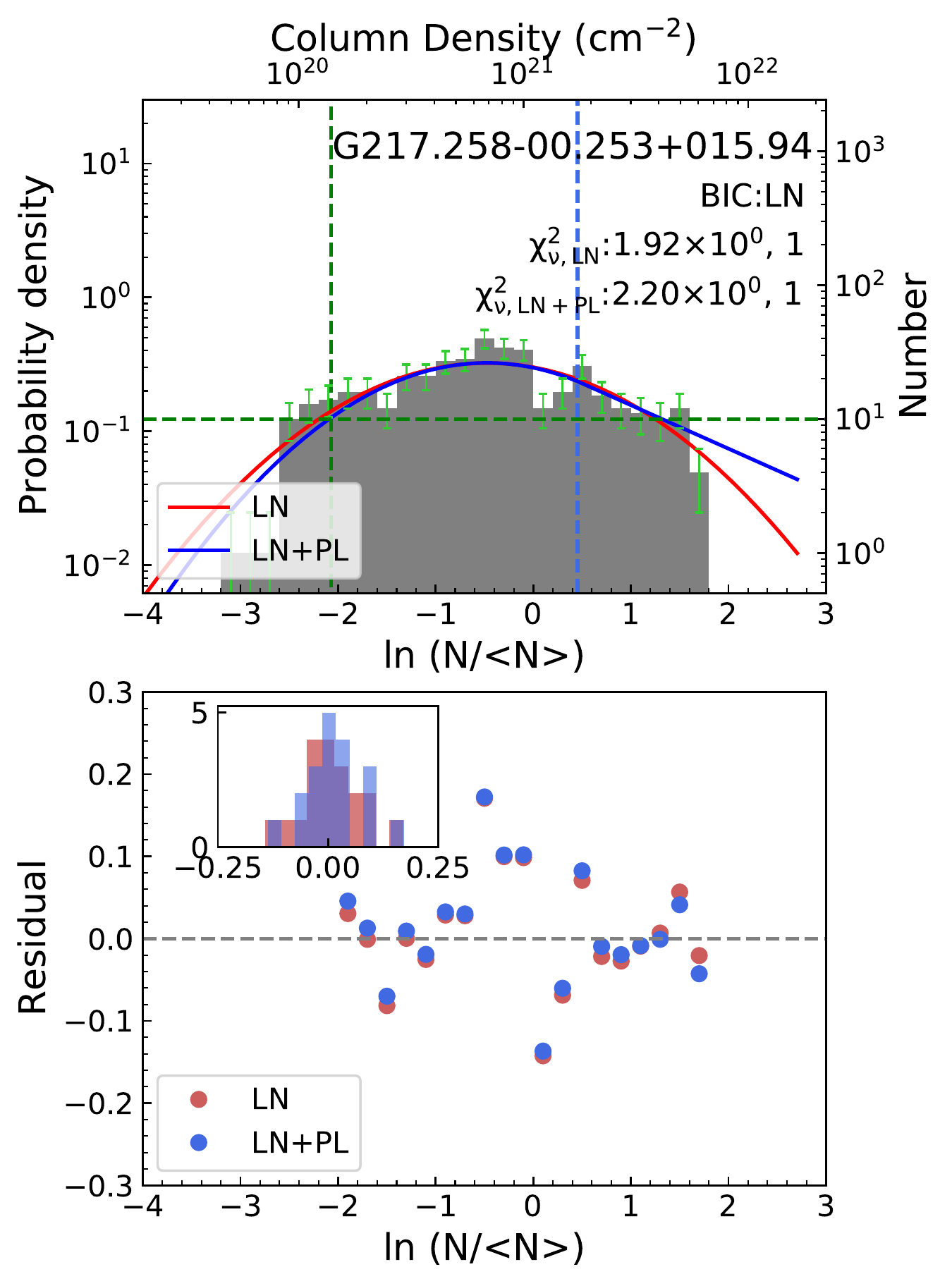}}
\subfigure{\includegraphics[trim=0cm 0cm 0cm 0cm, width= 0.23\linewidth, clip]{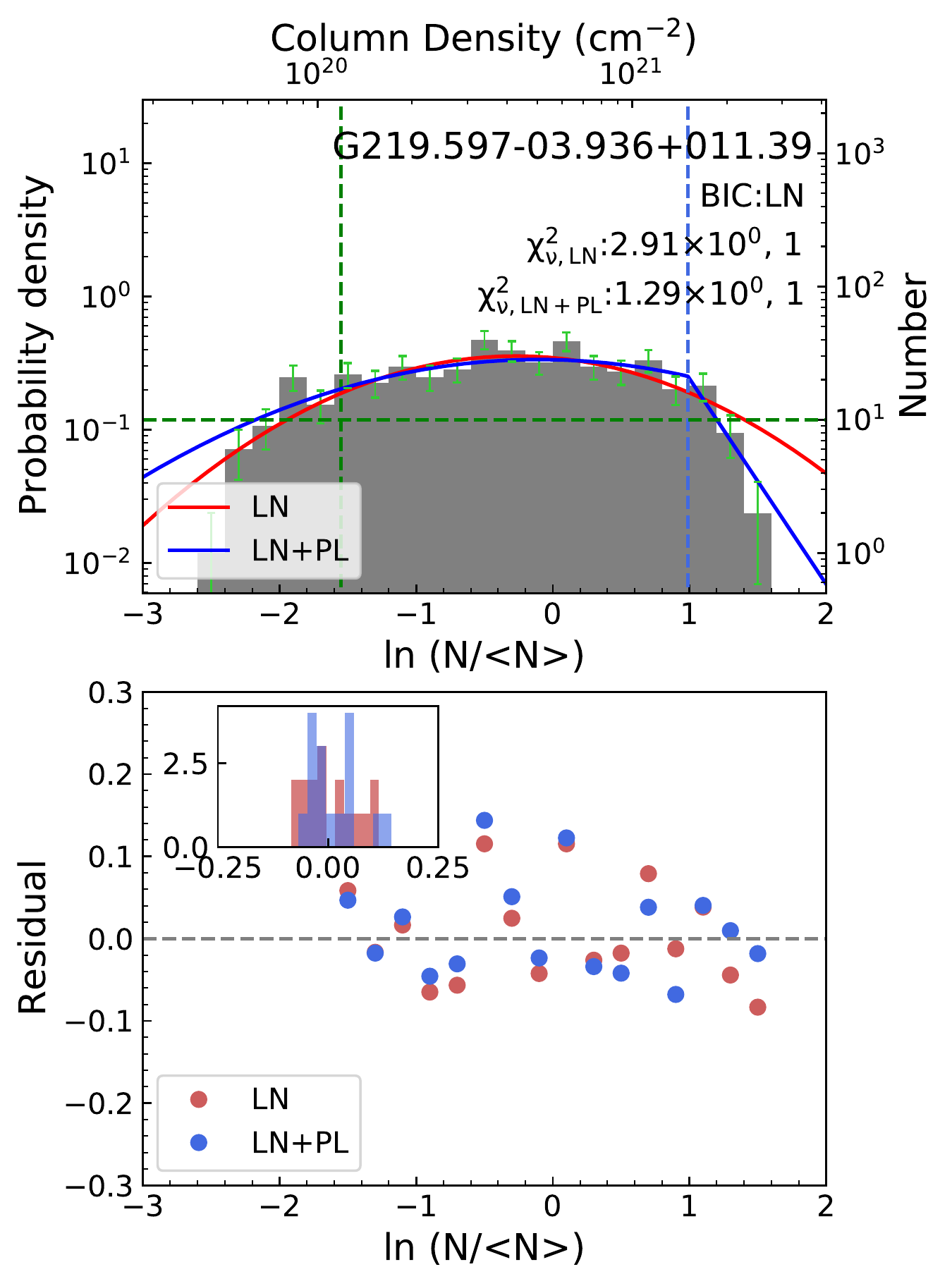}}
\subfigure{\includegraphics[trim=0cm 0cm 0cm 0cm, width= 0.23\linewidth, clip]{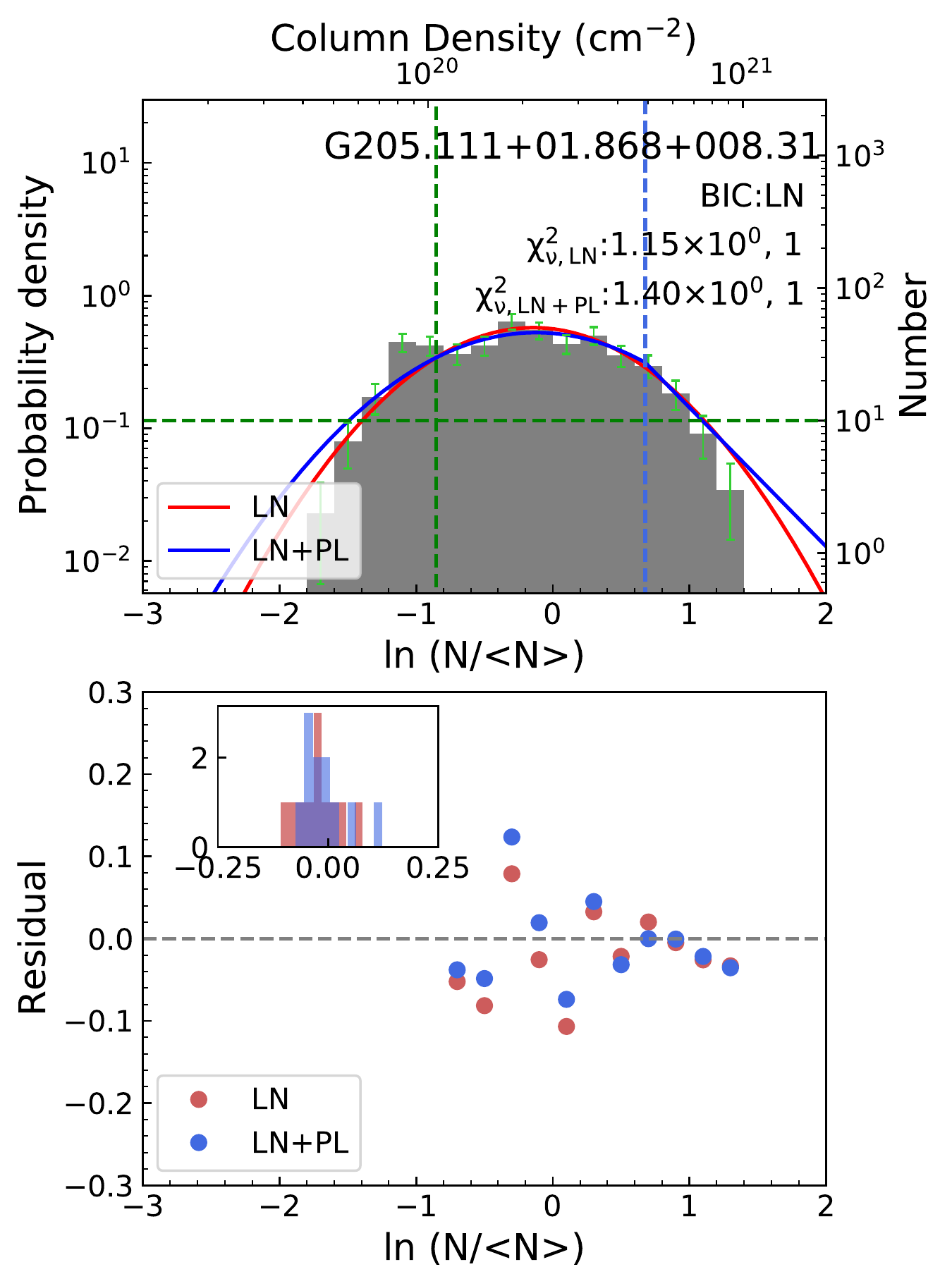}}
\subfigure{\includegraphics[trim=0cm 0cm 0cm 0cm, width= 0.23\linewidth, clip]{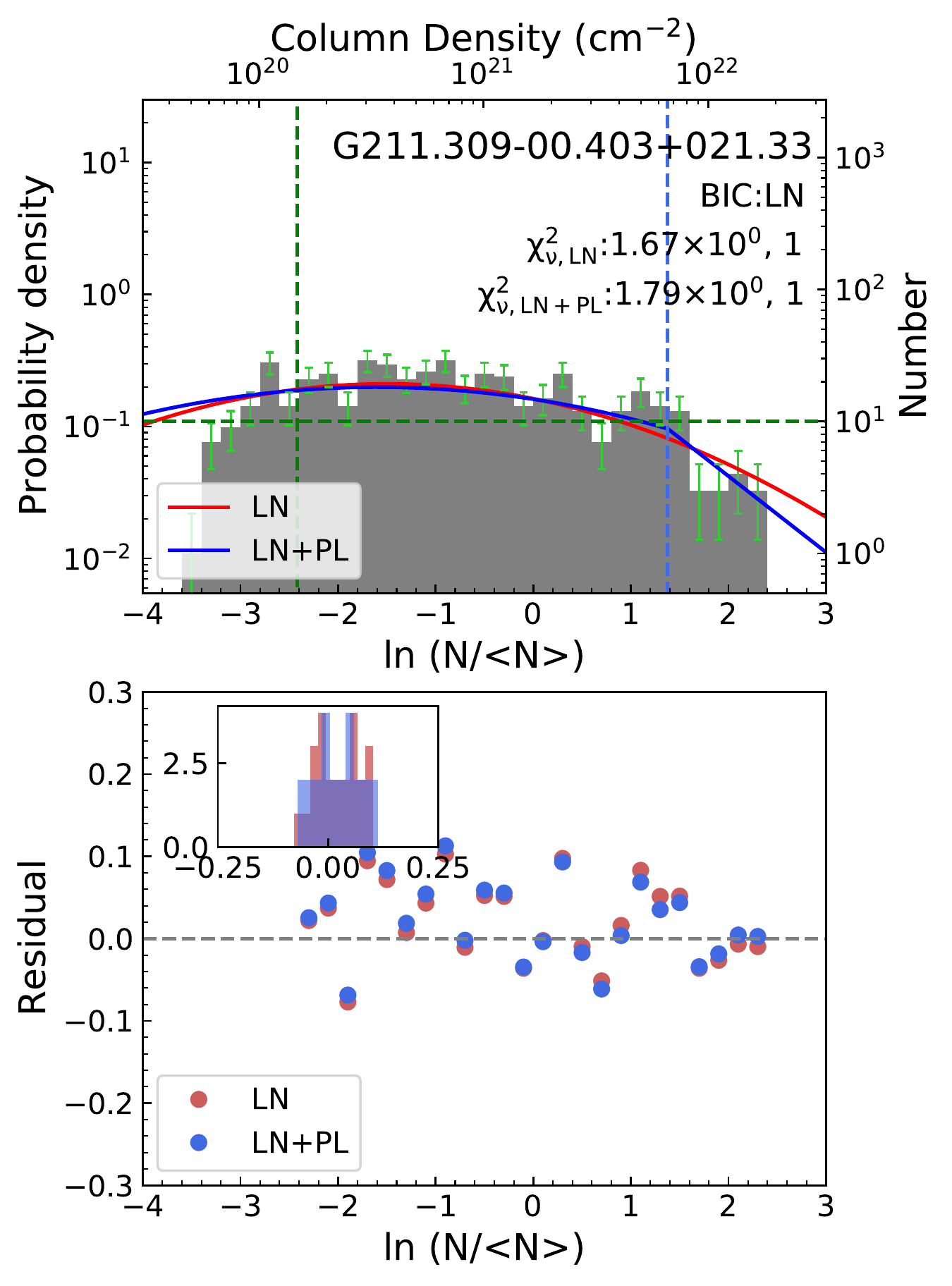}}

\subfigure{\includegraphics[trim=0cm 0cm 0cm 0cm, width= 0.23\linewidth, clip]{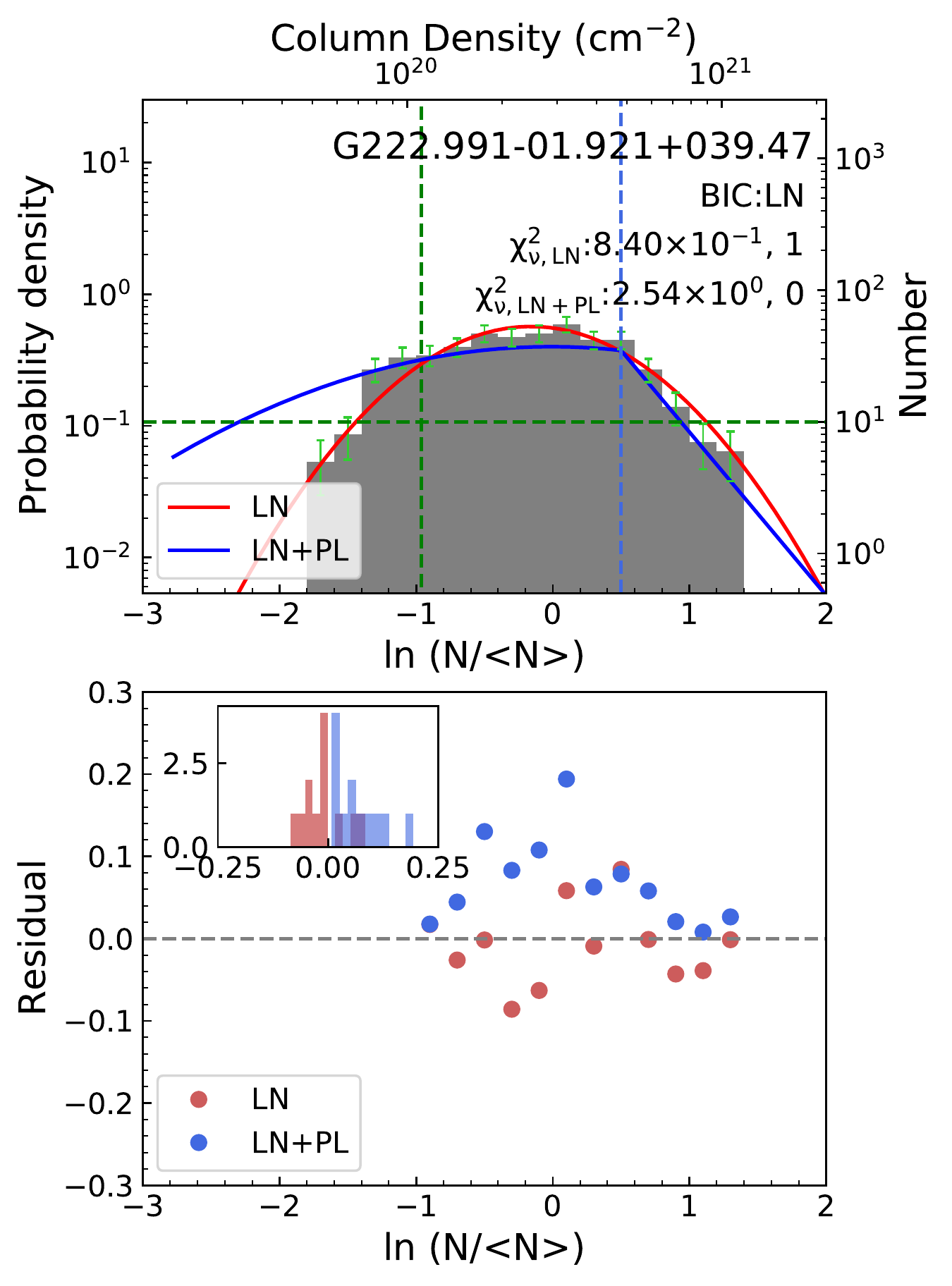}}
\subfigure{\includegraphics[trim=0cm 0cm 0cm 0cm, width= 0.23\linewidth, clip]{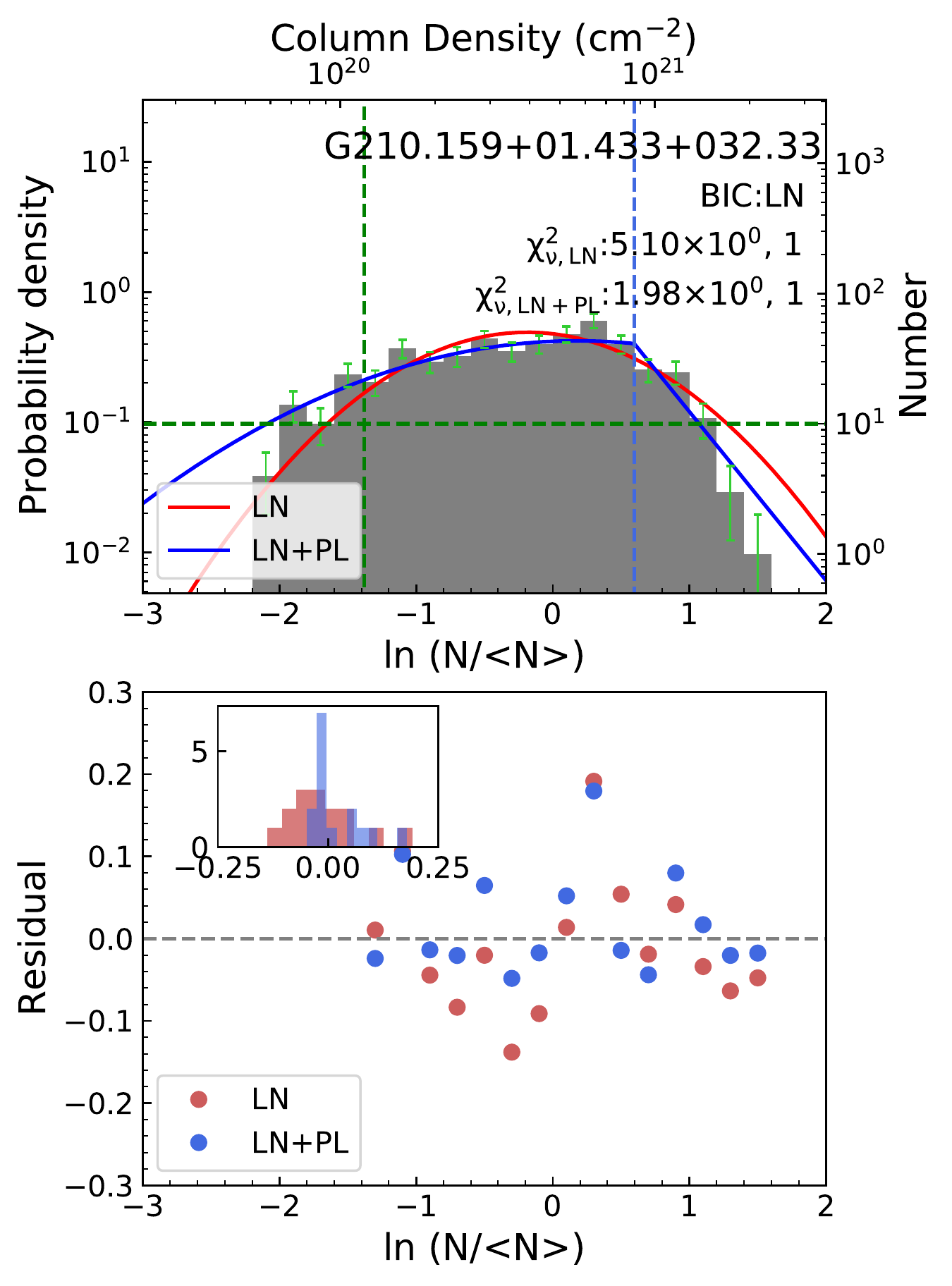}}
\subfigure{\includegraphics[trim=0cm 0cm 0cm 0cm, width= 0.23\linewidth, clip]{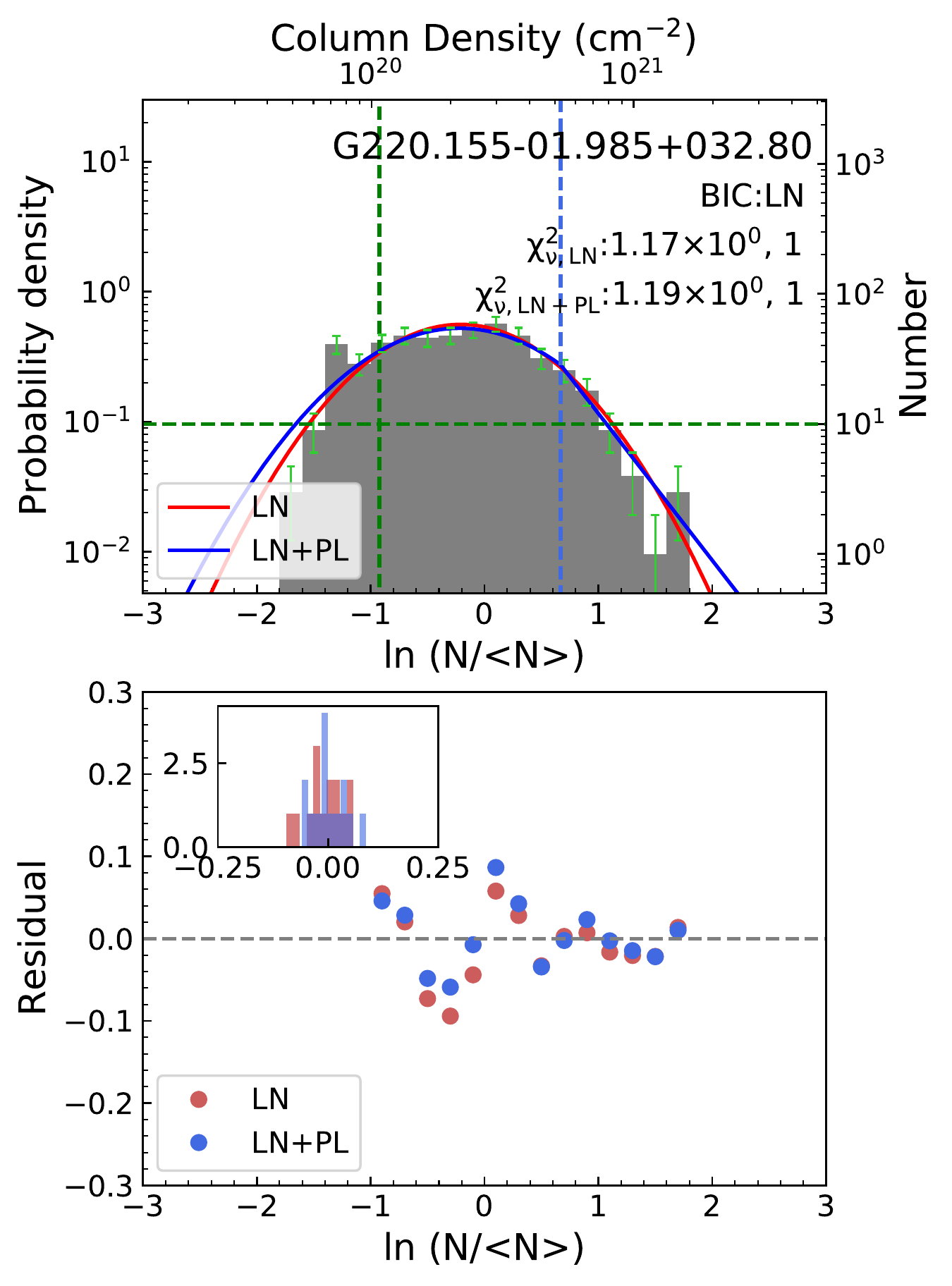}}
\subfigure{\includegraphics[trim=0cm 0cm 0cm 0cm, width= 0.23\linewidth, clip]{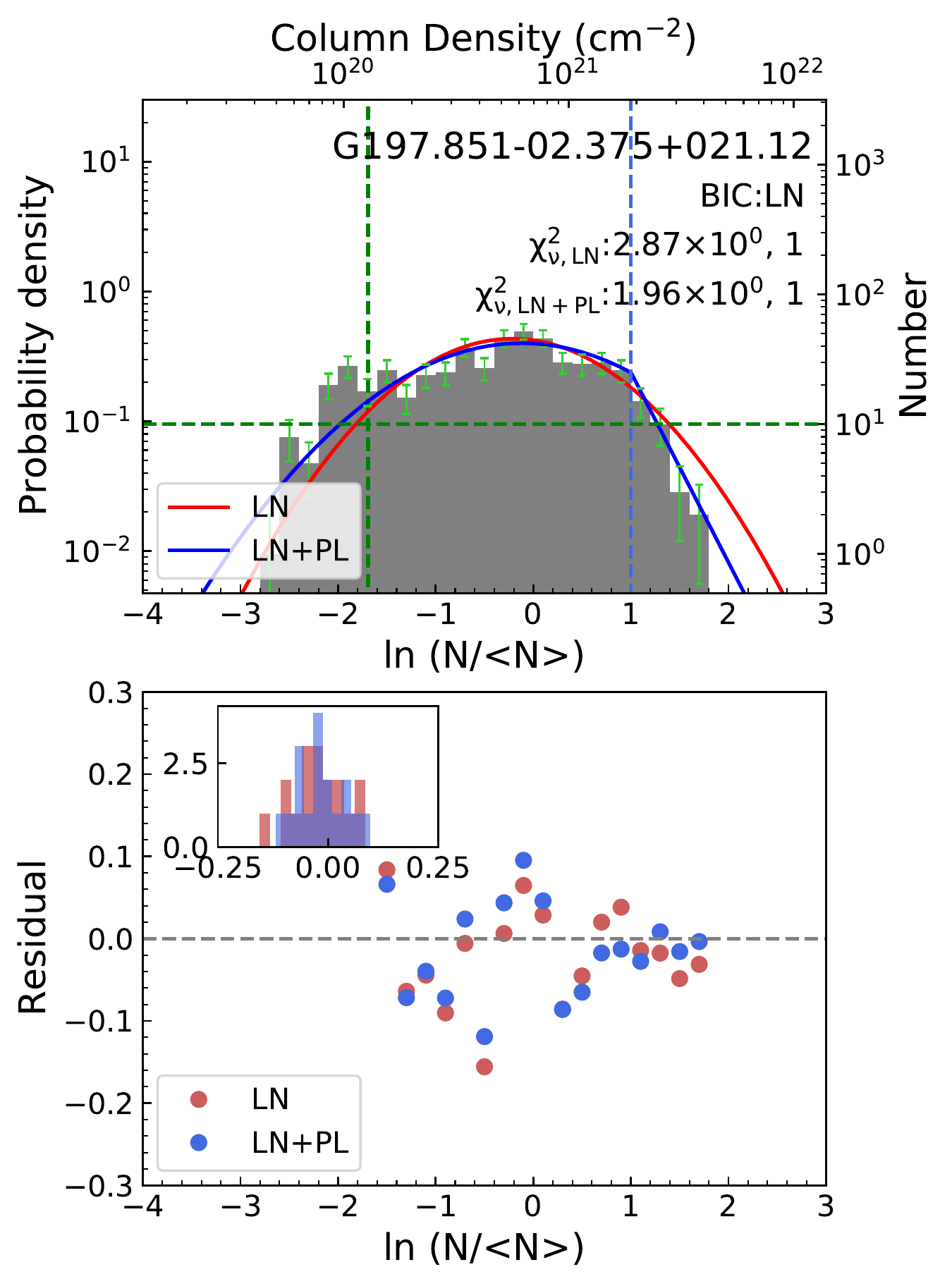}}

\subfigure{\includegraphics[trim=0cm 0cm 0cm 0cm, width= 0.23\linewidth, clip]{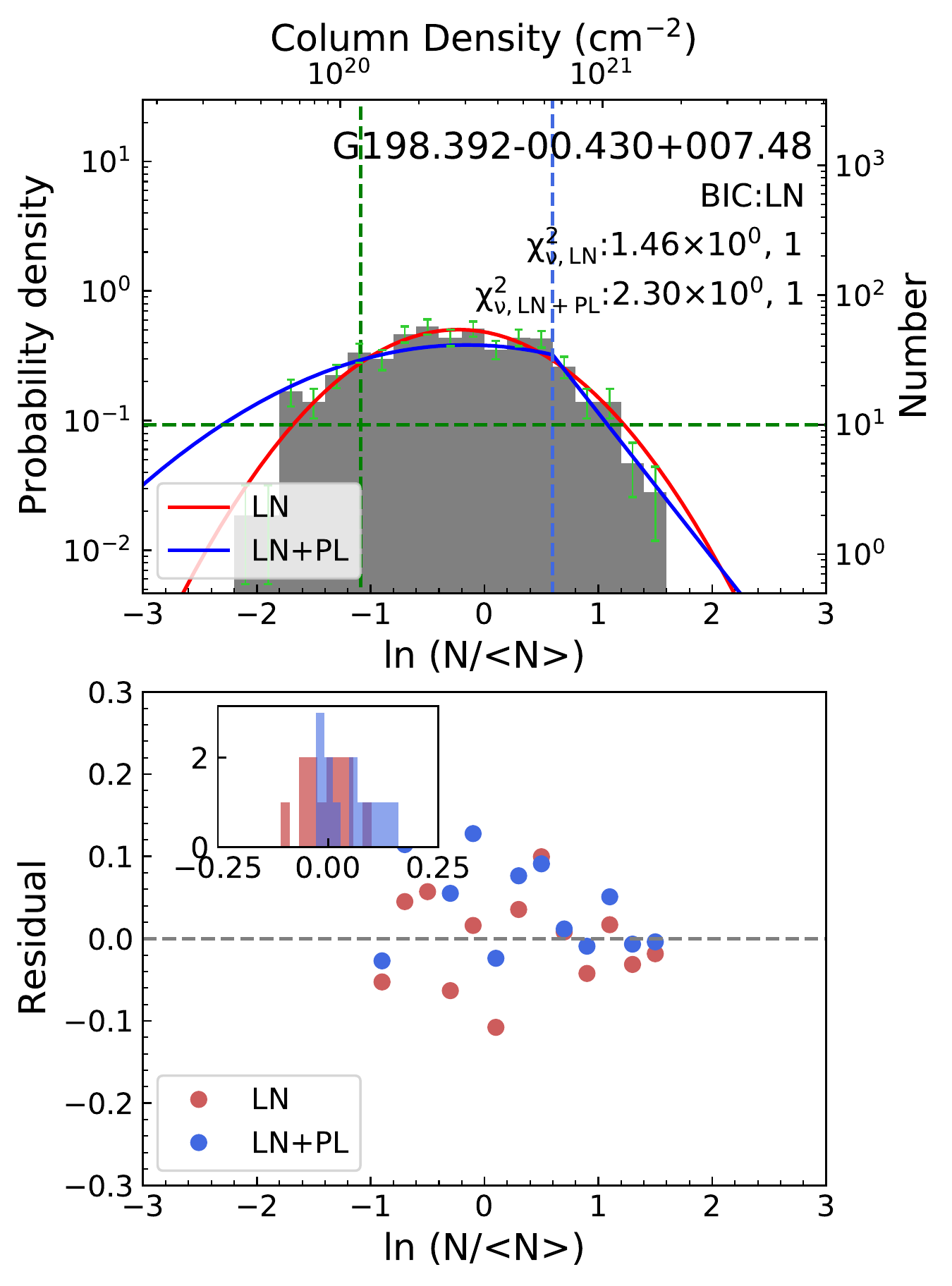}}
\subfigure{\includegraphics[trim=0cm 0cm 0cm 0cm, width= 0.23\linewidth, clip]{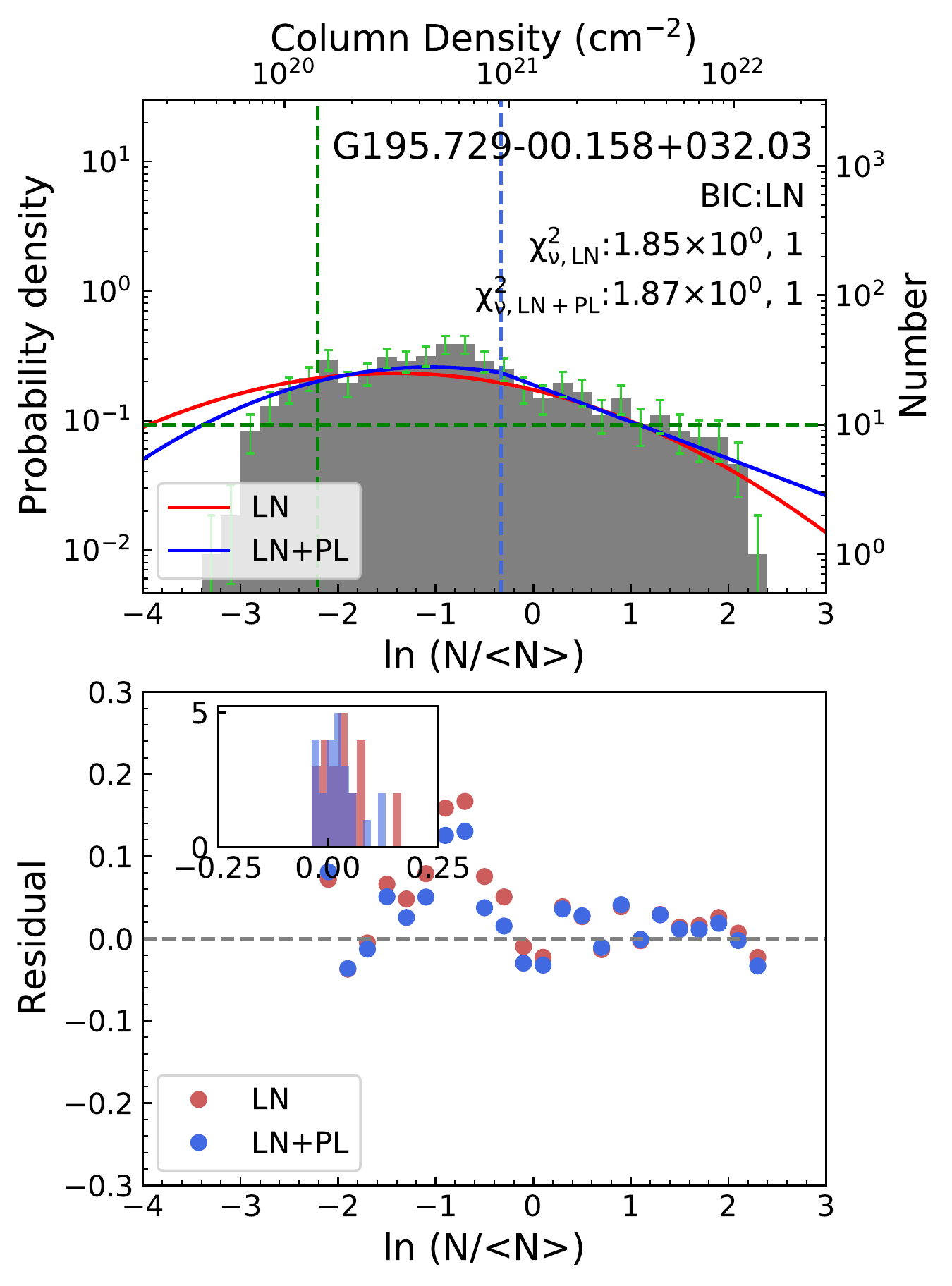}}
\subfigure{\includegraphics[trim=0cm 0cm 0cm 0cm, width= 0.23\linewidth, clip]{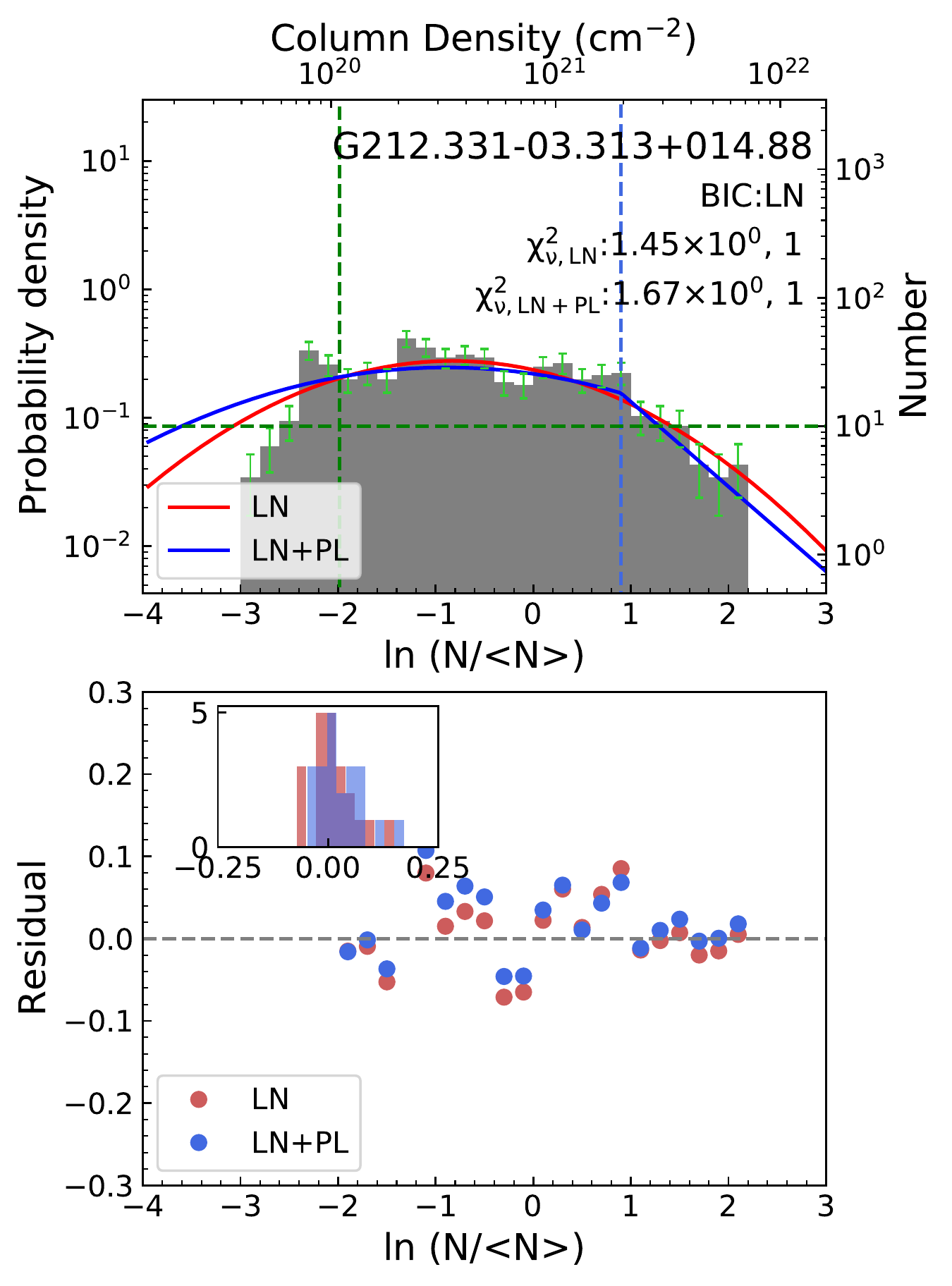}}
\subfigure{\includegraphics[trim=0cm 0cm 0cm 0cm, width= 0.23\linewidth, clip]{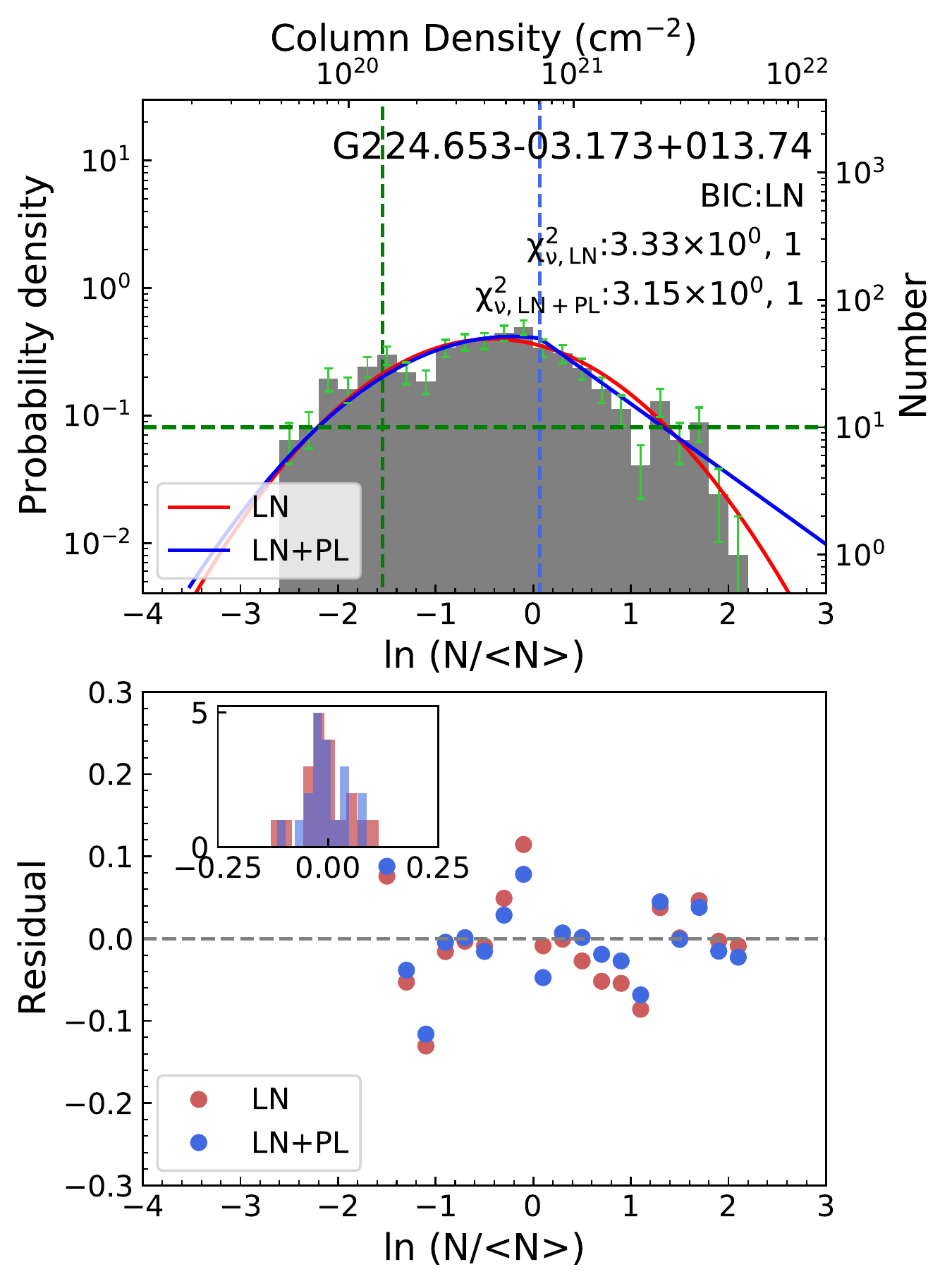}}

\subfigure{\includegraphics[trim=0cm 0cm 0cm 0cm, width= 0.23\linewidth, clip]{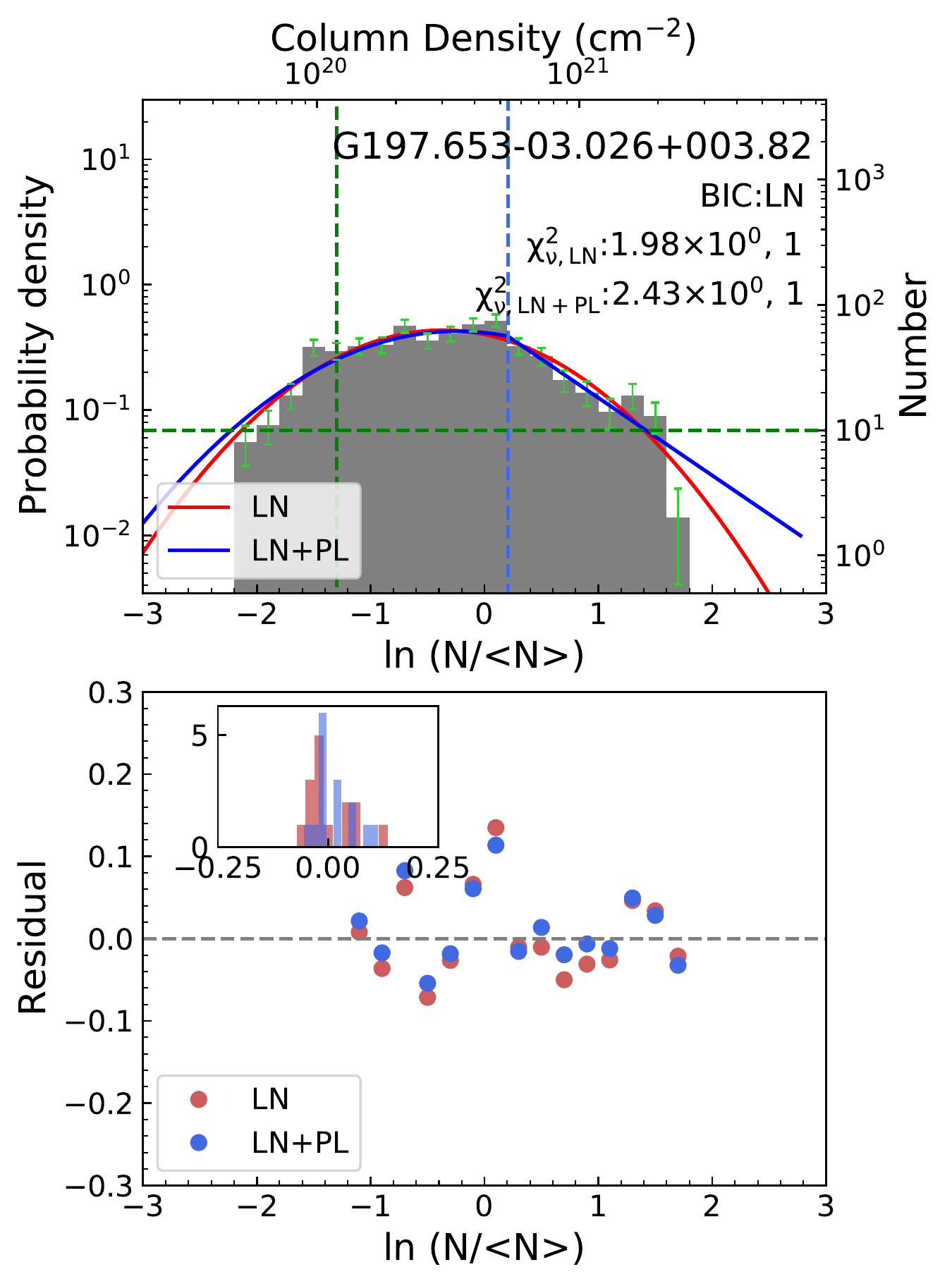}}
\subfigure{\includegraphics[trim=0cm 0cm 0cm 0cm, width= 0.23\linewidth, clip]{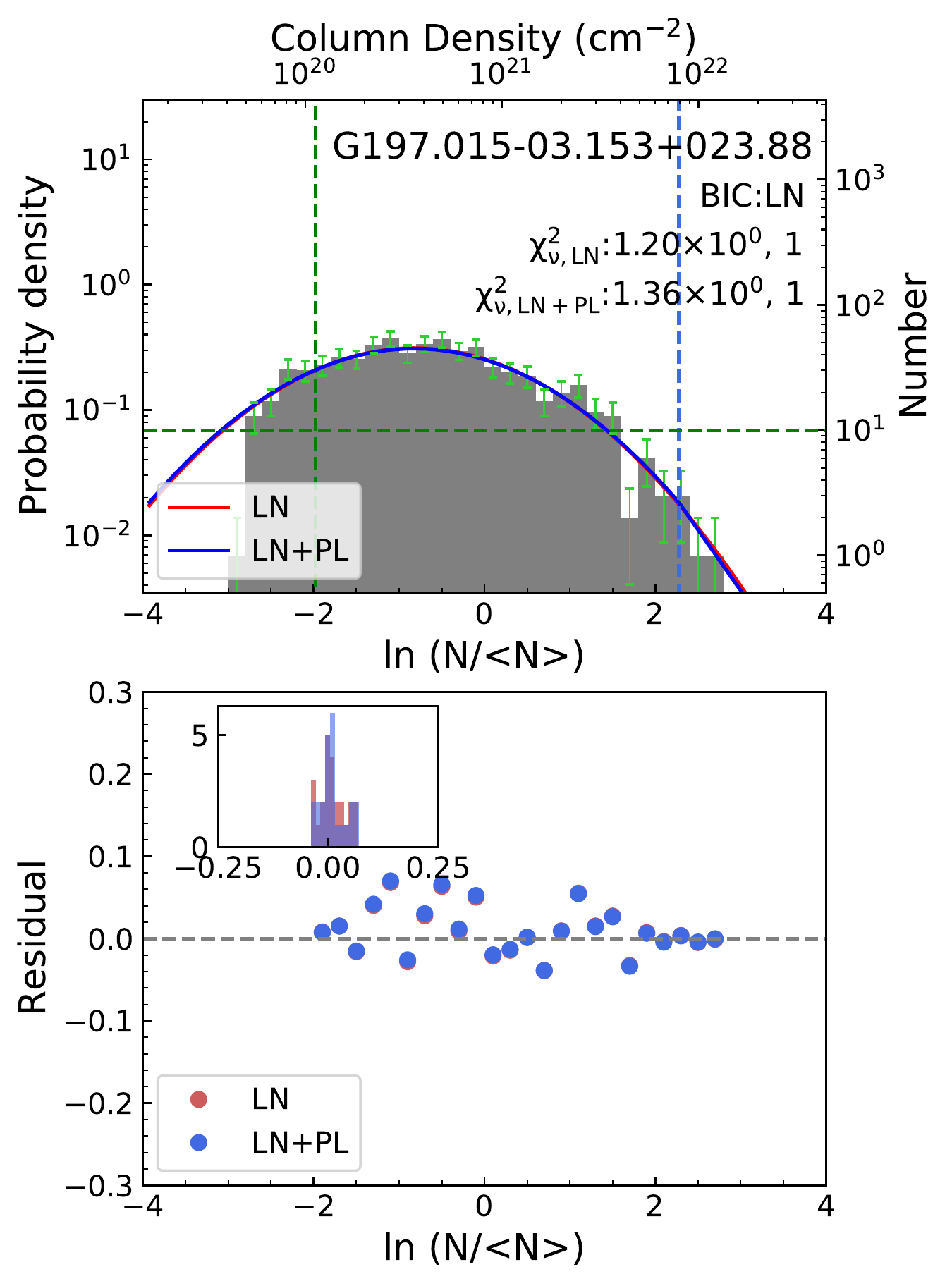}}
\subfigure{\includegraphics[trim=0cm 0cm 0cm 0cm, width= 0.23\linewidth, clip]{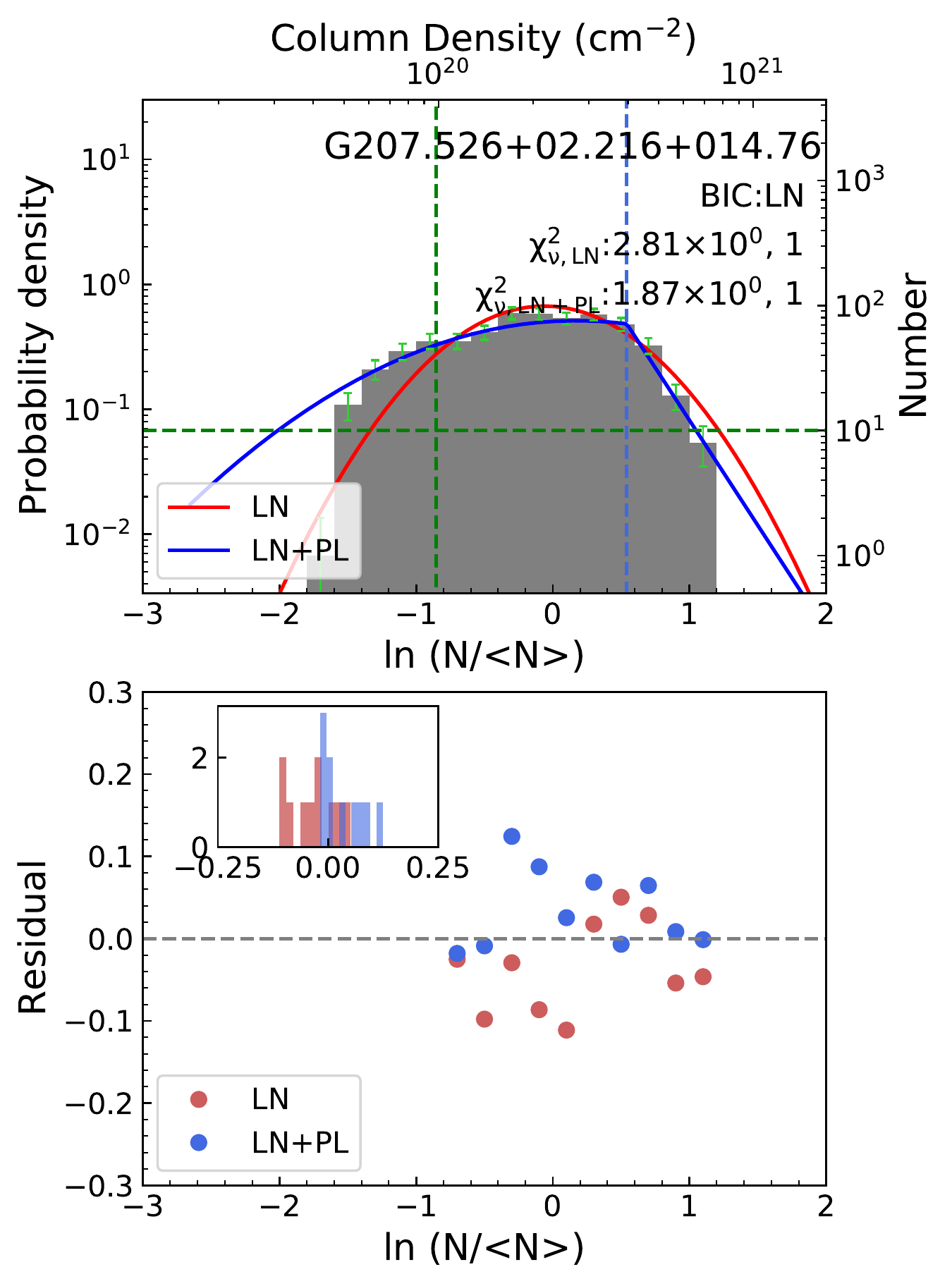}}
\subfigure{\includegraphics[trim=0cm 0cm 0cm 0cm, width= 0.23\linewidth, clip]{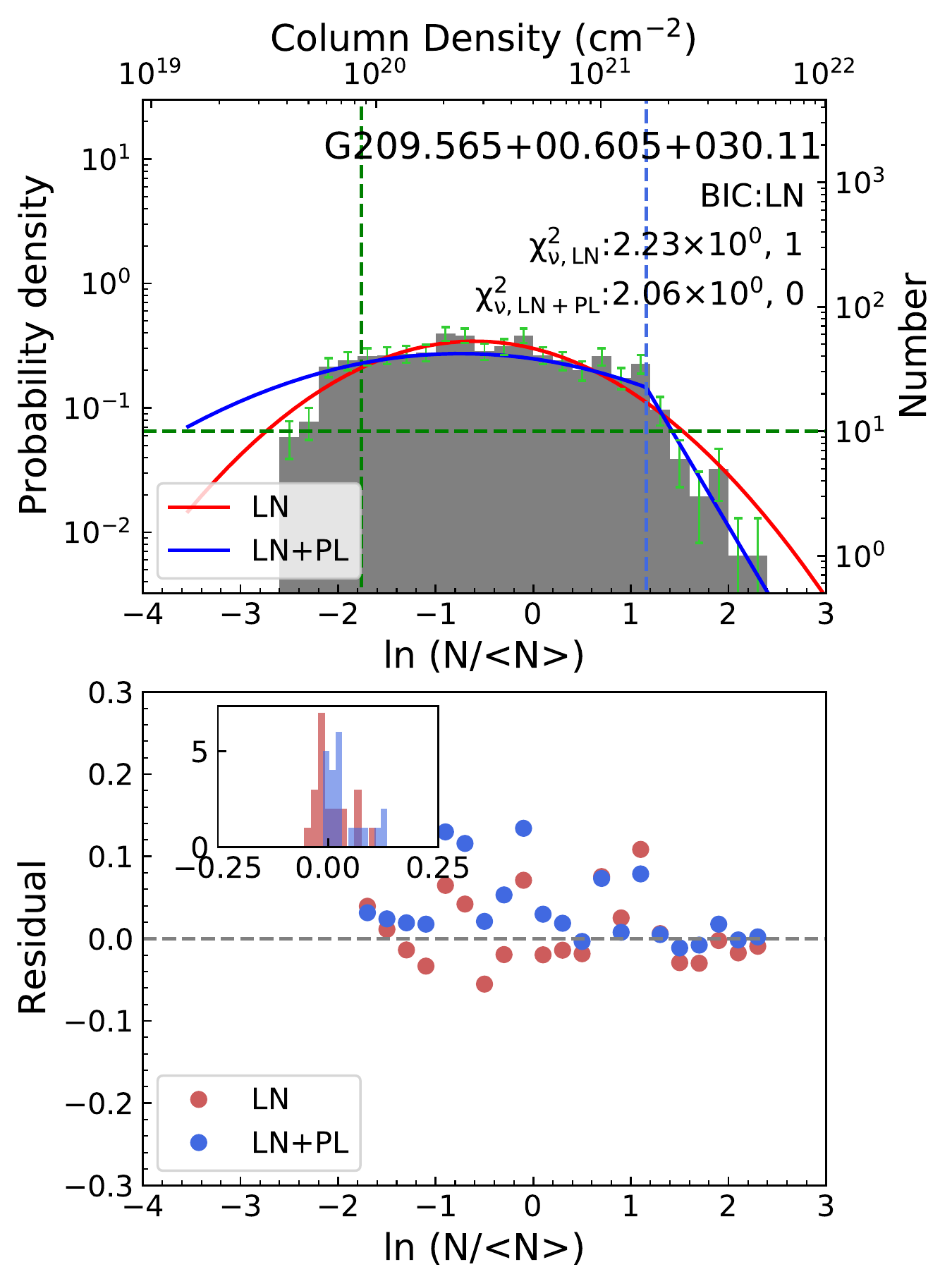}}

\end{figure}
\begin{figure}
\subfigure{\includegraphics[trim=0cm 0cm 0cm 0cm, width= 0.23\linewidth, clip]{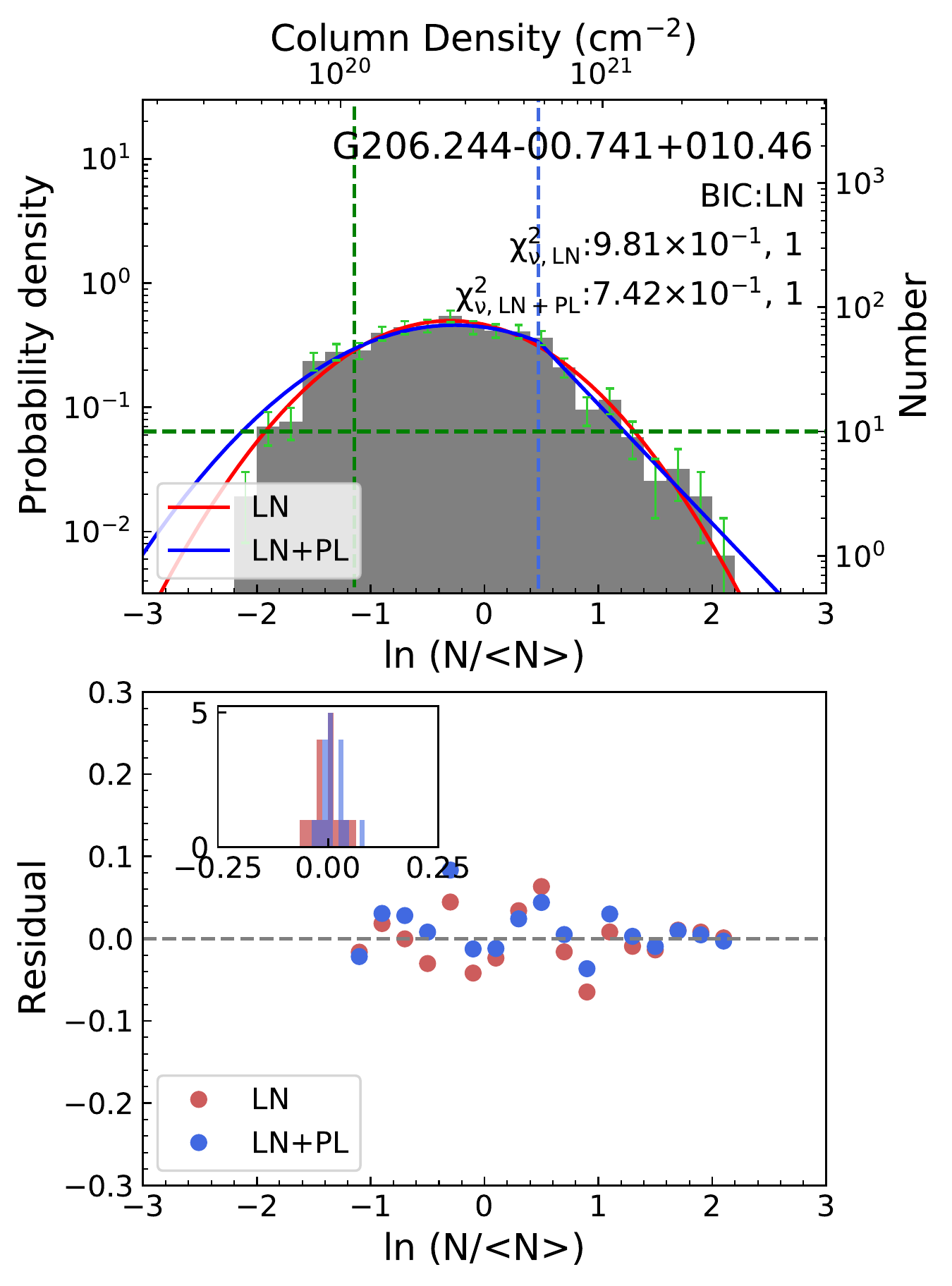}}
\subfigure{\includegraphics[trim=0cm 0cm 0cm 0cm, width= 0.23\linewidth, clip]{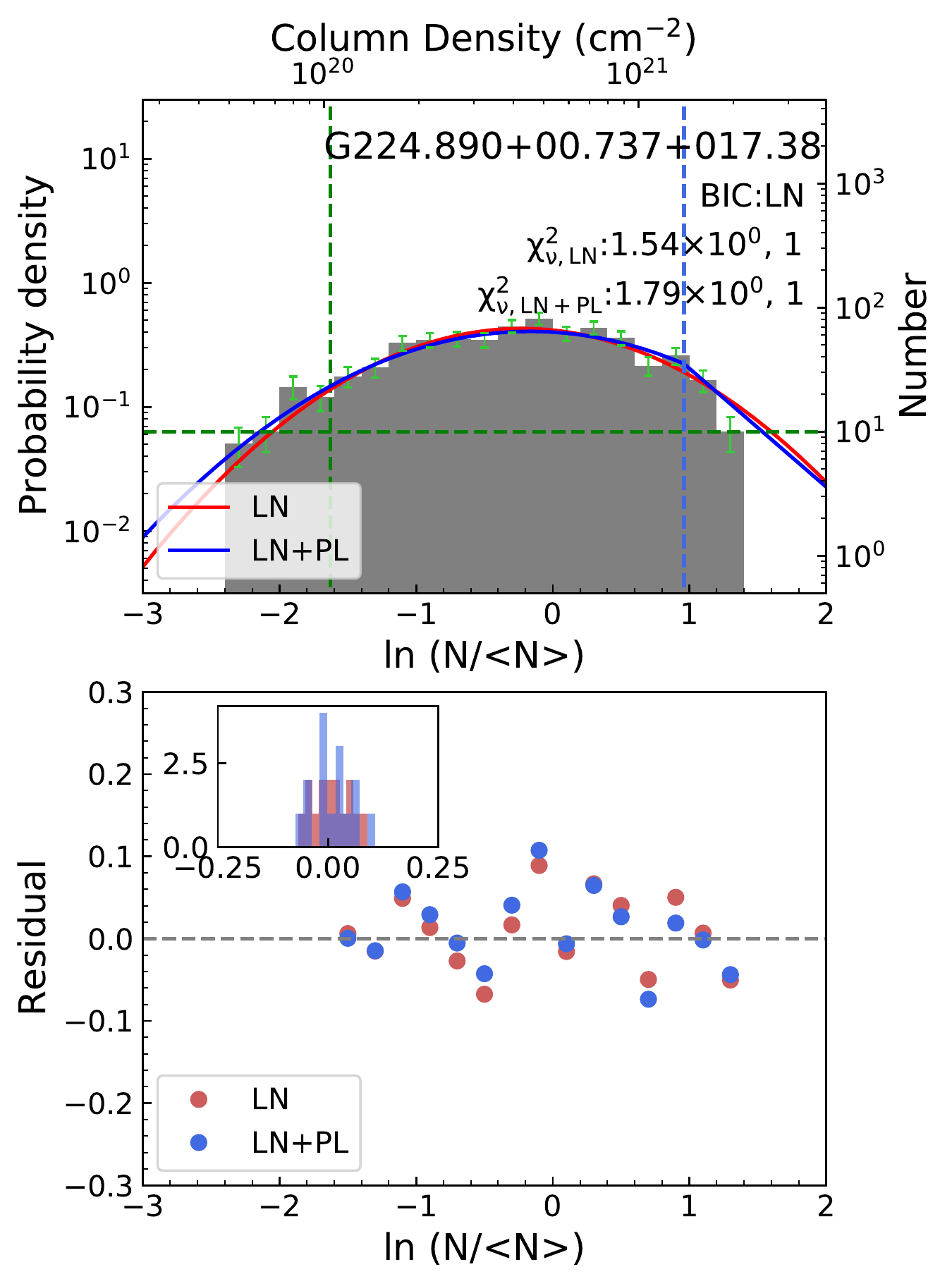}}
\subfigure{\includegraphics[trim=0cm 0cm 0cm 0cm, width= 0.23\linewidth, clip]{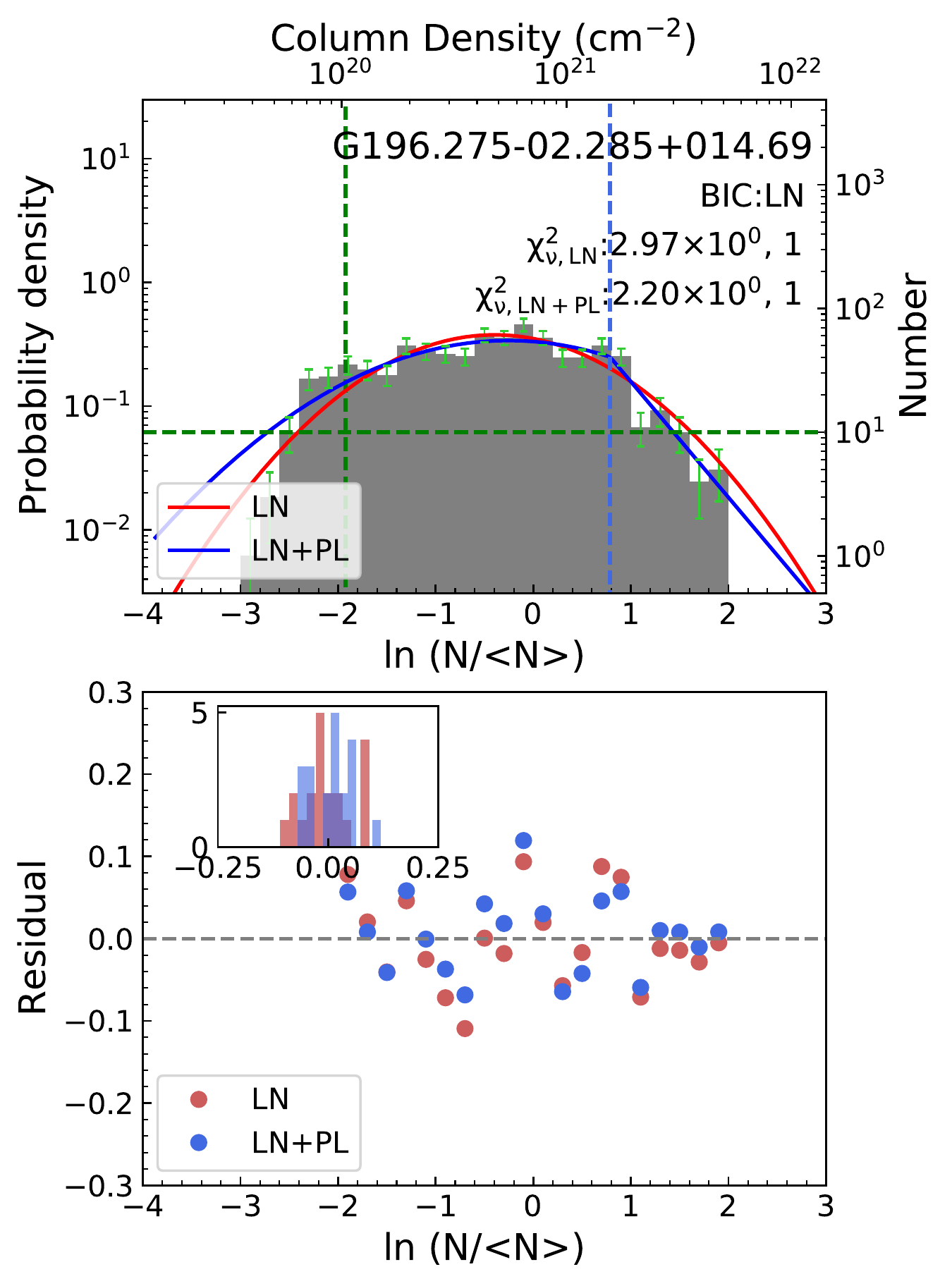}}
\subfigure{\includegraphics[trim=0cm 0cm 0cm 0cm, width= 0.23\linewidth, clip]{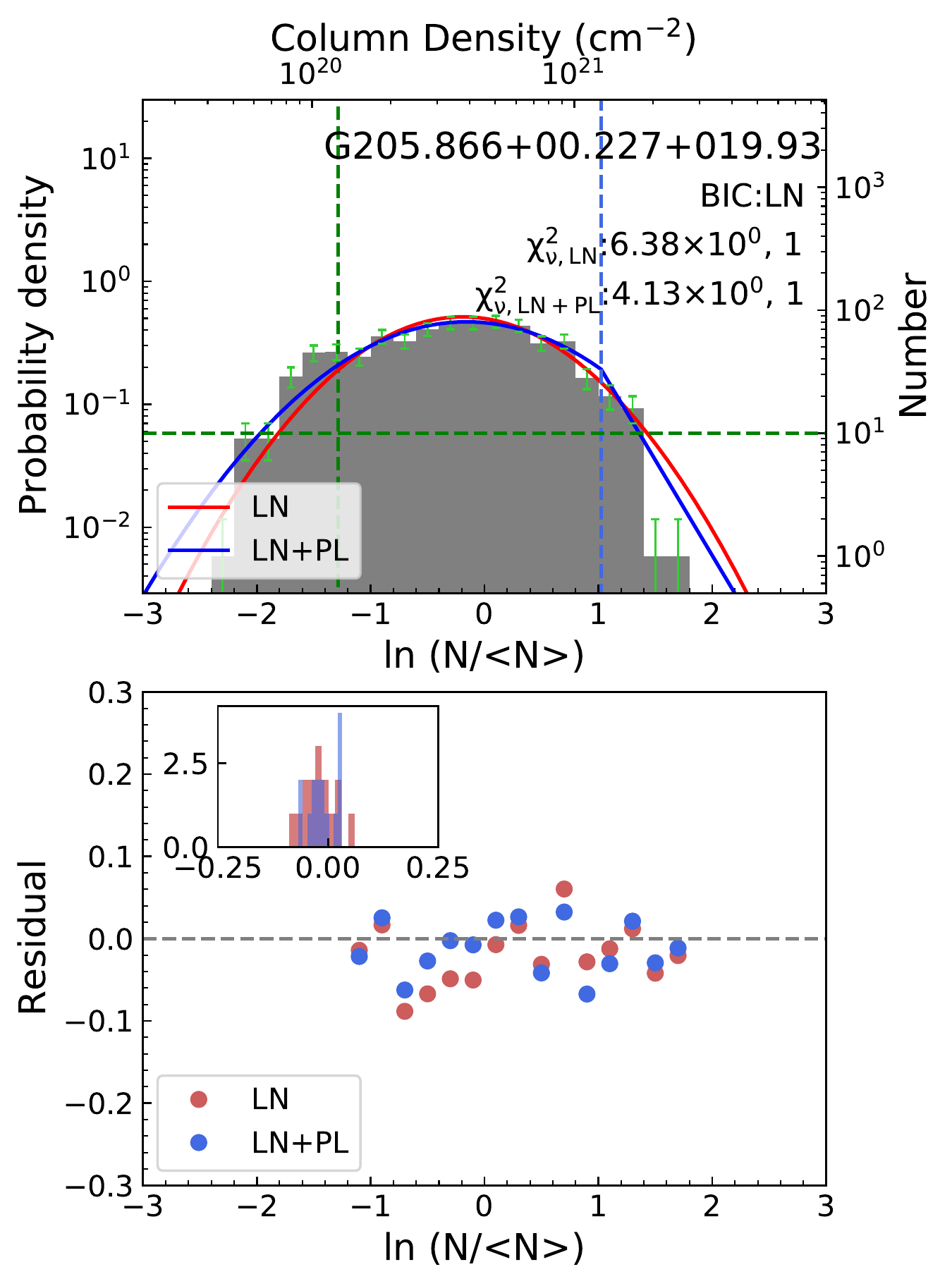}}

\subfigure{\includegraphics[trim=0cm 0cm 0cm 0cm, width= 0.23\linewidth, clip]{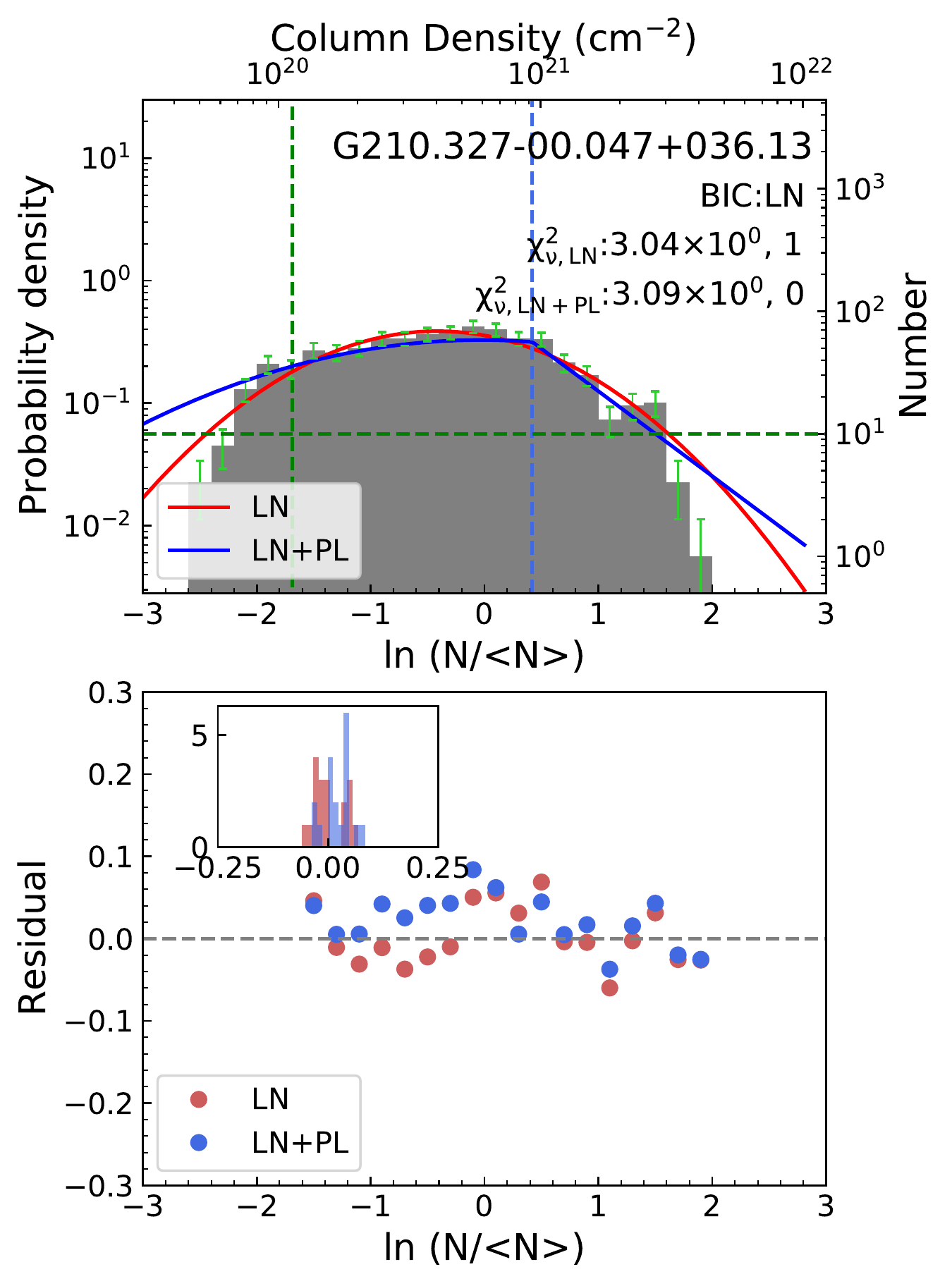}}
\subfigure{\includegraphics[trim=0cm 0cm 0cm 0cm, width= 0.23\linewidth, clip]{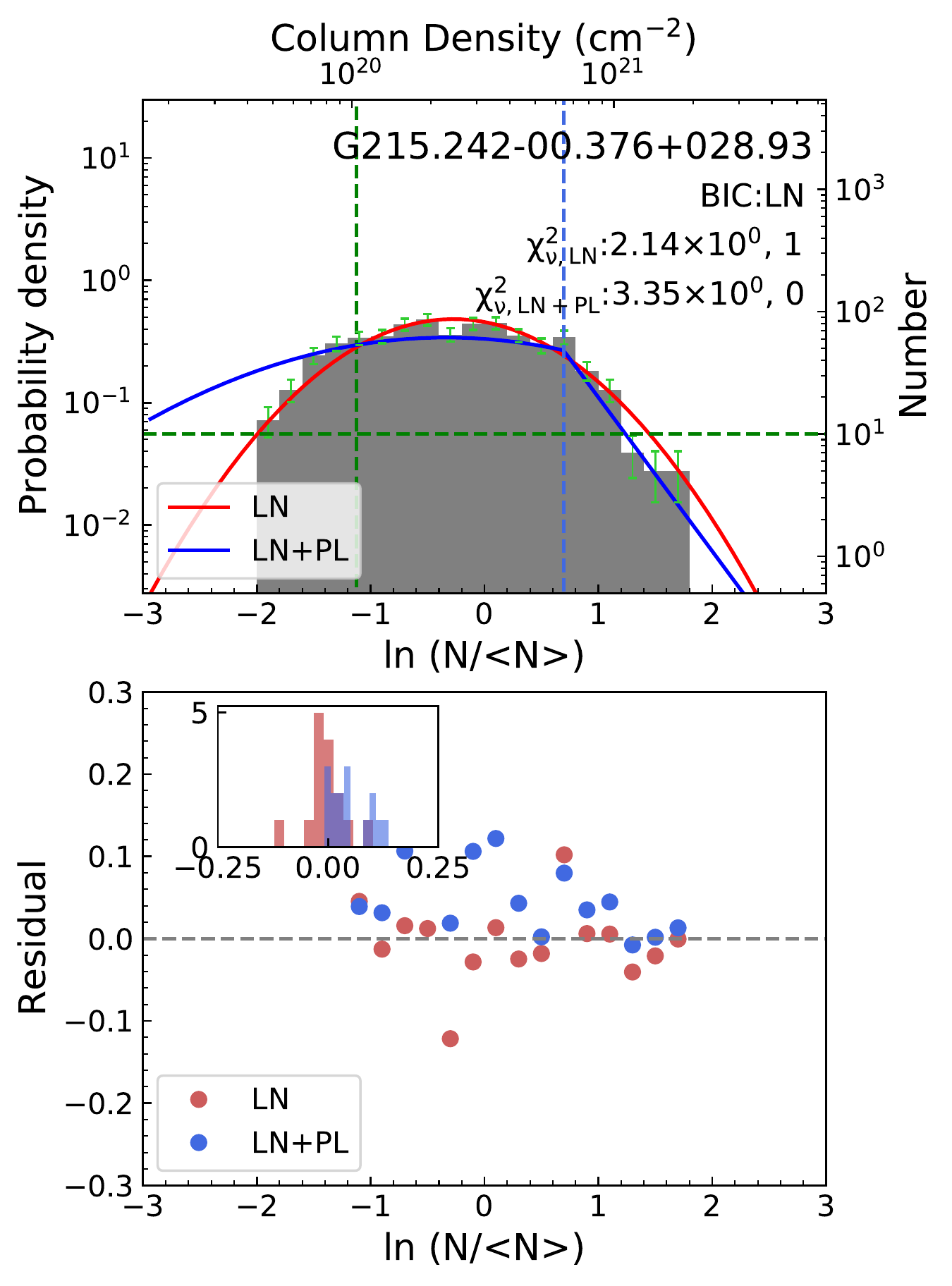}}
\subfigure{\includegraphics[trim=0cm 0cm 0cm 0cm, width= 0.23\linewidth, clip]{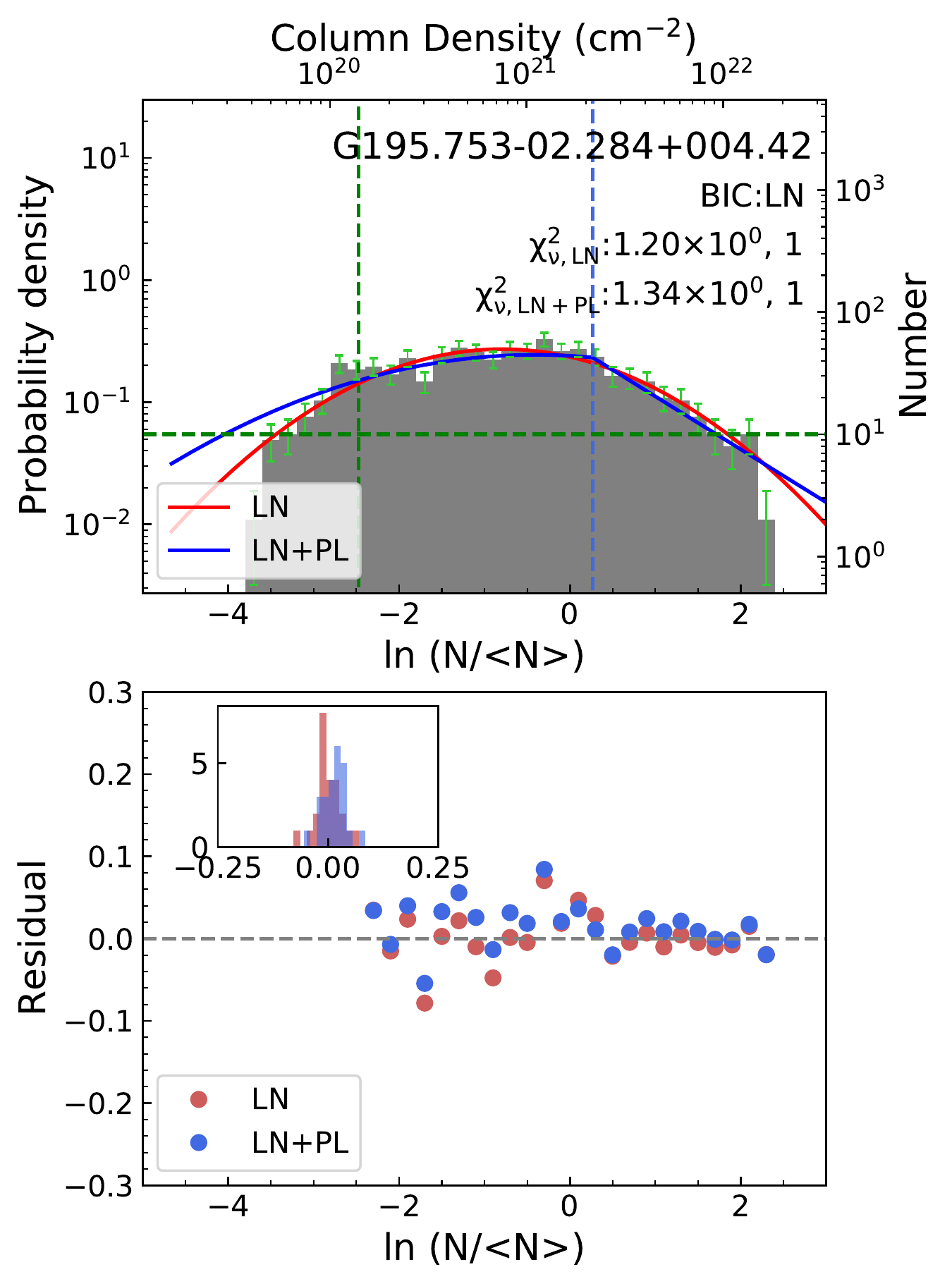}}
\subfigure{\includegraphics[trim=0cm 0cm 0cm 0cm, width= 0.23\linewidth, clip]{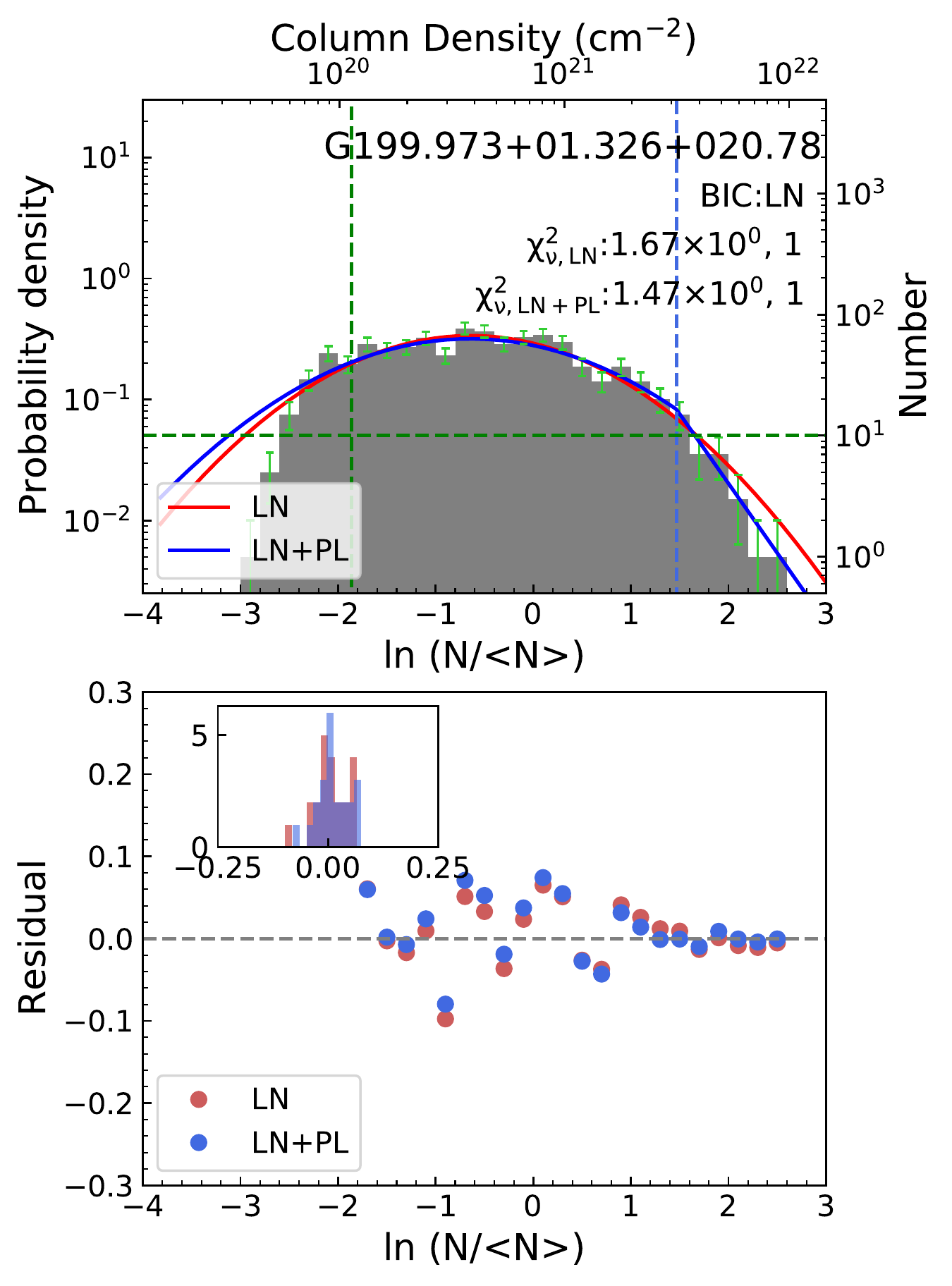}}

\subfigure{\includegraphics[trim=0cm 0cm 0cm 0cm, width= 0.23\linewidth, clip]{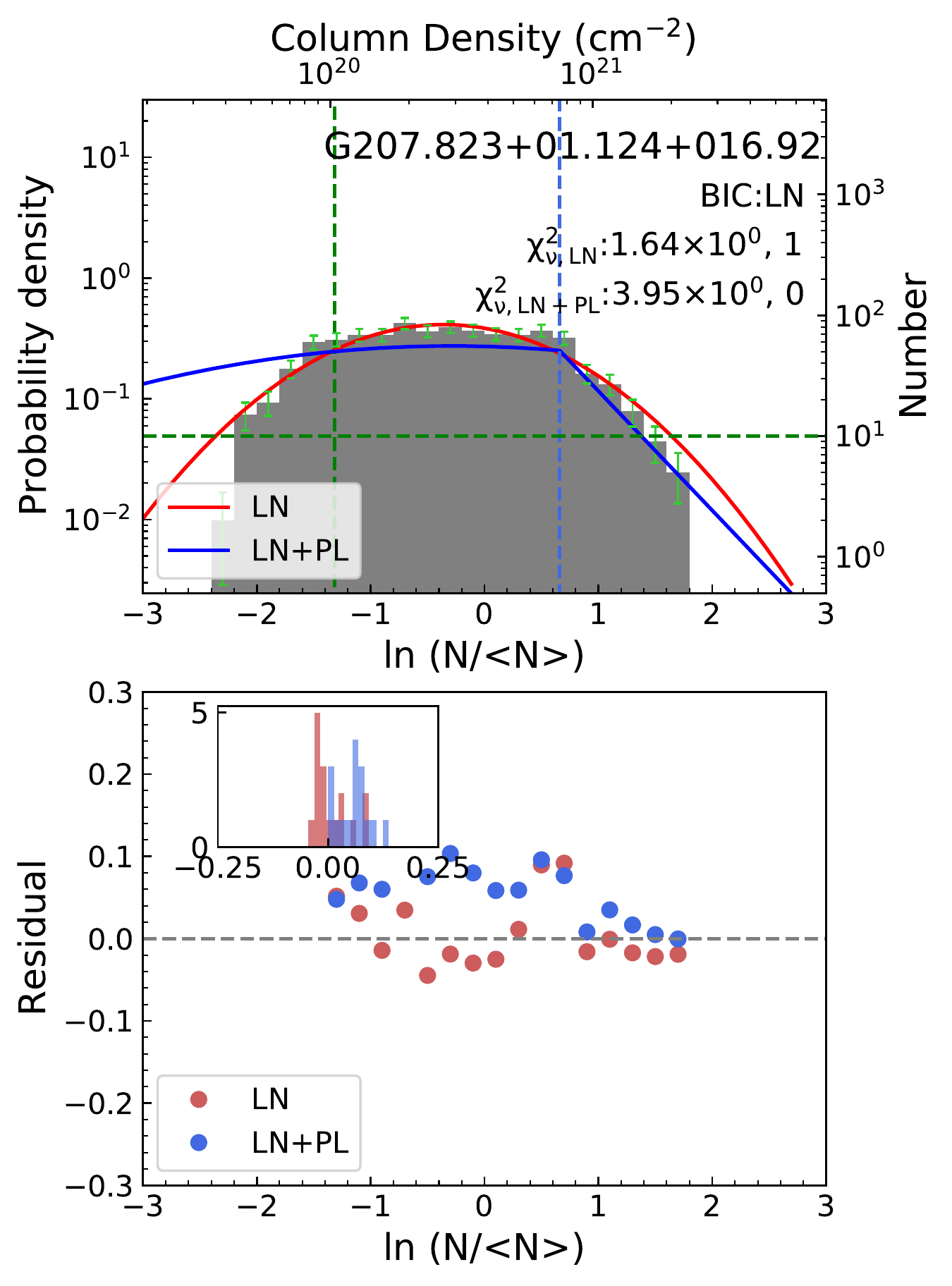}}
\subfigure{\includegraphics[trim=0cm 0cm 0cm 0cm, width= 0.23\linewidth, clip]{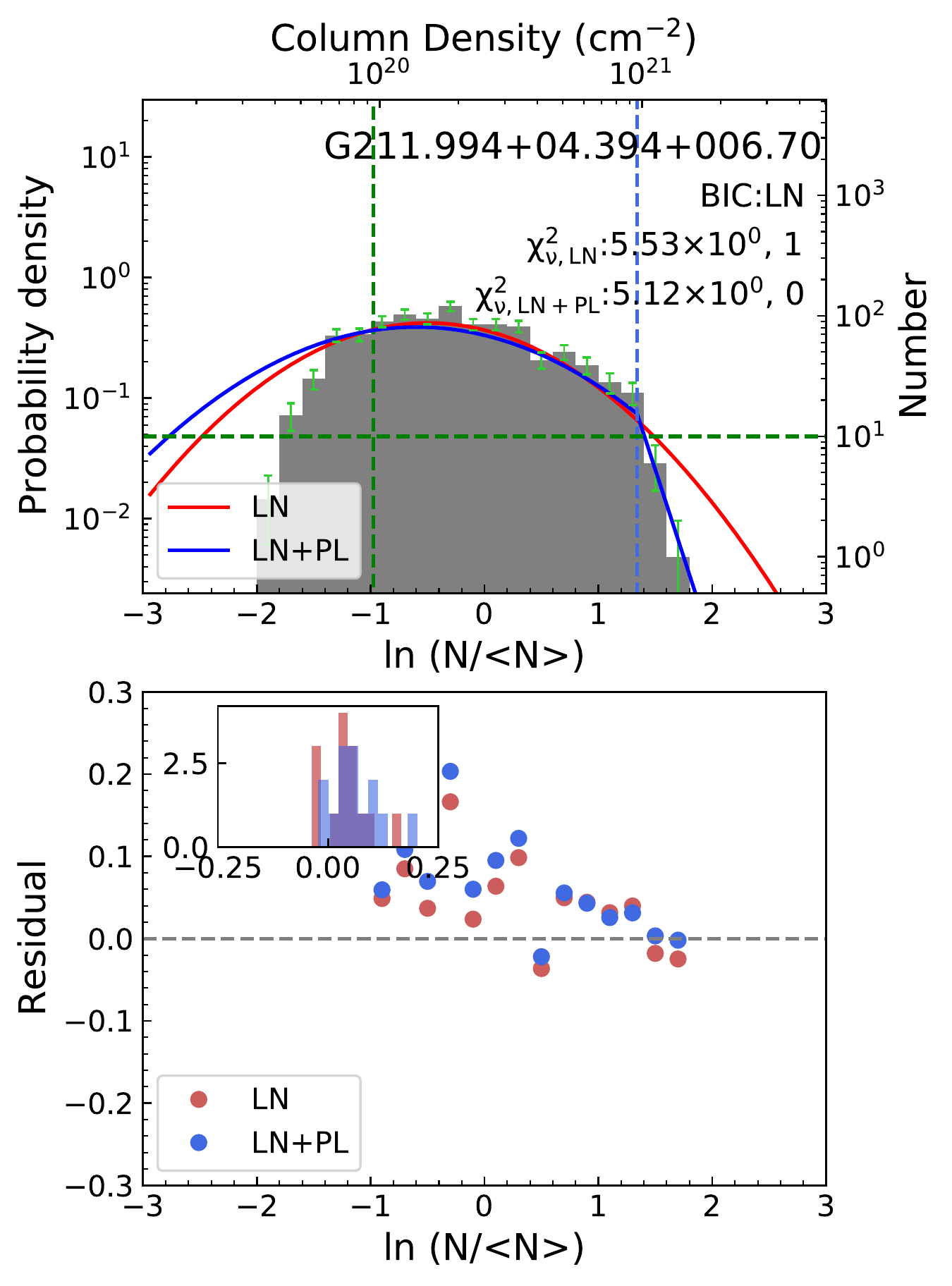}}
\subfigure{\includegraphics[trim=0cm 0cm 0cm 0cm, width= 0.23\linewidth, clip]{{G212.141-01.044+044.18_pdf_residual}.pdf}}
\subfigure{\includegraphics[trim=0cm 0cm 0cm 0cm, width= 0.23\linewidth, clip]{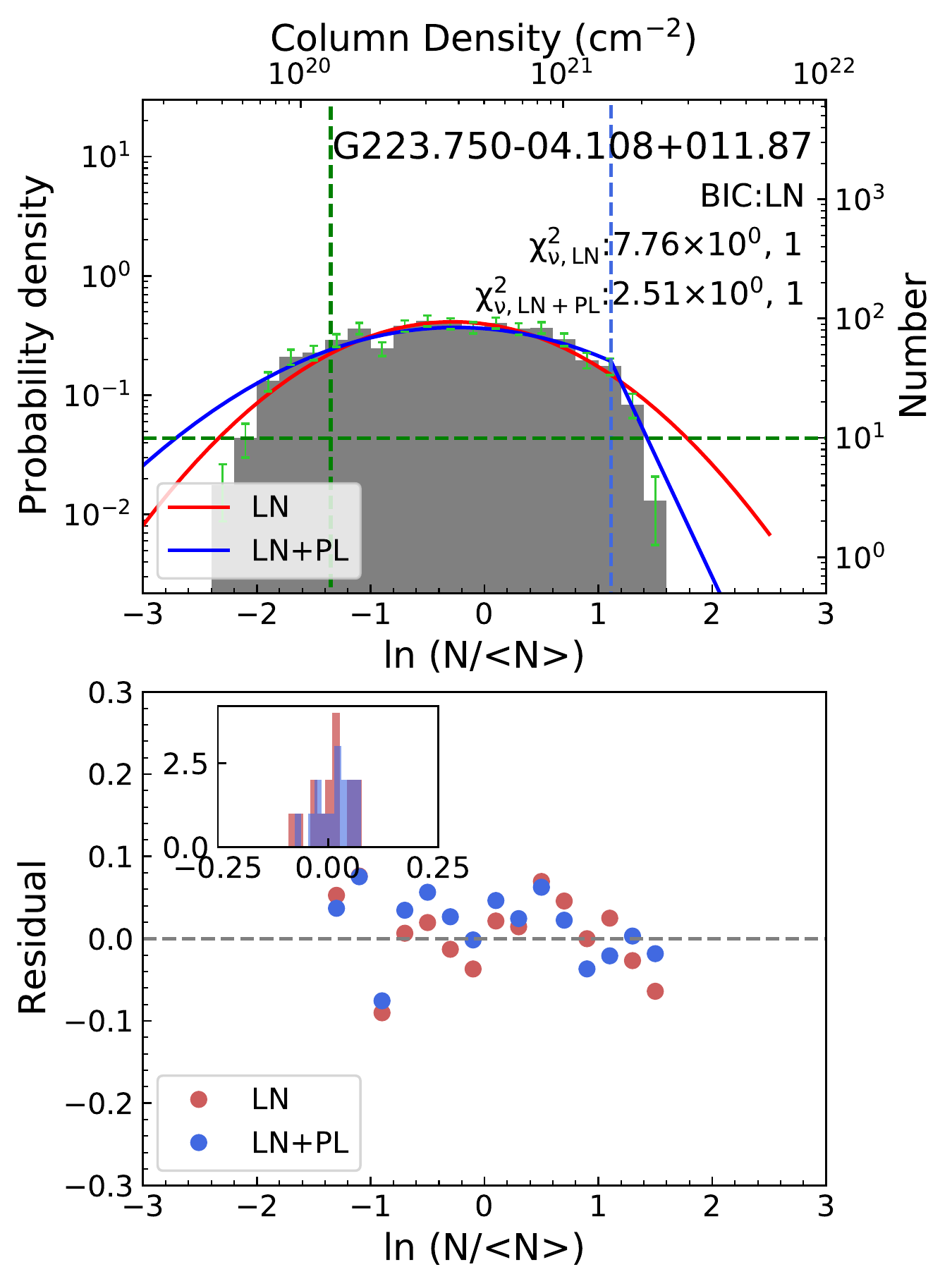}}

\subfigure{\includegraphics[trim=0cm 0cm 0cm 0cm, width= 0.23\linewidth, clip]{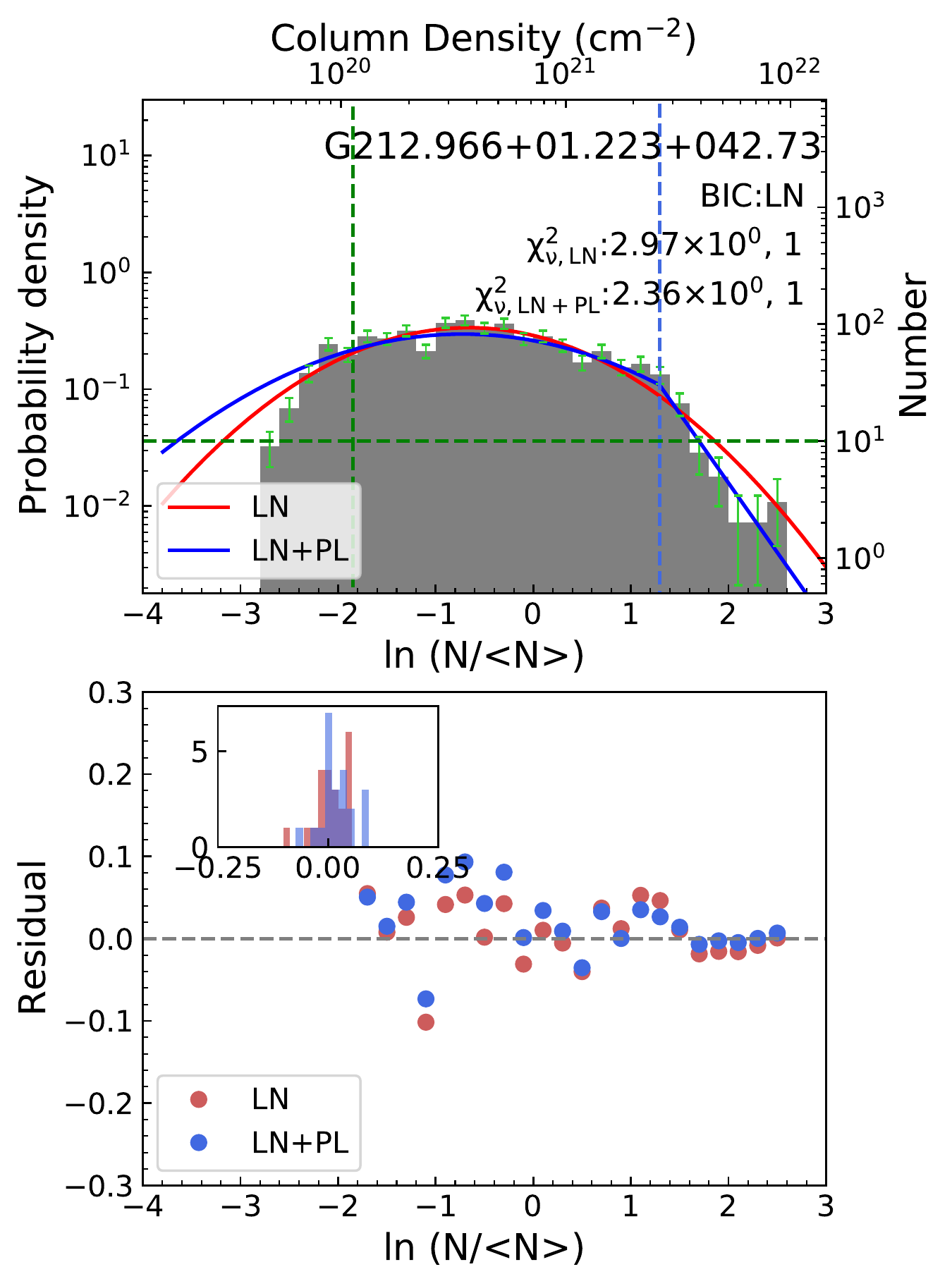}}
\subfigure{\includegraphics[trim=0cm 0cm 0cm 0cm, width= 0.23\linewidth, clip]{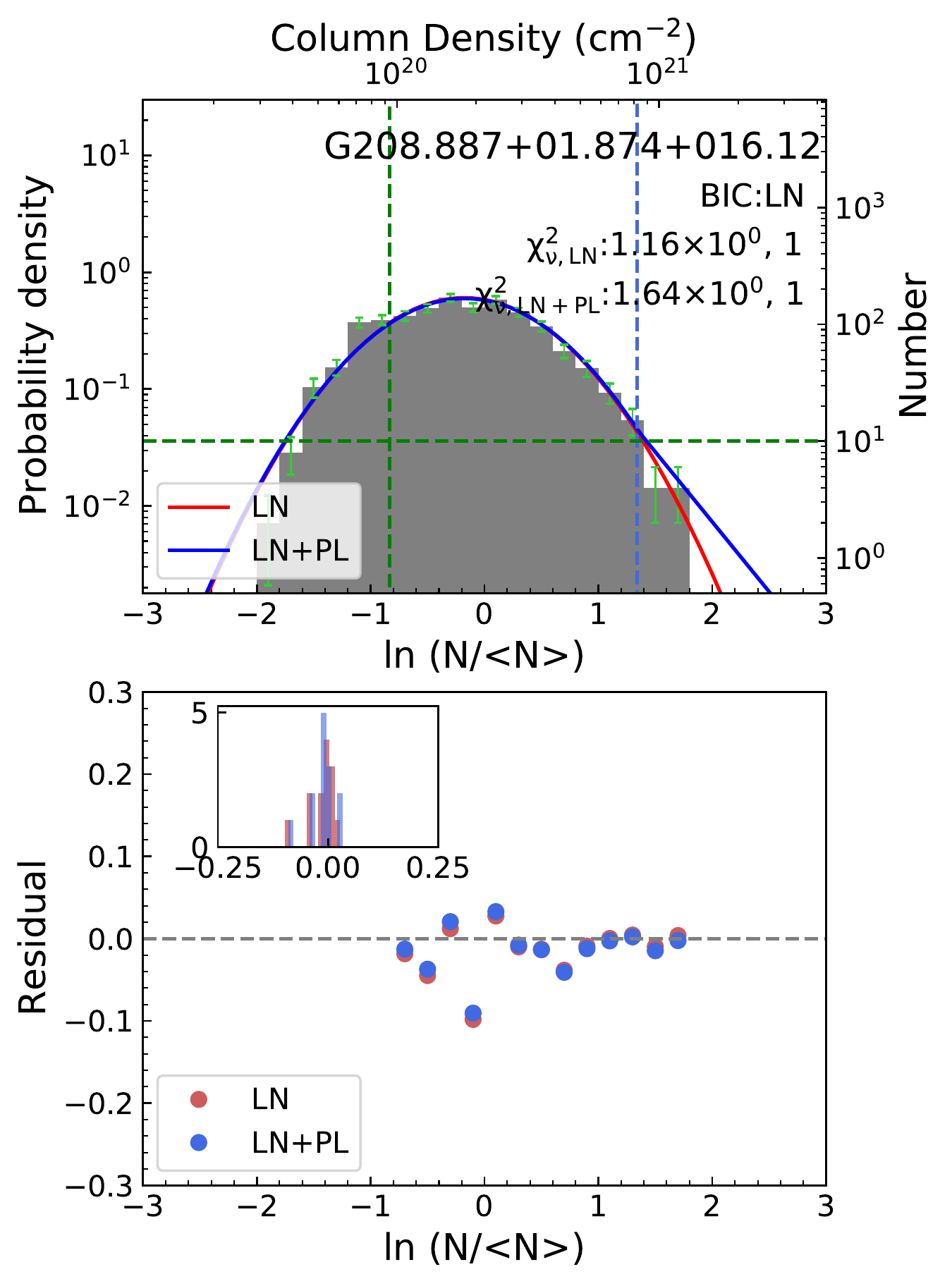}}
\subfigure{\includegraphics[trim=0cm 0cm 0cm 0cm, width= 0.23\linewidth, clip]{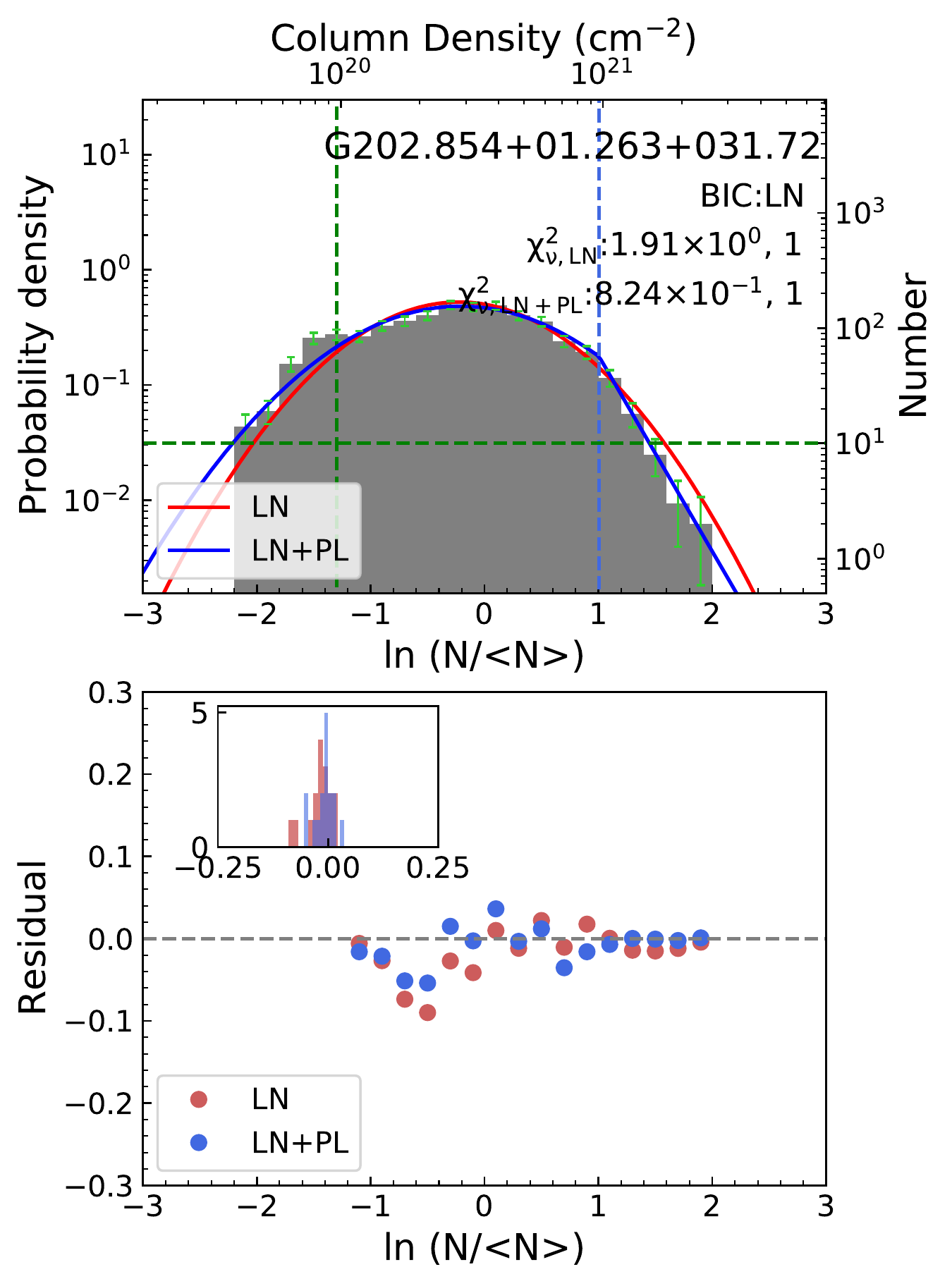}}
\subfigure{\includegraphics[trim=0cm 0cm 0cm 0cm, width= 0.23\linewidth, clip]{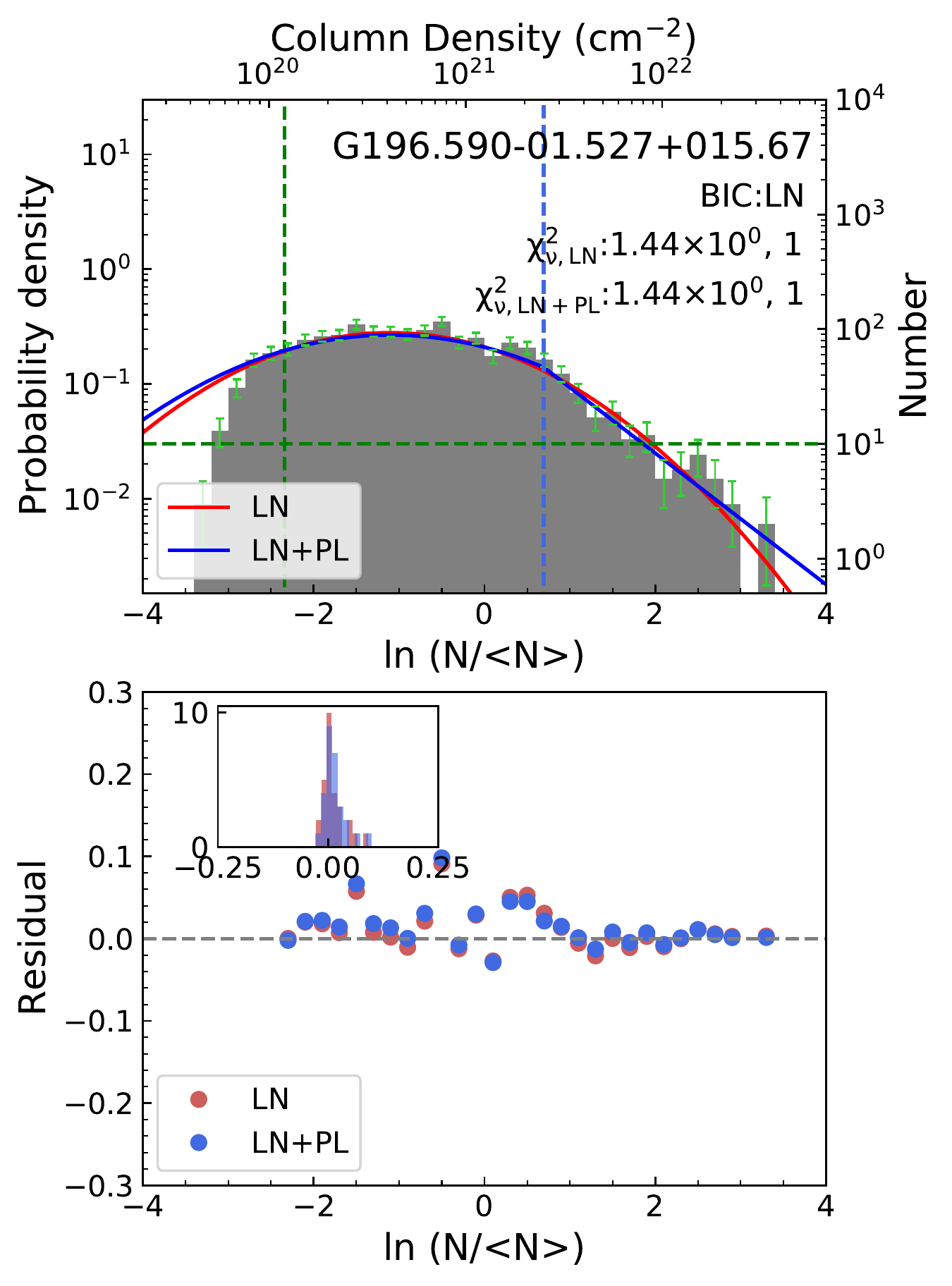}}

\end{figure}
\begin{figure}
\subfigure{\includegraphics[trim=0cm 0cm 0cm 0cm, width= 0.23\linewidth, clip]{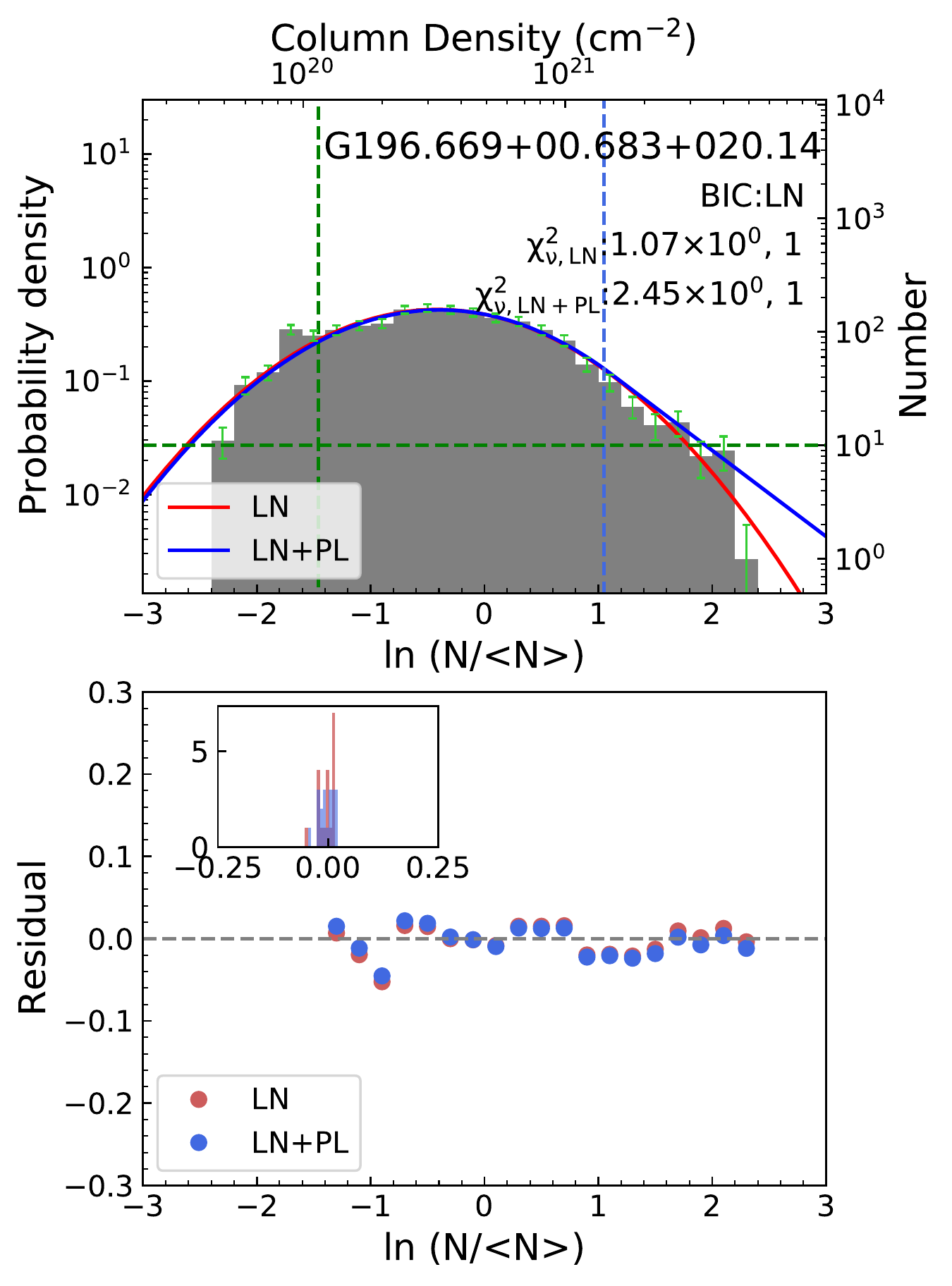}}
\subfigure{\includegraphics[trim=0cm 0cm 0cm 0cm, width= 0.23\linewidth, clip]{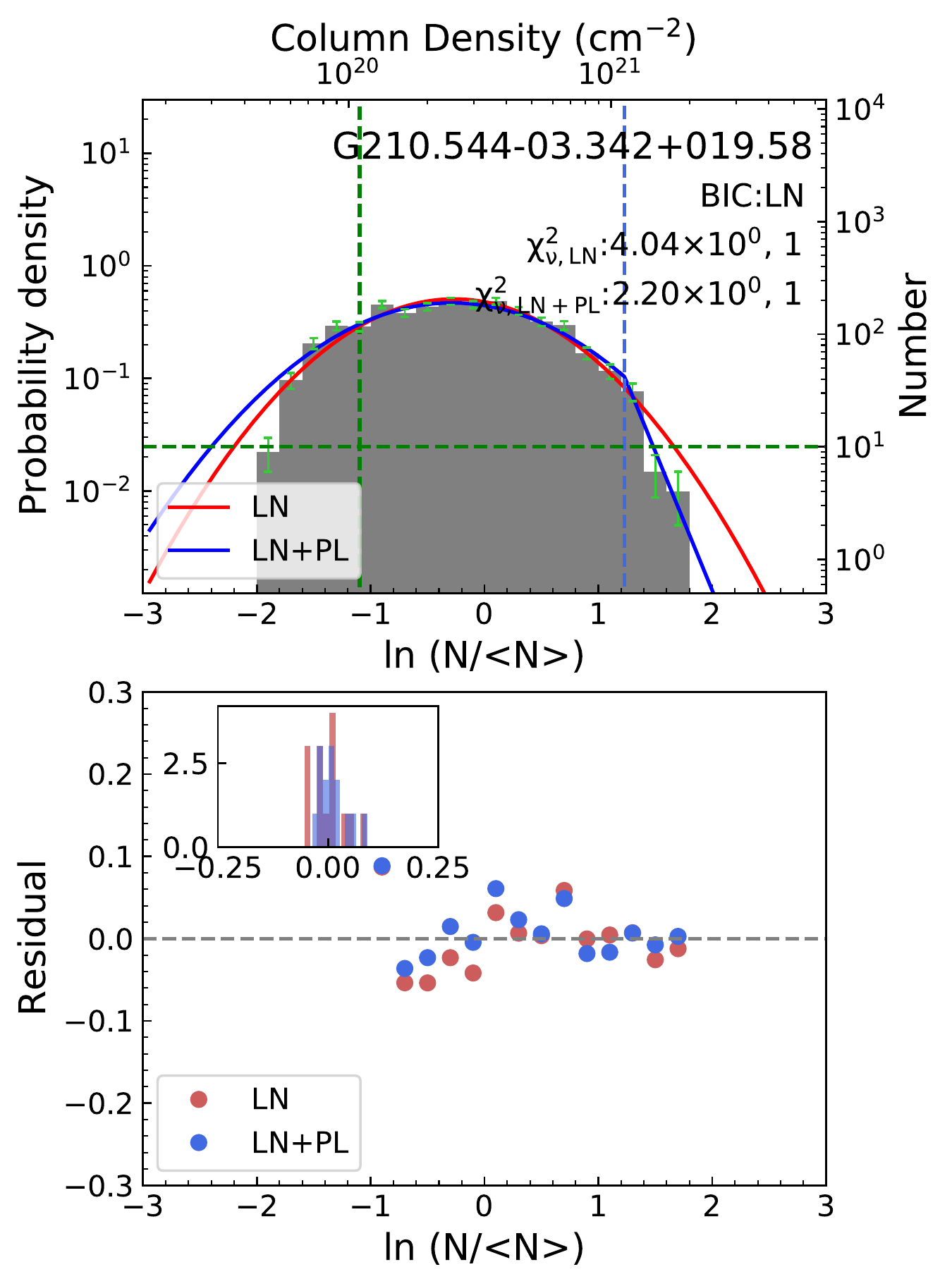}}
\subfigure{\includegraphics[trim=0cm 0cm 0cm 0cm, width= 0.23\linewidth, clip]{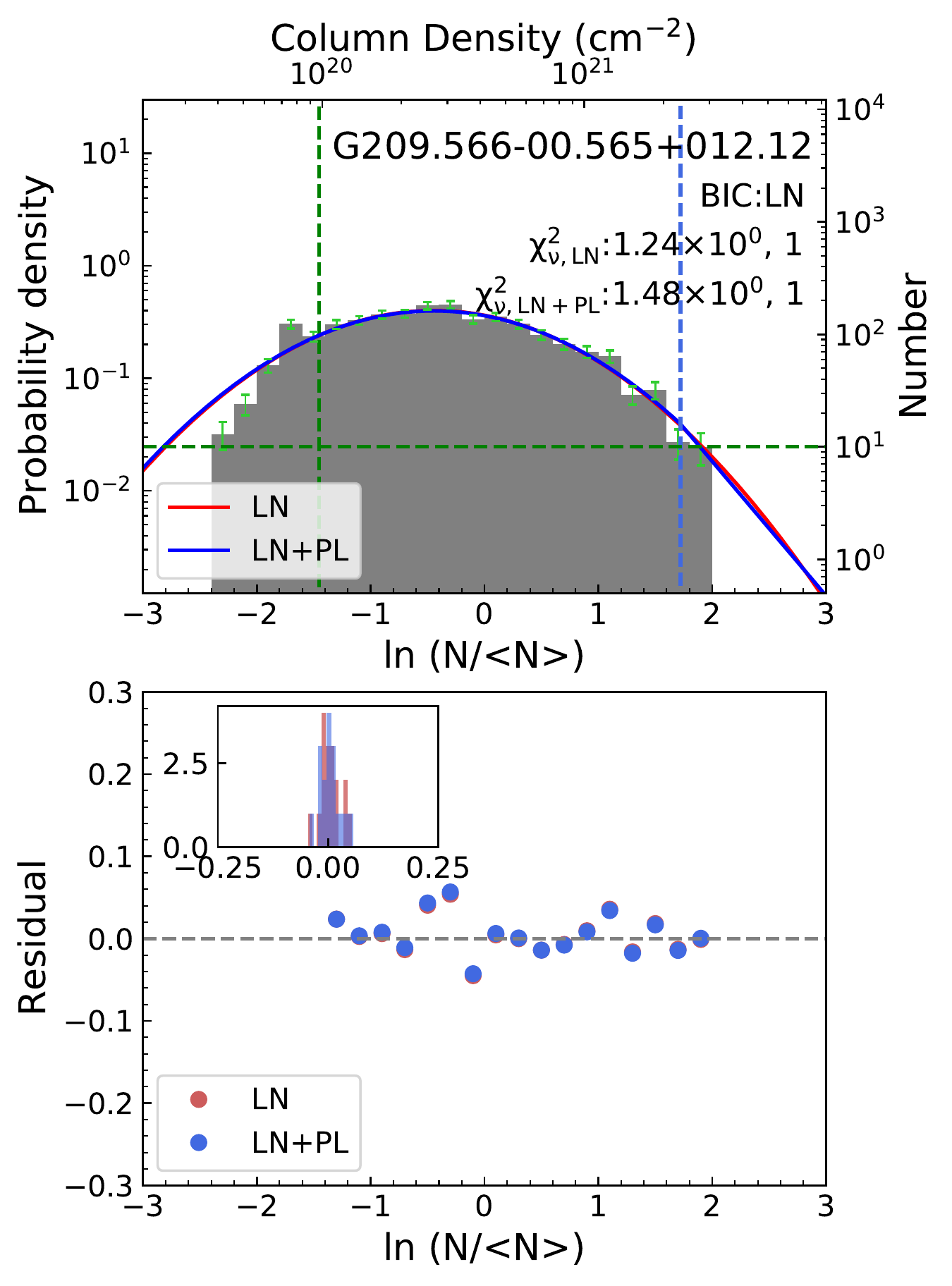}}
\subfigure{\includegraphics[trim=0cm 0cm 0cm 0cm, width= 0.23\linewidth, clip]{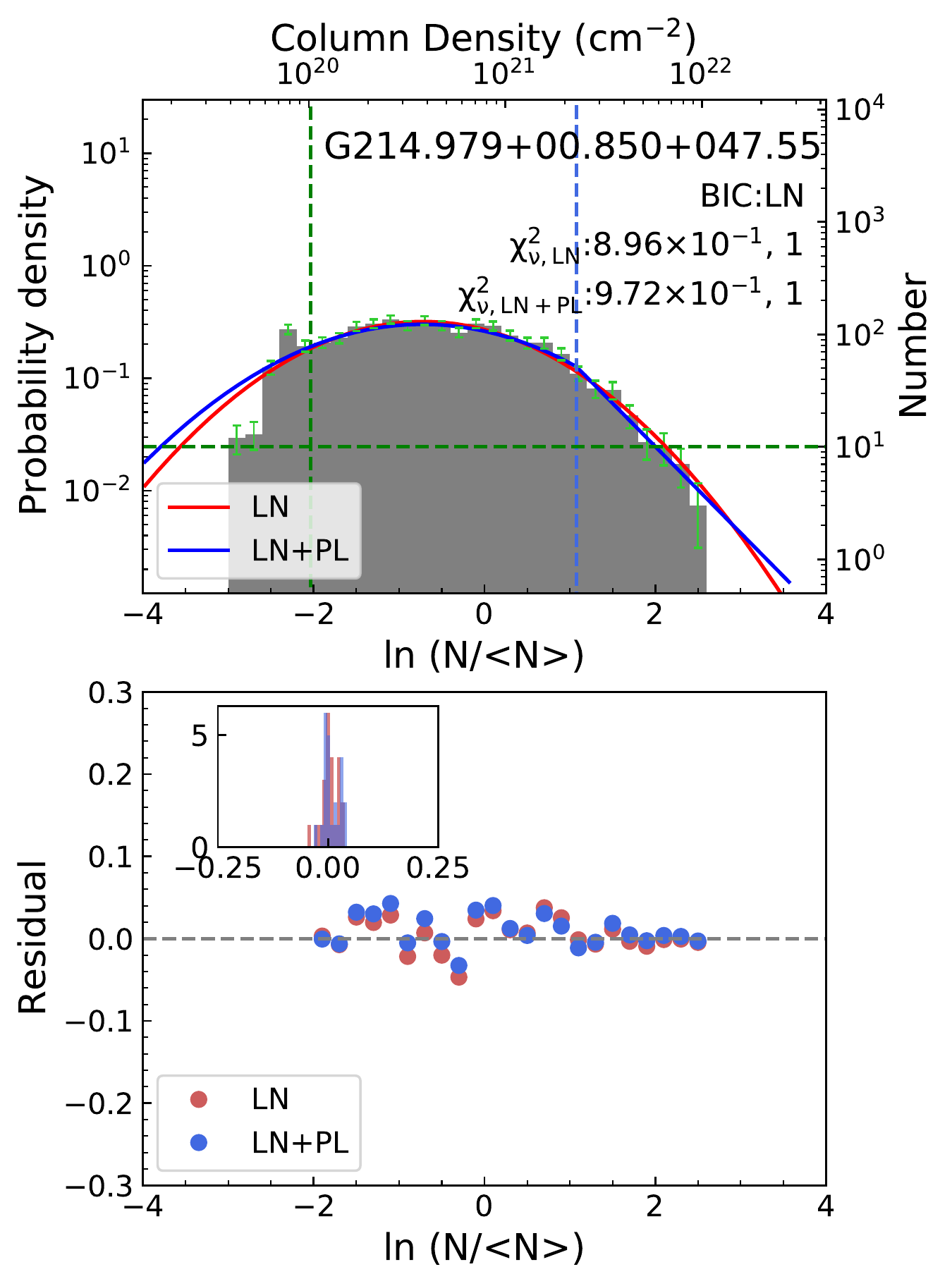}}

\subfigure{\includegraphics[trim=0cm 0cm 0cm 0cm, width= 0.23\linewidth, clip]{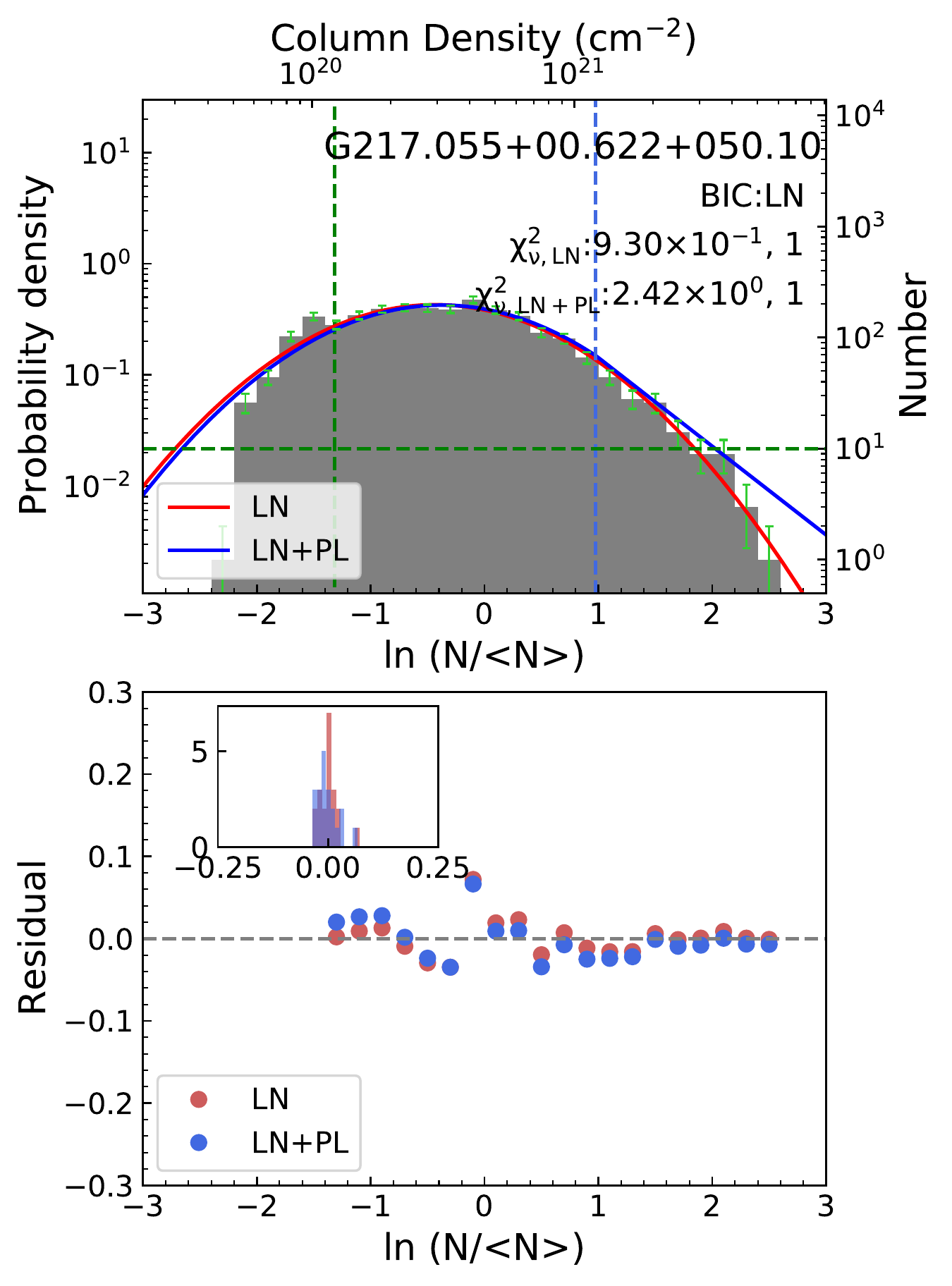}}
\subfigure{\includegraphics[trim=0cm 0cm 0cm 0cm, width= 0.23\linewidth, clip]{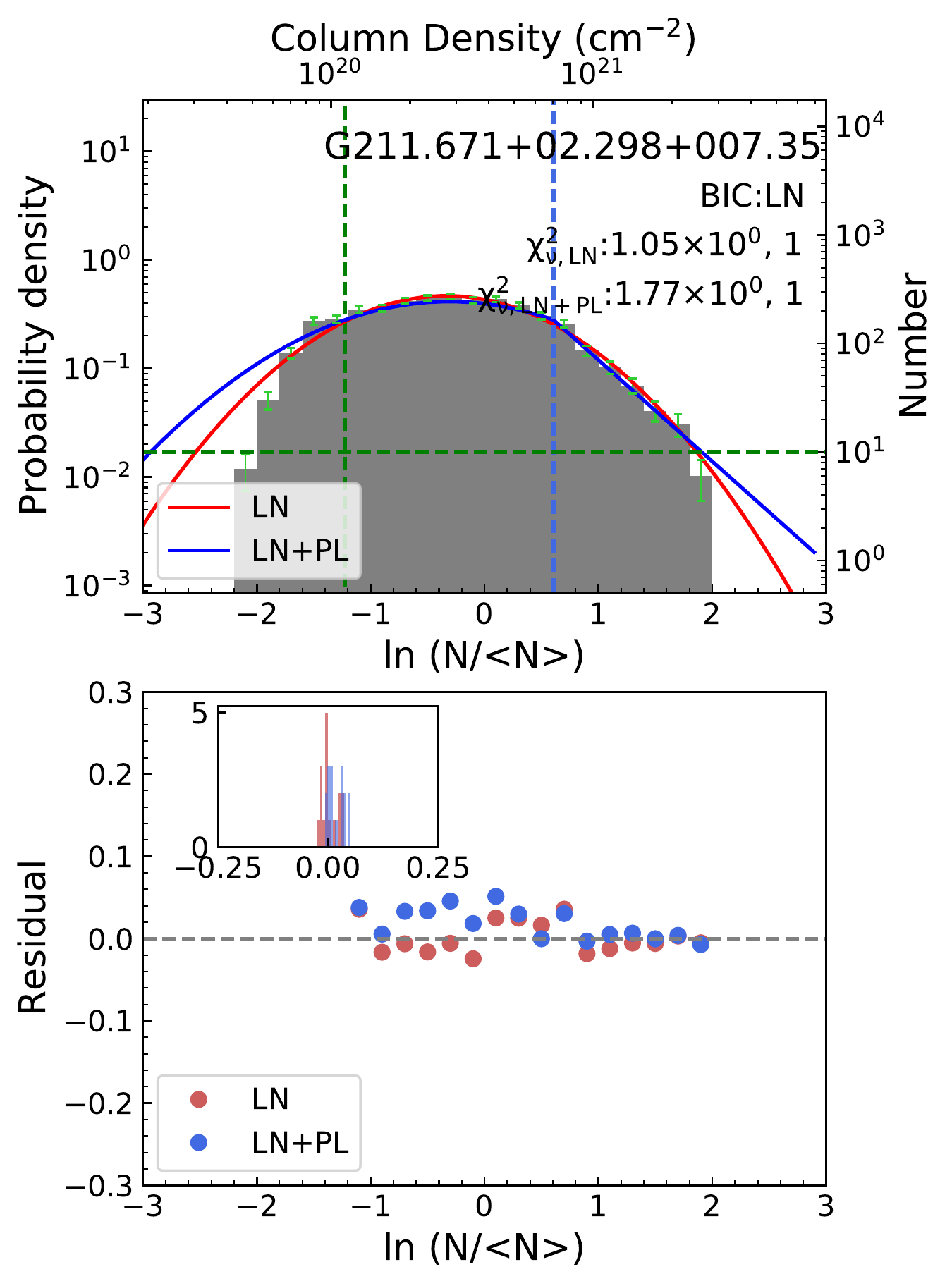}}

\caption{LN PDFs. Symbols and legends have the same meaning as those in Figure \ref{fig3}.}
\label{fig16}
\end{figure}

%% file: pdf_residual_LN+PL.tex
\begin{figure}
\subfigure{\includegraphics[trim=0cm 0cm 0cm 0cm, width= 0.23\linewidth, clip]{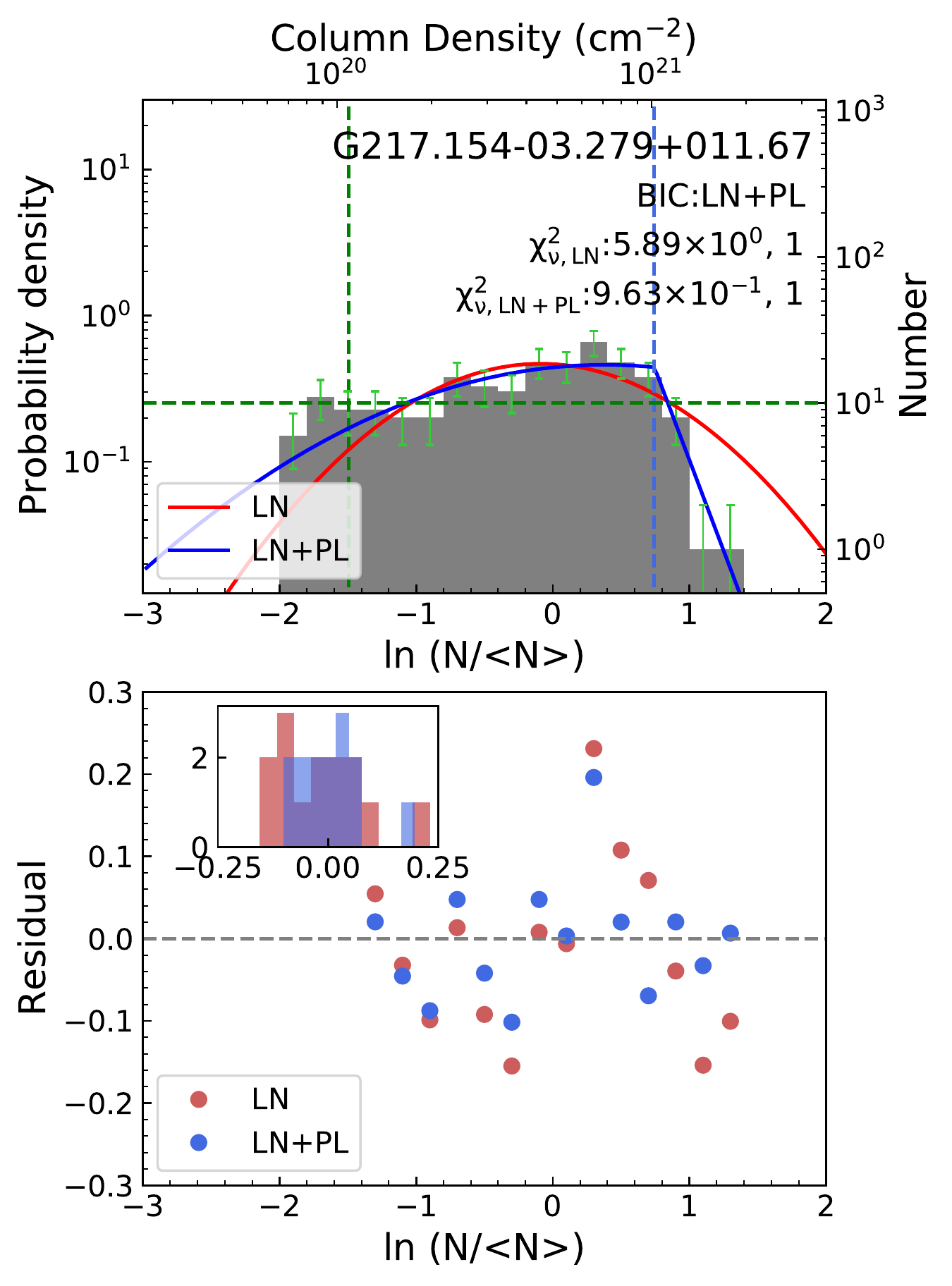}}
\subfigure{\includegraphics[trim=0cm 0cm 0cm 0cm, width= 0.23\linewidth, clip]{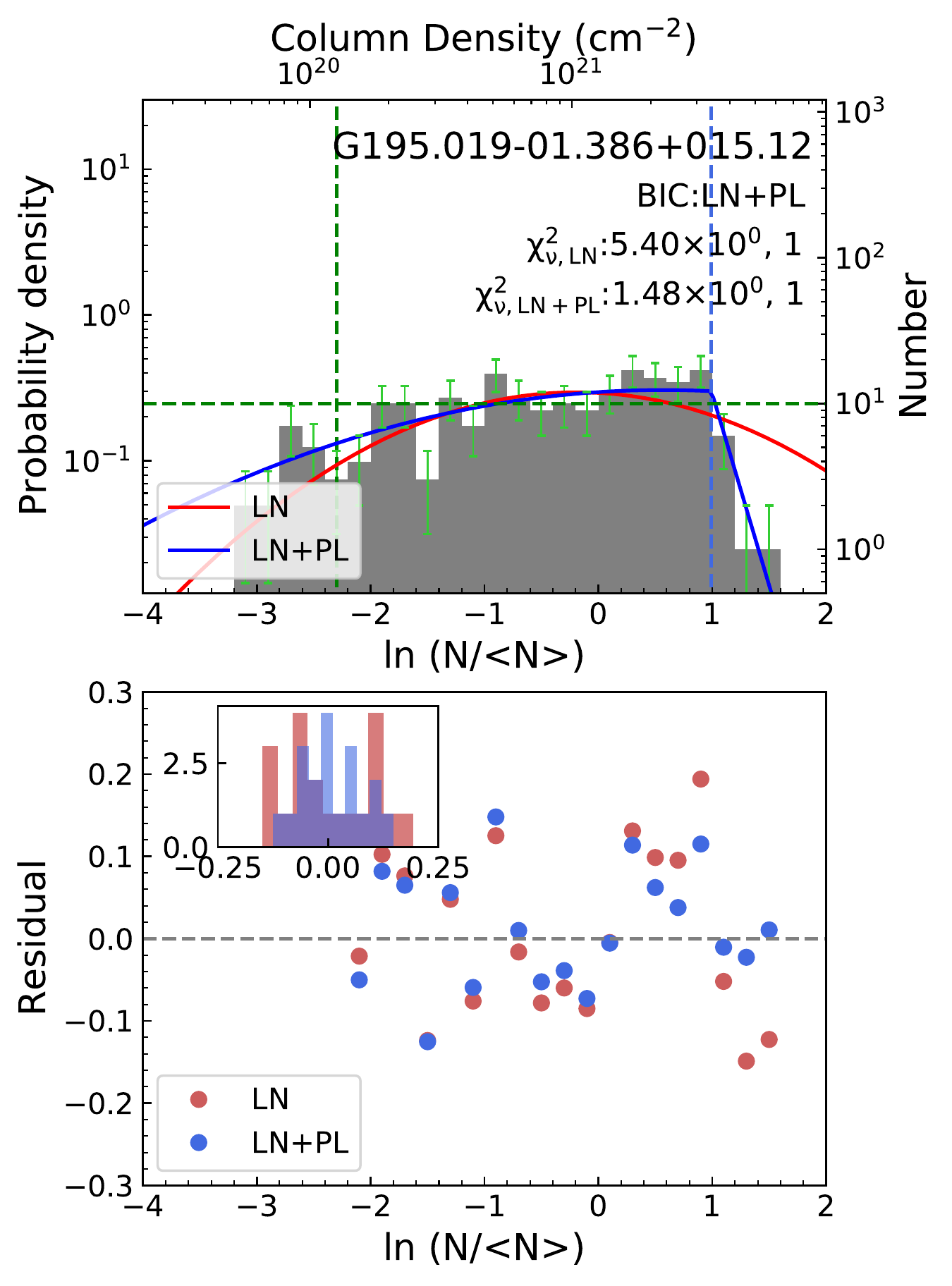}}
\subfigure{\includegraphics[trim=0cm 0cm 0cm 0cm, width= 0.23\linewidth, clip]{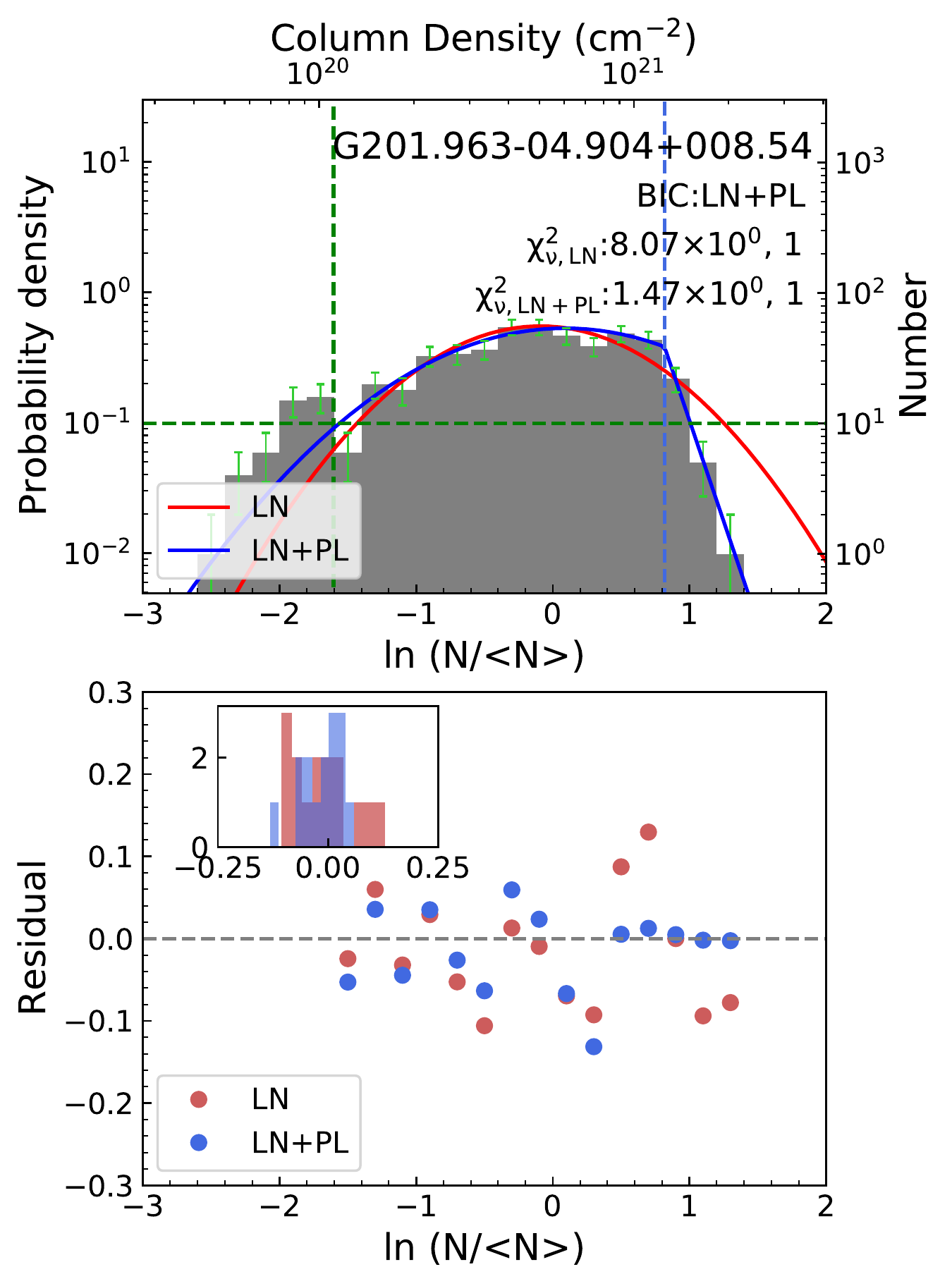}}
\subfigure{\includegraphics[trim=0cm 0cm 0cm 0cm, width= 0.23\linewidth, clip]{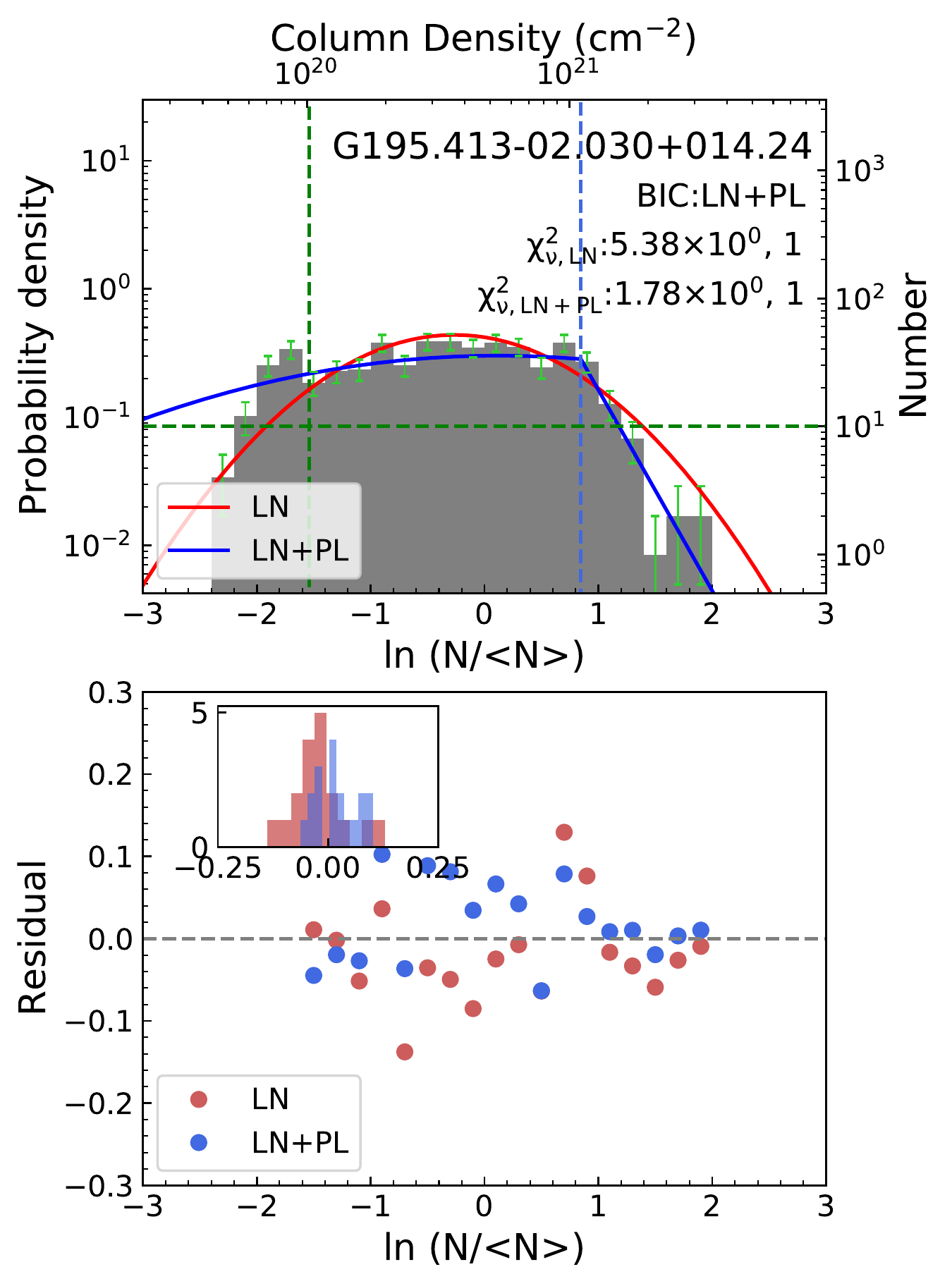}}

\subfigure{\includegraphics[trim=0cm 0cm 0cm 0cm, width= 0.23\linewidth, clip]{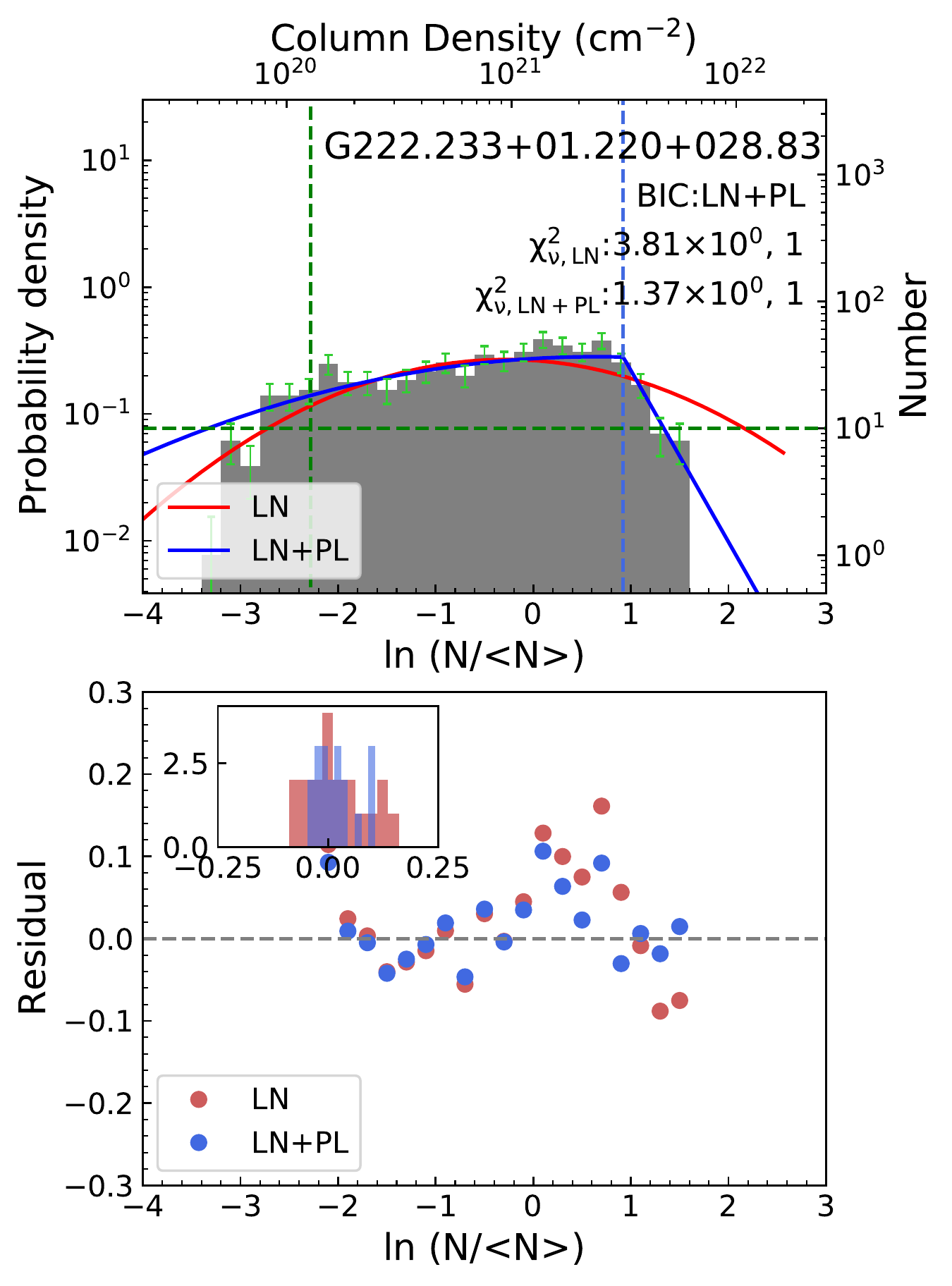}}
\subfigure{\includegraphics[trim=0cm 0cm 0cm 0cm, width= 0.23\linewidth, clip]{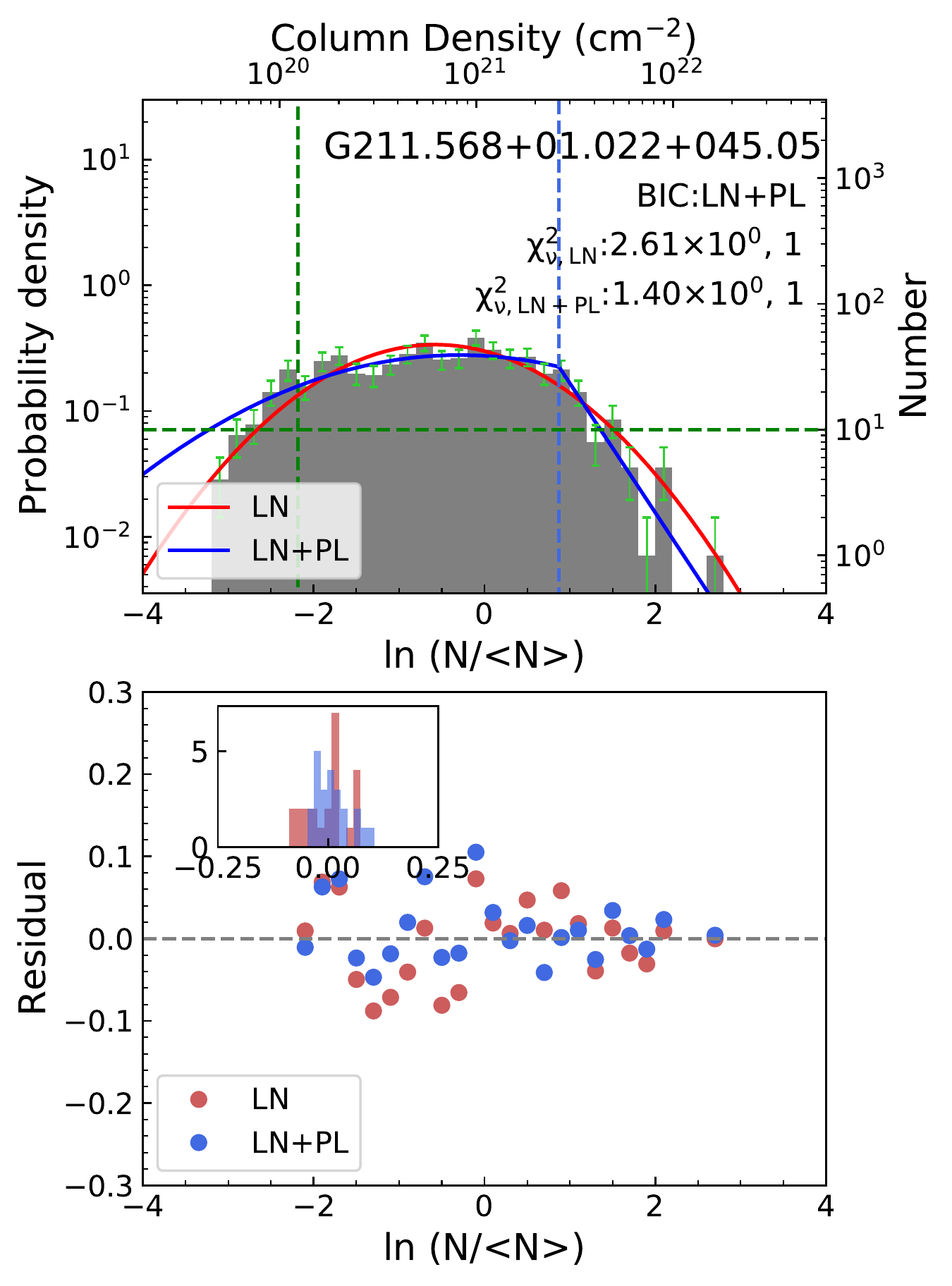}}
\subfigure{\includegraphics[trim=0cm 0cm 0cm 0cm, width= 0.23\linewidth, clip]{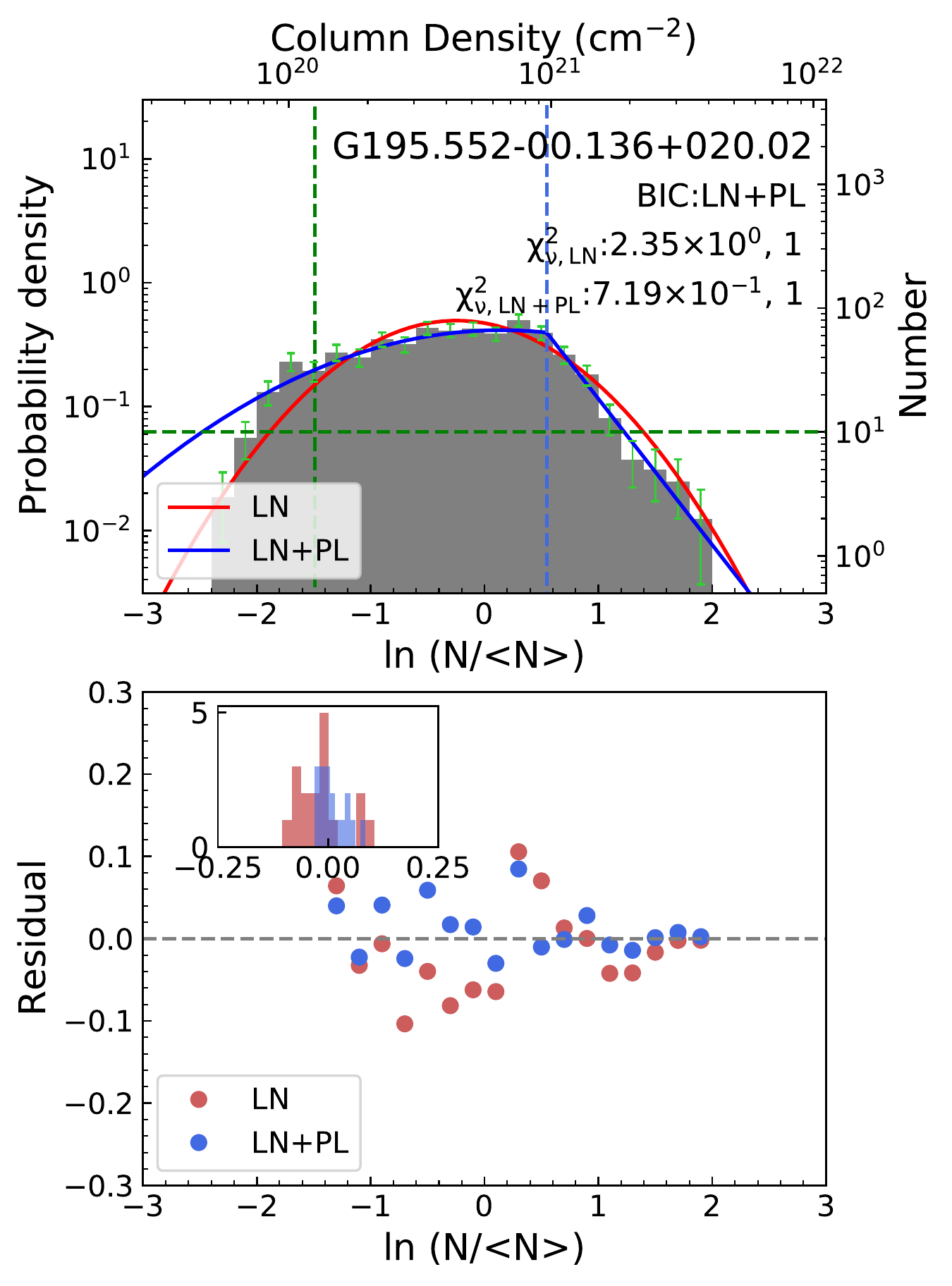}}
\subfigure{\includegraphics[trim=0cm 0cm 0cm 0cm, width= 0.23\linewidth, clip]{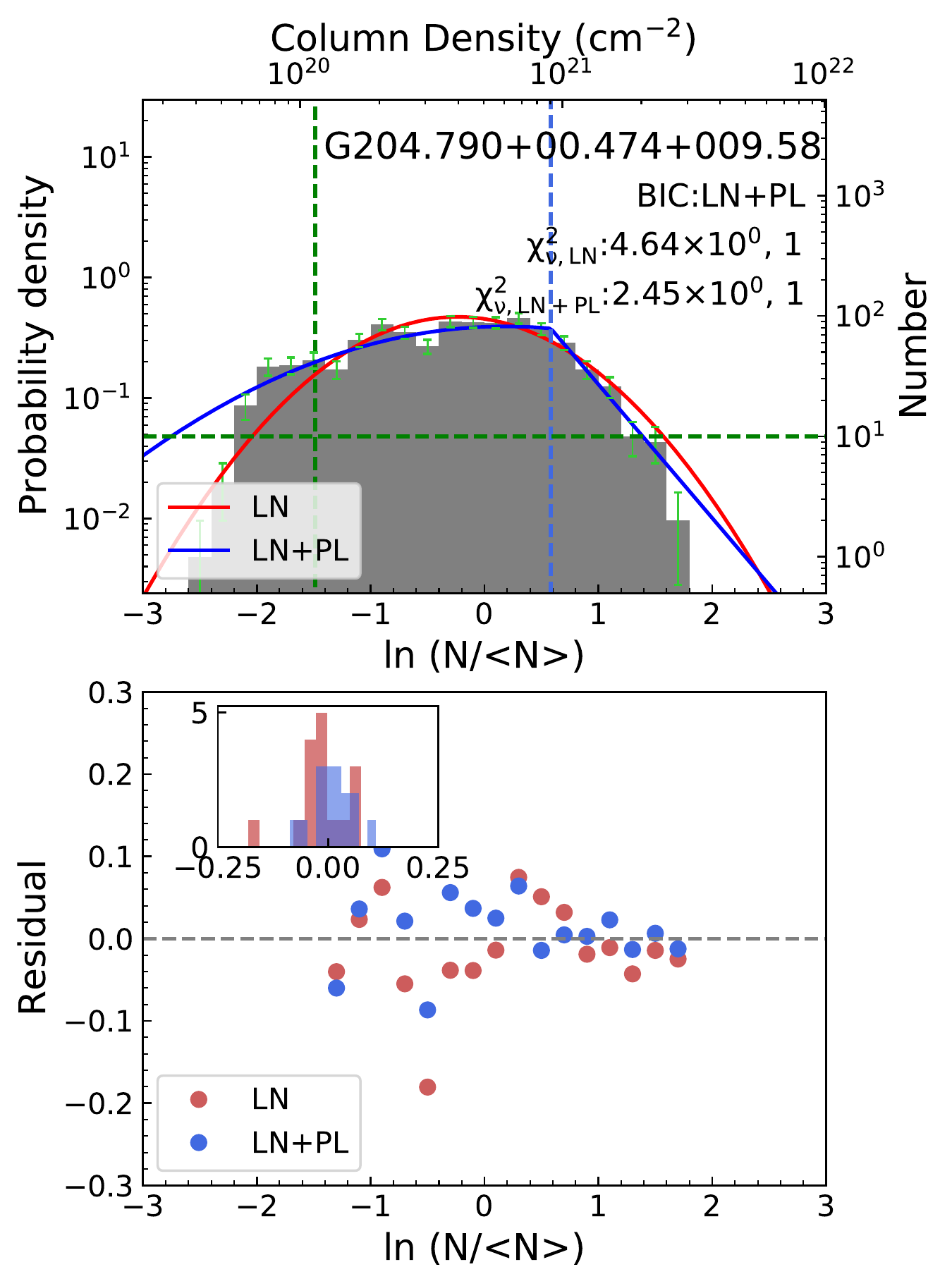}}

\subfigure{\includegraphics[trim=0cm 0cm 0cm 0cm, width= 0.23\linewidth, clip]{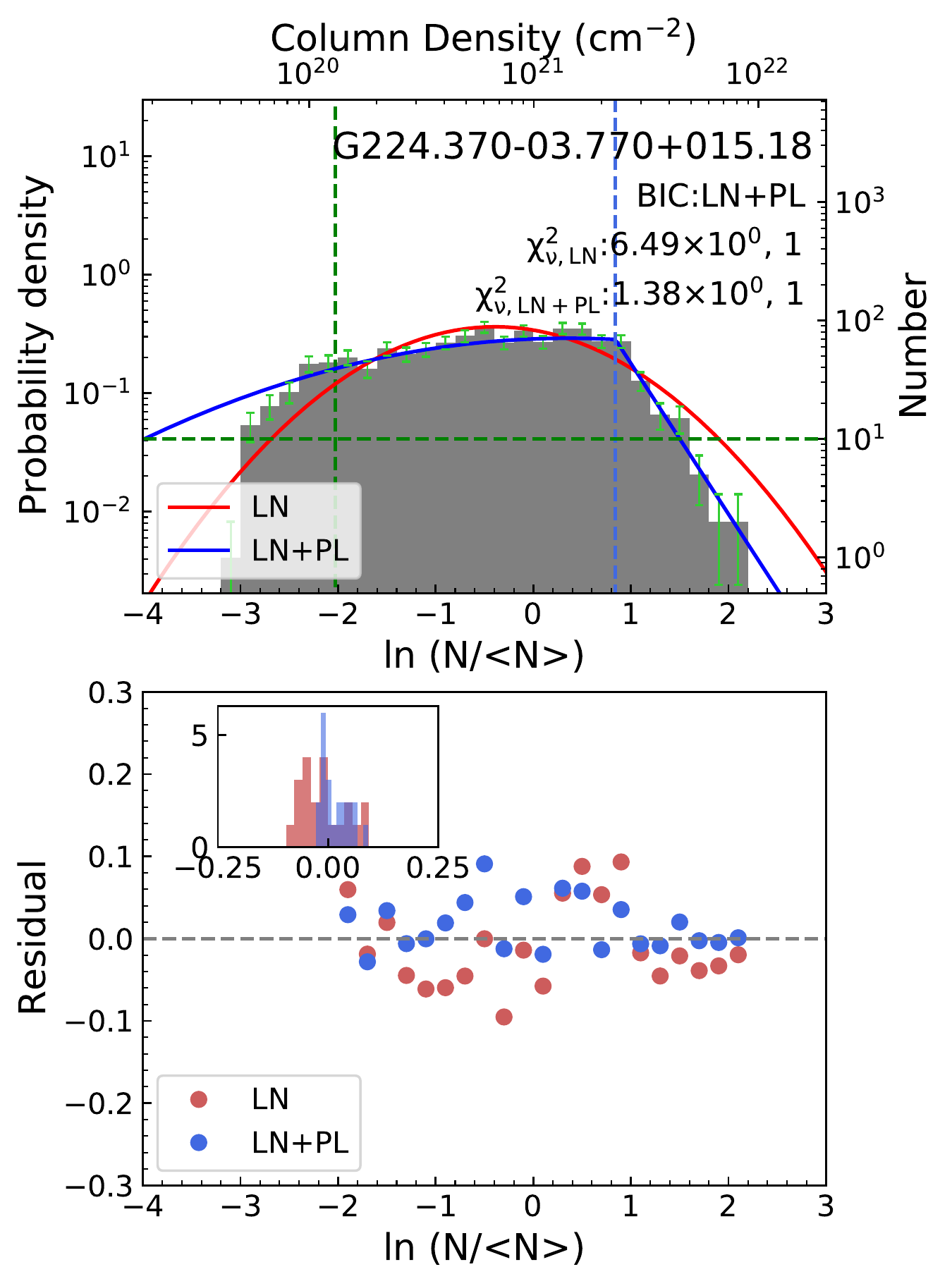}}
\subfigure{\includegraphics[trim=0cm 0cm 0cm 0cm, width= 0.23\linewidth, clip]{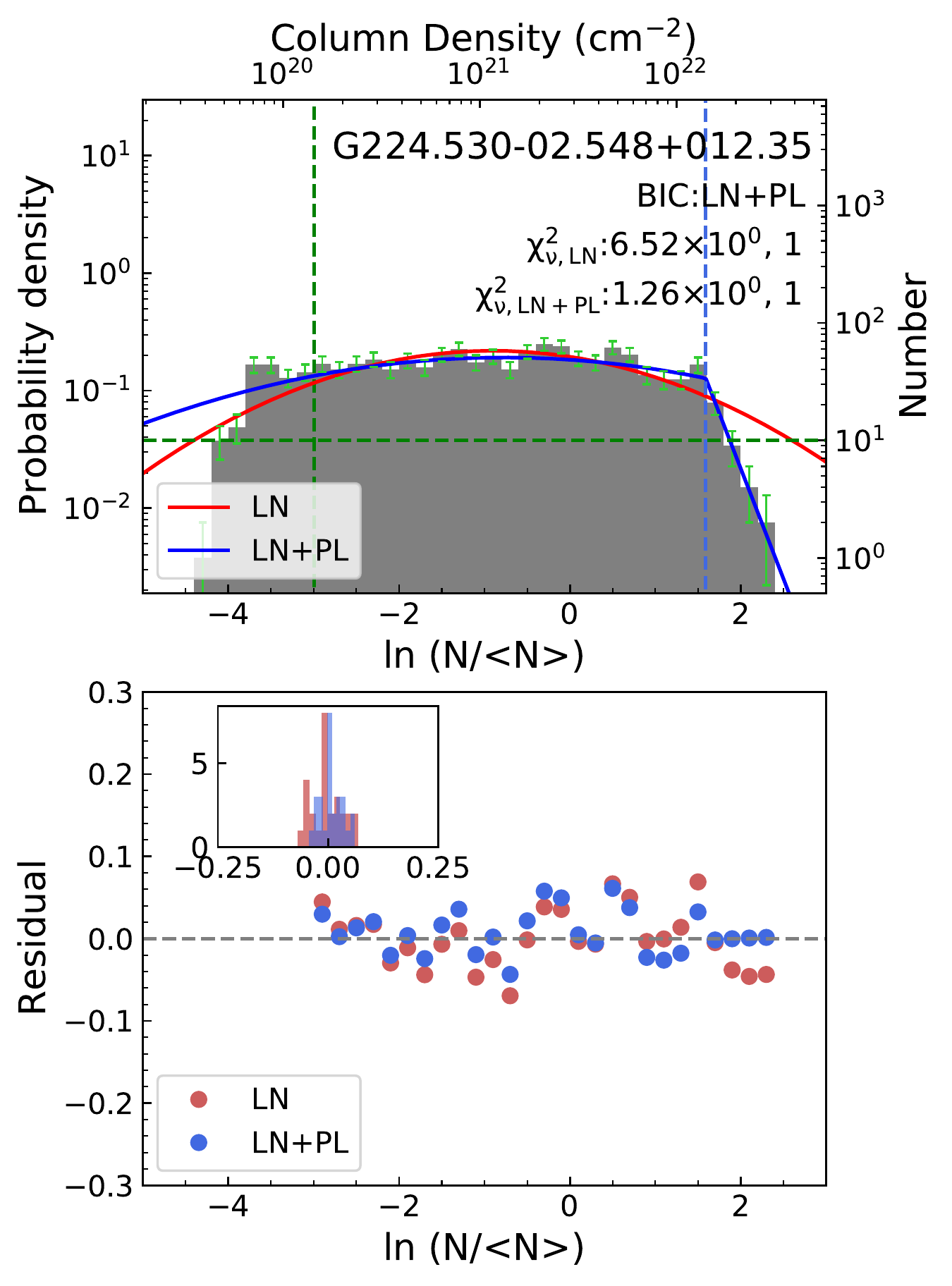}}
\subfigure{\includegraphics[trim=0cm 0cm 0cm 0cm, width= 0.23\linewidth, clip]{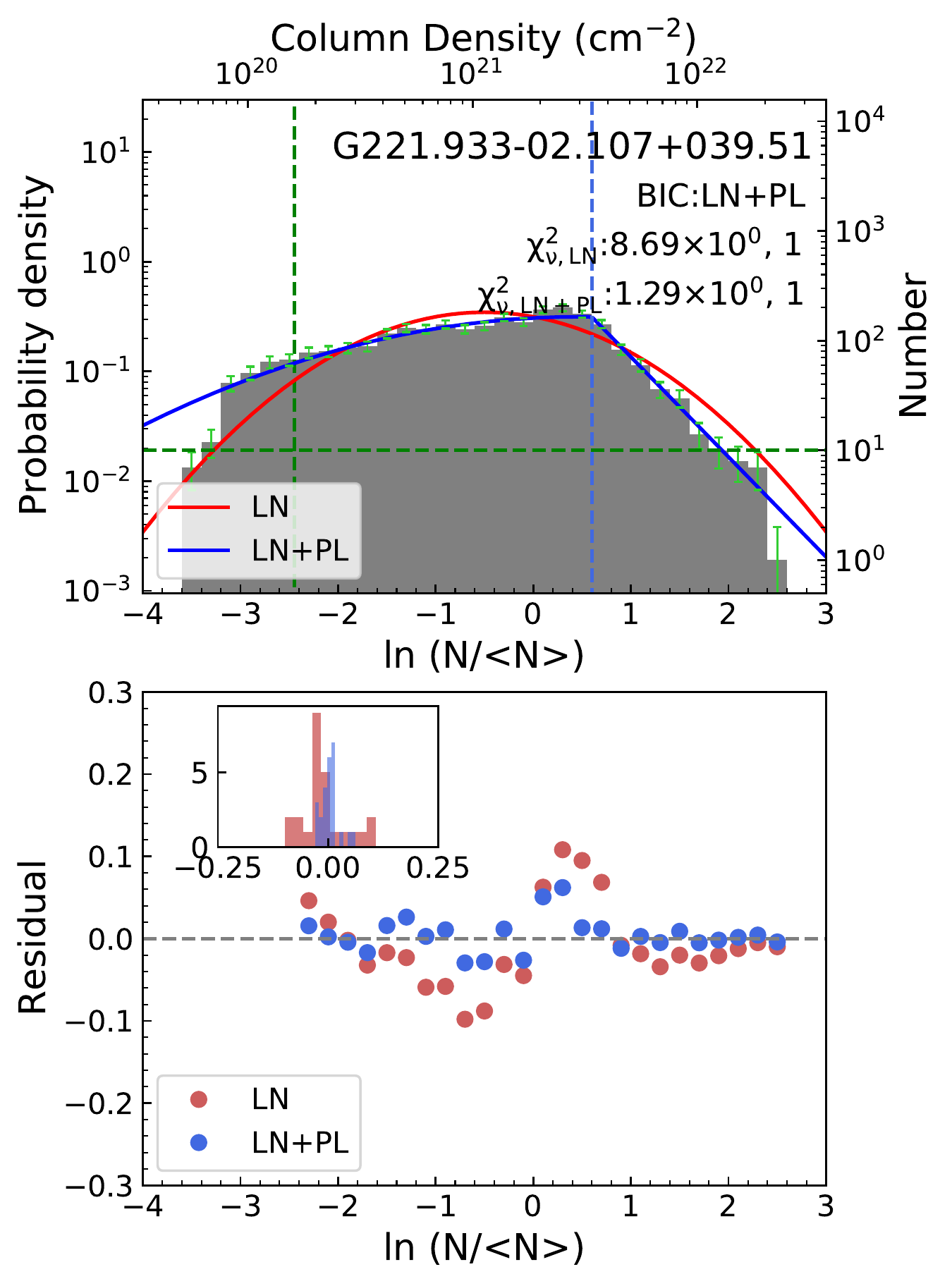}}
\subfigure{\includegraphics[trim=0cm 0cm 0cm 0cm, width= 0.23\linewidth, clip]{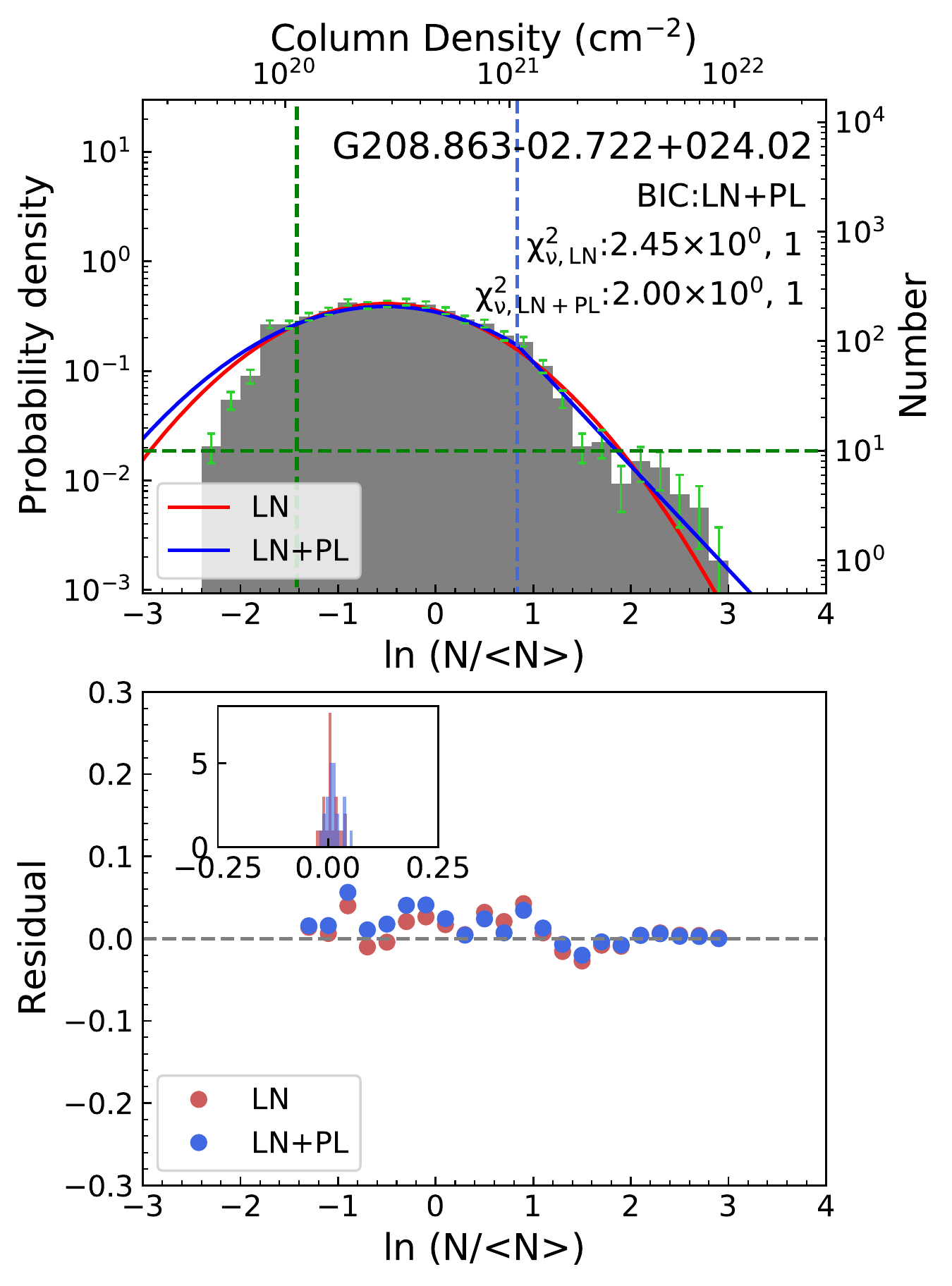}}

\subfigure{\includegraphics[trim=0cm 0cm 0cm 0cm, width= 0.23\linewidth, clip]{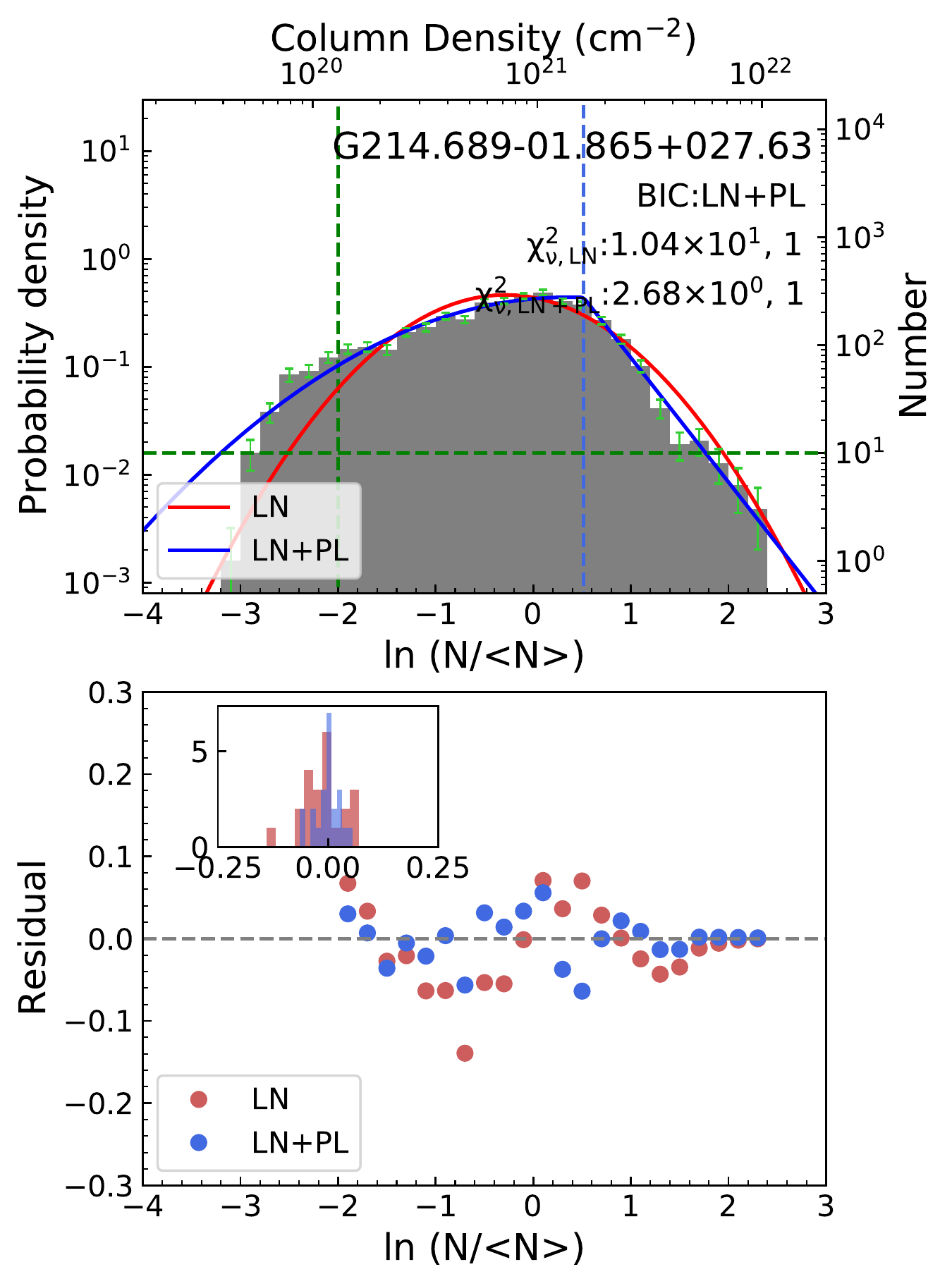}}
\subfigure{\includegraphics[trim=0cm 0cm 0cm 0cm, width= 0.23\linewidth, clip]{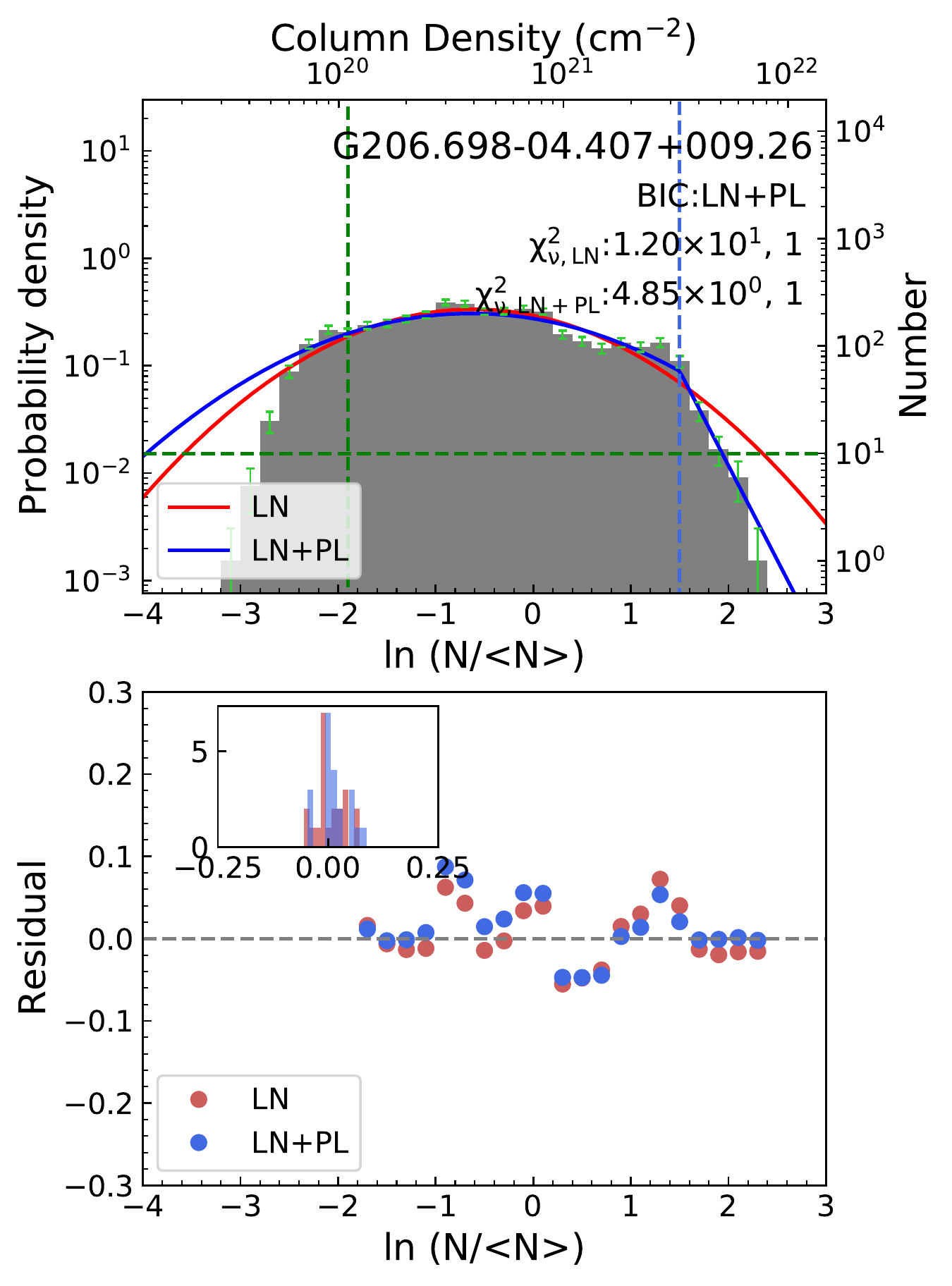}}
\subfigure{\includegraphics[trim=0cm 0cm 0cm 0cm, width= 0.23\linewidth, clip]{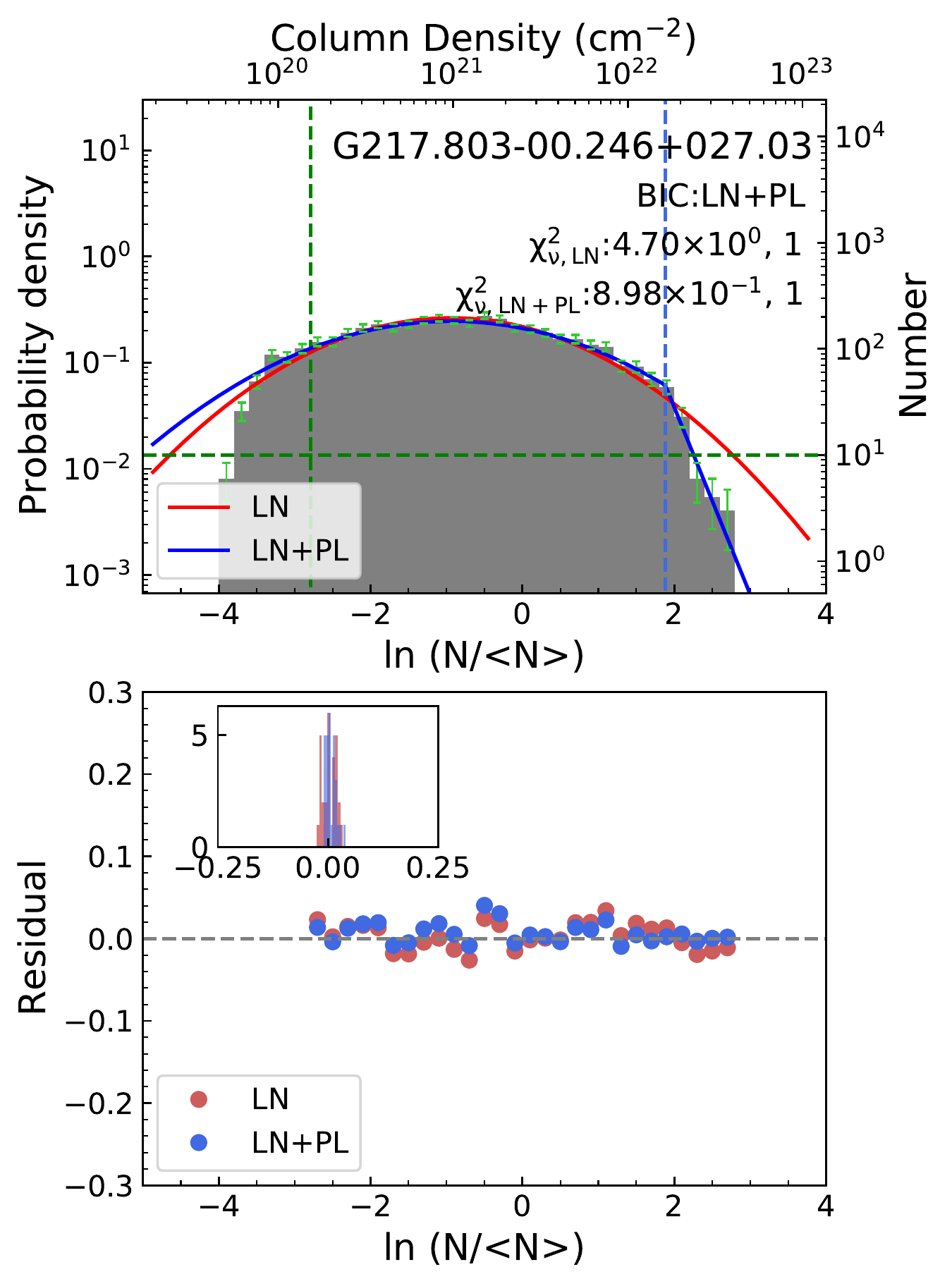}}
\subfigure{\includegraphics[trim=0cm 0cm 0cm 0cm, width= 0.23\linewidth, clip]{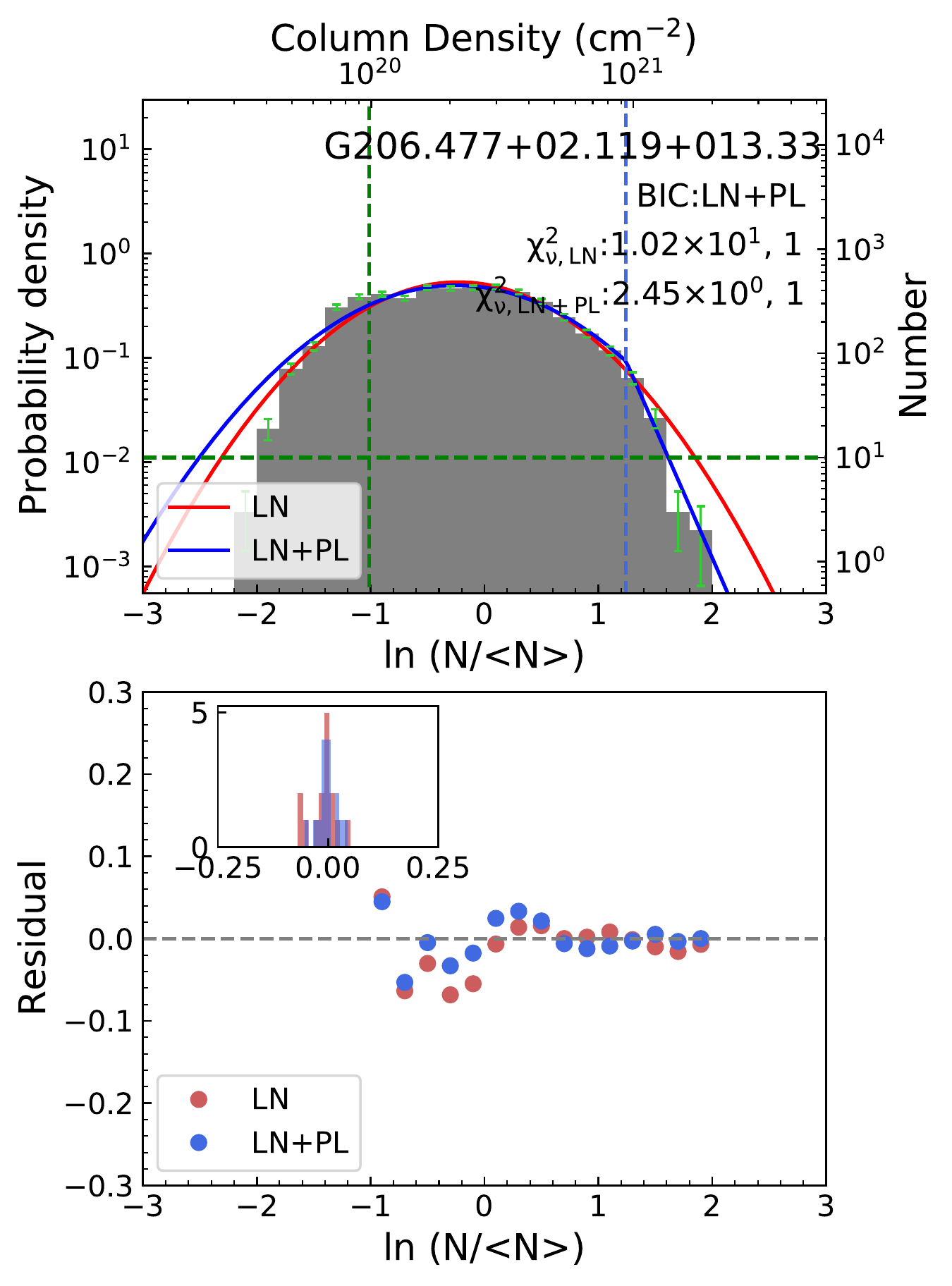}}

\end{figure}
\begin{figure}
\subfigure{\includegraphics[trim=0cm 0cm 0cm 0cm, width= 0.23\linewidth, clip]{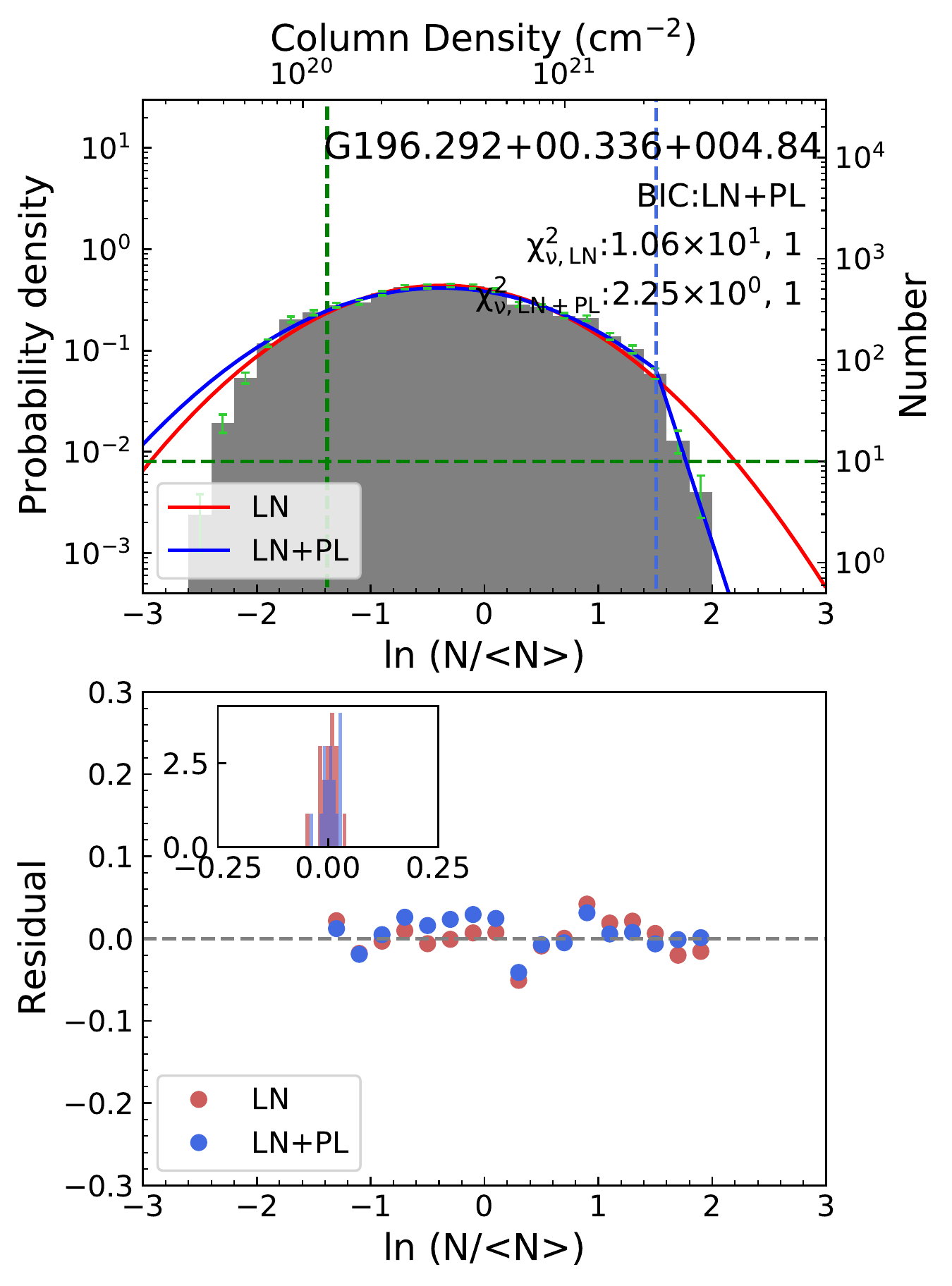}}
\subfigure{\includegraphics[trim=0cm 0cm 0cm 0cm, width= 0.23\linewidth, clip]{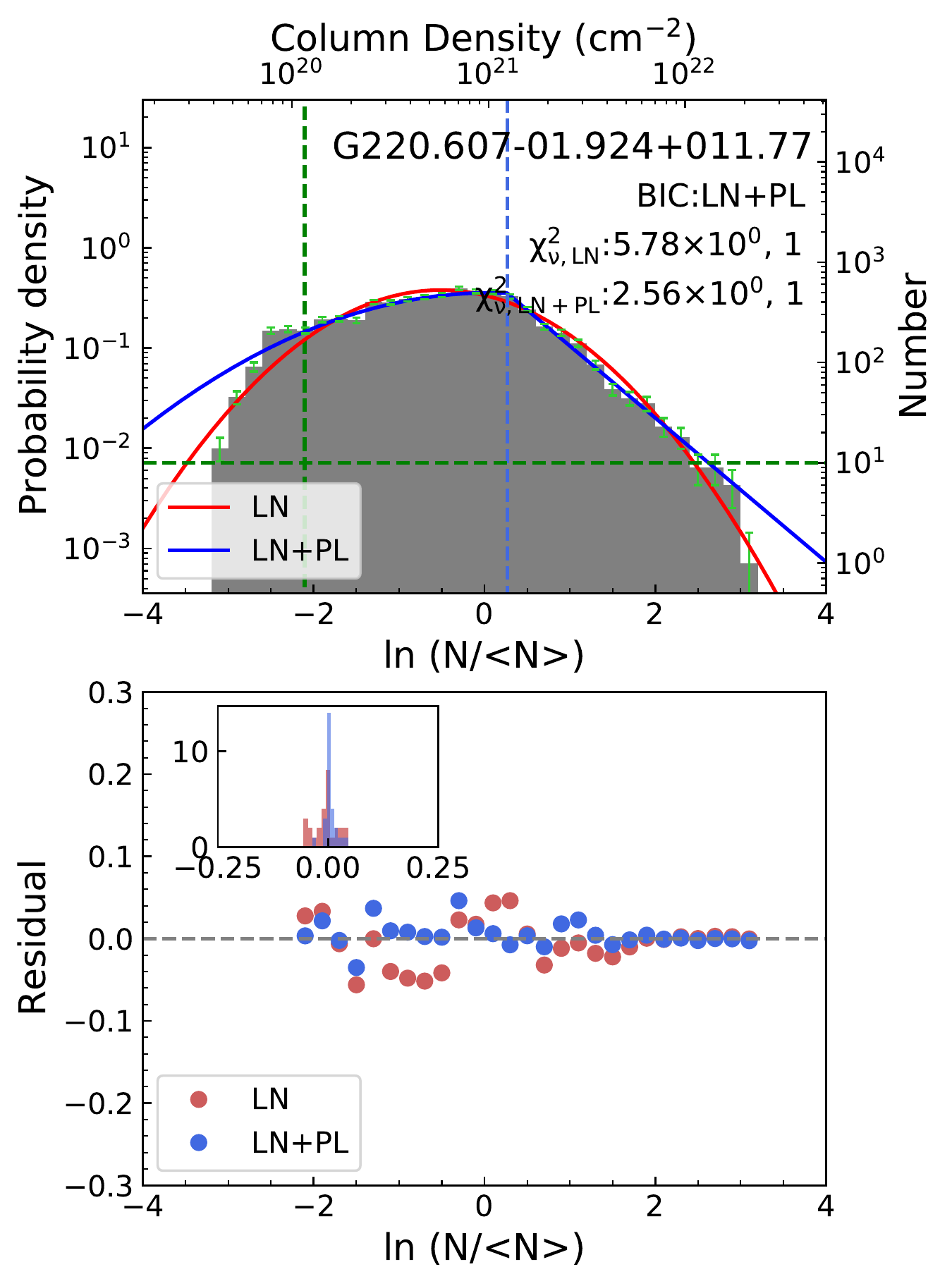}}
\subfigure{\includegraphics[trim=0cm 0cm 0cm 0cm, width= 0.23\linewidth, clip]{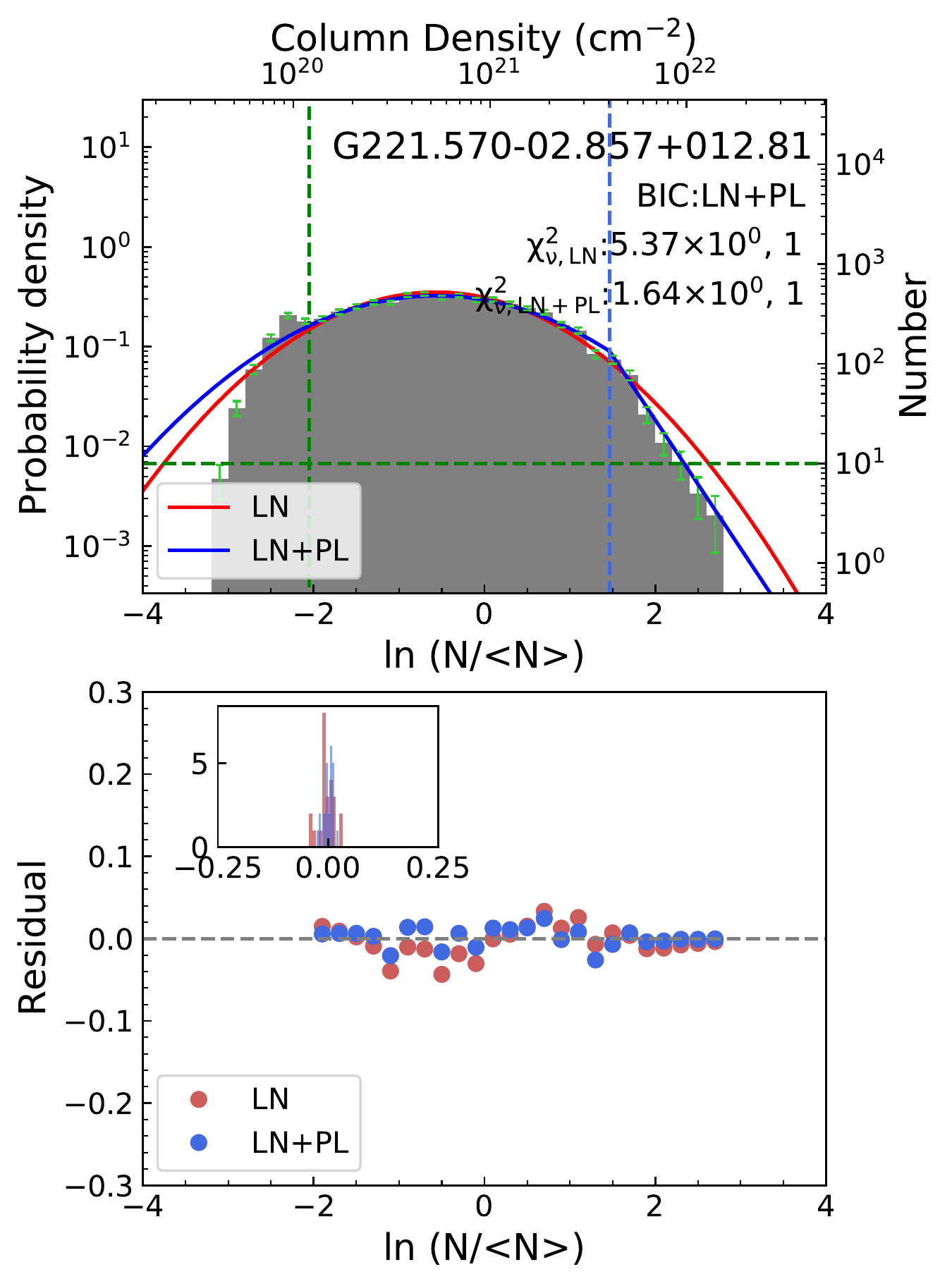}}
\subfigure{\includegraphics[trim=0cm 0cm 0cm 0cm, width= 0.23\linewidth, clip]{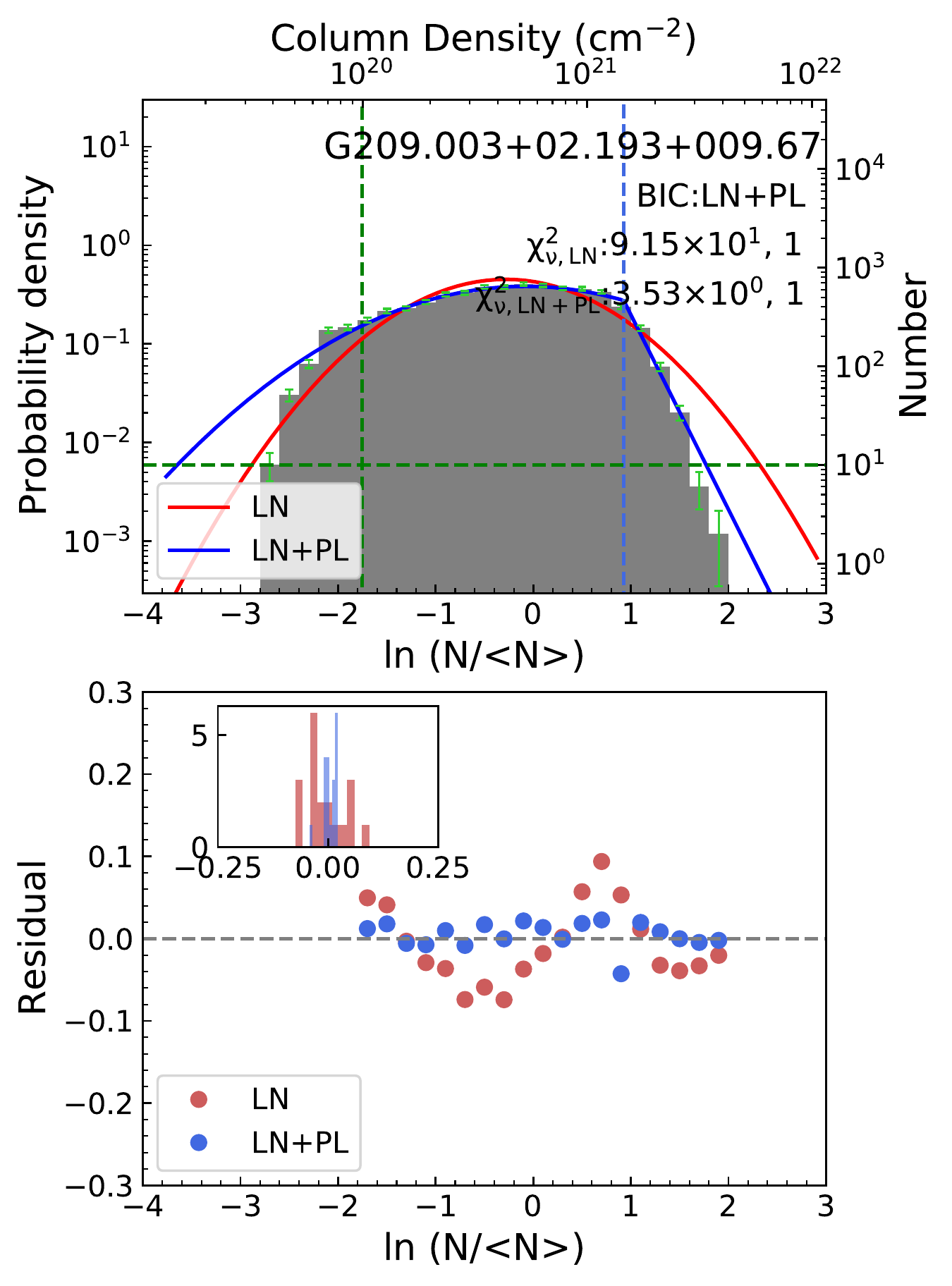}}

\subfigure{\includegraphics[trim=0cm 0cm 0cm 0cm, width= 0.23\linewidth, clip]{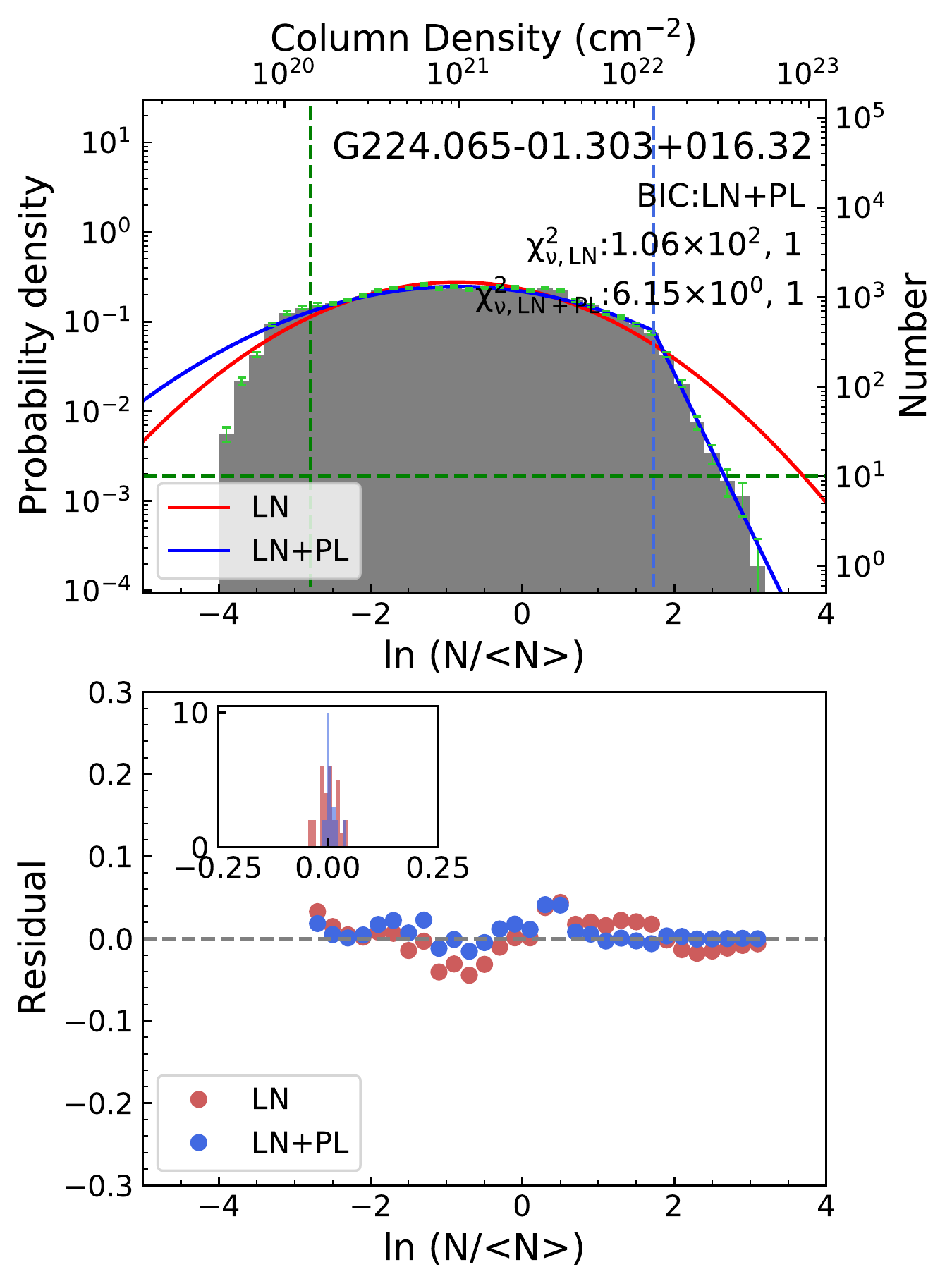}}
\subfigure{\includegraphics[trim=0cm 0cm 0cm 0cm, width= 0.23\linewidth, clip]{{G202.028+01.592+005.99_pdf_residual}.pdf}}

\caption{LN+PL PDFs. Symbols and legends have the same meaning as those in Figure \ref{fig3}.}

\label{fig17}
\end{figure}

%% file: pdf_residual_NC.tex
\begin{figure}
\subfigure{\includegraphics[trim=0cm 0cm 0cm 0cm, width= 0.23\linewidth, clip]{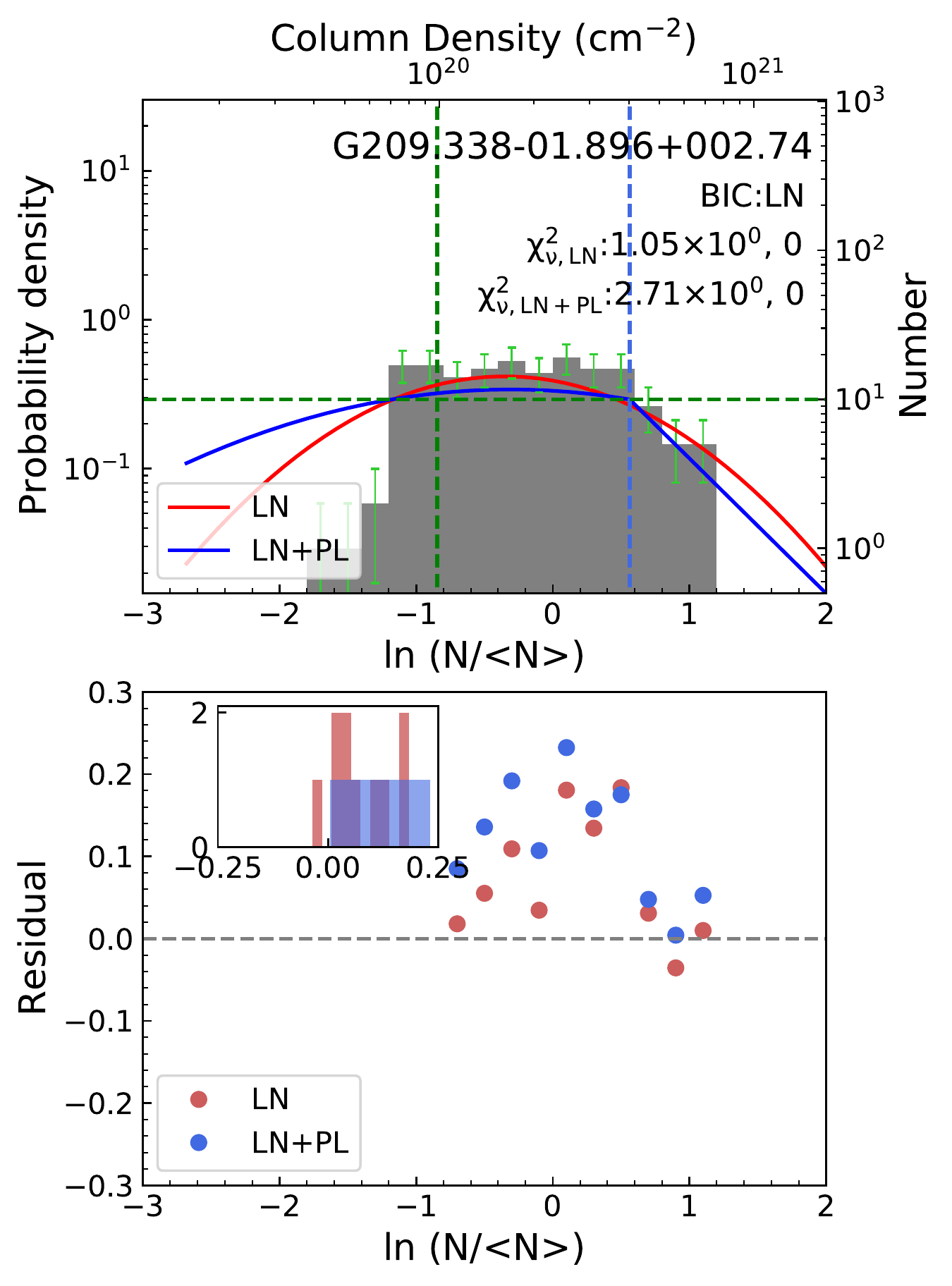}}
\subfigure{\includegraphics[trim=0cm 0cm 0cm 0cm, width= 0.23\linewidth, clip]{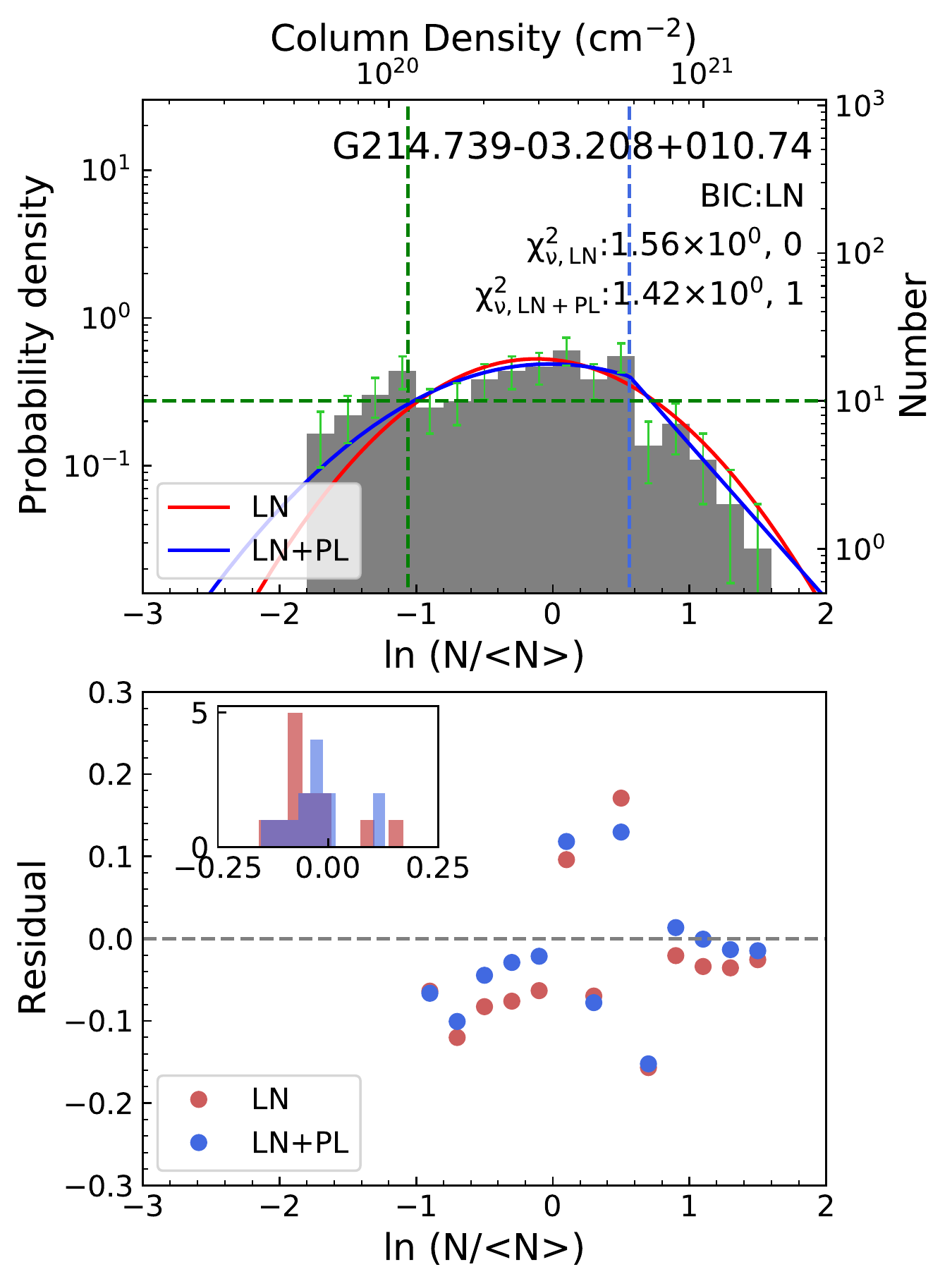}}
\subfigure{\includegraphics[trim=0cm 0cm 0cm 0cm, width= 0.23\linewidth, clip]{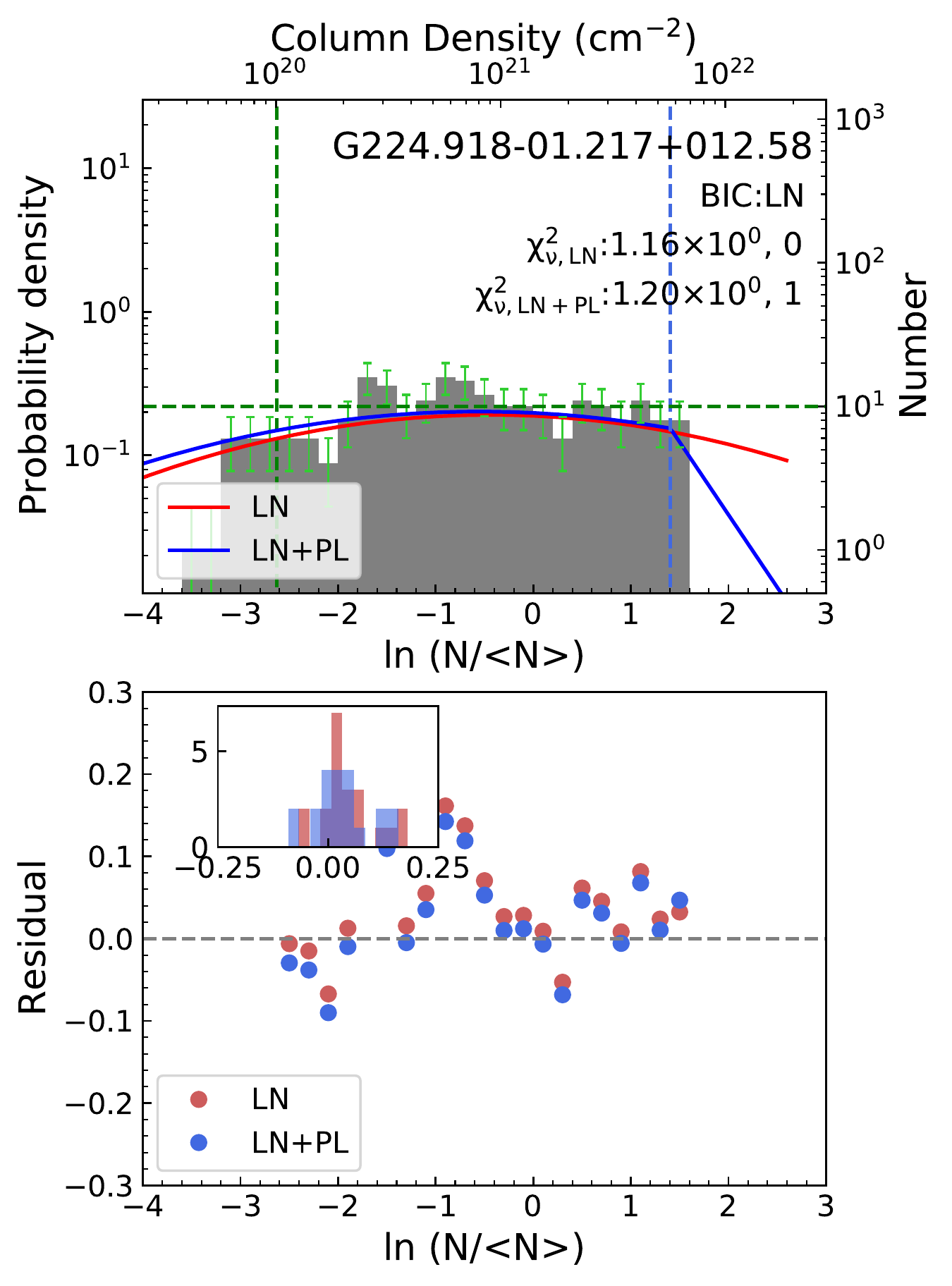}}
\subfigure{\includegraphics[trim=0cm 0cm 0cm 0cm, width= 0.23\linewidth, clip]{{G211.084-01.799+041.46_pdf_residual}.pdf}}

\subfigure{\includegraphics[trim=0cm 0cm 0cm 0cm, width= 0.23\linewidth, clip]{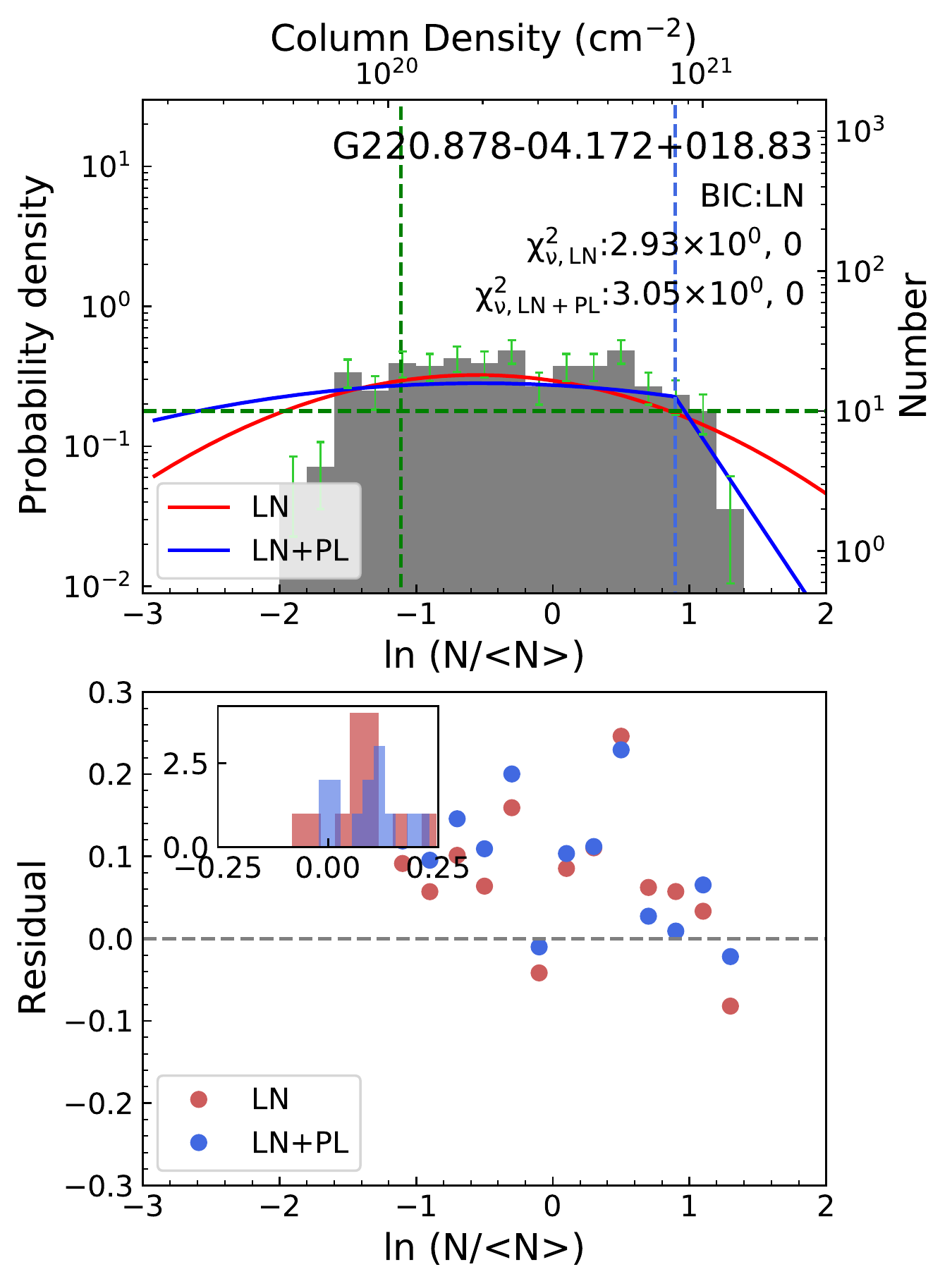}}
\subfigure{\includegraphics[trim=0cm 0cm 0cm 0cm, width= 0.23\linewidth, clip]{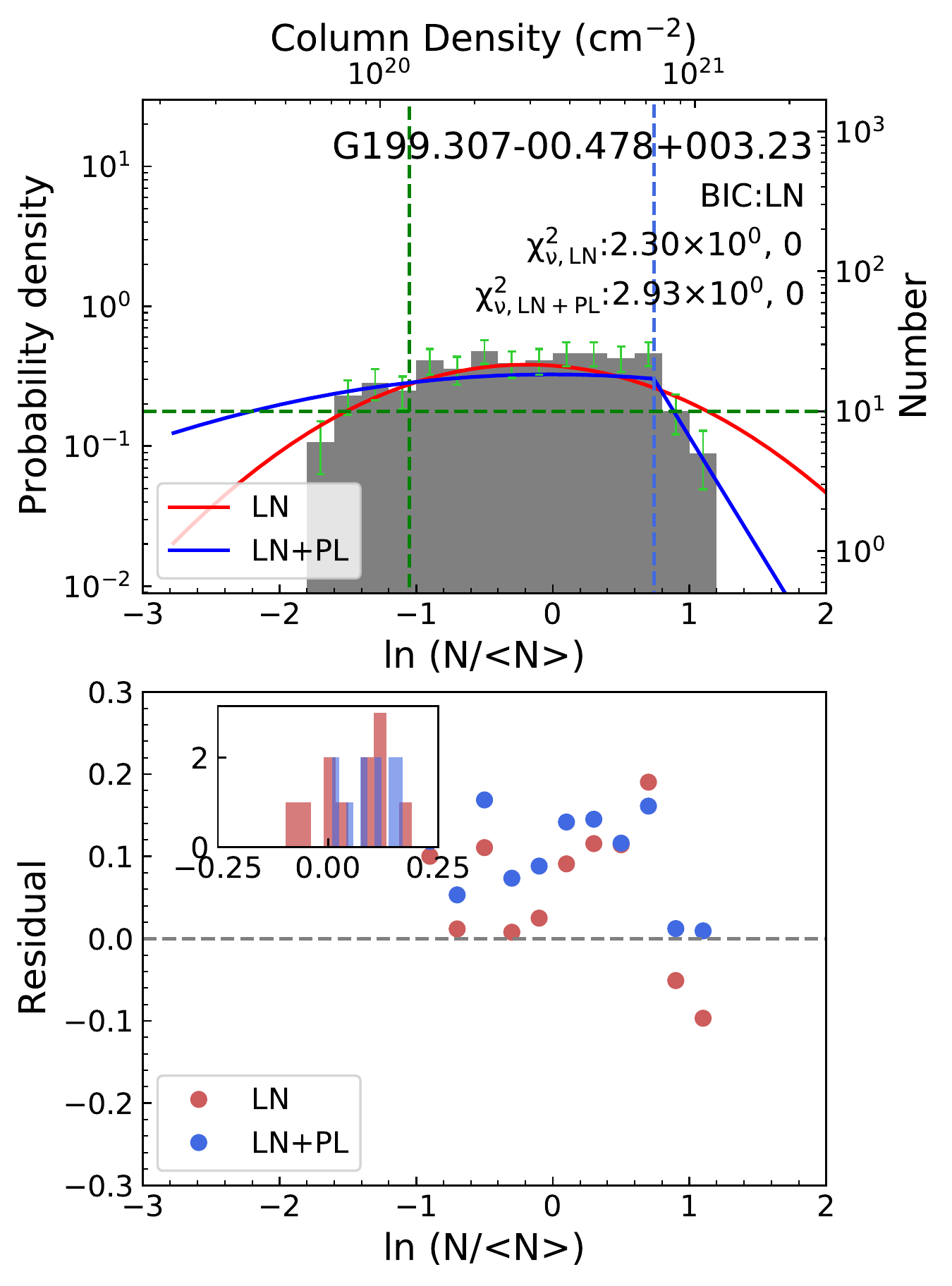}}
\subfigure{\includegraphics[trim=0cm 0cm 0cm 0cm, width= 0.23\linewidth, clip]{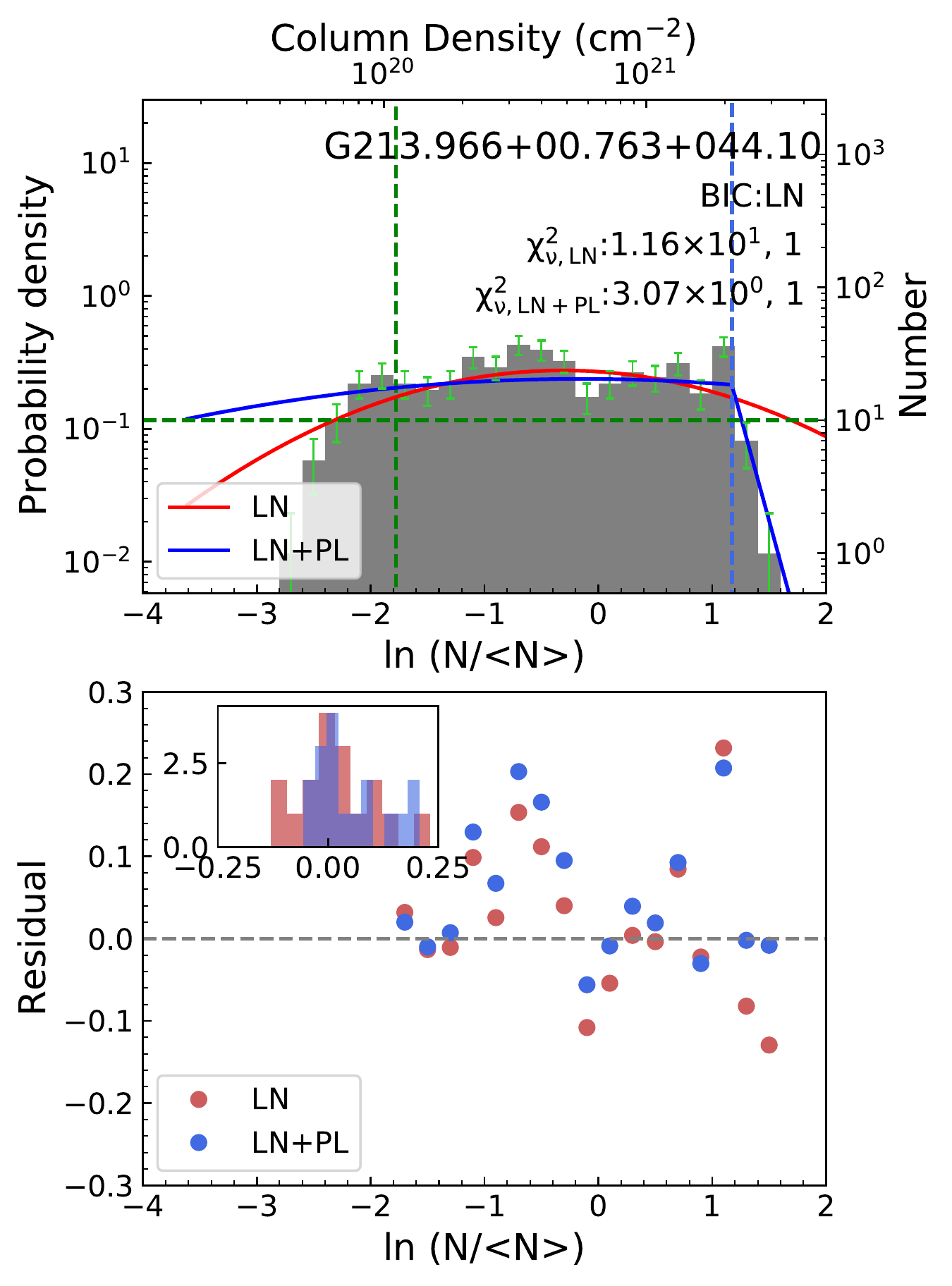}}
\subfigure{\includegraphics[trim=0cm 0cm 0cm 0cm, width= 0.23\linewidth, clip]{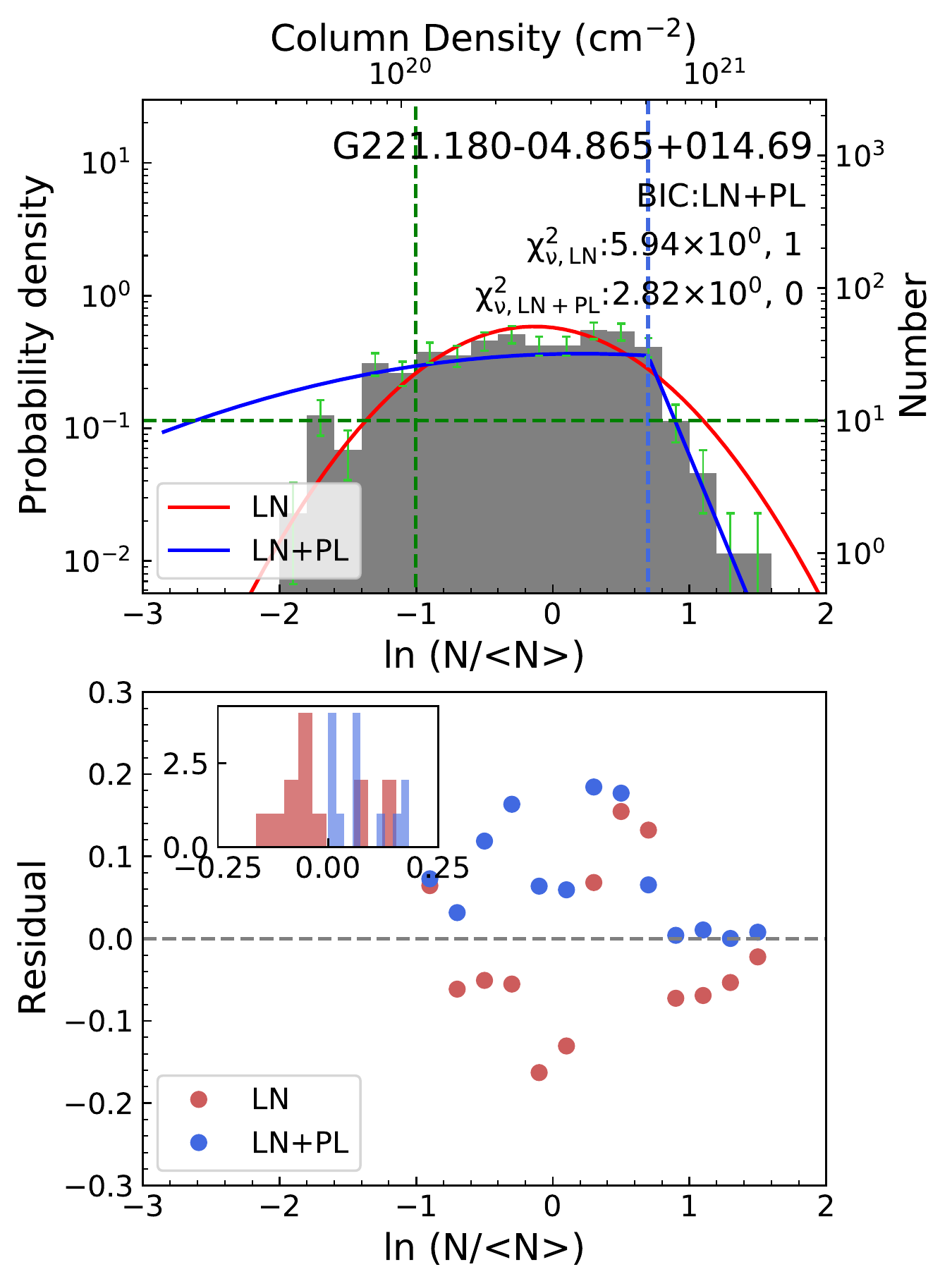}}

\subfigure{\includegraphics[trim=0cm 0cm 0cm 0cm, width= 0.23\linewidth, clip]{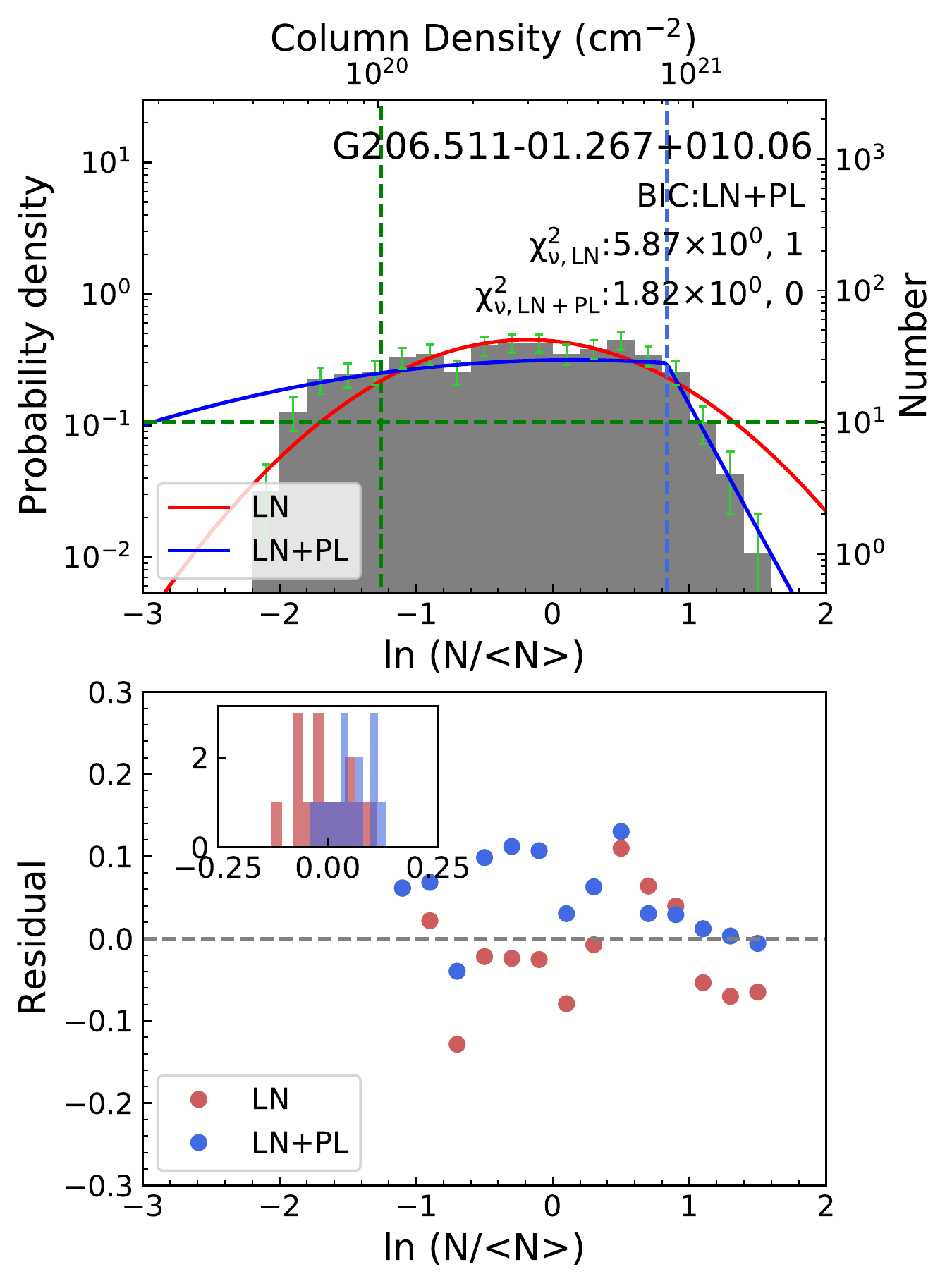}}
\subfigure{\includegraphics[trim=0cm 0cm 0cm 0cm, width= 0.23\linewidth, clip]{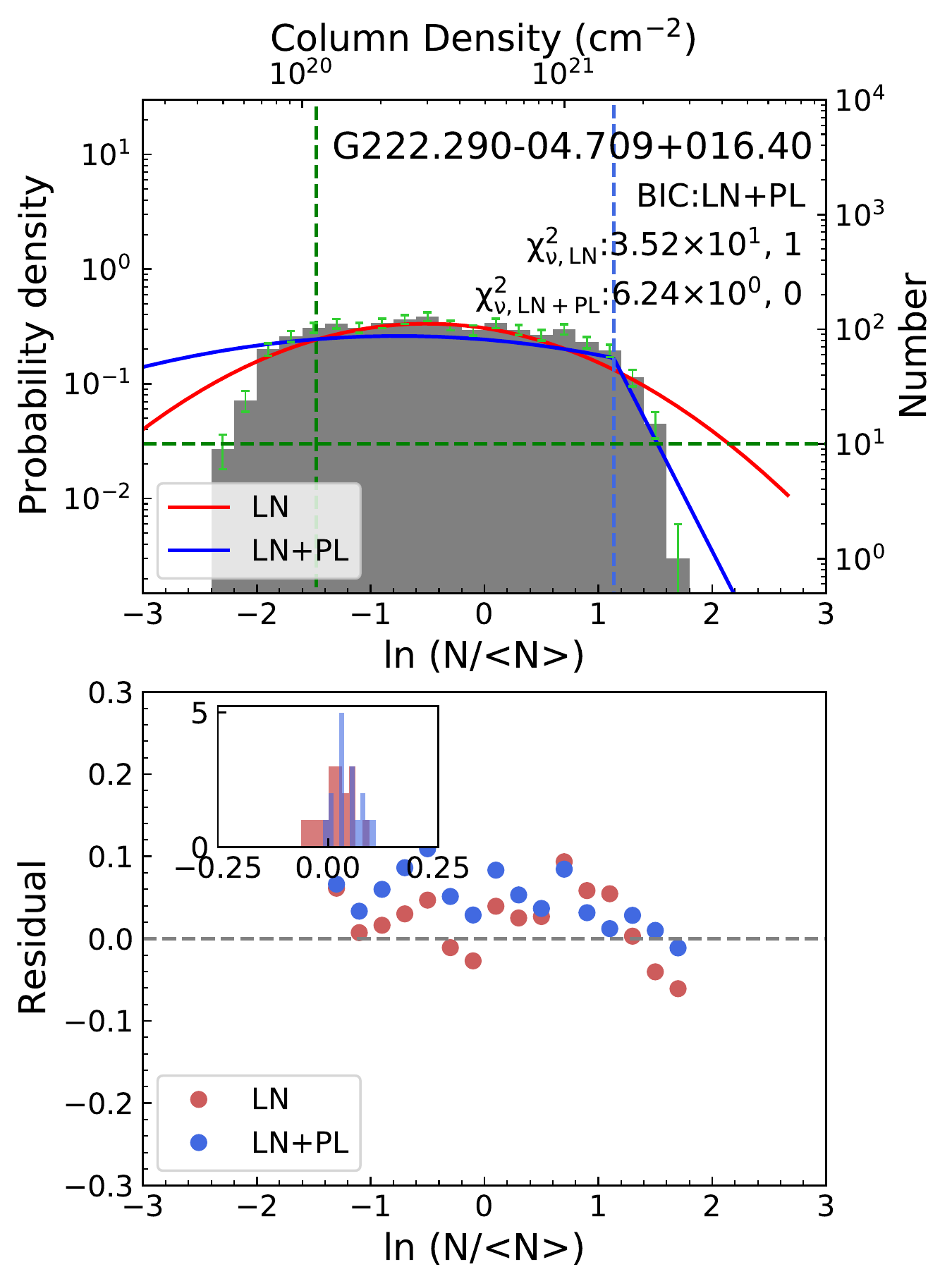}}
\subfigure{\includegraphics[trim=0cm 0cm 0cm 0cm, width= 0.23\linewidth, clip]{{G216.584-02.515+025.52_pdf_residual}.pdf}}
\subfigure{\includegraphics[trim=0cm 0cm 0cm 0cm, width= 0.23\linewidth, clip]{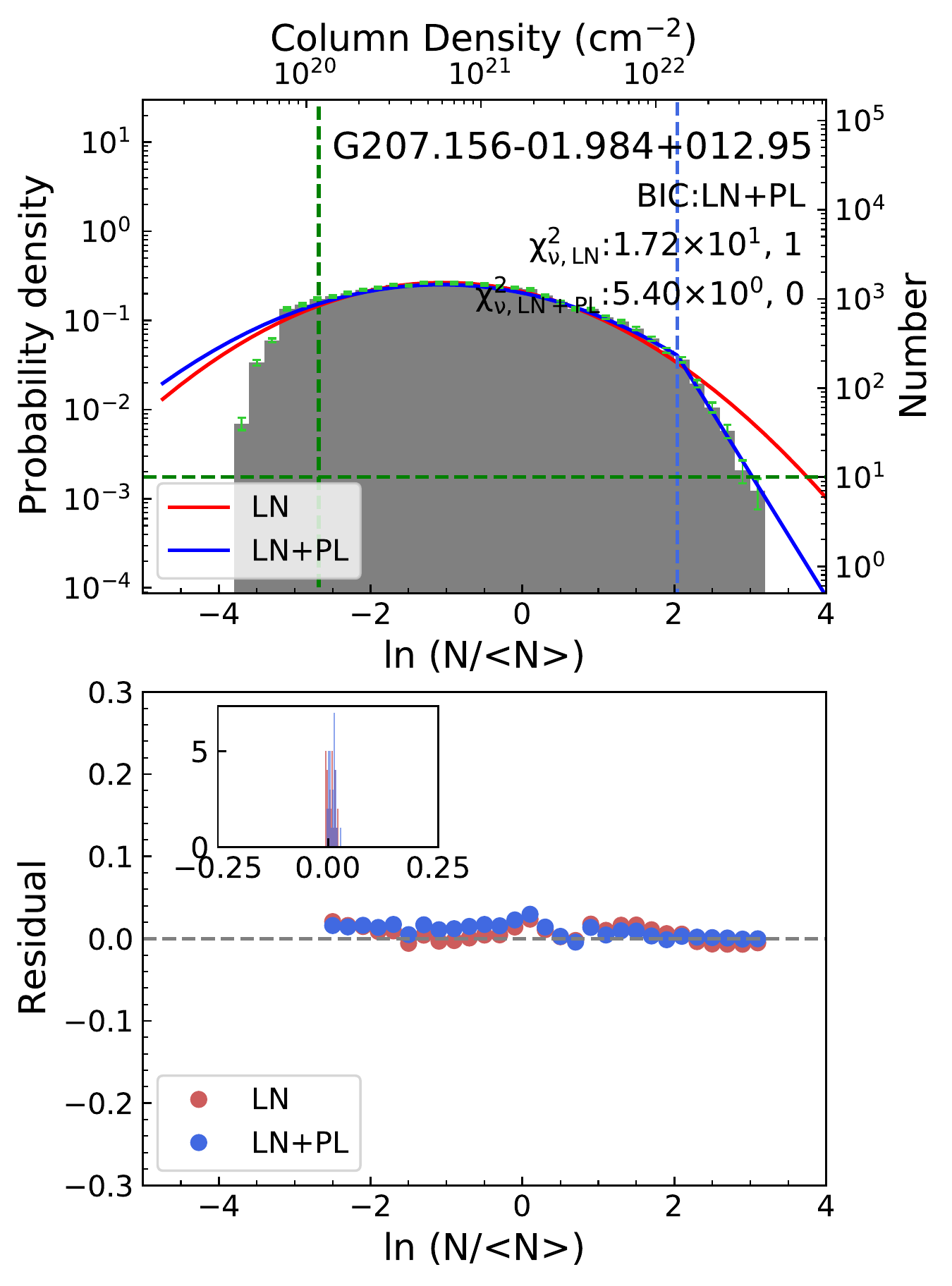}}

\caption{PDFs in the UN category. Symbols and legends have the same meaning as those in Figure \ref{fig3}.}
\label{fig18}
\end{figure}